\pdfoutput=1
\documentclass[10pt,journal,cspaper,compsoc,finalsubmission]{IEEEtran}
\ifCLASSOPTIONcompsoc
\else
\fi
\ifCLASSINFOpdf
\else
\fi

\usepackage{amsfonts}
\usepackage{amsmath}
\usepackage{units}
\usepackage{algorithmic}
\usepackage[boxed,lined,commentsnumbered]{algorithm2e}
\usepackage{grffile}
\usepackage{graphicx}
\usepackage[T1]{fontenc}
\usepackage{hyperref}
\hypersetup{
    colorlinks=false,
    pdfborder={0 0 0},
}

\usepackage{color}
\definecolor{gray}{rgb}{.8,.8,.8}

\newcommand{\R}{\mathbb{R}}
\newcommand{\x}{\mathbf{x}}
\newcommand{\N}{\mathbf{N}}
\newcommand{\y}{\mathbf{y}}
\newcommand{\p}{\mathbf{p}}
\newcommand{\q}{\mathbf{q}}

\usepackage[utf8]{inputenc}

\begin{document}
%
\title{A General Framework for\\ Bilateral and Mean Shift Filtering}
%
%
%
%

\author{Justin~Solomon, Keenan~Crane, Adrian~Butscher, and~Chris~Wojtan
\IEEEcompsocitemizethanks{\IEEEcompsocthanksitem J.\ Solomon is with the Department
of Computer Science, Stanford University, Stanford, CA, 94305.
\IEEEcompsocthanksitem K.\ Crane is with the Department of Computer Science, Columbia University, New York, NY, 10027.
\IEEEcompsocthanksitem A.\ Butscher is with the Max Planck Center for Visual Computing and Communication, Saarbr\"ucken, Germany.
\IEEEcompsocthanksitem C.\ Wojtan is with the Institute of Science and Technology Austria, Klosterneuburg, Austria.}
\thanks{}}

%
%

\markboth{}%
{Solomon \MakeLowercase{\textit{et al.}}: Bilateral and Mean Shift Filtering}
%


\IEEEcompsoctitleabstractindextext{%
\begin{abstract}
We present a generalization of the bilateral filter that can be applied to feature-preserving smoothing of signals on images, meshes, and other domains within a single unified framework.  Our discretization is competitive with state-of-the-art smoothing techniques in terms of both accuracy and speed, is easy to implement, and has parameters that are straightforward to understand.  Unlike previous bilateral filters developed for meshes and other irregular domains, our construction reduces \emph{exactly} to the image bilateral on rectangular domains and comes with a rigorous foundation in both the smooth and discrete settings.  These guarantees allow us to construct unconditionally convergent mean-shift schemes that handle a variety of extremely noisy signals.  We also apply our framework to geometric edge-preserving effects like feature enhancement and show how it is related to local histogram techniques.
\end{abstract}

}

\maketitle

\IEEEdisplaynotcompsoctitleabstractindextext

%
\IEEEpeerreviewmaketitle

\begin{figure*}[t]
\centering
\begin{tabular}{c@{}c@{}c}
\begin{tabular}{c@{}c}
  \includegraphics[height=0.16\linewidth]{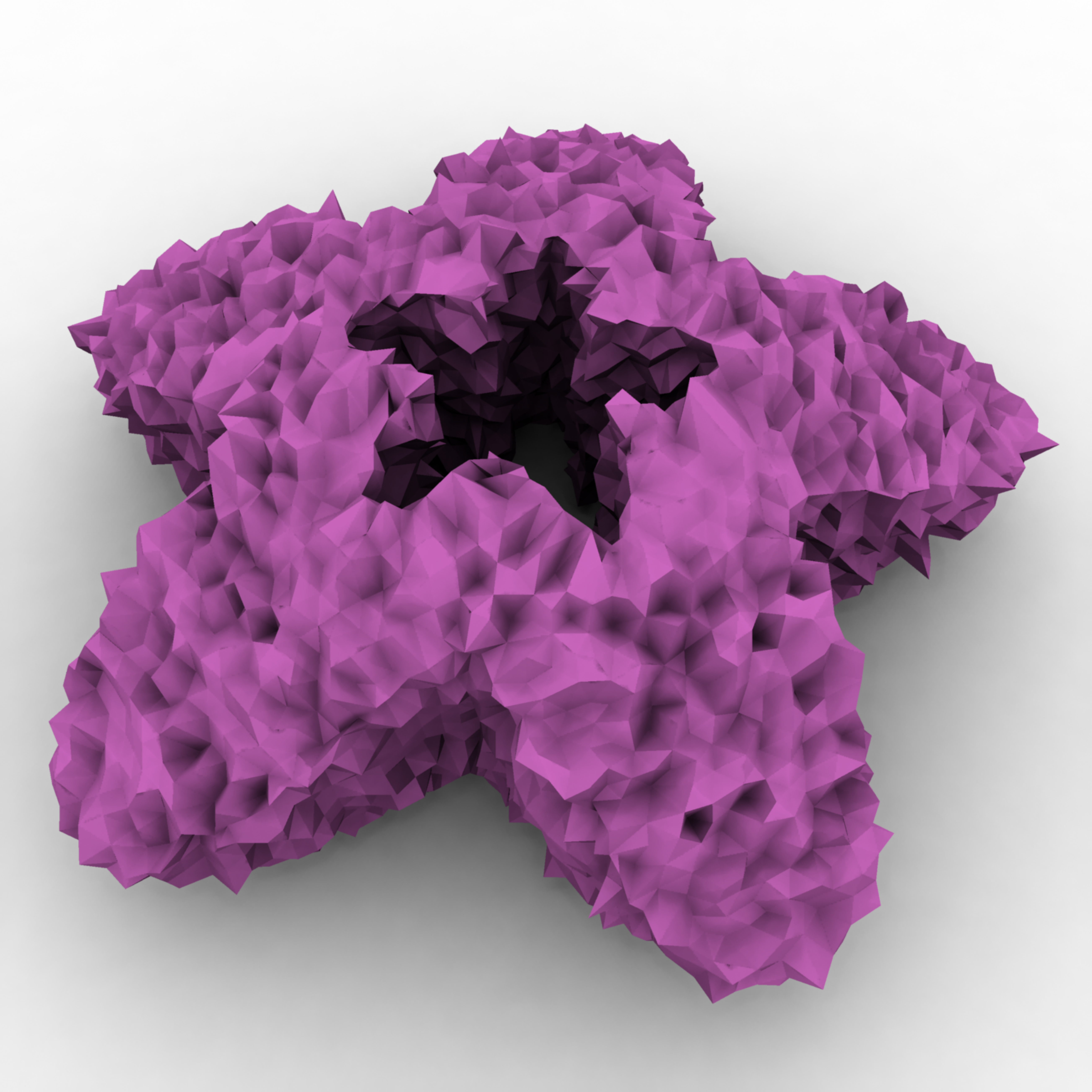}&
  \includegraphics[height=0.16\linewidth]{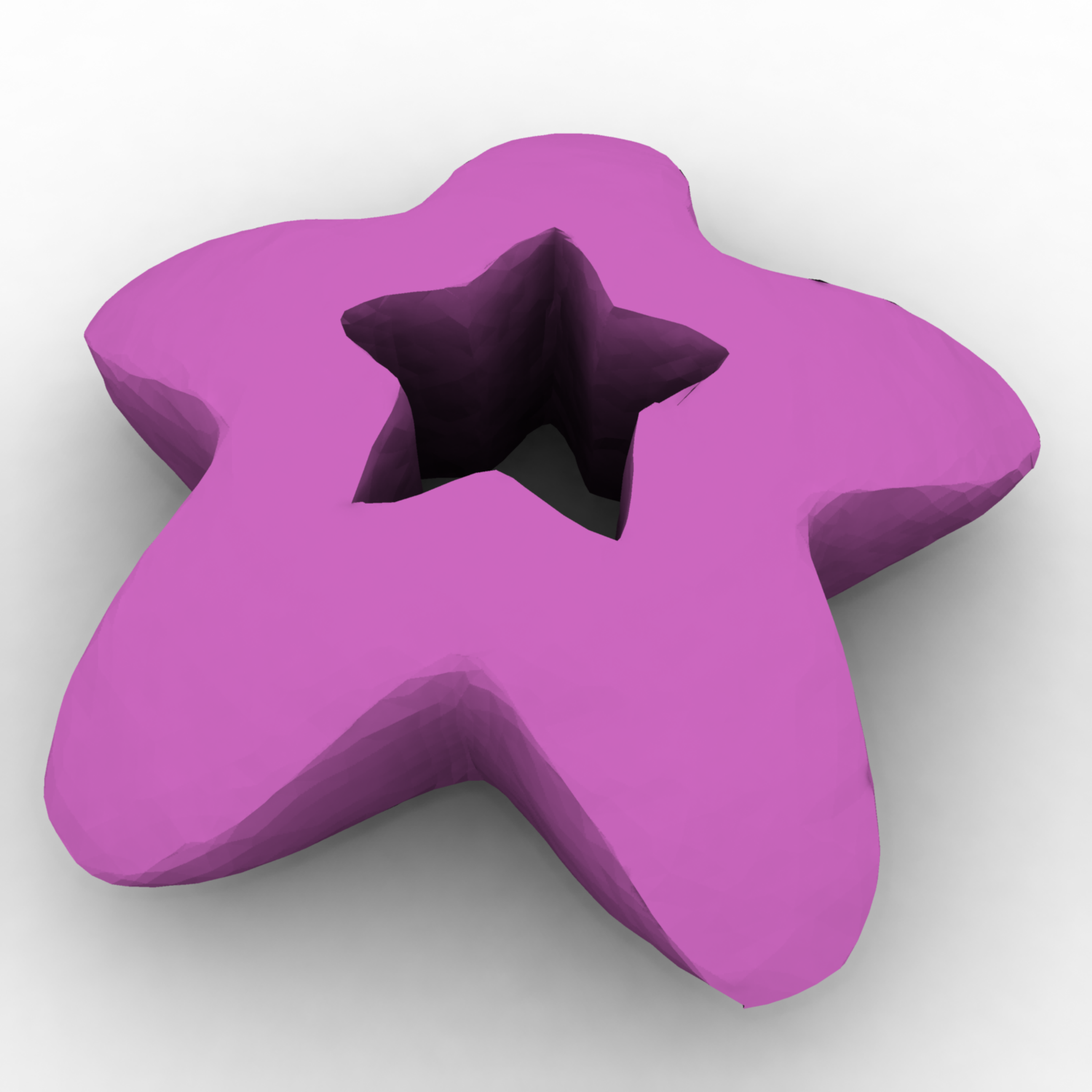}
\end{tabular}&
\begin{tabular}{c@{}c}
  \includegraphics[height=0.16\linewidth]{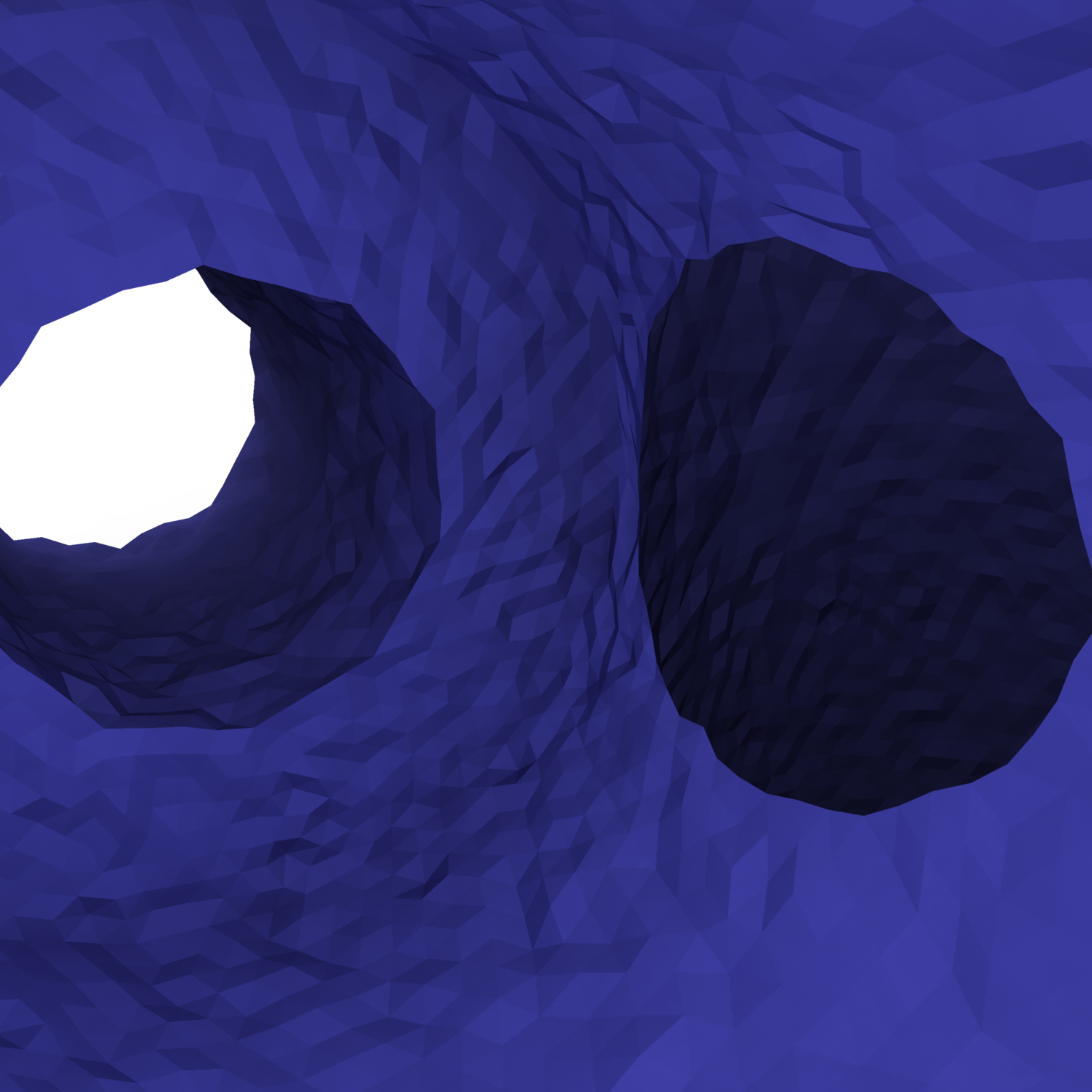}&
  \includegraphics[height=0.16\linewidth]{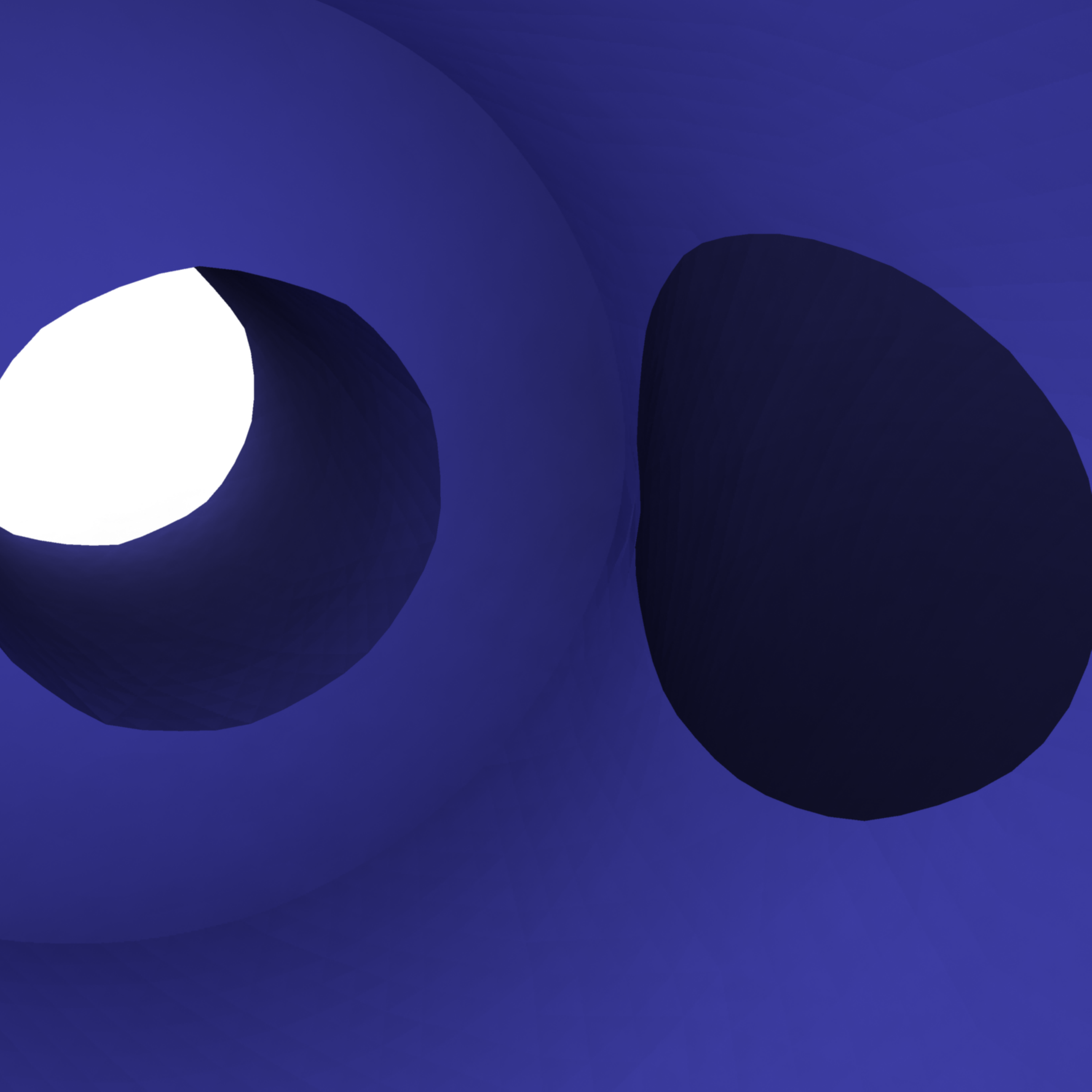}
\end{tabular} &
\begin{tabular}{c}
\begin{tabular}{c@{}c}
  \includegraphics[height=0.08\linewidth]{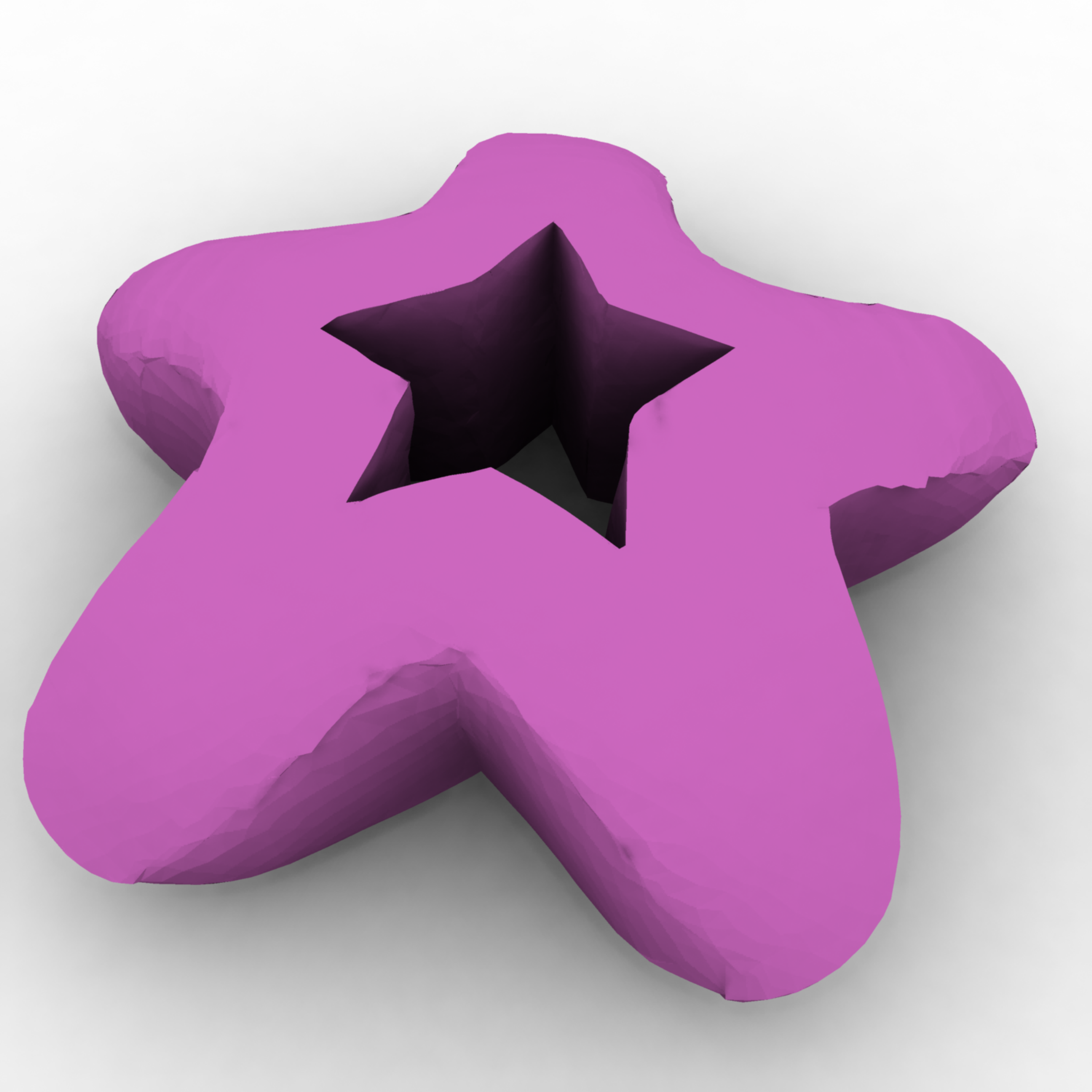}& 
  \includegraphics[height=0.08\linewidth]{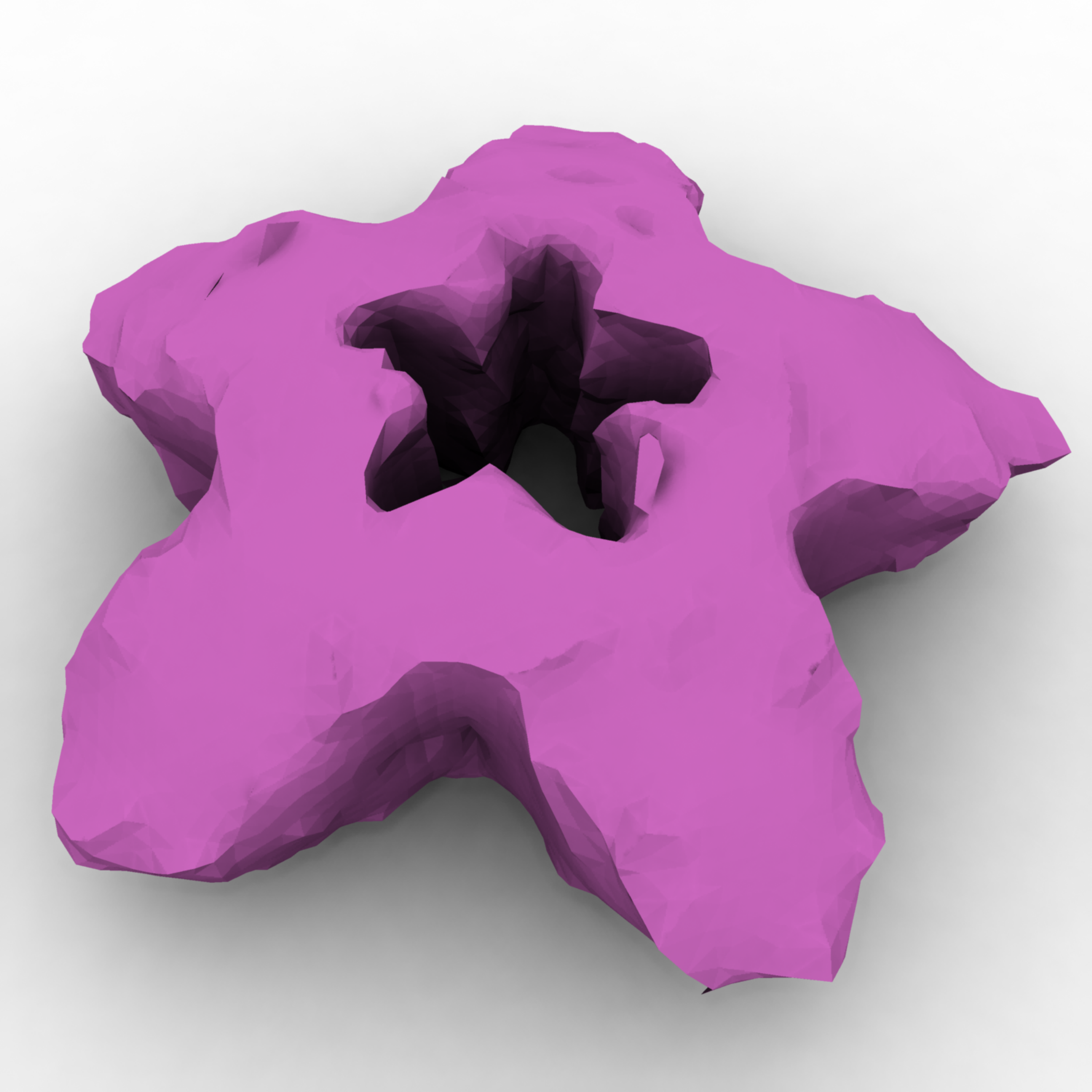}
\end{tabular}\\
\includegraphics[height=0.08\linewidth]{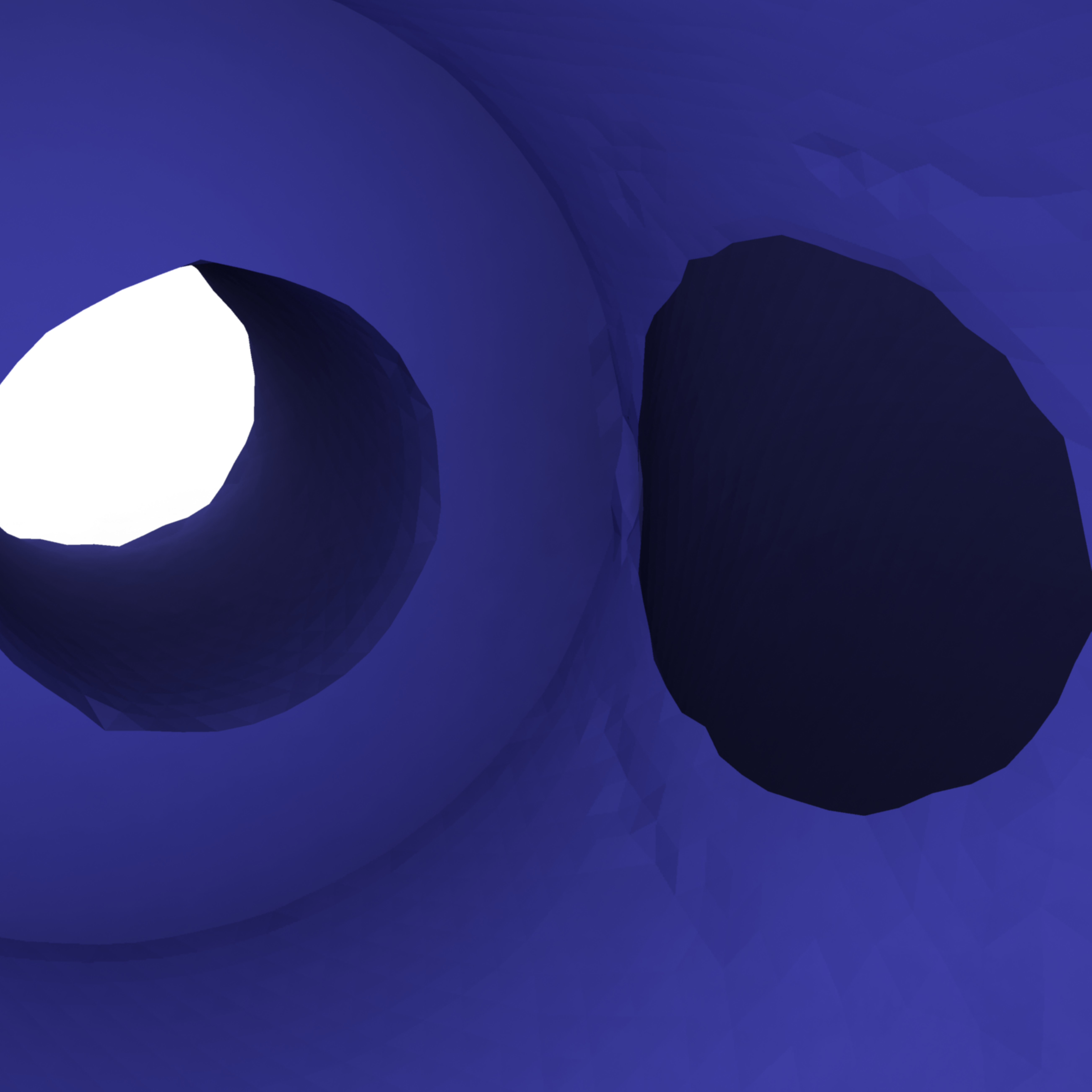}
\end{tabular}\\
(a) & (b) & (c)
\end{tabular}
  \caption{(a,b) Examples of edge-preserving mesh smoothing using our mean-shift filter; noise is removed without mollifying sharp edges, and in (b) the circular holes are rounded; (c) comparisons with~\protect\cite{zheng11},~\protect\cite{hildebrandt04}, and~\protect\cite{fan10}, resp.}
  \label{fig:teaser}
\end{figure*}

\section{Introduction}

Signals on images, surfaces, and other domains rarely obey the smoothness assumptions imposed by methods from classical signal processing.  Even when these methods are successful with respect to formal measures like smoothness and continuity, the resulting signal may fail to meet basic aesthetic or perceptual criteria.  For instance, Gaussian convolution is arguably an ideal image denoising filter, yet it ignores object boundaries and other semantic features.

As a result, a variety of nonlinear filters have been developed to take priors on signal content into account.  In particular, an effective replacement for Gaussian convolution is the \emph{bilateral filter}: rather than blindly averaging pixels that are near each other, the bilateral blends pixels that are nearby in both location and intensity.  The result is a filter that behaves like Gaussian convolution within object boundaries but prevents pixels on opposite sides of a boundary from averaging together.

Due to the success of the bilateral in image processing and computational photography, many attempts have been made to adapt it to geometric domains like meshes.  This transition is not straightforward, however: existing discretizations rely on local operations that are sensitive to the triangulation or use a distortion-inducing parameterization.  In some sense these methods are only ``inspired'' by the bilateral filter and provide few guarantees in the limit of refinement.

We introduce a bilateral filtering technique for signals  on \emph{any} domain admitting a diffusion operator.  This filter coincides with the image bilateral in the planar case but can also be used to process signals on meshes, point clouds, and other domains with minimal modification.  We can also process \emph{geometric} signals such as $xyz$ positions or mesh normals, enabling applications such as mesh smoothing.  Our discretization is a faithful interpretation of the continuous formulation and naturally extends to a larger class of filtering tasks.  More generally, our formulation builds upon and generalizes many previous image filtering (\cite{tomasi98,paris06}), mesh smoothing (\cite{fleishman03,sun07}), and distributional mode-finding (\cite{comaniciu02,kobayashi10}) techniques.

Iterative application of the bilateral leads to the \emph{mean shift} filter, introduced in~\cite{weijer01} and elsewhere, which has stronger denoising and edge-sharpening properties.  We show that the standard formulation of the mean shift translates directly into our framework and can be used to filter signals like surface normals, which are naturally treated as signals with values on the sphere $S^2$.  The result is a strong geometry filter illustrated in Figure~\ref{fig:teaser}.

Our method applies to several tasks from geometry processing including mesh smoothing, normal filtering on oriented point clouds, and curvature smoothing, all while respecting sharp edges.  We also explore how modifications of our filter can be used to achieve interesting feature enhancement effects that respect sharp edges and prove that a slight modification of our method generates a smooth analog of a recently-introduced mesh vertex descriptor.

\subsection{Contributions}

The basic contribution of this paper is a framework for bilateral filtering of signals with \emph{arbitrary} domain and distance manifolds in Section~\ref{sec:generalized_bilateral}.  Section~\ref{sec:mean_shift} develops schemes for mean-shift filtering using the generalized bilateral as a base, including proof that these methods are unconditionally convergent.  We describe a stable, easy-to-implement, and convergent discretization in Section~\ref{sec:discretization} and apply it to signals encountered in computer graphics in Sections~\ref{sec:scalar_processing} and~\ref{sec:normal_signal}, including geometric signals.  Section~\ref{sec:additional_applications} suggests additional applications and non-smoothing uses of our method.

\section{Background}\label{sec:background}

\cite{sun07,botsch10} survey work on mesh smoothing and fairing; we focus on bilateral geometry filtering schemes, which are the closest to our method.

\subsection{Scalar Bilateral Filtering}

The bilateral filter was introduced in~\cite{tomasi98} for filtering signals $f:I\rightarrow\R^n$ on an image $I$ using a kernel that is the product of a spatial term $W_s$ and an intensity term $W_c$:
\begin{equation}
\bar{f}(\x) = \frac{\int_I f(\y)W_s(\|\x-\y\|)W_c(\|f(\x)-f(\y)\|)\,d\y}{\int_I W_s(\|\x-\y\|)W_c(\|f(\x)-f(\y)\|)\,d\y}
\end{equation}
Pixels are combined only when they are nearby both in space \emph{and} in intensity.  The \emph{cross bilateral} filters a signal $f_1$ using intensity distances from another signal $f_2$~\cite{petschnigg04,eisemann04}:
\begin{equation}
\bar{f}(\x) = \frac{\int_I f_1(\y)W_s(\|\x-\y\|)W_c(\|f_2(\x)-f_2(\y)\|)\,d\y}{\int_I W_s(\|\x-\y\|)W_c(\|f_2(\x)-f_2(\y)\|)\,d\y}\label{eq:cross_bilateral}
\end{equation}
For instance, $f_1$ may be too noisy to have well-defined features, but it can instead be smoothed using features from $f_2$.  Considerable work has been put into accelerating these filters; see~\cite{paris06,adams09,adams10} for recent examples.

Several methods apply bilateral filtering on non-image domains.  Mostly, they map the domain to a regular grid and apply image processing methods; for instance,~\cite{miropolsky04} uses the bilateral on a voxel grid for surface reconstruction.  \cite{adams09} can be used to process signals that are not on grids, but distances for $f_1$ and $f_2$ must be measured using the Euclidean norm $\|\cdot\|_2$.  \cite{eigensatz08} makes use of a bilateral on scalar mesh curvature signals, but their focus is on shape editing rather than evaluation of the bilateral itself.

\subsection{Mesh Bilateral Filtering}

\begin{table*}
\centering
\begin{tabular}{|l|p{5in}|}\hline
\textbf{Paper} & \textbf{Description}\\\hline
\cite{fleishman03} &Bilaterally filters the height function of the surface over vertex tangent planes\\
\cite{jones03}&Combines vertices with their projections onto nearby tangent planes; bilateral weights take into account distances to the tangent plane projection and to the tangent plane center\\
\cite{hu04} &Uses bilateral filtering as part of a multi-pass approach to modify Laplacian smoothing using weights inspired by those in~\cite{fleishman03}\\
\cite{jones04} & Iteratively applies a modification of~\cite{jones03} to improve surface normals for rendering.\\
\cite{duguet04} & Bilaterally filters jets on point clouds for reconstruction\\
\cite{hou05}& Bilaterally filters mesh normals and then adjusts surface; weights are Gaussians in normal difference and an approximation of geodesic distance\\
\cite{shimizu05} & Explicitly filters sharp edges and then faces separately using extrinsic distances, edge directions, normal difference, and projections as in~\cite{jones03}\\
\cite{lee05} & Filters face normals using Euclidean distance between centroids and normal differences\\
\cite{wang06} & Filters non-manifold surfaces by iteratively applying a bilateral similar to~\cite{jones03} and remeshing\\
\cite{adams09}& Filters the difference between a mesh and its Laplace-smoothed counterpart in principal curvature coordinates using spin-images~\cite{johnson99} for weights without a distance term\\
\cite{fan10} & Denoises quadric surface approximations by extending~\cite{fleishman03}\\
\cite{nociar10} & Applies~\cite{lee05} with automatic parameter choice to normals and fits a new surface\\
\cite{zheng11}& Locally filters face normals using one-ring information; derives alternative implicit normal smoothing scheme using one-ring bilateral weights to change Laplacian operator\\
\cite{vialaneix11}&Approximates mesh bilateral filtering using separable filters along curvature directions\\\hline
\end{tabular}
\caption{A summary of previous attempts to adapt bilateral filtering to mesh domains.}\label{table:other_papers}
\vspace{-.2in}
\end{table*}

One domain in which applications of the bilateral extend beyond grid-based methods is mesh fairing and smoothing.   Table~\ref{table:other_papers} lists several past approaches to extend the bilateral to mesh domains in this fashion.  Despite the considerable amount of research devoted to mesh bilateral filtering, we find that \emph{none} of the prior contributions exhibits the following desirable properties simultaneously, and most methods do not exhibit more than one at a time:
\begin{enumerate}
\item Use of intrinsic and smooth distance weights respecting the domain's metric without resorting to parameterization\label{item:no_parameterization}
\item Convergence in the limit of refinement or theory identifying the effects of the filter on an abstract surface\label{item:convergence}
\item Applicability to multiple signal types and domains\label{item:applicability}
\item Reduction to~\cite{tomasi98} for image signals\label{item:image_reduction}
\end{enumerate}
These desiderata characterize desirable behavior and convergence of generalized bilateral filtering techniques.  For example,~\ref{item:no_parameterization}) ensures that the algorithm is tailored for mesh processing rather than adapting image-based strategies to local neighborhoods; avoiding local parameterization also contributes to algorithmic efficiency.  Item~\ref{item:convergence}) helps ensure that discretizations of filter integrals converge to their continuous counterparts; ad-hoc methods considering ring-based vertex neighborhoods on meshes do not satisfy this criterion.  We include~\ref{item:applicability}) to ensure that filters support multiple applications without tuning for a narrow set of domains, and~\ref{item:image_reduction}) confirms our intuition that a filter is truly ``bilateral'' and thus can be understood using intuition from image processing.  Our algorithm satisfies all these criteria and still performs comparably to the methods in Table~\ref{table:other_papers}.  

\subsection{Mean Shift Filtering}

Mean shift filtering, introduced for image segmentation in~\cite{comaniciu02}, was shown to be equivalent to iterated cross bilateral filtering in~\cite{weijer01}--before the bilateral filter formally was introduced.  Given this connection,~\cite{paris07} and others make use of bilateral filter accelerations to accomplish mean shift.  It produces strong feature-preserving denoising for images, but few attempts have been made to apply it to mesh domains.  \cite{yamauchi05} mean shifts mesh normals for segmentation; \cite{shamir06} proposes a mesh mean shift operator requiring local geodesic parameterizations.  While attempts to mean shift signals \emph{on} meshes or surfaces have been limited, mean shift filtering has been applied to different manifold-\emph{valued} signals; for instance,~\cite{kobayashi10,subbarao06,subbarao09} propose mean shift methods for filtering sphere-, analytic manifold-, and Riemannian manifold-valued signals, resp.  Our framework bridges the gaps among a variety of existing methods in this domain.

\section{Generalized Bilateral Filtering}\label{sec:generalized_bilateral}

Take $\Sigma$ to be the domain of a signal $f_1:\Sigma\rightarrow\R^n$ equipped with a nonnegative symmetric kernel $K_\Sigma:\Sigma\times\Sigma\rightarrow\R$.  Intuitively, we can think of $K_\Sigma(\x,\y)$ as measuring the proximity between $\x$ and $\y$ on $\Sigma$.  For instance, signal processing on an image might take $\Sigma\subseteq\R^2$ as the image plane, $n=3$ for RGB channels, and $K_\Sigma(\x,\y)=e^{-\|\x-\y\|^2/\sigma^2}$, the usual Gaussian blur kernel.  More generally, if $\Sigma$ is any domain  admitting a Laplacian operator $L$, such as a graph, surface, mesh, or point cloud, we can take $K_\Sigma$ to be the kernel corresponding to a solution at some fixed $t>0$ of the heat equation $\frac{\partial u}{\partial t} = Lu$, where $u(\x,t): \Sigma\times[0,\infty)\rightarrow\R$; that is, $K_\Sigma(\x,\y)$ measures how much a unit of heat diffuses from $\x$ to $\y$ along $\Sigma$ in $t$ time.

We can define a blurred version of $f_1$ as the convolution
\begin{equation}
\hat{f}_1(\x) = \frac{1}{z(\x)}\int_\Sigma f_1(\y) K_\Sigma(\x,\y)\,d\y\label{eq:blur}
\end{equation}
where $z(\x)$ is the normalizing value $\int_\Sigma K_\Sigma(\x,\y)\,d\y$.  Let $\mathbf{T}(f)$ be the linear operator on square-integrable functions taking $f_1$ to $\hat f_1$; in other words, $\mathbf{T}$ blurs functions $f$ with kernel $K_\Sigma$.

In parallel with the cross bilateral~\eqref{eq:cross_bilateral}, take $f_2:\Sigma\rightarrow\Gamma$ to be a function designed so that if $f_2(\x)$ and $f_2(\y)$ are distant, the signal $f_1$ at $\x$ and $\y$ should not be blended during filtering.  We assume that $\Gamma$ is a compact manifold with or without boundary; for instance, using RGB colors would yield $\Gamma=[0,1]^3$, while using surface normals yields $\Gamma=S^2$, the unit sphere.  We equip $\Gamma$ with its own kernel $K_\Gamma:\Gamma\times\Gamma\rightarrow\R$.

\begin{figure}[thb]
\centering
\includegraphics[width=2.8in]{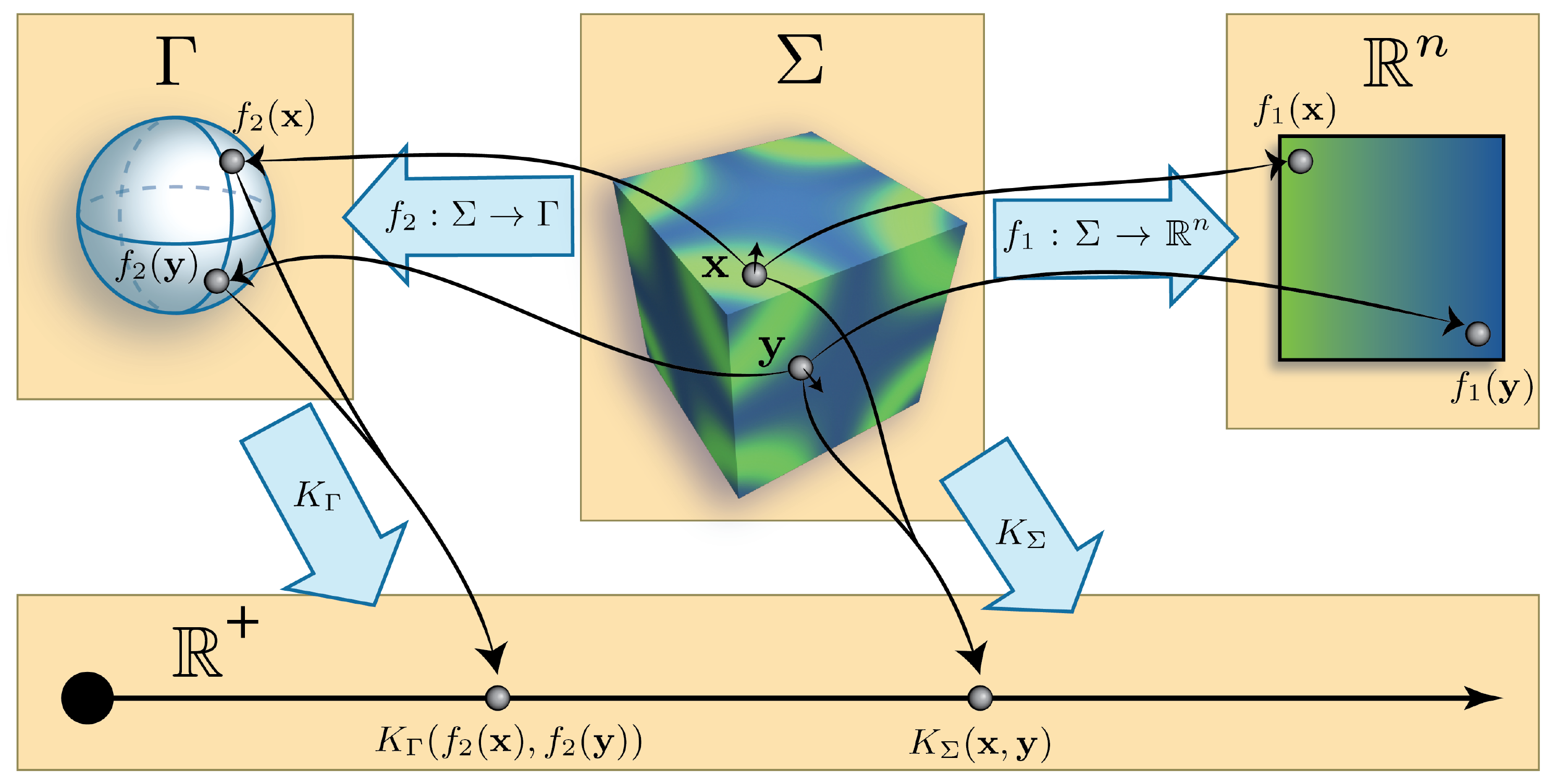}\\
\includegraphics[width=2in]{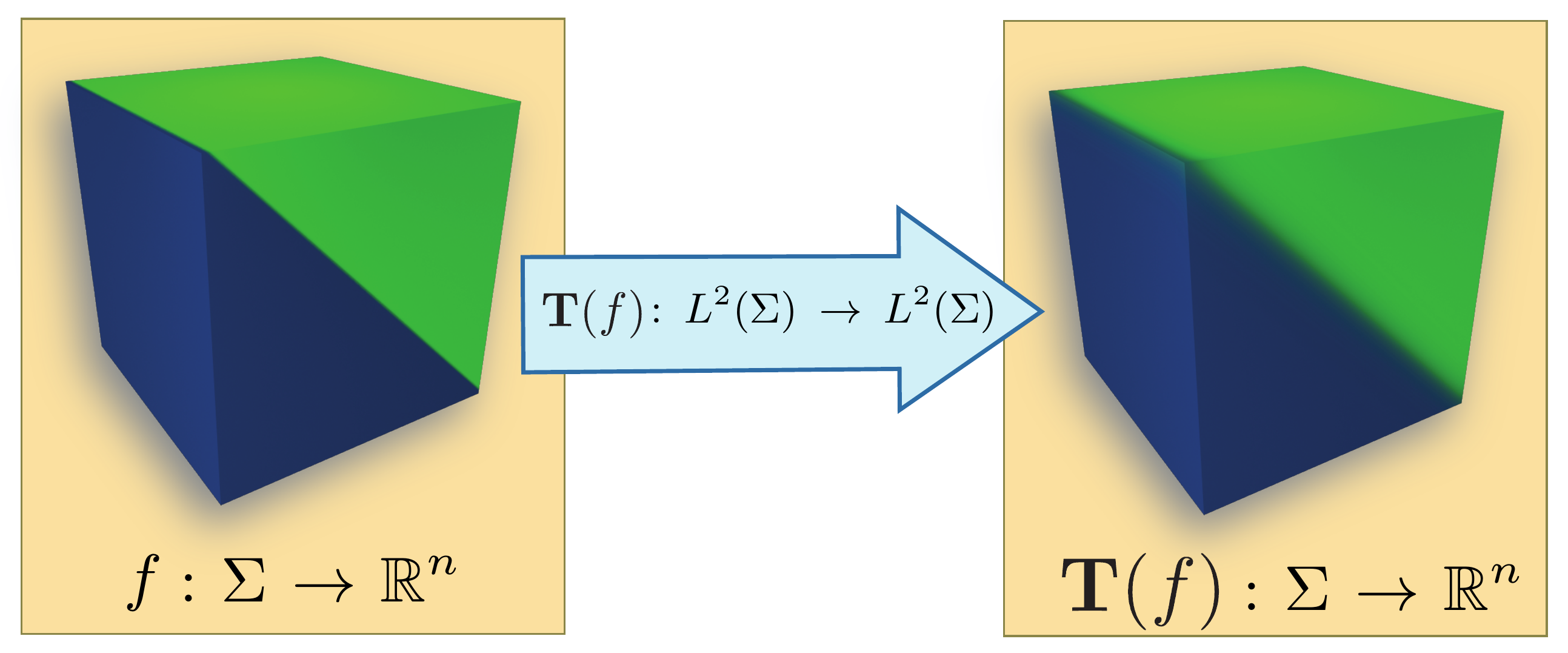}
\caption{Notation.}\label{fig:notation}
\end{figure}

With this notation (illustrated in Figure~\ref{fig:notation}) in place, we can introduce the \emph{generalized cross-bilateral filter} as follows:
\begin{equation}
\bar{f}(\x) = \frac{\int_\Sigma f_1(\y) K_\Sigma(\x,\y)K_\Gamma(f_2(\x),f_2(\y))\,d\y}{\int_\Sigma K_\Sigma(\x,\y)K_\Gamma(f_2(\x),f_2(\y))\,d\y}\label{eq:generalized_bilateral}
\end{equation}
Note the similarity to the image cross bilateral filter~\eqref{eq:cross_bilateral}.  The main difference is that we allow our kernel functions to take into account $\x$ and $\y$ (as well as $f_2(\x)$ and $f_2(\y)$) directly rather than just the norms $\|\x-\y\|$ and $\|f_2(\x)-f_2(\y)\|$.

We can re-express the cross bilateral using the diffusion operator $\mathbf{T}$ defined above.  In particular, for fixed $\p\in\Gamma$ define numerator and denominator functions as:
\begin{align}
f^{num}_{\p}(\y) &= f_1(\y) K_\Gamma(\p,f_2(\y))\label{eq:numerator}\\
f^{den}_{\p}(\y) &= K_\Gamma(\p,f_2(\y))\label{eq:denominator}
\end{align}
Then, we have
\begin{equation}
\bar{f}(\x) = \frac{\mathbf{T}[f^{num}_{f_2(\x)}(\cdot)](\x)}{\mathbf{T}[f^{den}_{f_2(\x)}(\cdot)](\x)}\label{eq:alternative_form}
\end{equation}

\section{Generalized Mean Shift Filtering}\label{sec:mean_shift}

Bilateral filtering is reliable for minor denoising but is less effective on highly-noisy signals.  In particular, the $K_\Gamma$ term combines values only when they are similar; outliers thus will be influenced only slightly by their nearby counterparts.  Furthermore, in certain scenarios it is desired not only to smooth signals but also to sharpen edges.  For these purposes we propose a \emph{generalized mean shift} filter below.

For fixed $\x\in\Sigma$, we can rewrite the denominator of the bilateral~\eqref{eq:generalized_bilateral} as a probability distribution $h:\Gamma\rightarrow\R$ over $\Gamma$:
\begin{equation}
h_{\x}(\p) = \frac{1}{z(\x)}\int_\Sigma K_\Sigma(\x,\y) K_\Gamma(\p,f(\y))\,d\y\label{eq:local_histogram}
\end{equation}
where $z(\x)$ is a normalizing constant so that $\int_\Gamma h_{\x}(\p)\ d\p=1$.  This function, constructed using the same technique as~\cite{kass10}, represents the distribution of values of $f$ near $\x$.

If $\Gamma=\R^n$ with $K_\Gamma(\p,\q)=e^{-\|\p-\q\|^2/\sigma^2}$, taking the gradient with respect to $\p$ we find that peaks $\p^*$ of $h_{\x}(\p)$ satisfy
\begin{equation}
\p^* = \frac{\int_\Sigma f(\y) K_\Sigma(\x,\y) K_\Gamma(\p^*, f(\y))\,d\y}{\int_\Sigma K_\Sigma(\x,\y) K_\Gamma(\p^*, f(\y))\,d\y}
\end{equation}
This relationship suggests a fixed-point iteration scheme for finding peaks of $h_{\x}(\p)$ at all $\x$:
\begin{align}
f^{(0)}(\x) &= f(\x)\\
f^{(k+1)}(\x) &= \frac{\int_\Sigma f(\y) K_\Sigma(\x,\y) K_\Gamma(f^{(k)}(\x), f(\y))\,d\y}{\int_\Sigma K_\Sigma(\x,\y) K_\Gamma(f^{(k)}(\x), f(\y))\,d\y}
\end{align}
Each iteration applies a slightly modified cross bilateral~\eqref{eq:generalized_bilateral}.  This scheme is an instance of the mean-shift filter~\cite{comaniciu02}, which converges unconditionally to peaks of $h_{\x}$~\cite{li07}.

The derivation above assumes that $\Gamma=\R^n$.  This restriction to $\R^n$ reflects a general drawback of bilateral filters and related integral operators, that they can take inputs on a manifold $\Gamma$ but give outputs in the ambient $\R^n$; we are unaware of a bilateral filter that does not have this property without postprocessing.  In particular, filters including~\cite{zheng11} modify surface normals (on the sphere $S^2$) but result in filtered versions without unit length; these filters can be difficult to understand and control.  The description of the mean shift as a mode-finding technique, however, is valid for any $\Gamma$ independent of its embedding, and we can take advantage of this observation to build denoising methods that are \emph{intrinsic} to $\Gamma$.

More formally, our construction of $h$ remains valid when $\Gamma\neq\R^n$.  For instance, we can equip $\Gamma=S^2$ with the Von Mises--Fisher kernel $K_\Gamma(\p,\q)=e^{\p\cdot\q/\sigma}$ for unit vectors $\p$ and $\q$, used to represent isotropic distributions on the unit sphere~\cite{fisher53}.  In this case, a similar argument to the one above yields the mean-shift iteration:
\begin{align}
f^{(0)}(\x) &= f(\x)\\
f^{(k+1)}(\x) &= \frac{\int_\Sigma f(\y) K_\Sigma(\x,\y) K_\Gamma(f^{(k)}(\x), f(\y))\,d\y}{\|\int_\Sigma f(\y) K_\Sigma(\x,\y) K_\Gamma(f^{(k)}(\x), f(\y))\,d\y\|}\label{eq:sphere_mean_shift}
\end{align}
Each iterate has unit length and thus remains on $S^2$.  This new iterative scheme is an instance of the spherical mean shift algorithm in~\cite{kobayashi10} being carried out in parallel at each $\x\in\Sigma$, proving its convergence and its qualitative similarity to the Euclidean case.  Iterations of~\eqref{eq:sphere_mean_shift} are effectively averaging unit vectors; while this is the mathematically correct operation to carry out according to the Von Mises--Fisher kernel, there is some potential for numerical instability when $\sigma$ is large.  We have not observed such issues in the applications we propose for reasonable choices of $\sigma$; particular values are documented in the supplementary material.

We have concentrated above on two simple domains $\Gamma$: subsets of $\R^n$ and the sphere $S^2$.  These are by no means the only choices of $\Gamma$ that yield convergent mode-finding schemes.  \cite{subbarao06} and~\cite{subbarao09} provide mean shift methods when data is on analytic or Riemannian manifolds, resp., that can be adapted to our framework on $\Sigma$ in a similar manner.

\section{Discretization}\label{sec:discretization}

We employ a signal processing technique similar to that in~\cite{paris06} to evaluate the bilateral filter on discrete domains $\Sigma$ (Algorithm~\ref{alg:generalized}).  Our method applies essentially the same computations to $f^{den}_\p$ as $f^{num}_\p$, so for ease of notation during its development denote $f_\p$ as one of $f^{num}_\p$ or $f^{den}_\p$.

Suppose that we choose samples $\p_1,\ldots,\p_m\in\Gamma$ and a corresponding partition of unity $\phi_1,\ldots,\phi_m:\Gamma\rightarrow\R$ such that a function $g:\Gamma\rightarrow\R$ can be approximated as $g(\p)\approx \sum_i g(\p_i) \phi_i(\p)$.  Note that under mild continuity and compactness conditions, we can construct sequences of partitions such that the approximation converges to $g(\p)$ as $m\rightarrow\infty$.  This discretization is similar to the use of finite element bases to express functions on surfaces~\cite{johnson12}; for instance, on a triangle mesh, piecewise linear ``hat'' functions can serve as an appropriate partition of unity.

\SetAlCapSkip{.5em}
\makeatletter
\newcommand{\nosemic}{\renewcommand{\@endalgocfline}{\relax}}
\newcommand{\dosemic}{\renewcommand{\@endalgocfline}{\algocf@endline}}
\makeatother
\begin{algorithm}[t]
\SetCommentSty{textit}
\SetKwComment{tcc}{}{} 
\SetSideCommentRight
\SetKwInOut{Input}{Input}\SetKwInOut{Output}{Output}
\Input{Signal to be filtered $f_1:\Sigma\rightarrow\R^n$\\Cross bilateral function $f_2:\Sigma\rightarrow\Gamma$\\Samples $\p_1,\ldots,\p_m\in\Gamma$\\Partition of unity $\phi_1,\ldots,\phi_m:\Gamma\rightarrow\R$}
\Output{Filtered signal $\bar{f}:\Sigma\rightarrow\R^n$}
\BlankLine
$\bar{f}^{num}(\x), \bar{f}^{den}(\x)\leftarrow0\,\forall\x\in\Sigma$\tcc*[r]{Initialization}
\For{$i=1\textrm{ to }m$}{
$g^{num}(\x)\leftarrow f_1(\x)K_\Gamma(f_2(\x),\p_i)$\tcc*[r]{Weight signals}
$g^{den}(\x)\leftarrow K_\Gamma(f_2(\x),\p_i)$\;
$\hat{g}^{num}(\x)\leftarrow \mathbf{T}[g^{num}](\x)$\tcc*[r]{Apply blur operator}
$\hat{g}^{den}(\x)\leftarrow \mathbf{T}[g^{den}](\x)$\;
$\bar{f}^{num}(\x)\leftarrow \bar{f}^{num}(\x)$\nosemic\tcc*[r]{Collect}
\dosemic$\hspace{.75in}+ \hat{g}^{num}(\x)\phi_i(f_2(\x))$\;
$\bar{f}^{den}(\x)\leftarrow \bar{f}^{den}(\x) + \hat{g}^{den}(\x)\phi_i(f_2(\x))$\;
}
$\bar{f}(\x)\leftarrow \nicefrac{\bar{f}^{num}(\x)}{\bar{f}^{den}(\x)}$\tcc*[r]{Normalize}\vspace{.025in}
\caption{Generalized bilateral filtering algorithm\vspace{-.2in}}\label{alg:generalized}
\end{algorithm}

Define $g_i(\x)=f_{\p_i}(\x)$; this function can be computed for all $x$ in $\Sigma$ by evaluating $f_1$ and $K_\Gamma$ as in~\eqref{eq:numerator} and~\eqref{eq:denominator}.  The blurring operation~\eqref{eq:blur} is then applied to obtain $\hat{g}_i(\x)=\mathbf{T}[g_i](\x)$.  For instance, if $\Sigma$ is an image then $\mathbf{T}$ will be a Gaussian blur, while mesh bilateral filters would implement $\mathbf{T}$ using diffusion.  Our bilateral filter is thus approximated as:
\begin{equation}
\bar{f}(\x) \approx \frac{\sum_i \hat{g}^{num}_i(\x) \phi_i(f_2(\x))}{\sum_i \hat{g}^{den}_i(\x) \phi_i(f_2(\x))}\label{eq:bilateral_approximation}
\end{equation}

We show several concrete applications of bilateral filtering simply by applying this formulation to various domains and kernels.  If $K_\Gamma$ is straightforward to evaluate, the only time-consuming step is generating the functions $\hat{g}_i$ from $g_i$; that is, the time complexity of this algorithm is essentially that of carrying out $2m$ blurs~\eqref{eq:blur}.

\section{Processing Scalar Signals}\label{sec:scalar_processing}

Before introducing novel domains and signals, we verify that our bilateral filter applied to grayscale images reduces to the one presented in~\cite{paris06}.  Here, we define our signal domain as $\Sigma = \{1,\ldots,w\}\times\{1,\ldots,h\}$, a $w\times h$ grid of pixel values, and our signal range of grayscale intensities is $\Gamma=[0,1]$.  We take our image and intensity kernels to be $K_\Sigma(\x,\y)\equiv W_s(\|\x-\y\|)$ and $K_\Gamma(p,q)=W_c(|p-q|)$.  It is easy to check that in this case~\eqref{eq:generalized_bilateral} and~\eqref{eq:cross_bilateral} coincide.

Now, suppose we divide $\Gamma=[0,1]$ into $m$ equally-spaced samples $p_1,\ldots,p_m$ of width $\nicefrac{1}{m-1}$.  Define $\phi_i:[0,1]\rightarrow\R$ to be the piecewise linear hat function centered at $p_i$ with width $\nicefrac{2}{m-1}$.  Then,~\eqref{eq:bilateral_approximation} coincides with the ``signal processing approximation'' in~\cite{paris06}.  The approximation is indistinguishable from the exact bilateral on most images for $m$ as low as $20$, and it can be carried out using down/up-sampling or methods like~\cite{burt83,deriche93} for $\mathbf{T}$ in~\eqref{eq:blur}.

Generalizing somewhat, suppose we take $\Sigma$ to be a mesh with vertices $V$, edges $E$, and triangular faces $F$.  We represent scalar functions on $\Sigma$ as vectors $\mathbf{v}\in\R^{|V|}$ and construct a ``cotangent Laplacian'' matrix $L\in\R^{|V|\times|V|}$ with diagonal mass matrix $A\in\R^{|V|\times|V|}$ imitating the Laplacian operator on the smooth surface approximated by $\Sigma$~\cite{macneal49}.  We compute $\mathbf{T}(\mathbf{v})$ using heat flow using a single implicit time step $\mathbf{T}(\mathbf{v})\approx (I+\Delta t A^{-1} L)^{-1}\mathbf{v}$.  Multiple time steps or a higher-order discretization yield closer approximations, but the damping effect of a single implicit step has few perceptual differences and is faster to carry out; furthermore, it can be viewed as an isotropic instance of the screened Poisson equation~\cite{chuang11}, which may suggest future research directions making bilateral filtering faster or more anisotropic.  Since we apply $\mathbf{T}$ several times, we pre-factor time time step matrix using the sparse LU method in~\cite{davis04}.  We keep $\Gamma=[0,1]$ with Gaussian kernel $K_\Gamma(x,y)=e^{-|x-y|^2/\sigma^2}$.

\begin{figure}[htb]
\centering
\begin{tabular}{cccc}
\includegraphics[width=.2\linewidth]{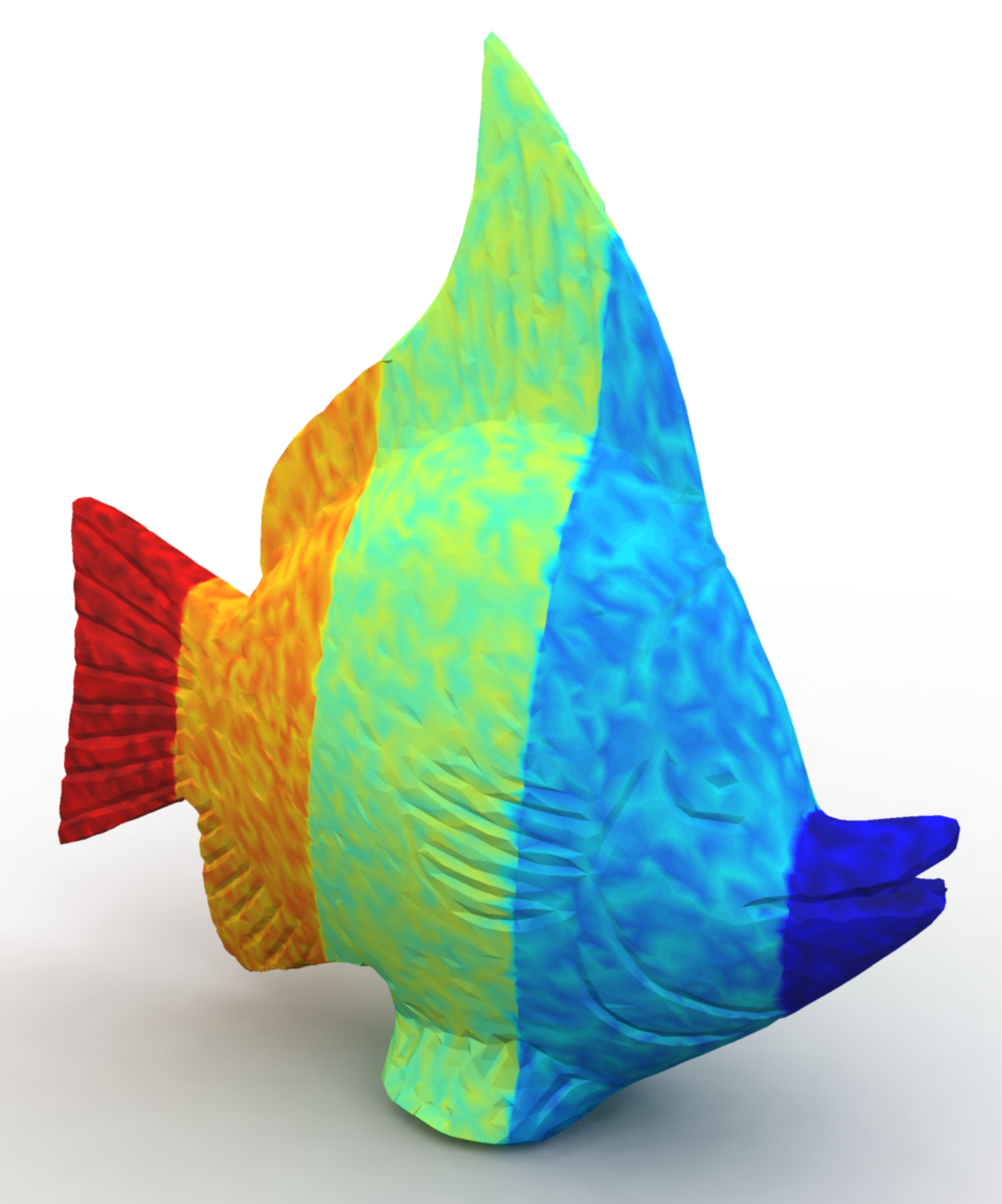}&
\includegraphics[width=.2\linewidth]{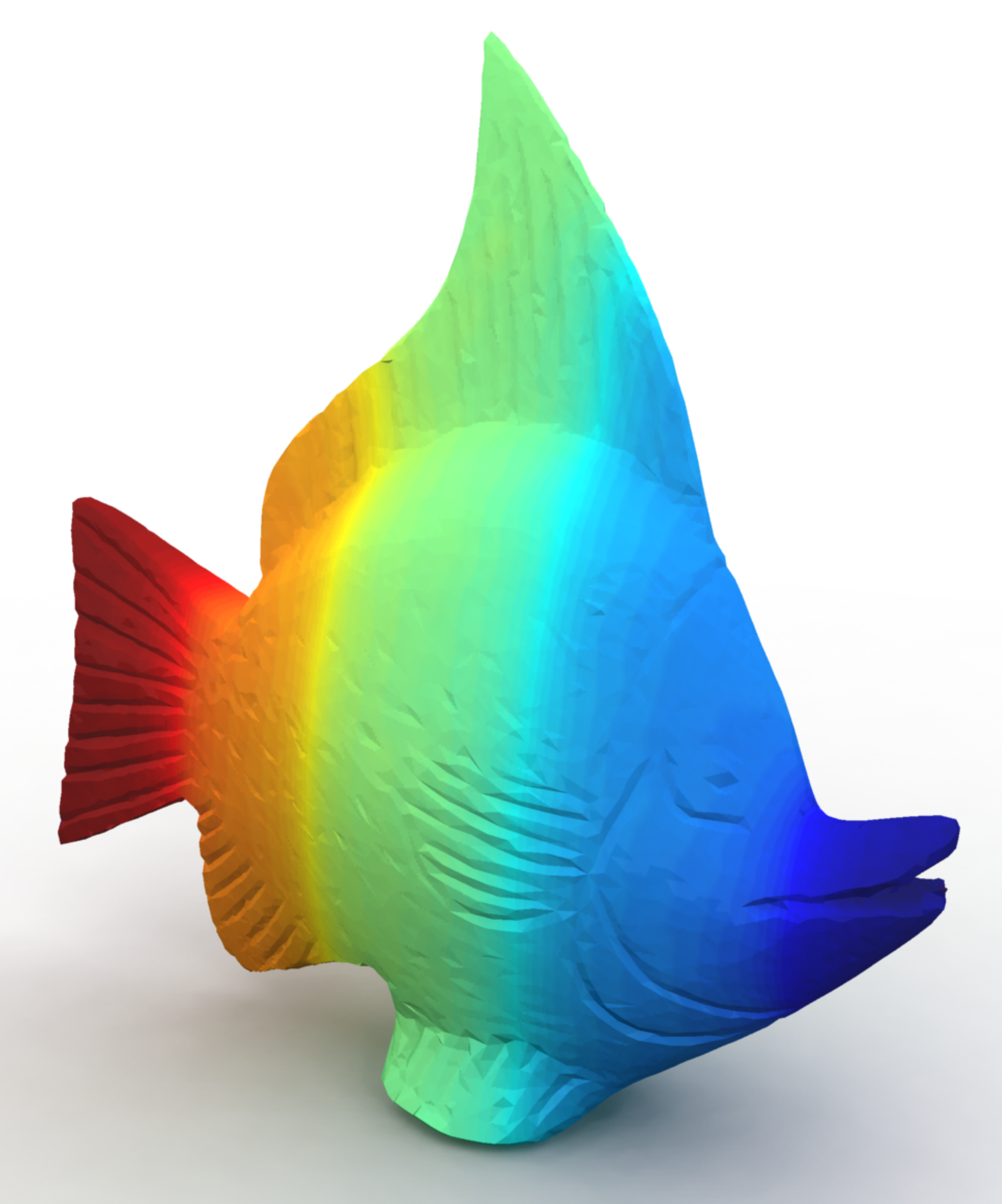}&
\includegraphics[width=.2\linewidth]{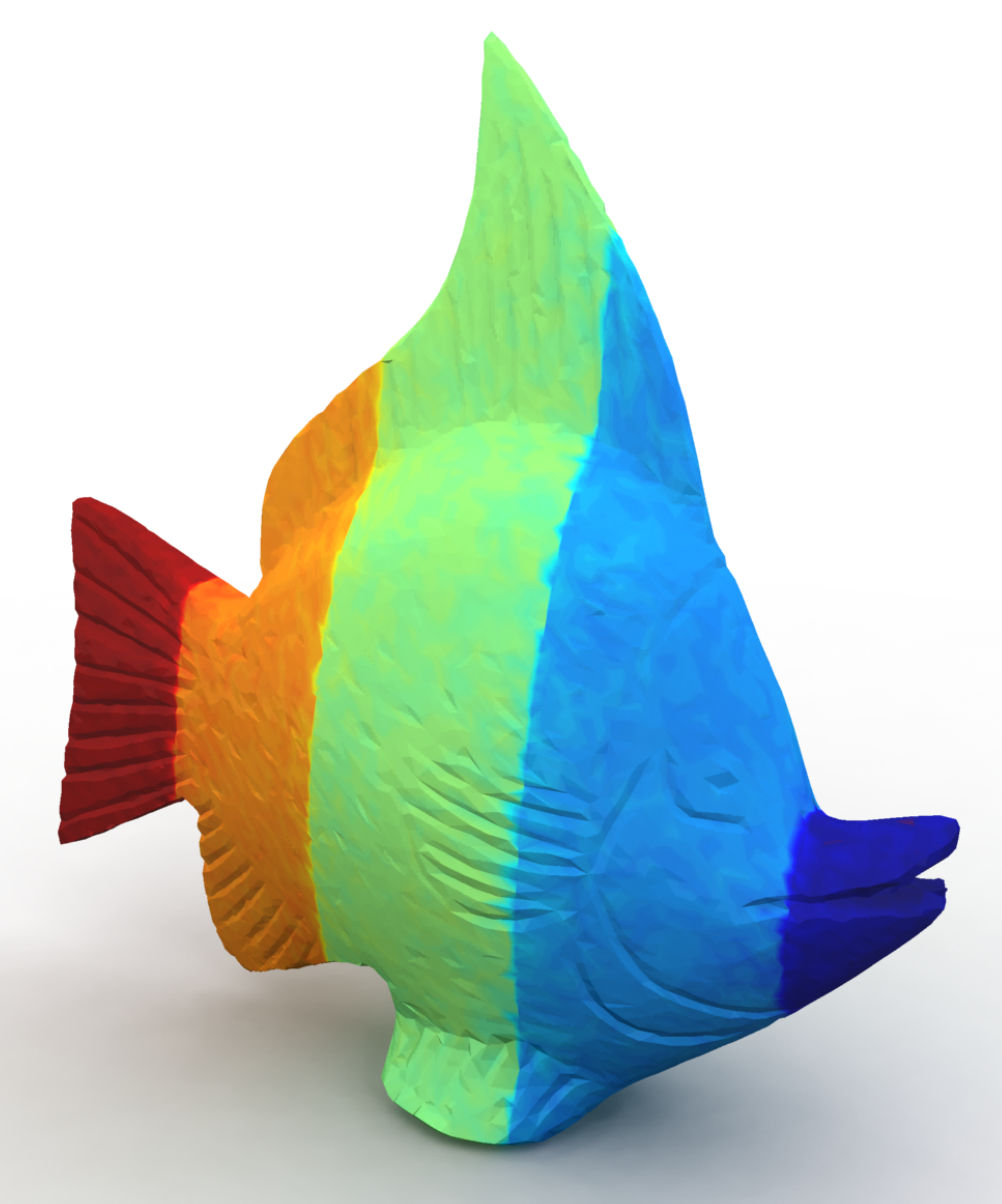}&
\includegraphics[width=.2\linewidth]{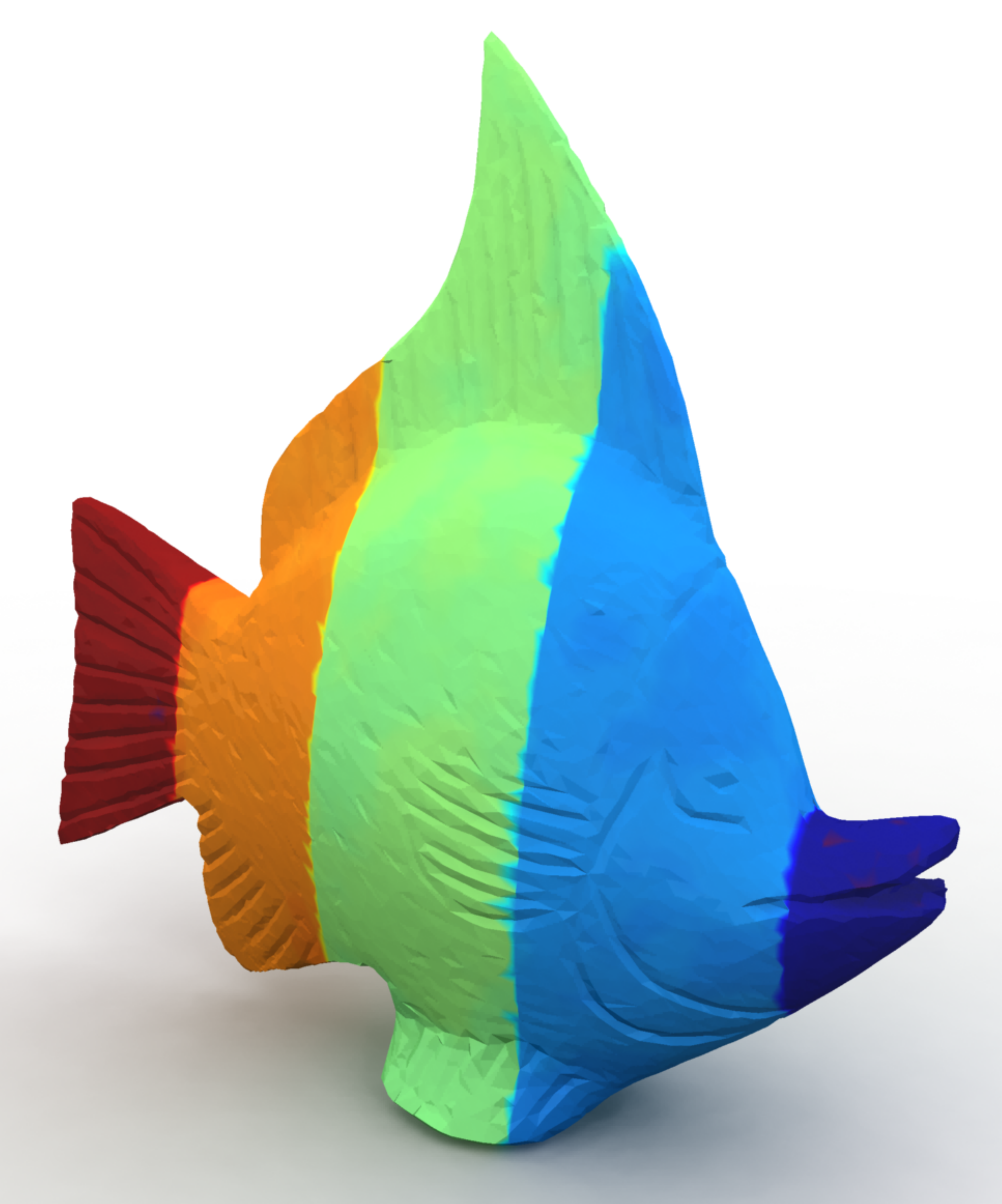}\\
(a) & (b) & (c) & (d)
\end{tabular}
\caption{A noisy function (a) smoothed using Laplacian diffusion (b), the generalized bilateral, (c), and the mean shift (d).  Diffusion does not preserve signal edges, the bilateral removes most of the noise while preserving edges, and the mean shift provides strong denoising.}\label{fig:scalar_bilateral}
\end{figure}

If we take $f_1=f_2\equiv f:\Sigma\rightarrow\R$, the generalized bilateral blurs $f$ while preserving its discontinuities.  Figure~\ref{fig:scalar_bilateral} shows the output of this method and the iterative mean shift on a noisy texture. Unlike the image bilateral and mesh methods relying on planar projection or parameterization, this bilateral respects the metric of $\Sigma$ regardless of the width of $K_\Sigma$.

\begin{figure}[thb]
\centering
\begin{tabular}{ccc}
\includegraphics[width=.27\linewidth]{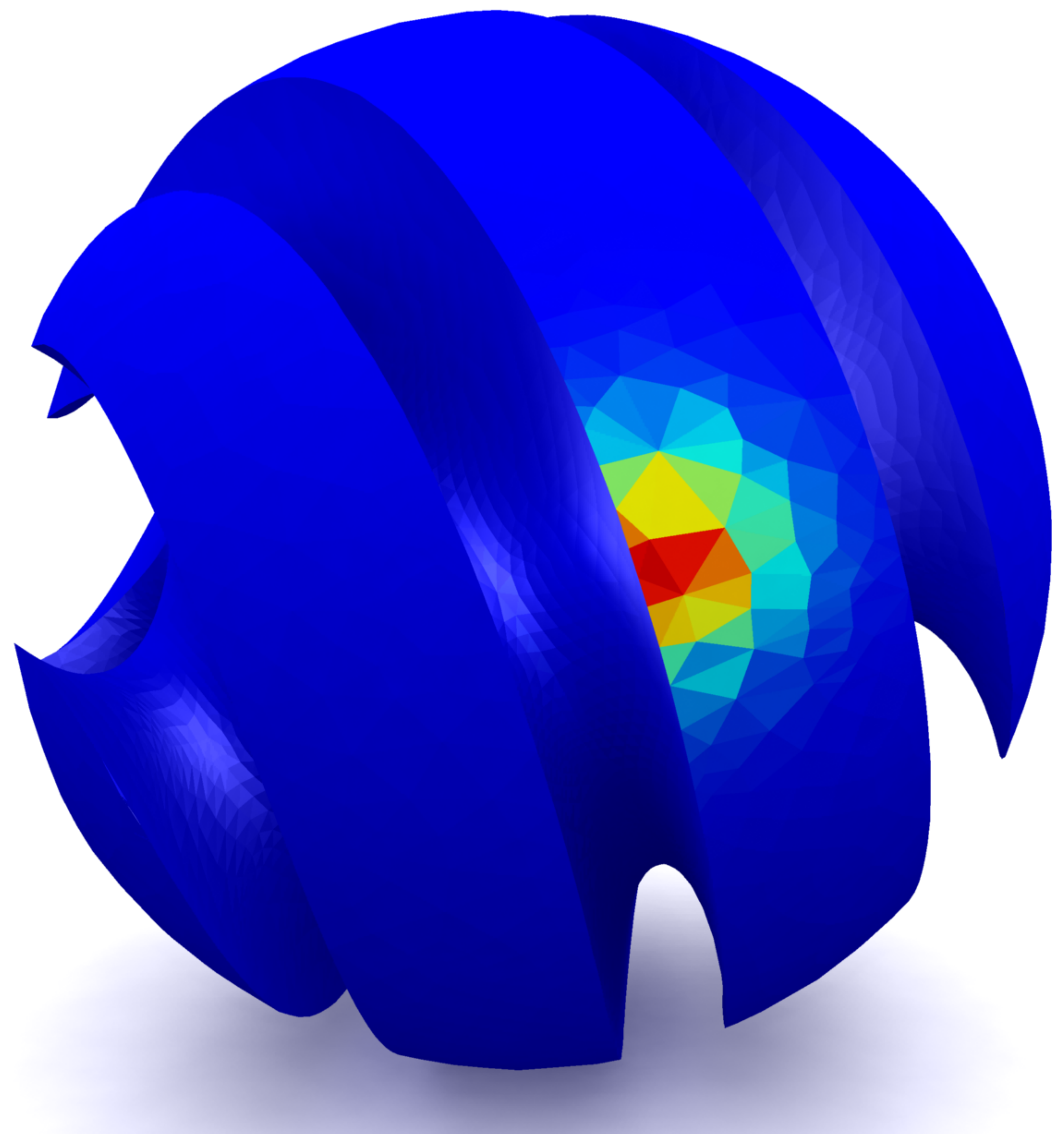}&
\includegraphics[width=.27\linewidth]{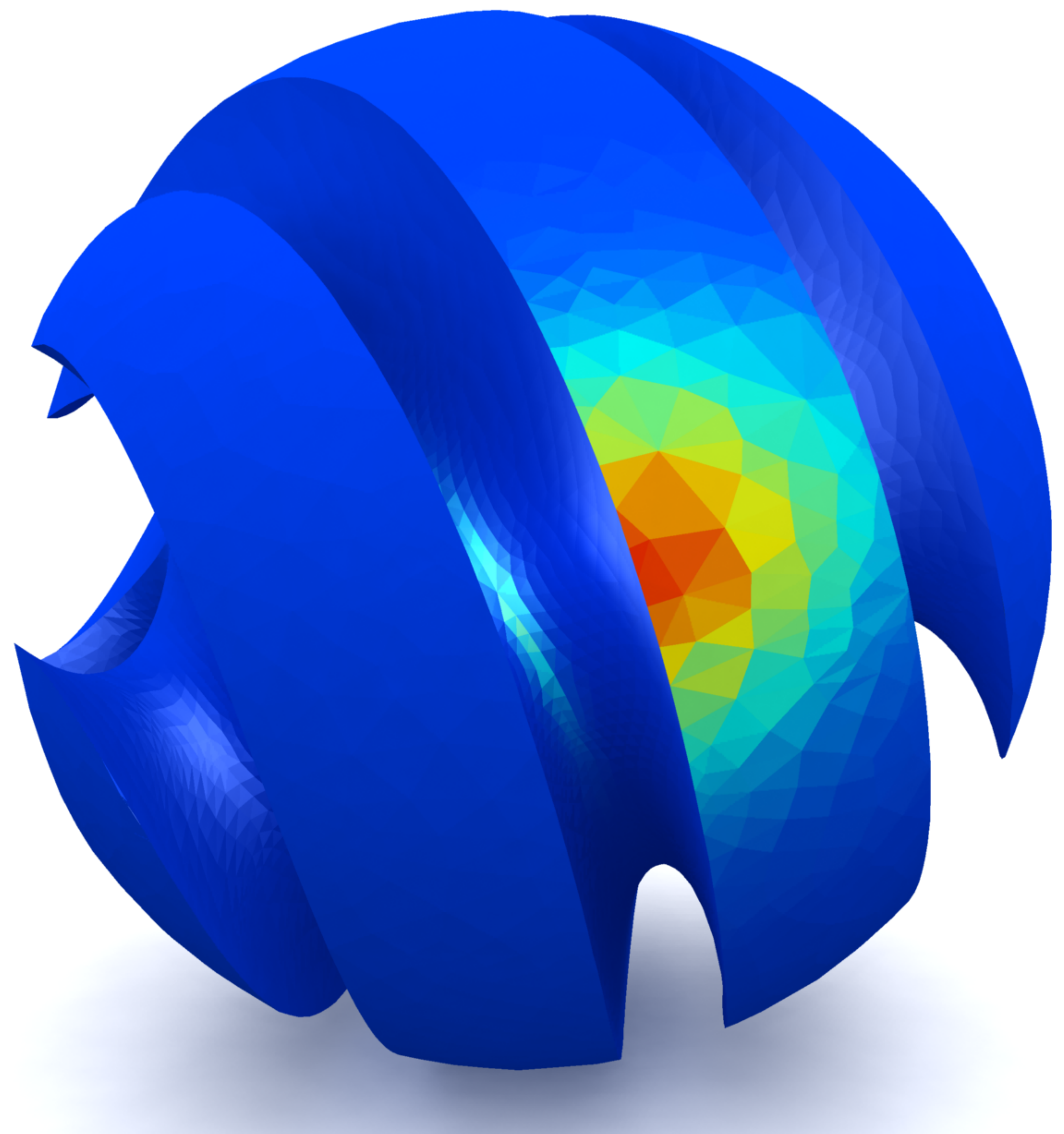}&
\includegraphics[width=.27\linewidth]{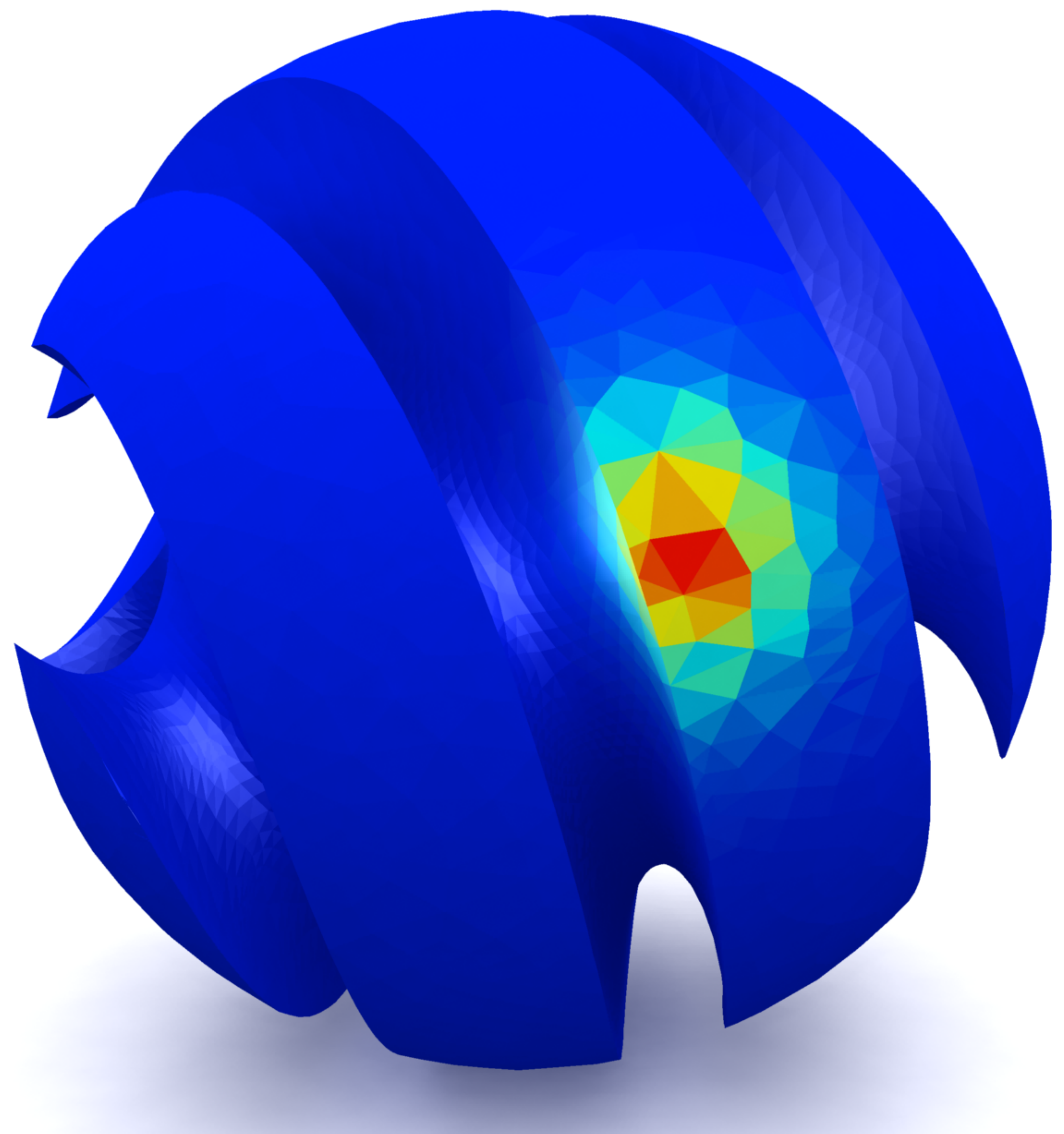}\\
(a) & (b) & (c)
\end{tabular}
\caption{Kernel of the normal cross bilateral (a).  Increasing the reach of $K_\Sigma$ widens the kernel (b), while increasing that of $K_\Gamma$ allows the kernel to continue over sharp edges (c).}\label{fig:normal_kernel}
\end{figure}

\section{Mesh Denoising}\label{sec:normal_signal}

We can extend the method in Section~\ref{sec:scalar_processing} by considering cross bilaterals for which $\Gamma$ is not $[0,1]$.  Most importantly, suppose $\Gamma=S^2$, the unit sphere, and take $f_2$ to be the signal $\N:F\rightarrow S^2$ given by unit face normals.  Our signal now is on mesh \emph{faces} rather than vertices to avoid ambiguous normals along sharp edges.  So, we replace $L$ from Section~\ref{sec:scalar_processing} with the dual $0$-form Laplacian $d \star d \star$ from discrete exterior calculus~\cite{hirani03}.  Figure~\ref{fig:normal_kernel} illustrates the bilateral kernel $K_\Sigma K_\Gamma$ in this context.

\begin{figure*}[thb]
\centering
\begin{tabular}{ccc}
\includegraphics[width=.3\linewidth]{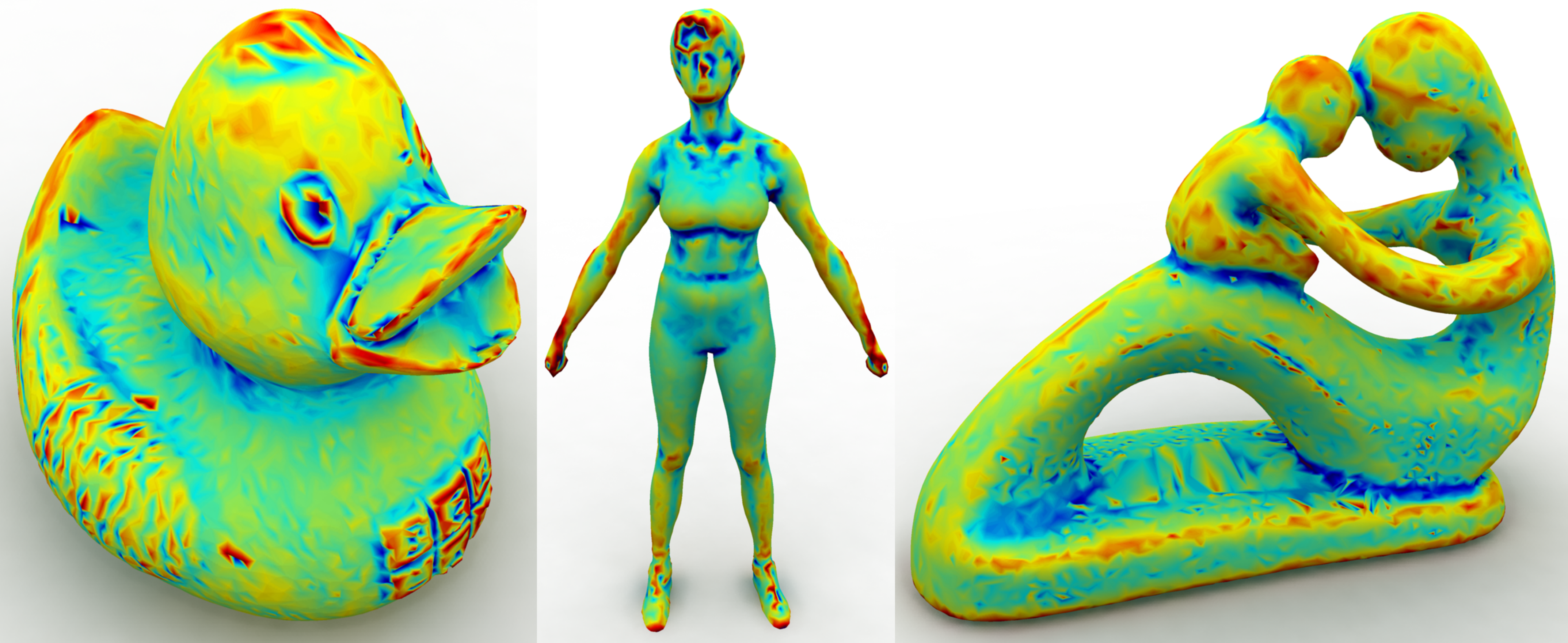} &
\includegraphics[width=.3\linewidth]{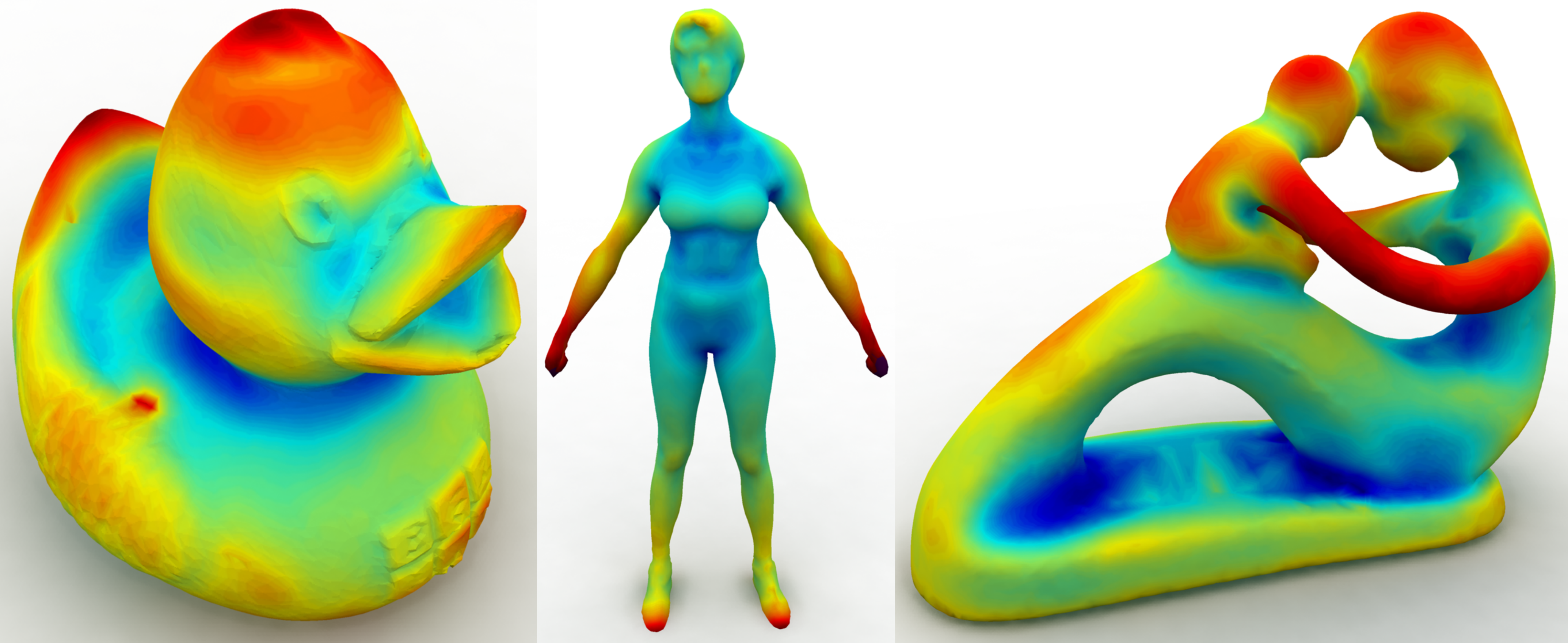} &
\includegraphics[width=.3\linewidth]{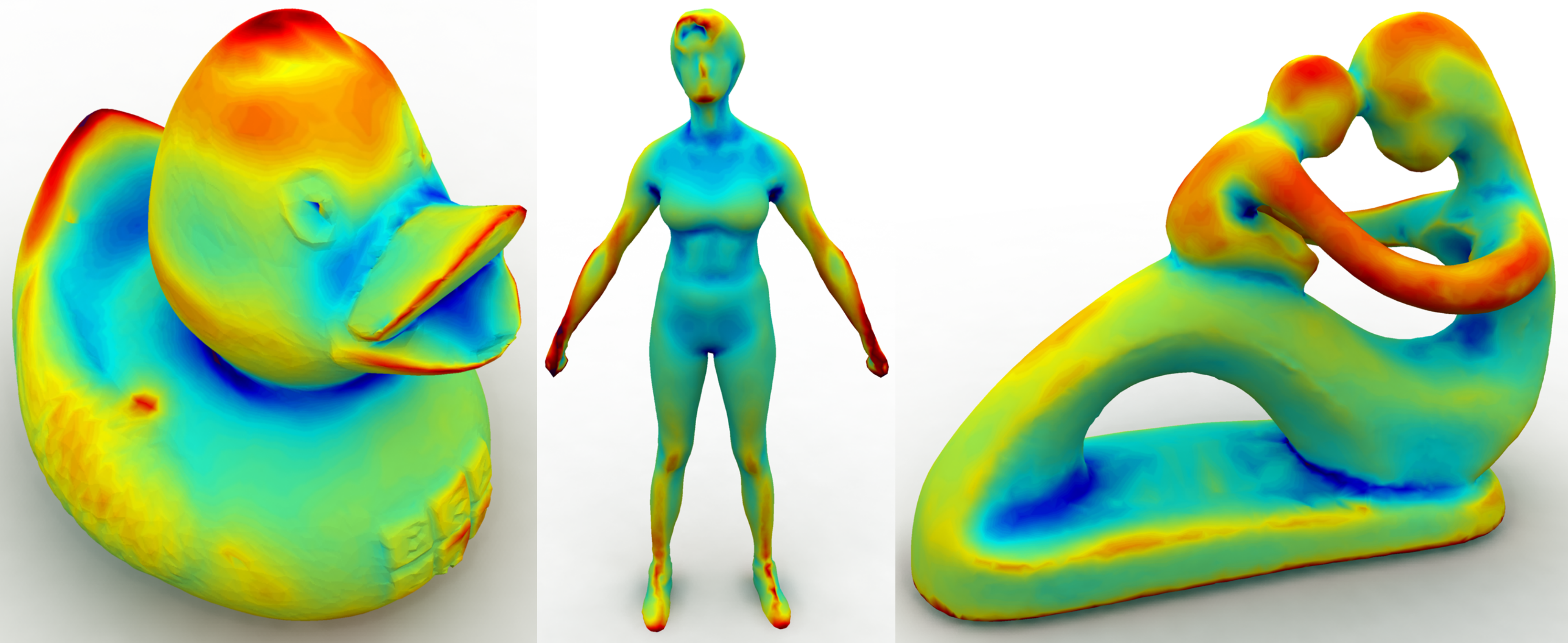} \\
(a) & (b) & (c)
\end{tabular}
\caption{Noisy mean curvature values obtained from a one-ring computation (a), Laplacian diffusion-smoothed mean curvatures (b), and bilateral-filtered mean curvatures (c).}\label{fig:normal_bilateral}
\end{figure*}

\begin{figure}[htb]
\centering
\begin{tabular}{cc}
\includegraphics[width=.45\linewidth]{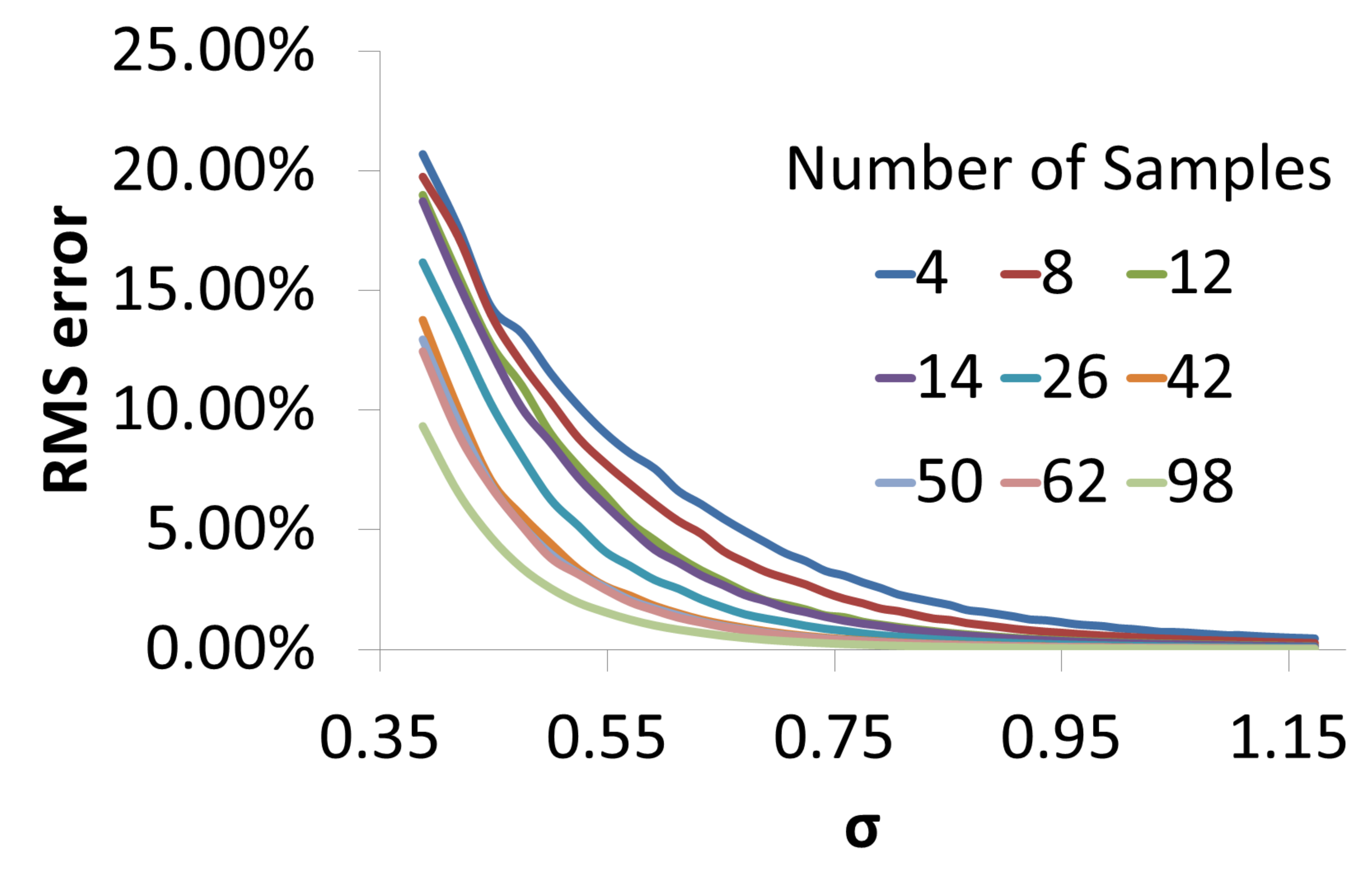}&
\includegraphics[width=.45\linewidth]{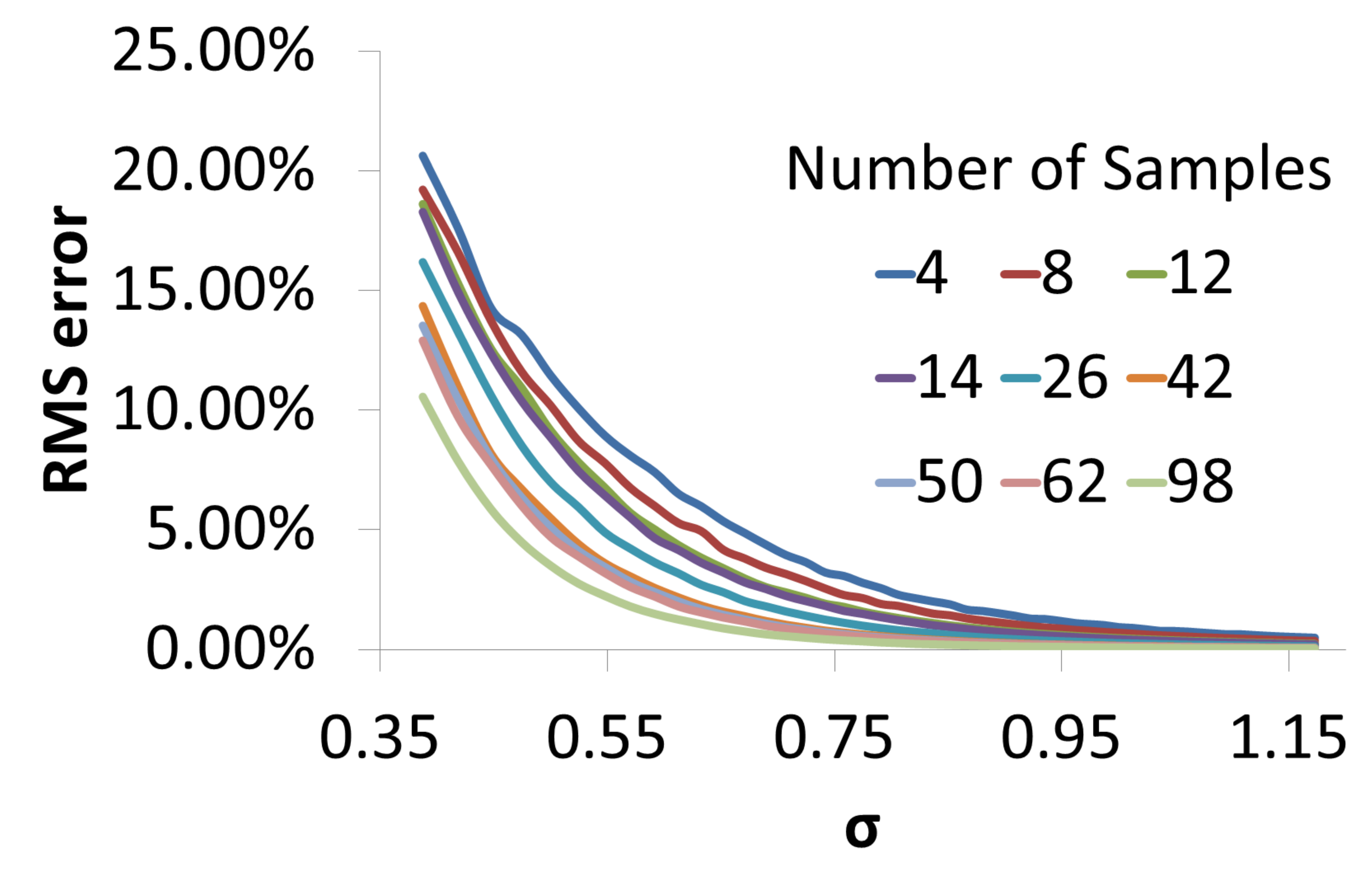}\\
(a)&(b)
\end{tabular}\\
\begin{tabular}{cccc}
(c)&
\includegraphics[height=.4in]{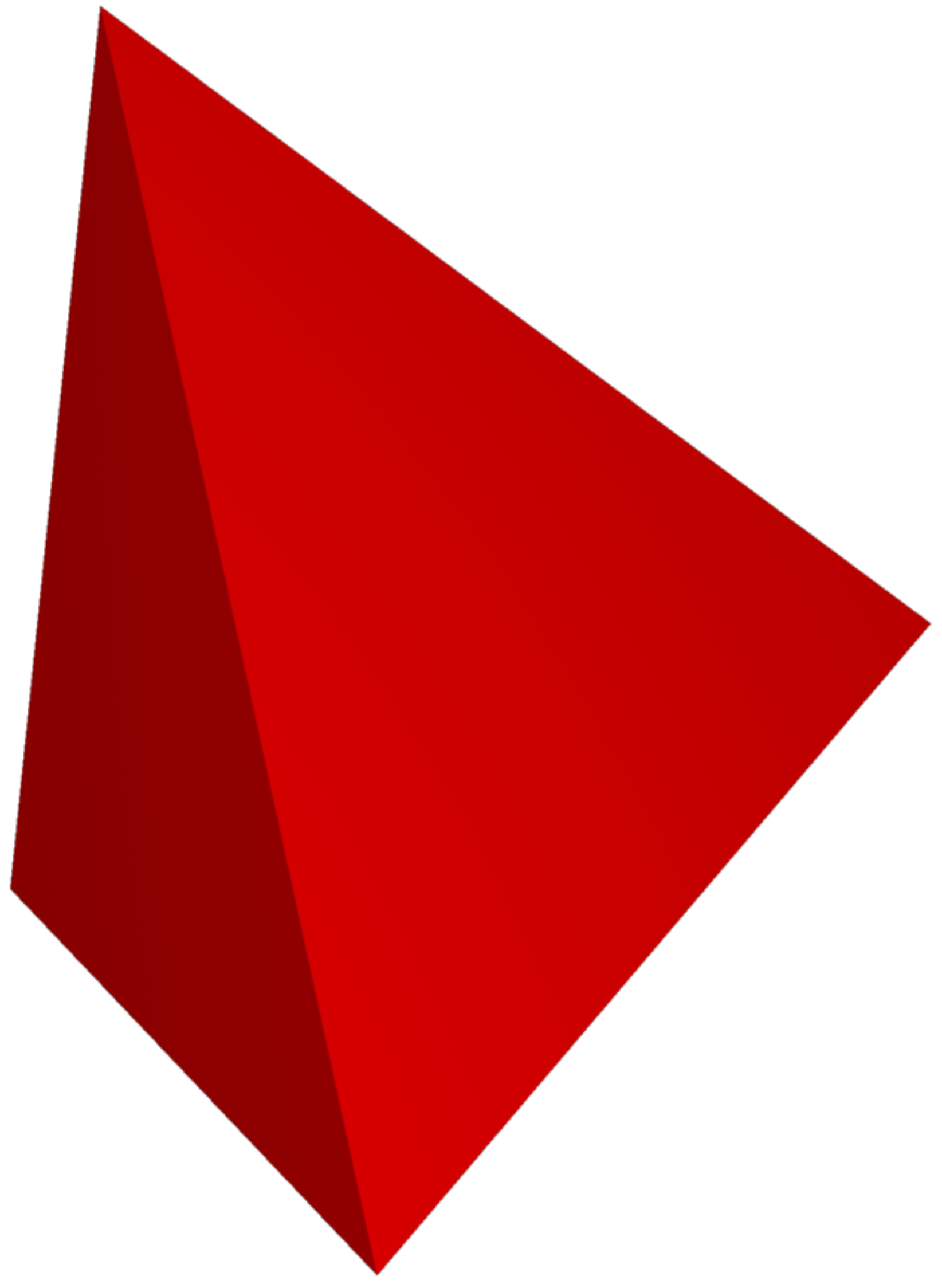} &
\includegraphics[height=.4in]{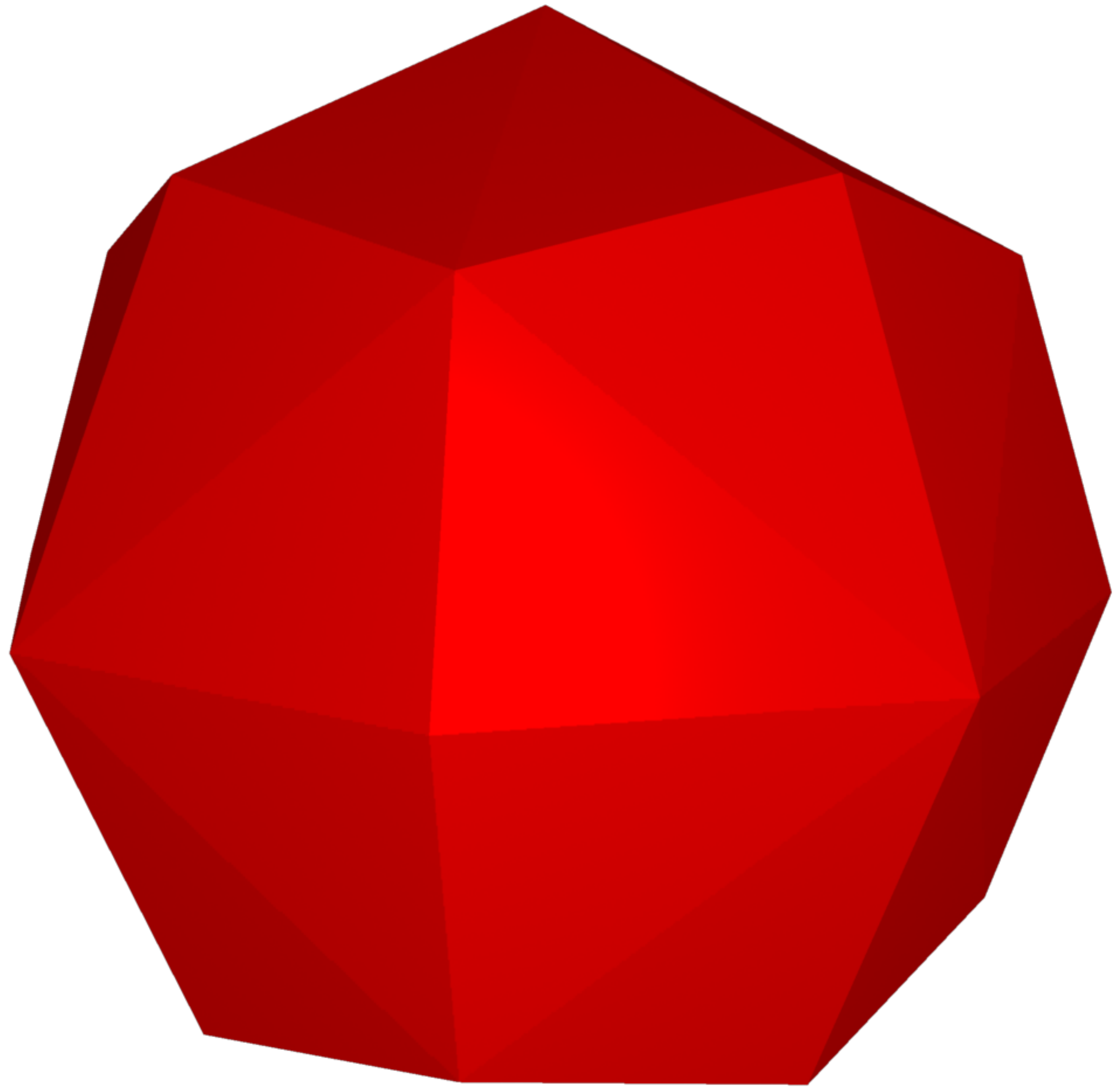} &
\includegraphics[height=.4in]{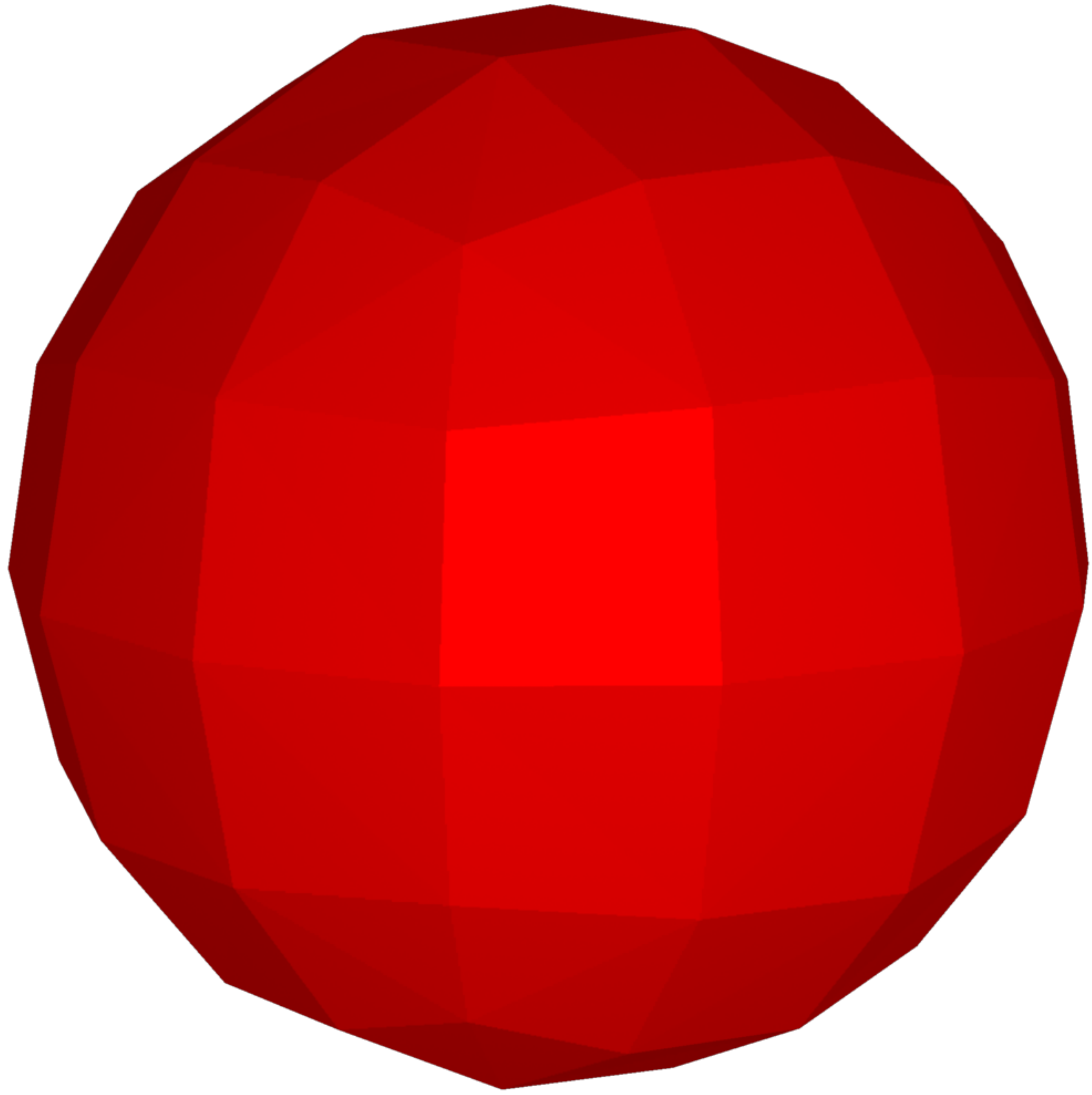} 
\end{tabular}
\caption{Mean-square reconstruction error of Von Mises--Fisher kernels of assorted sizes $\sigma$ using varying numbers of sample points and (a) piecewise-linear interpolation or (b) meshless interpolation; (c) approximations of the unit sphere for (a) with 4, 26, and 98 samples, resp.}\label{fig:sampling_rates}
\end{figure}

A partition of unity on $S^2$ is obtained using a regular polyhedron inscribed within $S^2$; each $\phi_i$ corresponds to a piecewise linear hat function centered at a vertex of the polyhedron projected to $S^2$.  An alternative more efficient and smoother partition of unity paralleling meshless integration is to use Von Mises--Fisher kernels centered at sample points on the unit sphere normalized to sum to 1; we choose the width of the kernels to be half the average distance from each sample to its closest neighbor.  We find little qualitative difference between these approaches and show experiments determining sufficient sampling rates for different kernel sizes in Figure~\ref{fig:sampling_rates}.  Applications of this filter to scalar functions on $\Sigma$ are shown in Figure~\ref{fig:normal_bilateral}; values are not combined over sharp edges since the normal $\N$ has a discontinuity there.

\def\bustStrongNoise{
\begin{tabular}{c@{}c@{}c@{}c@{}c}
\includegraphics[height=.8in]{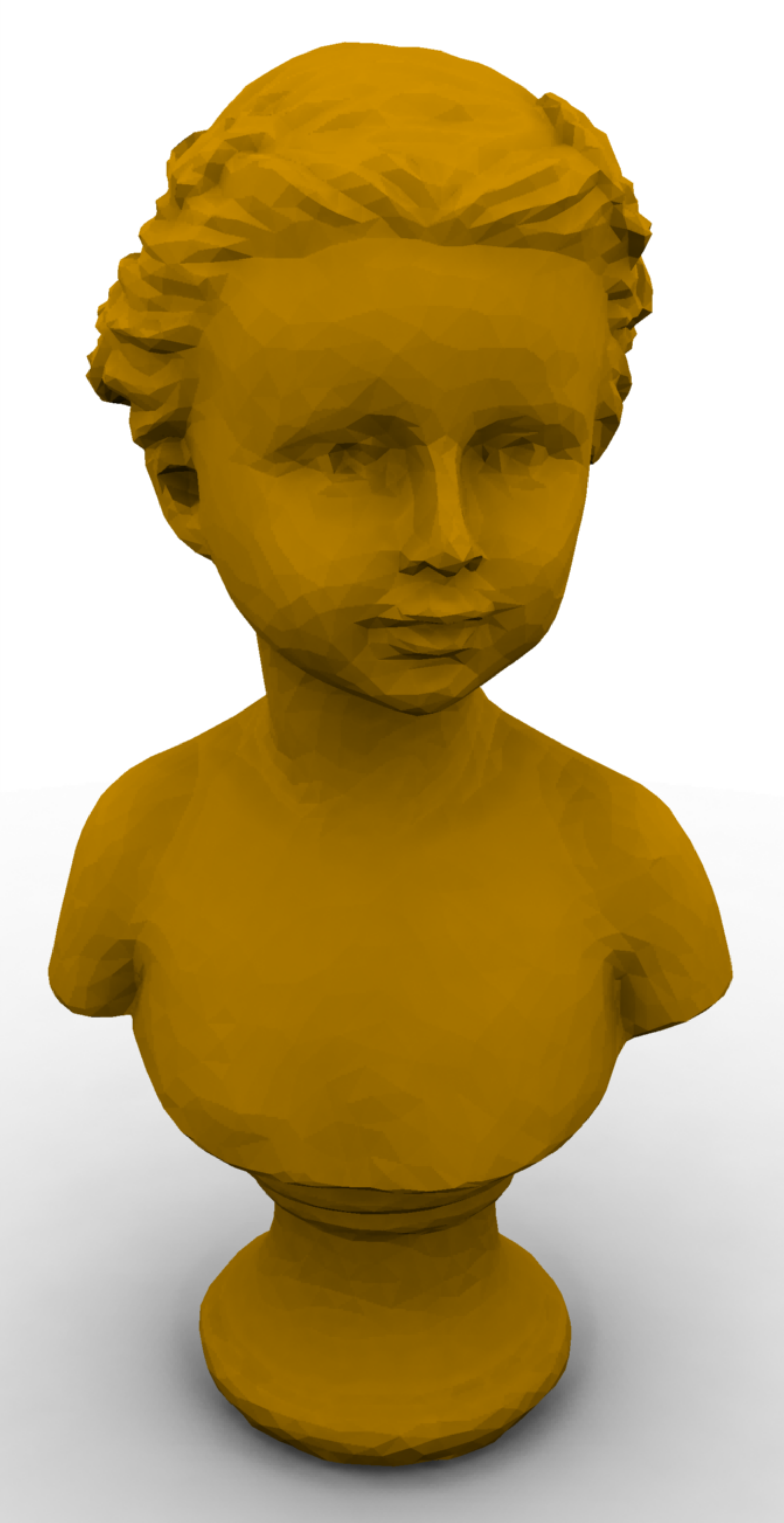} &
\includegraphics[height=.8in]{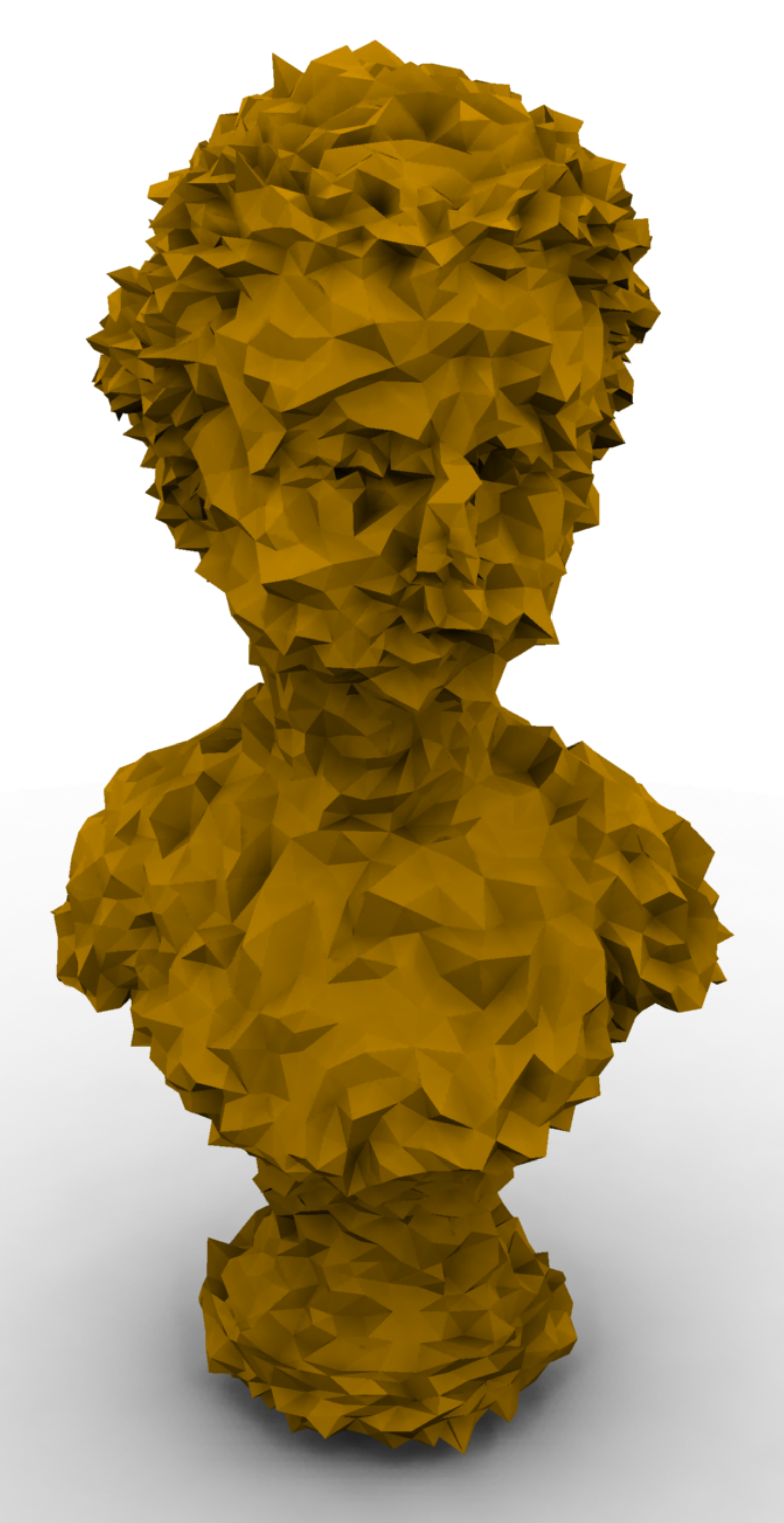} &
\includegraphics[height=.8in]{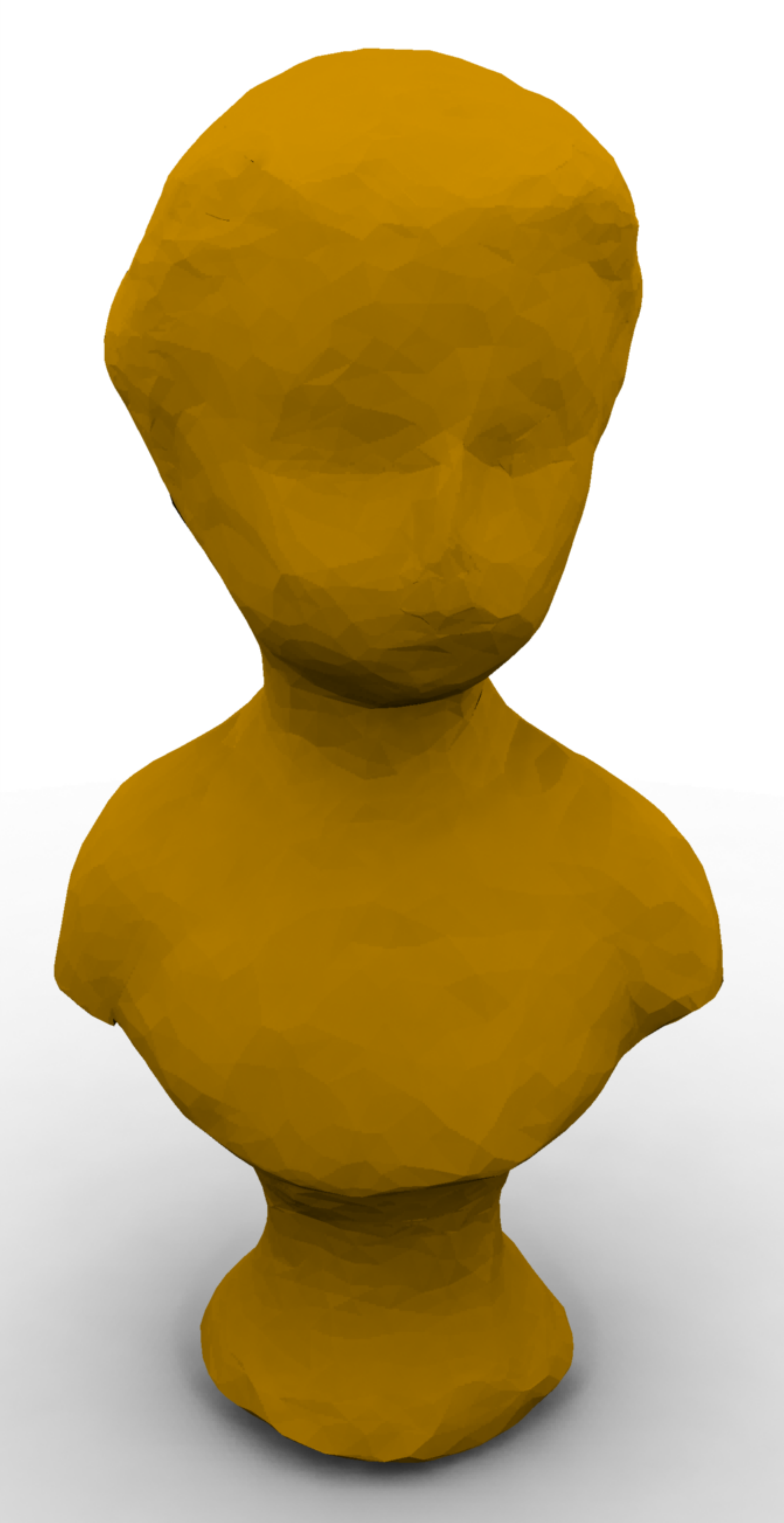} &
\includegraphics[height=.8in]{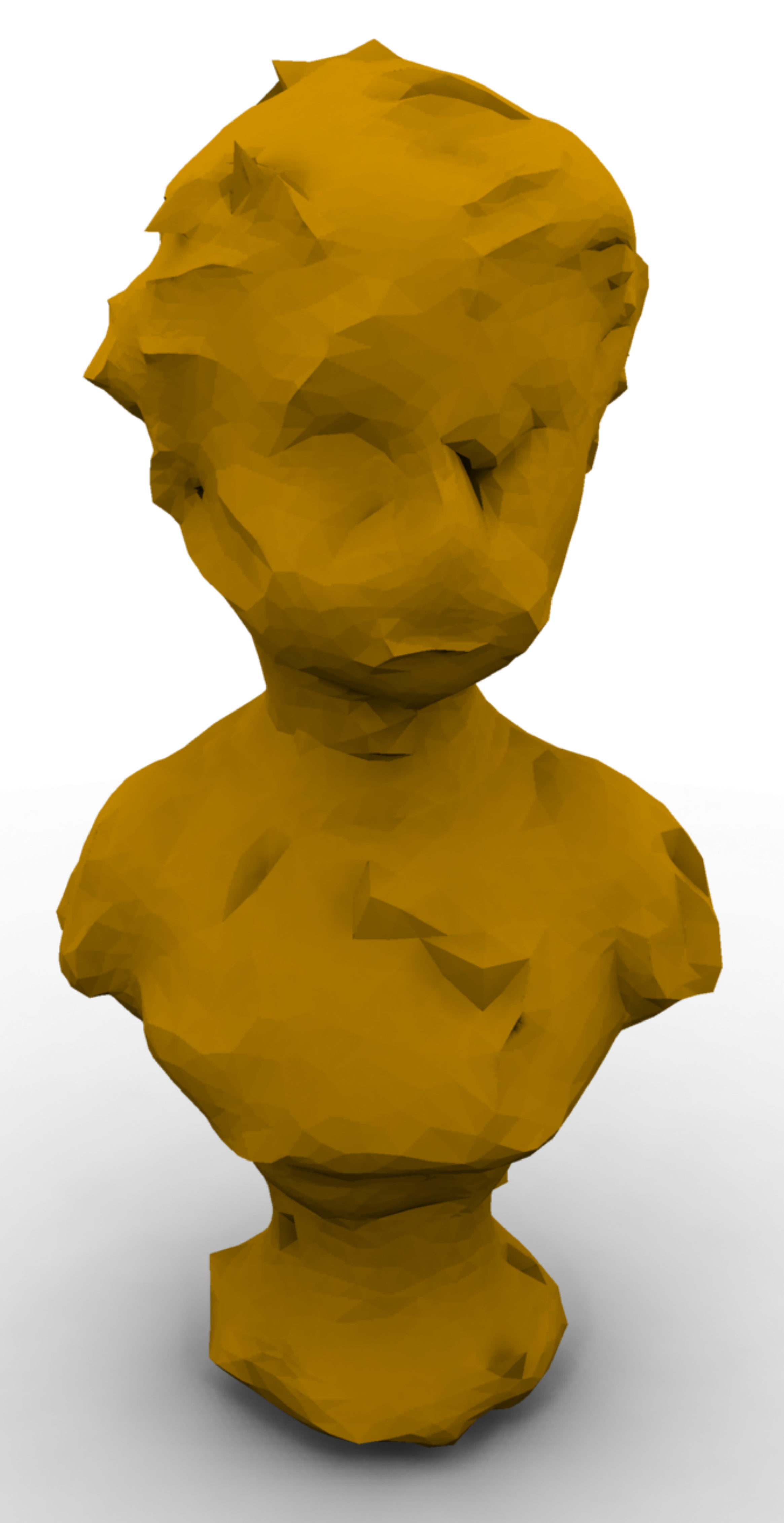} &
\includegraphics[height=.8in]{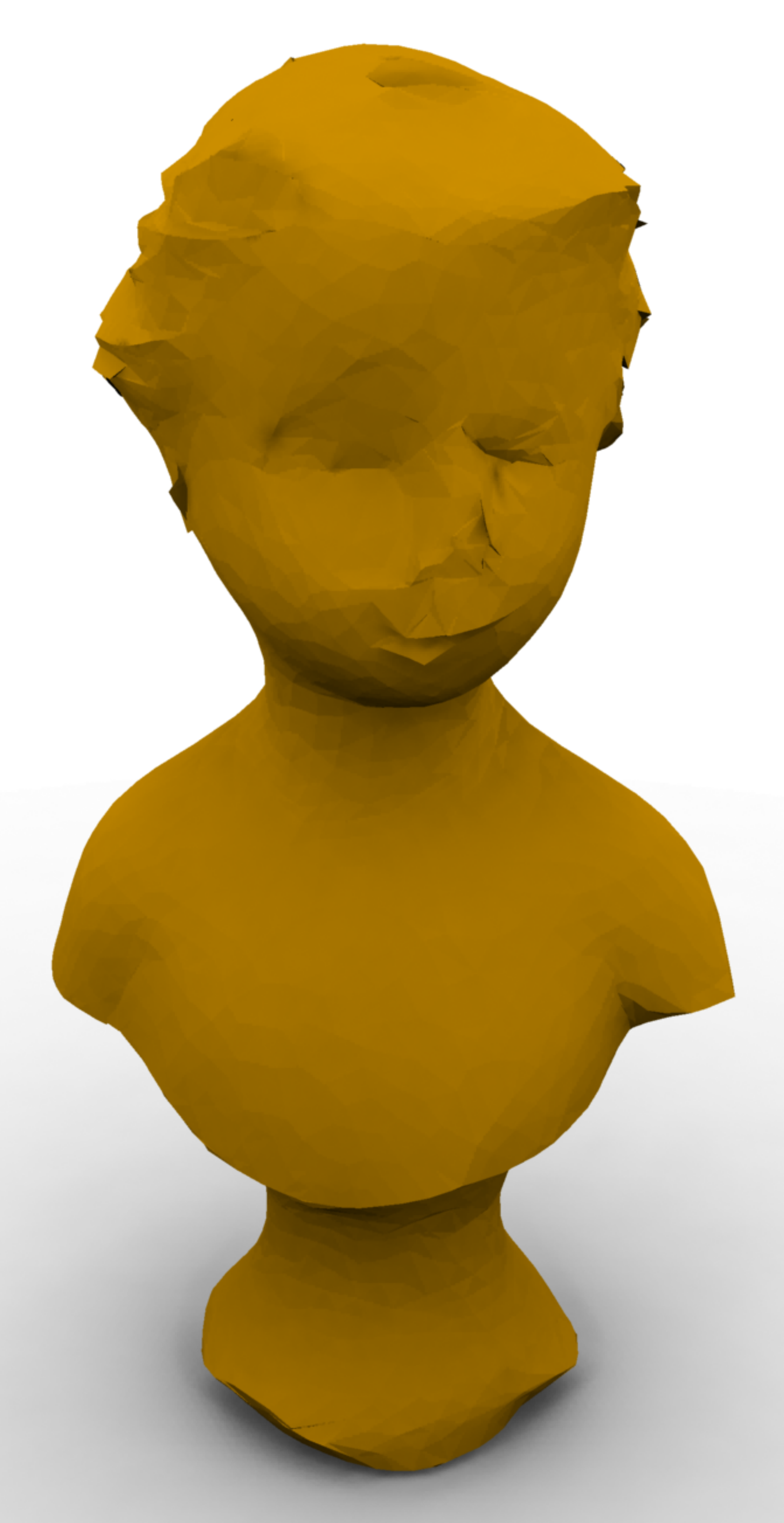}\\
(a)&(b) 0.296 &(d) 0.164 &(f) 0.212 &(g) 0.241
\end{tabular}
}

\def\bustWeakNoise{
\begin{tabular}{c@{}c@{}c@{}c@{}c@{}c}
\includegraphics[height=.8in]{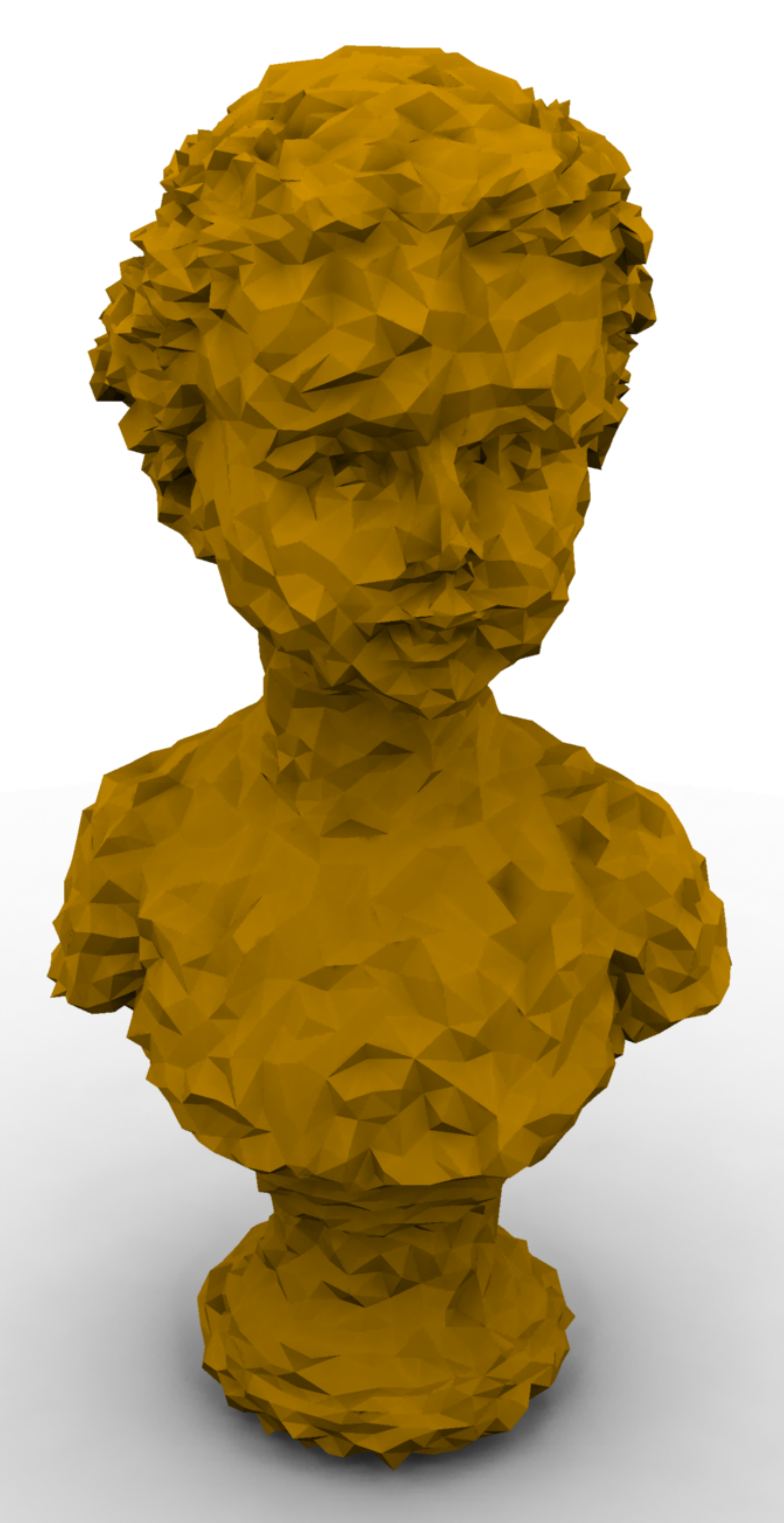} &
\includegraphics[height=.8in]{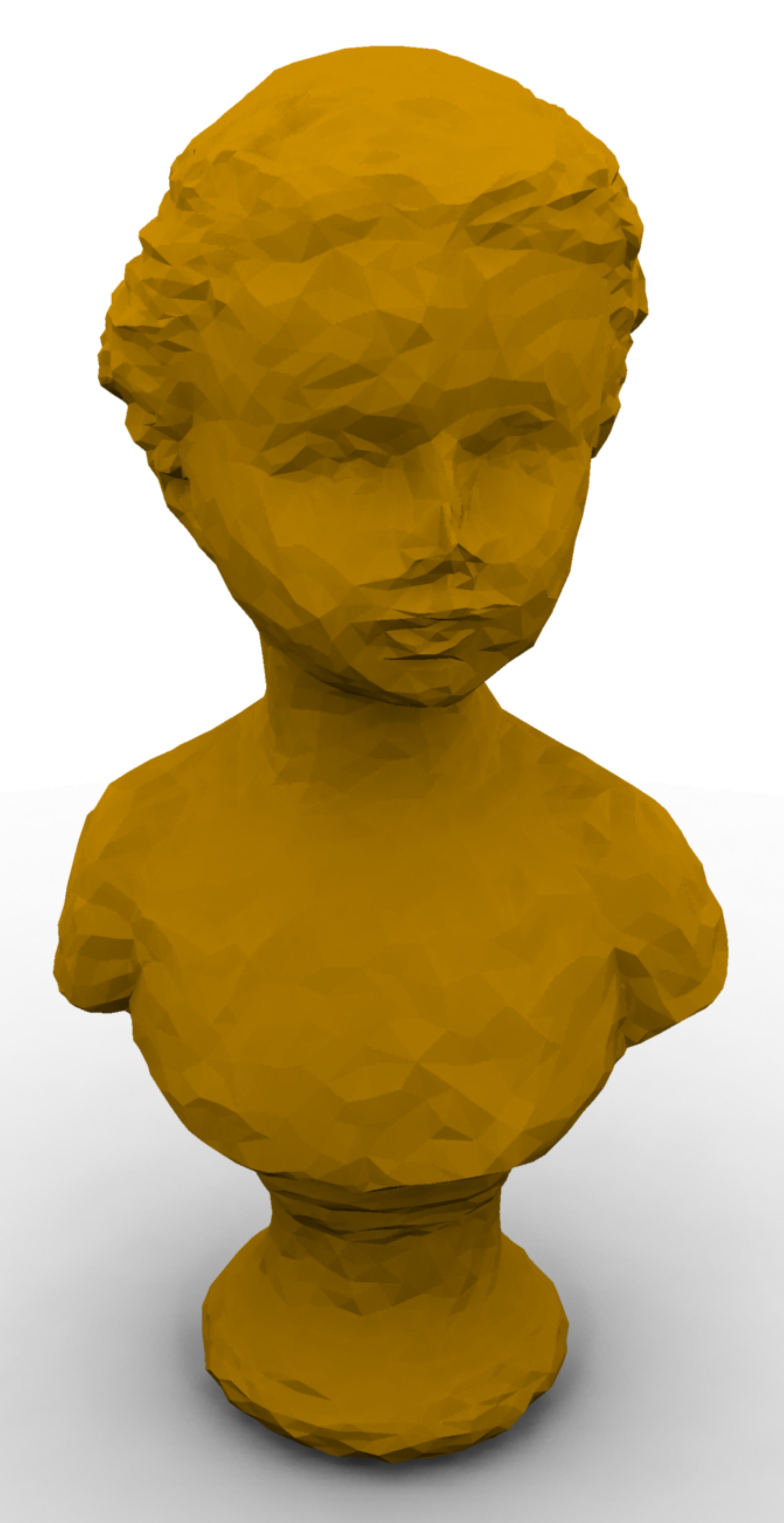} &
\includegraphics[height=.8in]{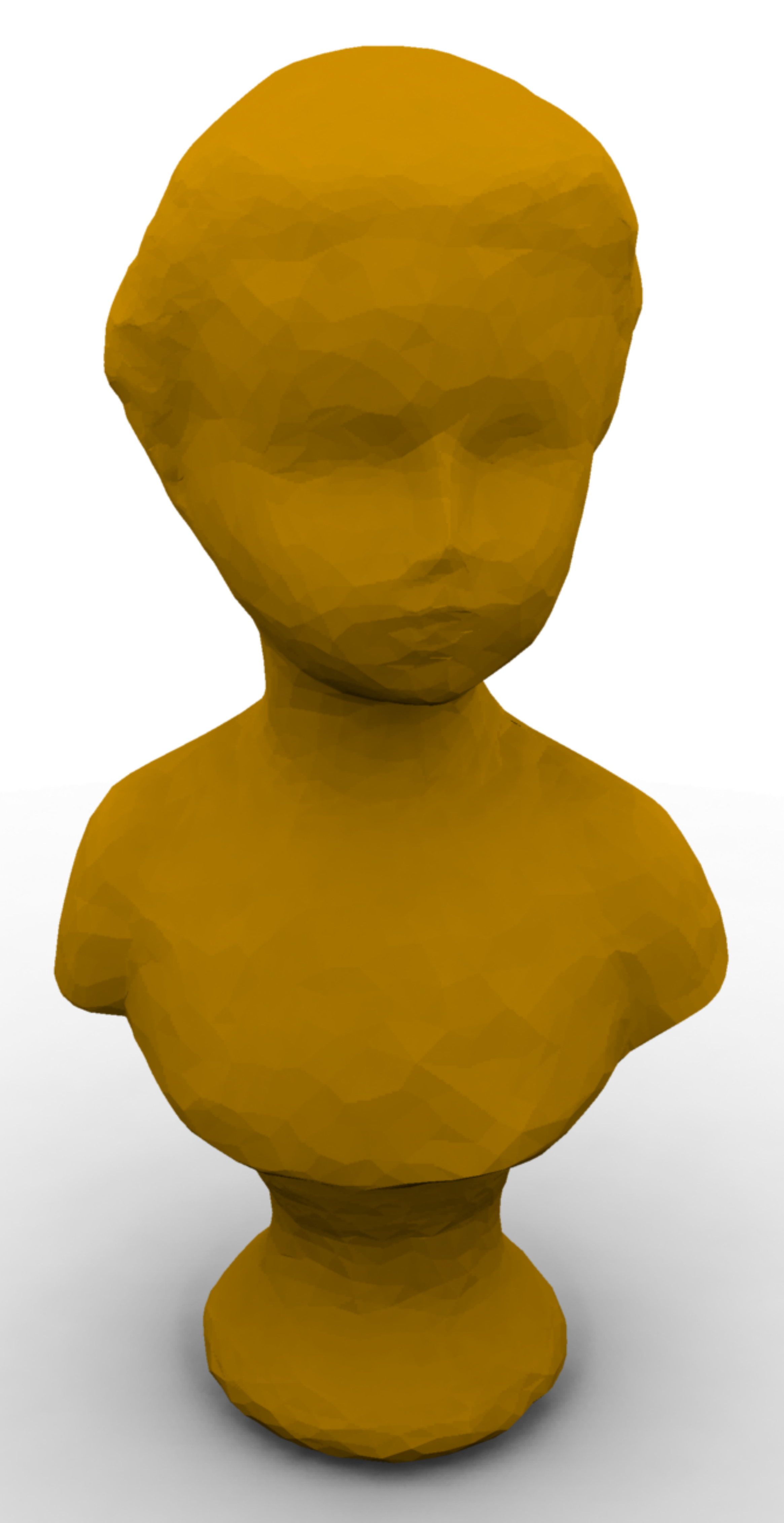} &
\includegraphics[height=.8in]{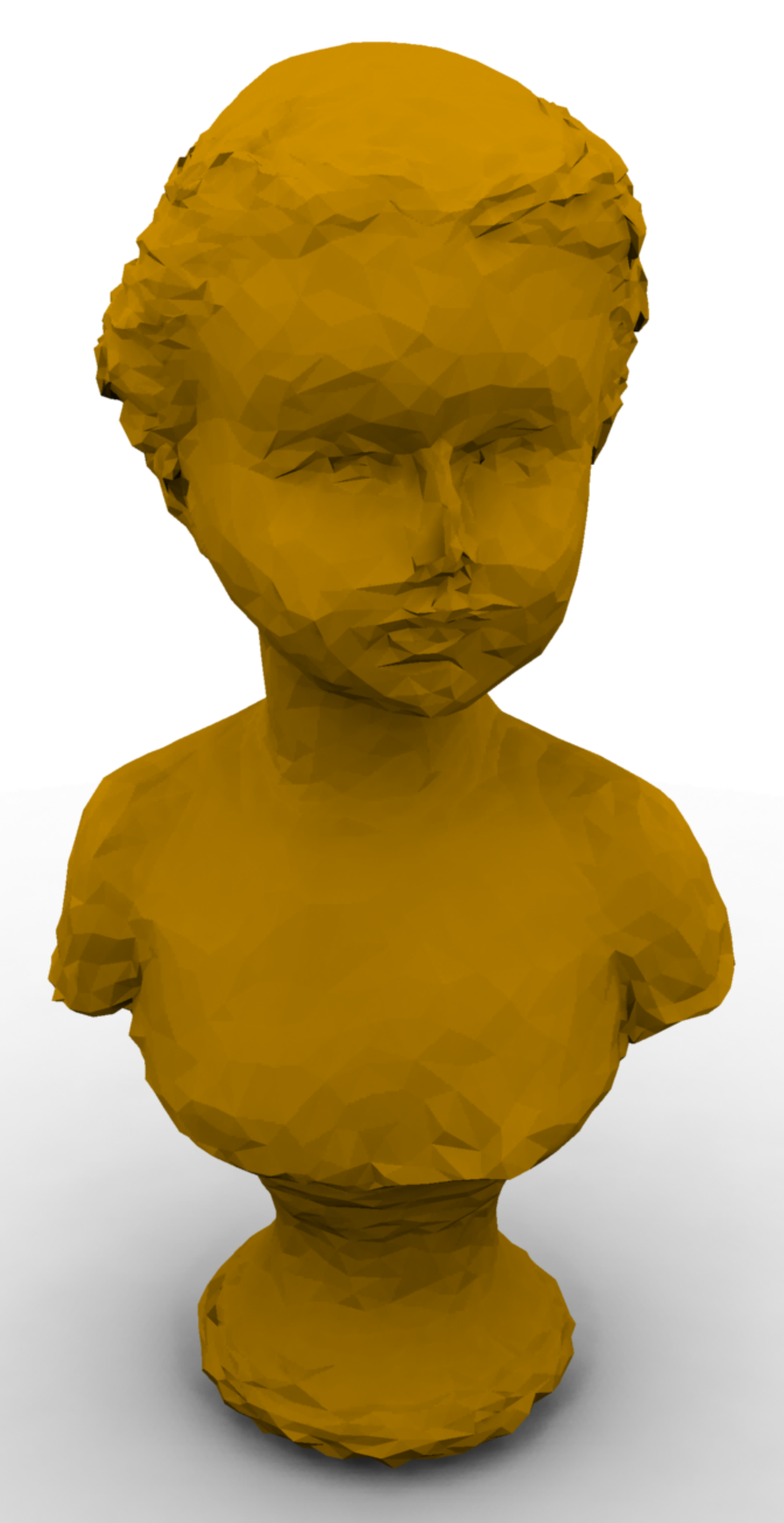} &
\includegraphics[height=.8in]{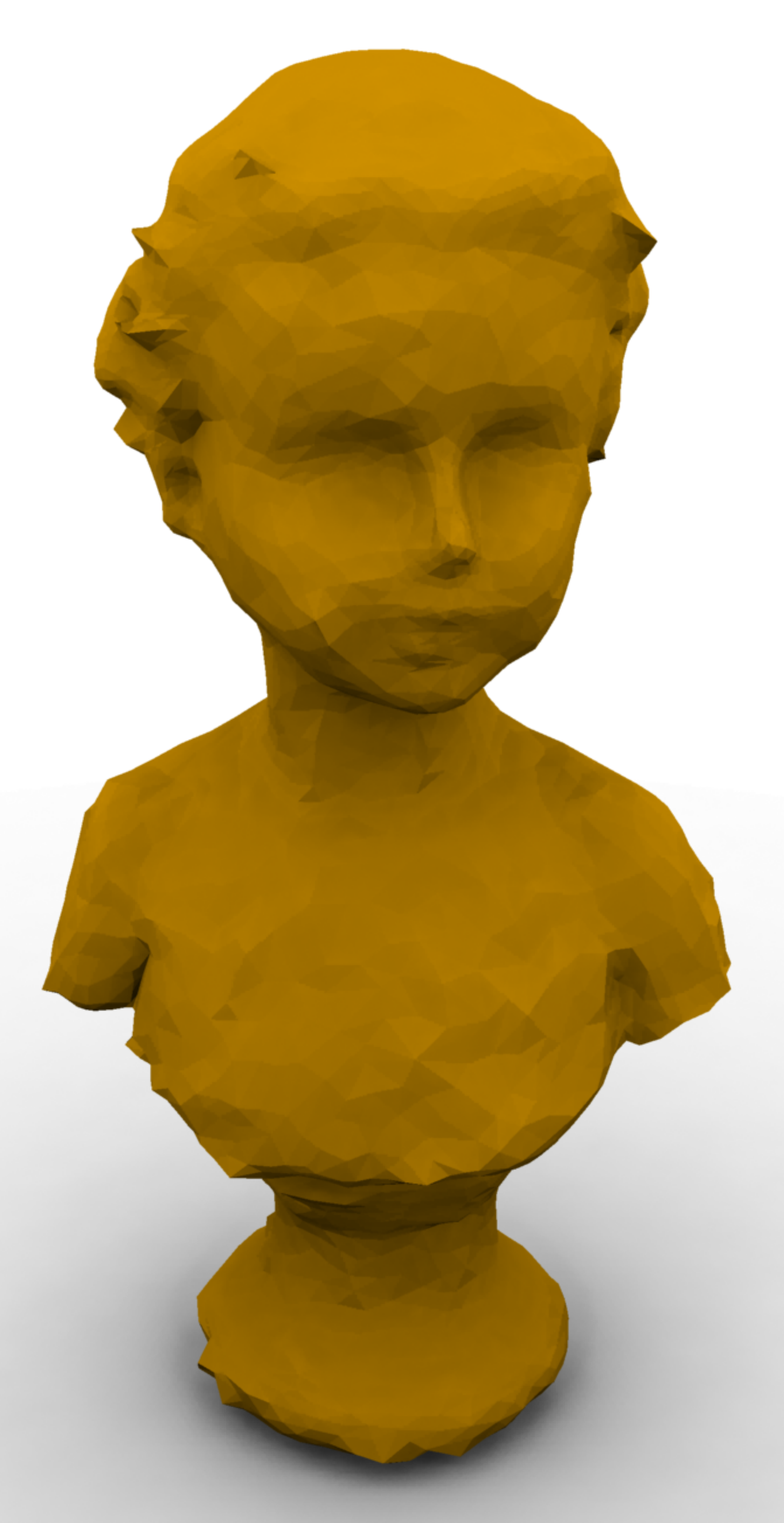} &
\includegraphics[height=.8in]{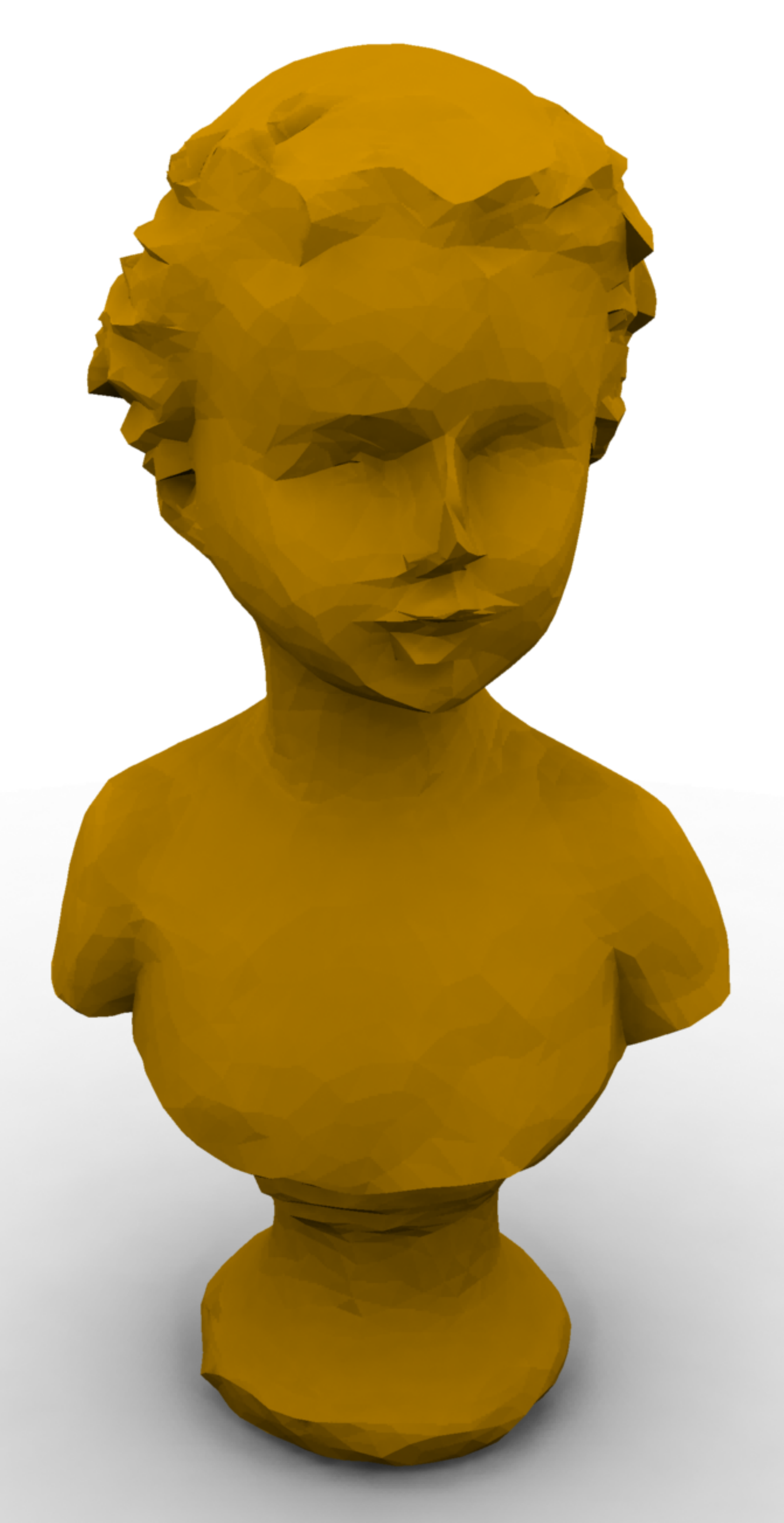}\\
(b) 0.114 &(c) 0.109 &(d) 0.104 &(e)&(f) 0.105&(g) 0.110
\end{tabular}
}

\def\circularBox{
\begin{tabular}{c@{}c}
\includegraphics[height=.8in]{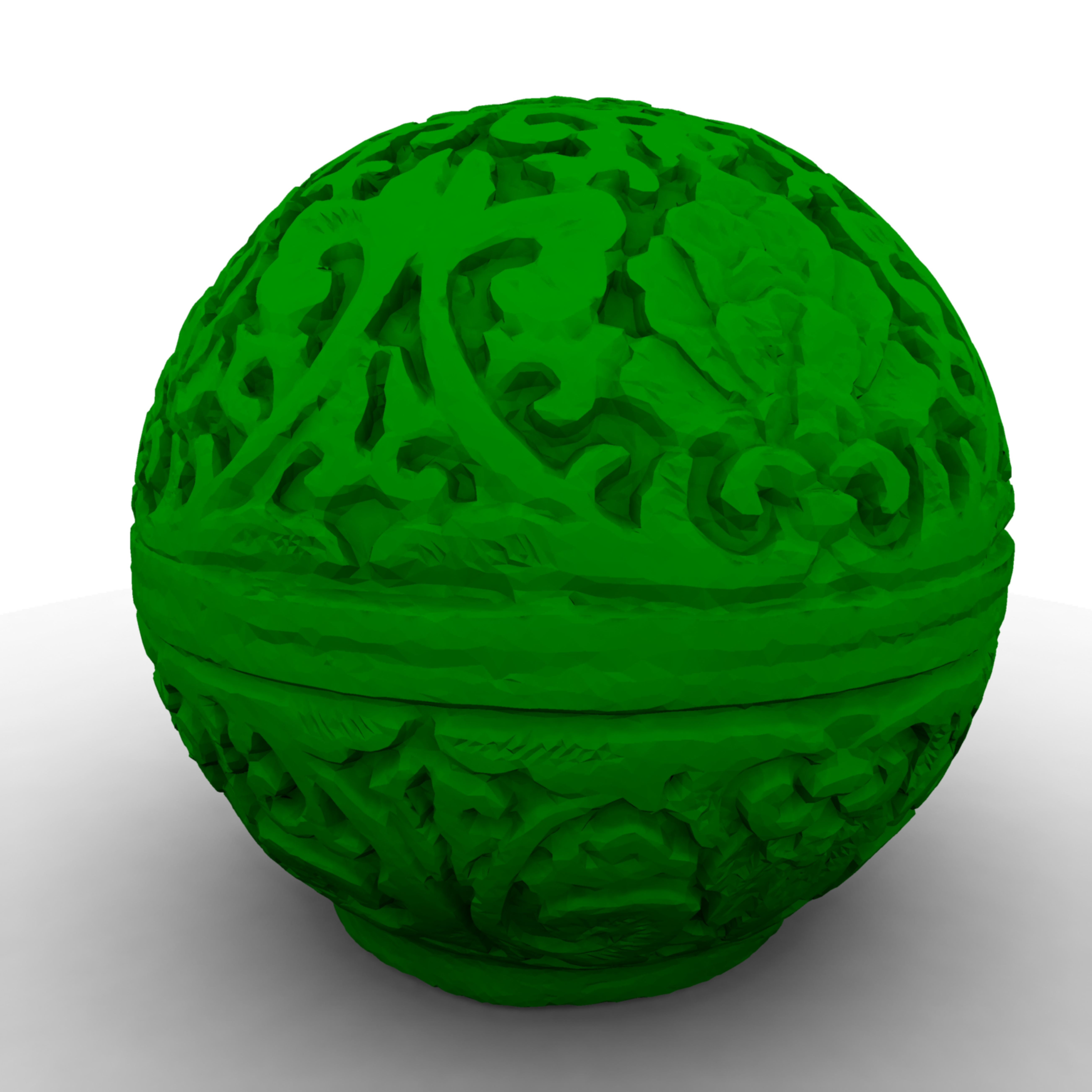} &
\includegraphics[height=.8in]{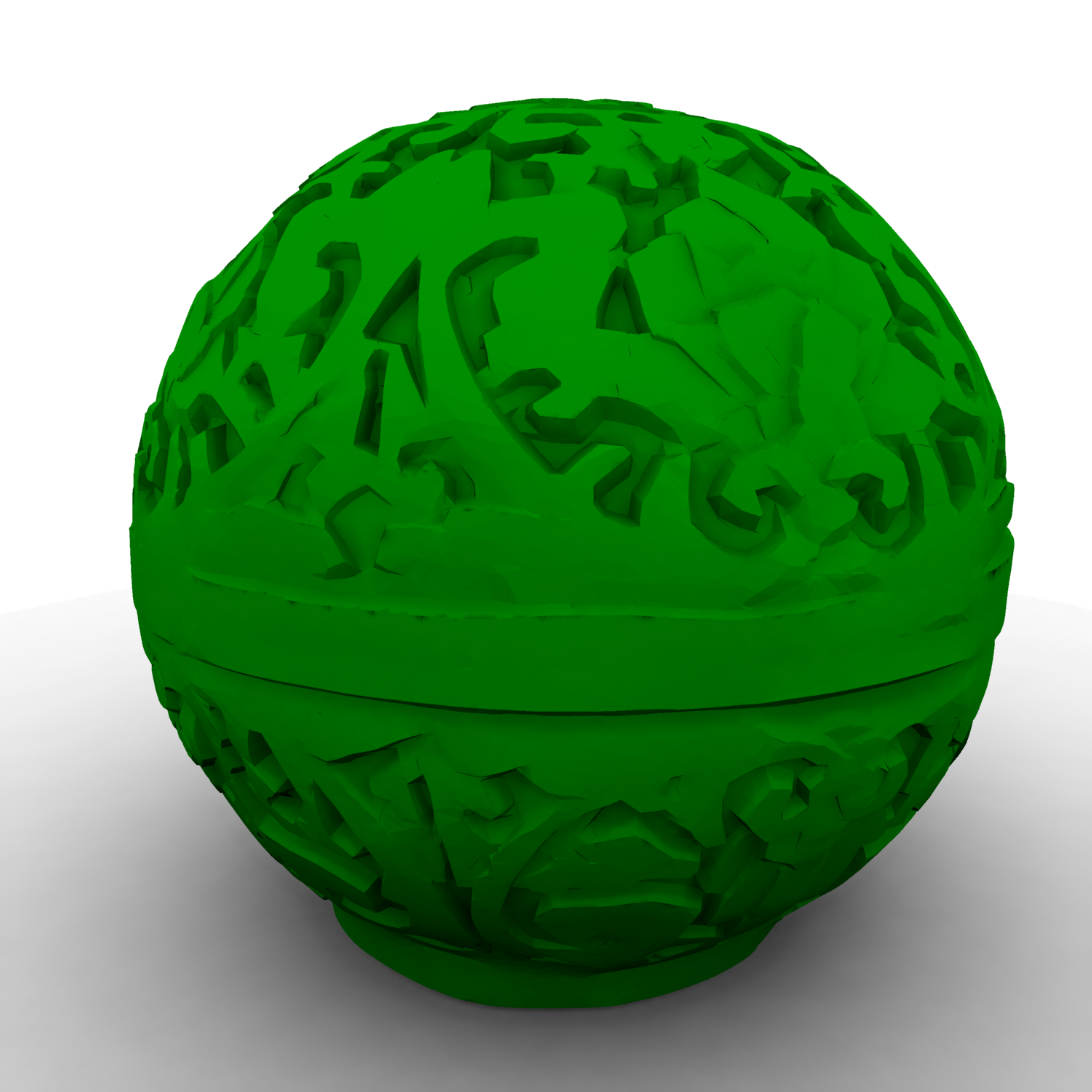} \\
(a)&(d)
\end{tabular}
}

\def\doubleTorus{
\begin{tabular}{c@{}c@{}c@{}c@{}c@{}c@{}c}
\includegraphics[height=.8in]{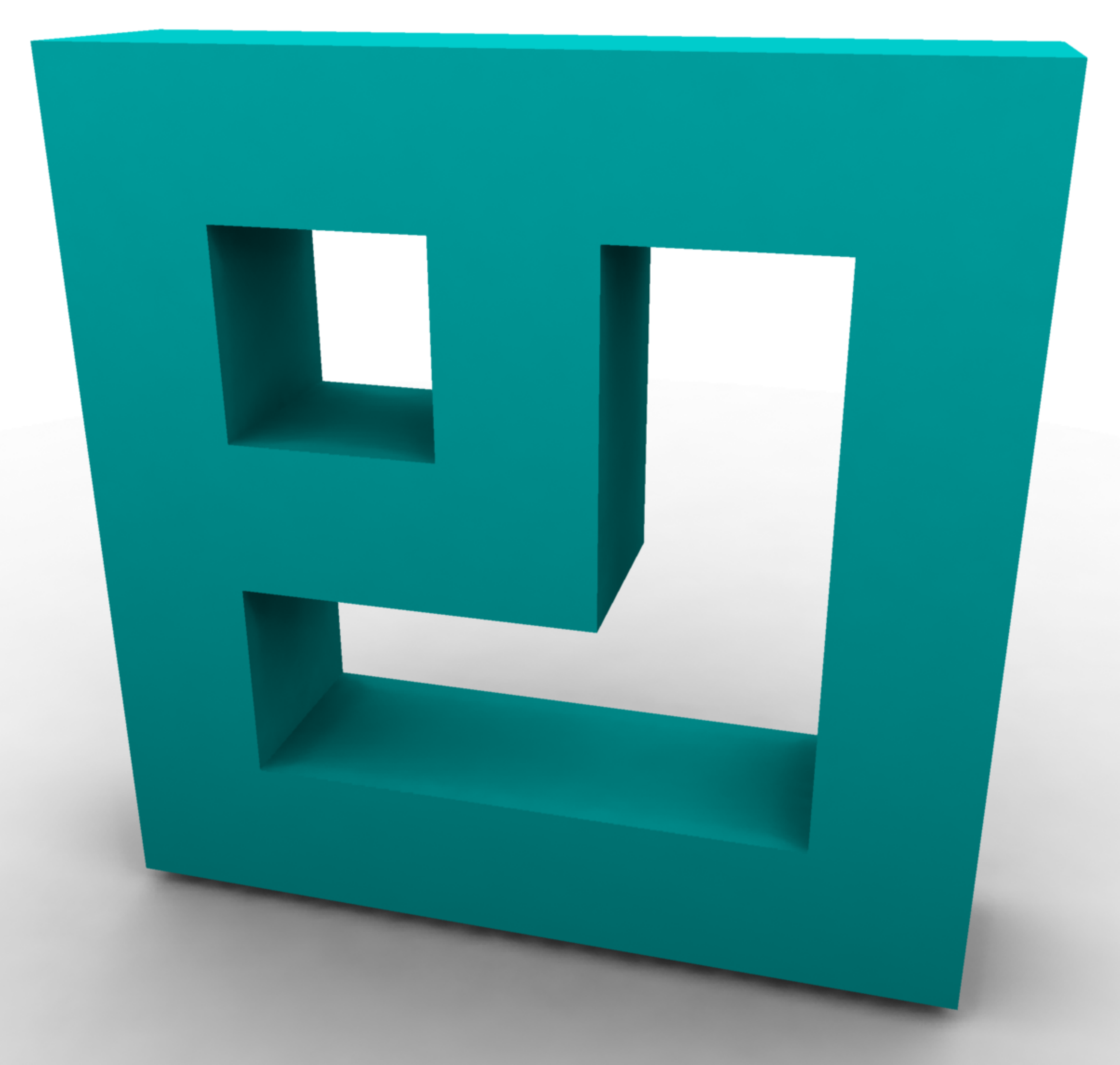} &
\includegraphics[height=.8in]{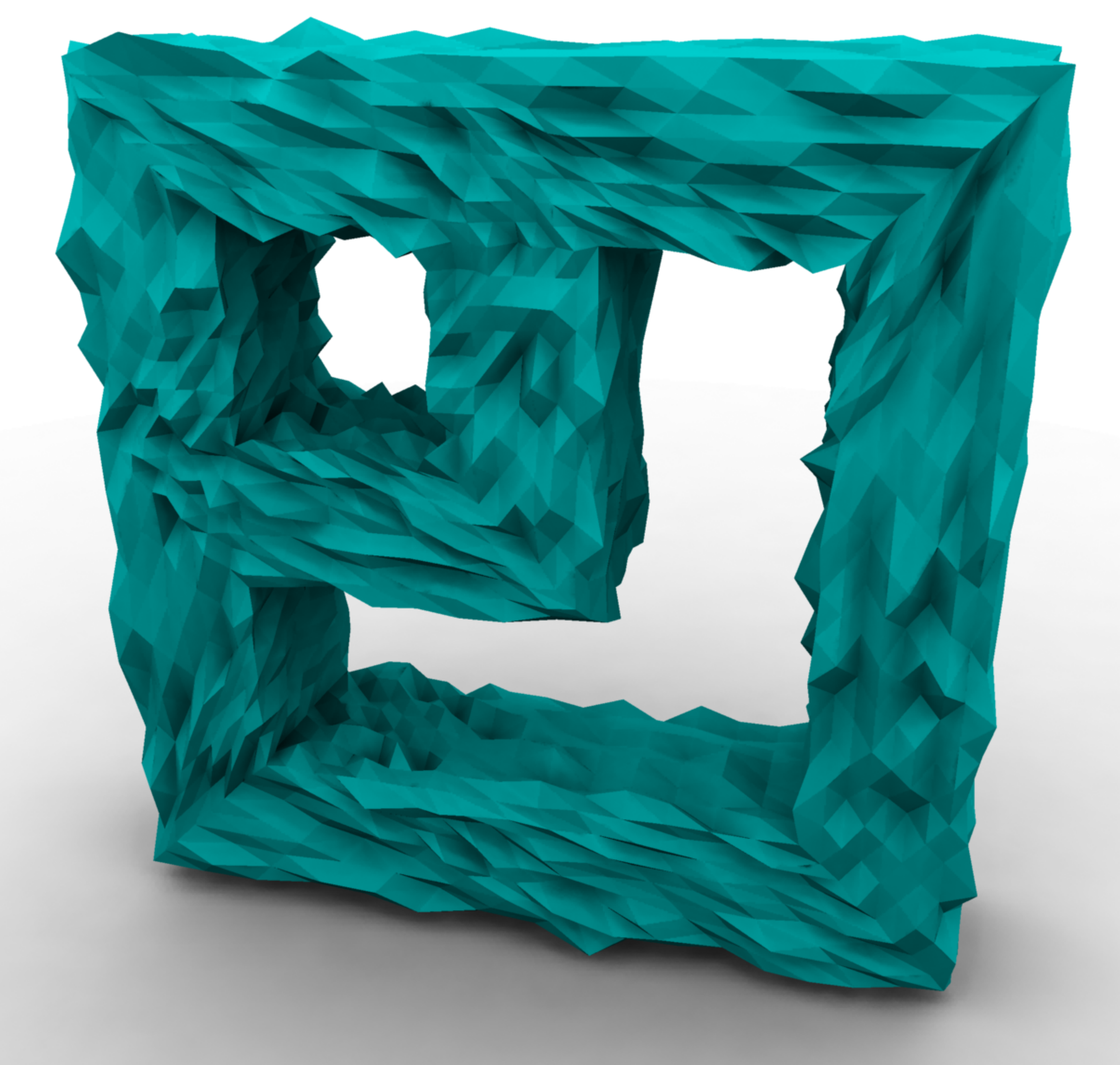} &
\includegraphics[height=.8in]{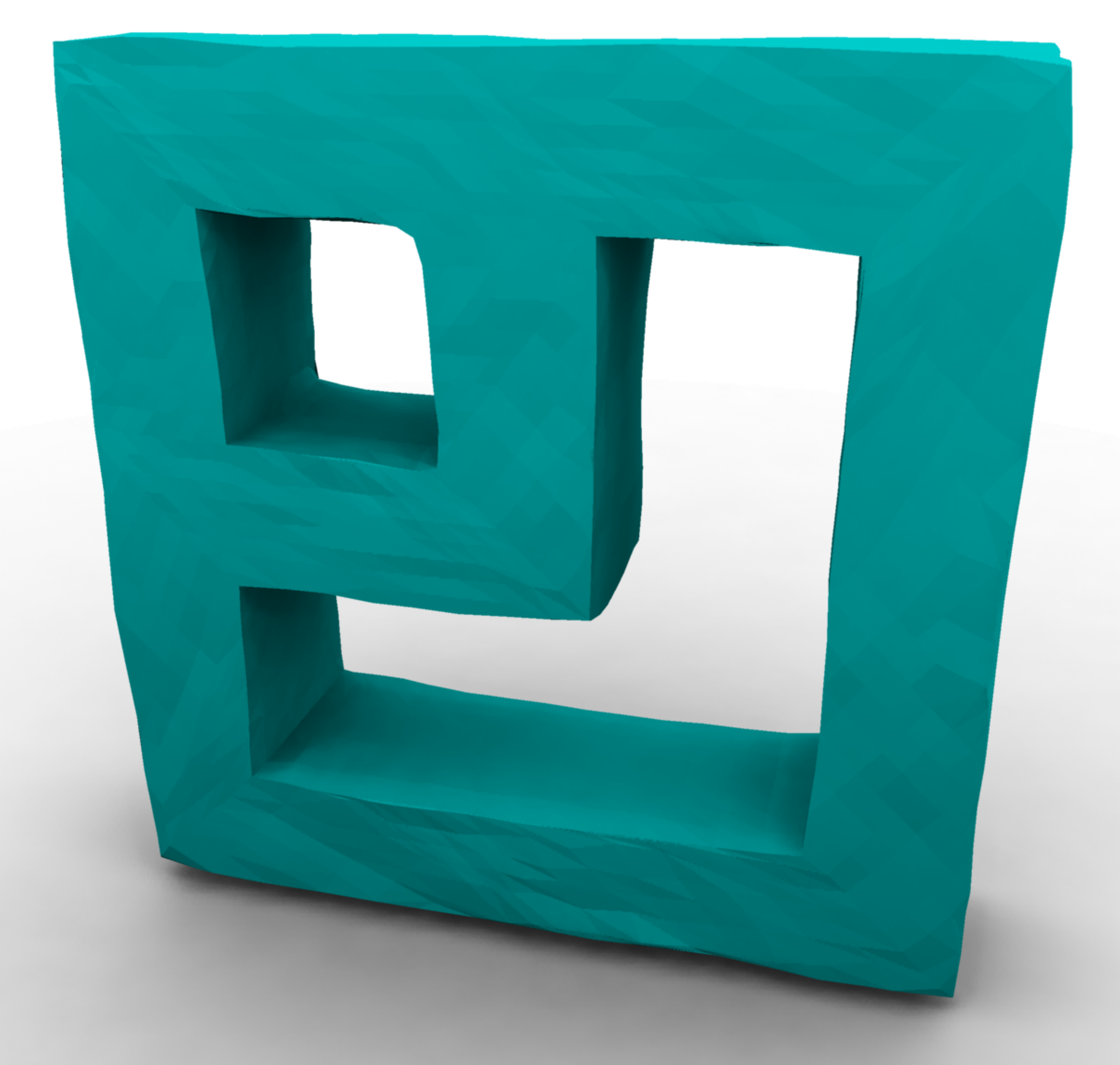} &
\includegraphics[height=.8in]{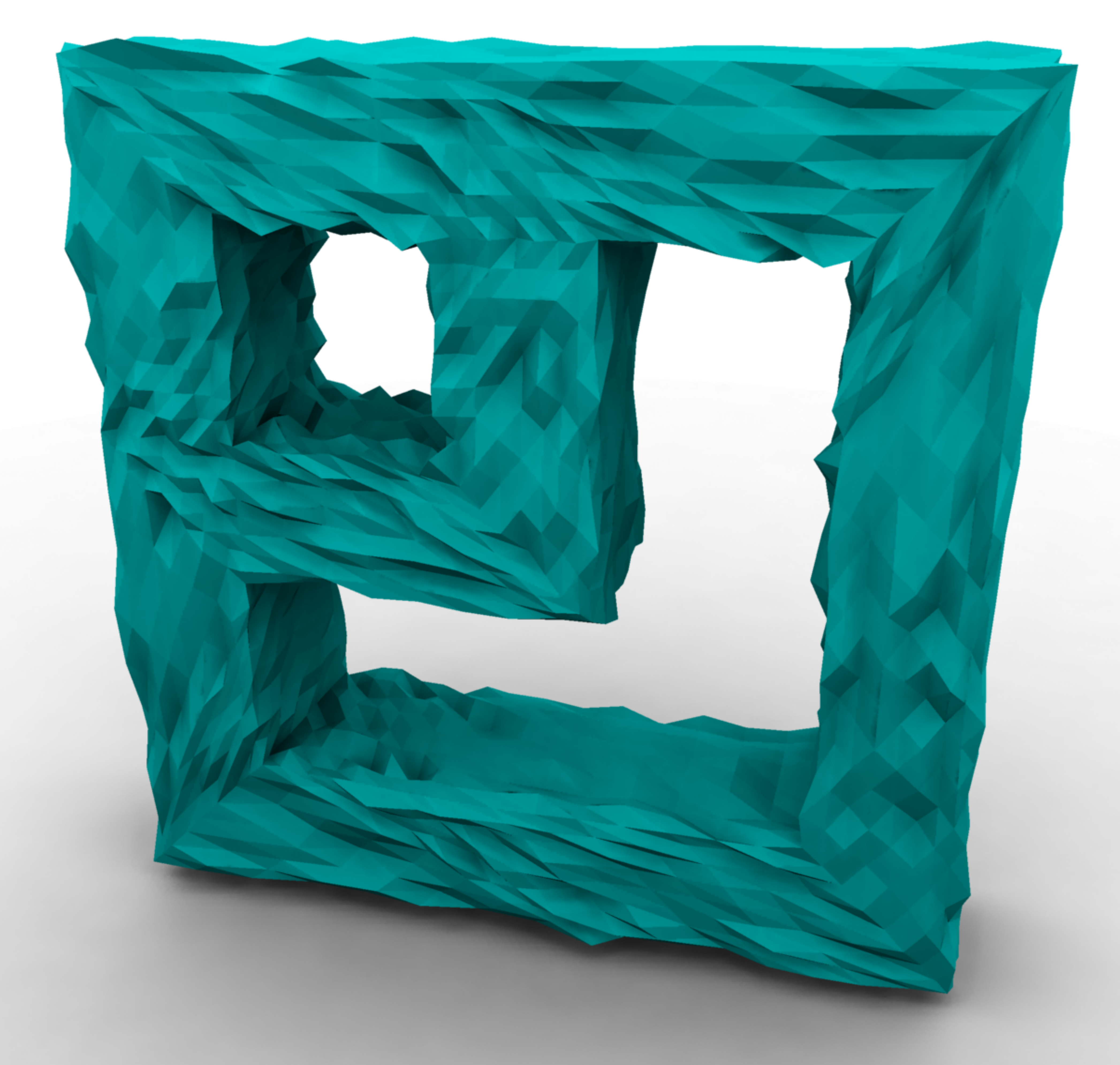} &
\includegraphics[height=.8in]{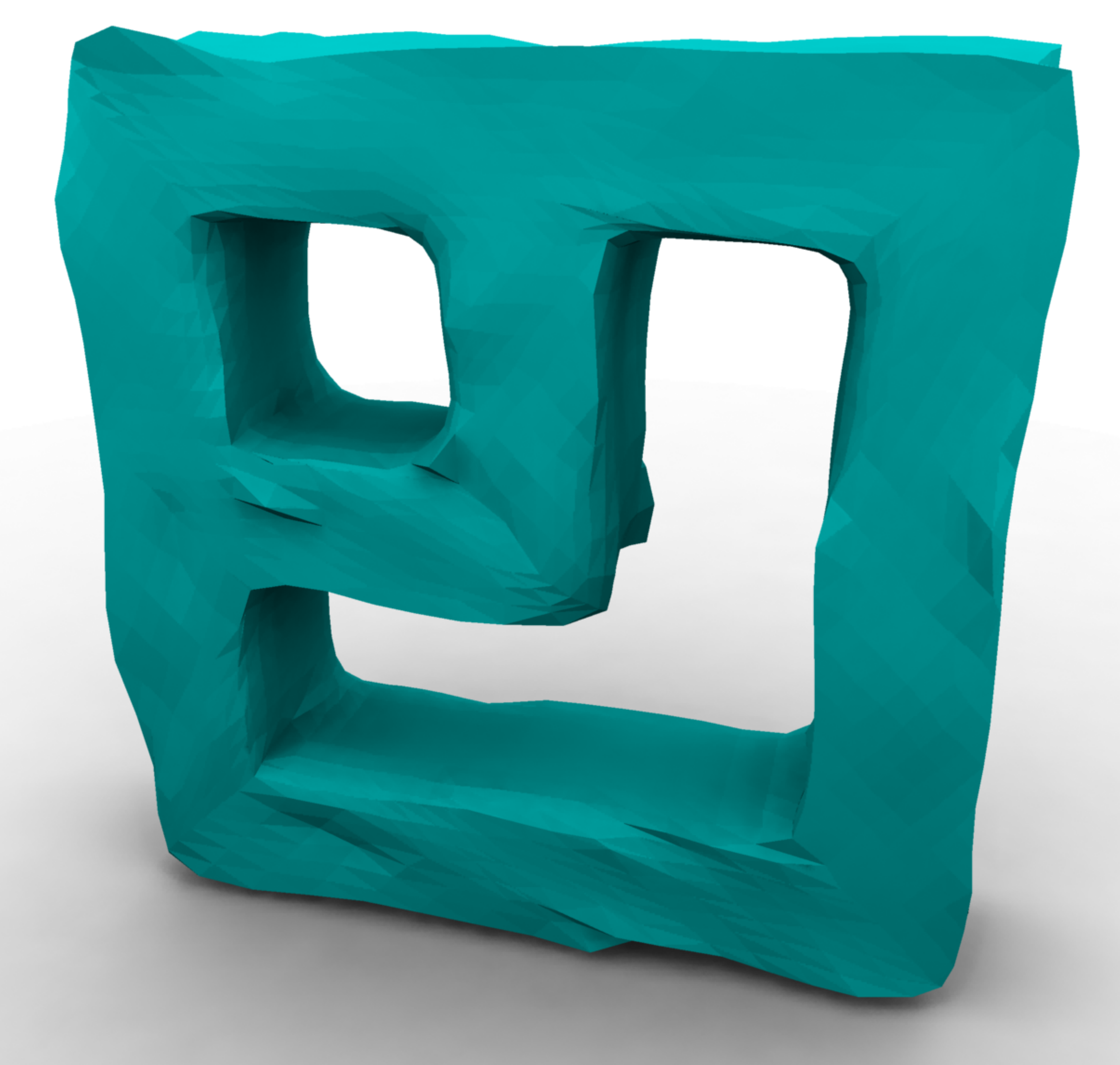} &
\includegraphics[height=.8in]{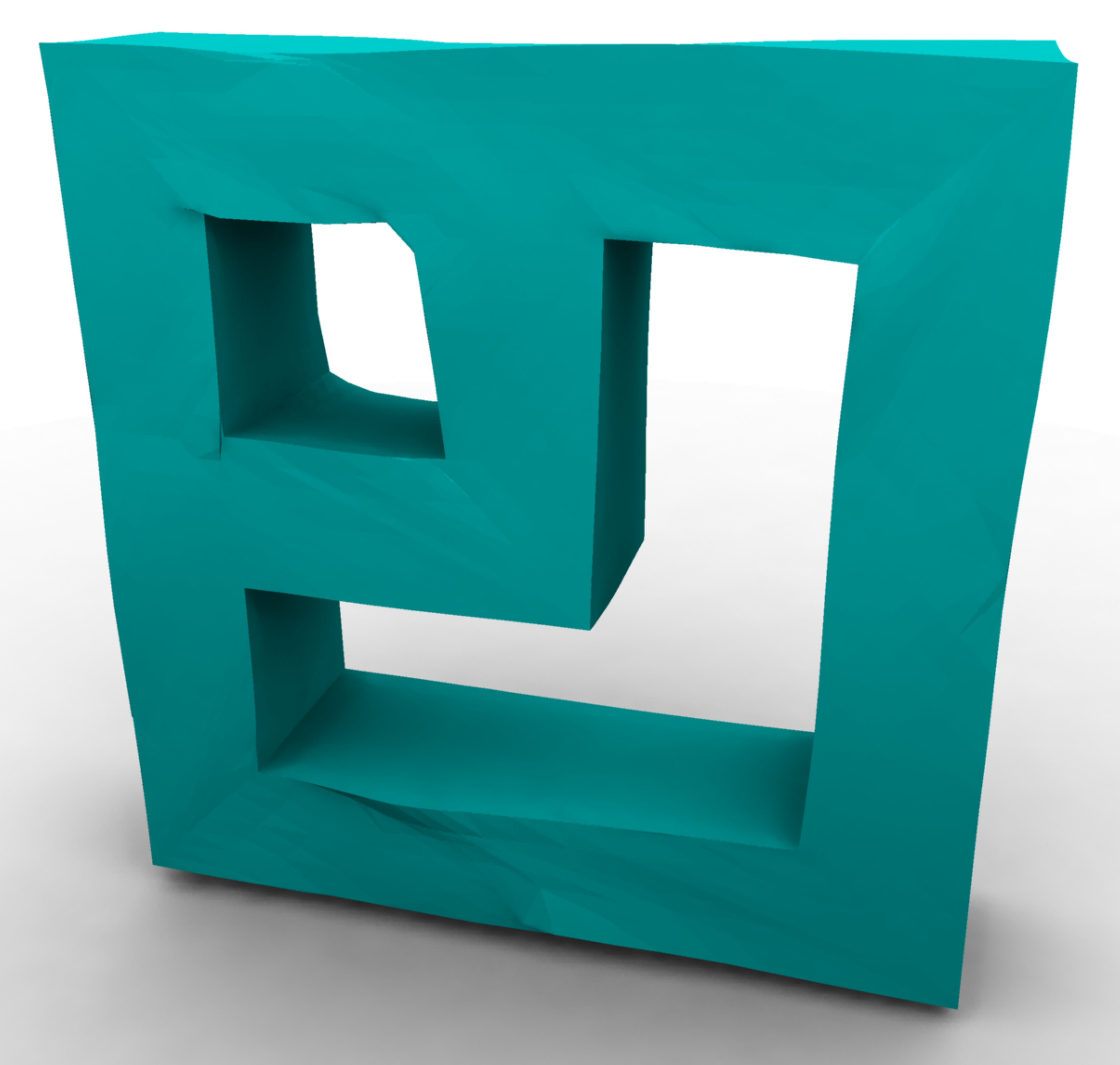}\\
(a)&(b) 0.111&(d) 0.074&(e)&(f) 0.094&(g) 0.087
\end{tabular}
}

\def\fandiskNoisier{
\begin{tabular}{c@{}c@{}c@{}c@{}c@{}c@{}c}
\includegraphics[height=.8in]{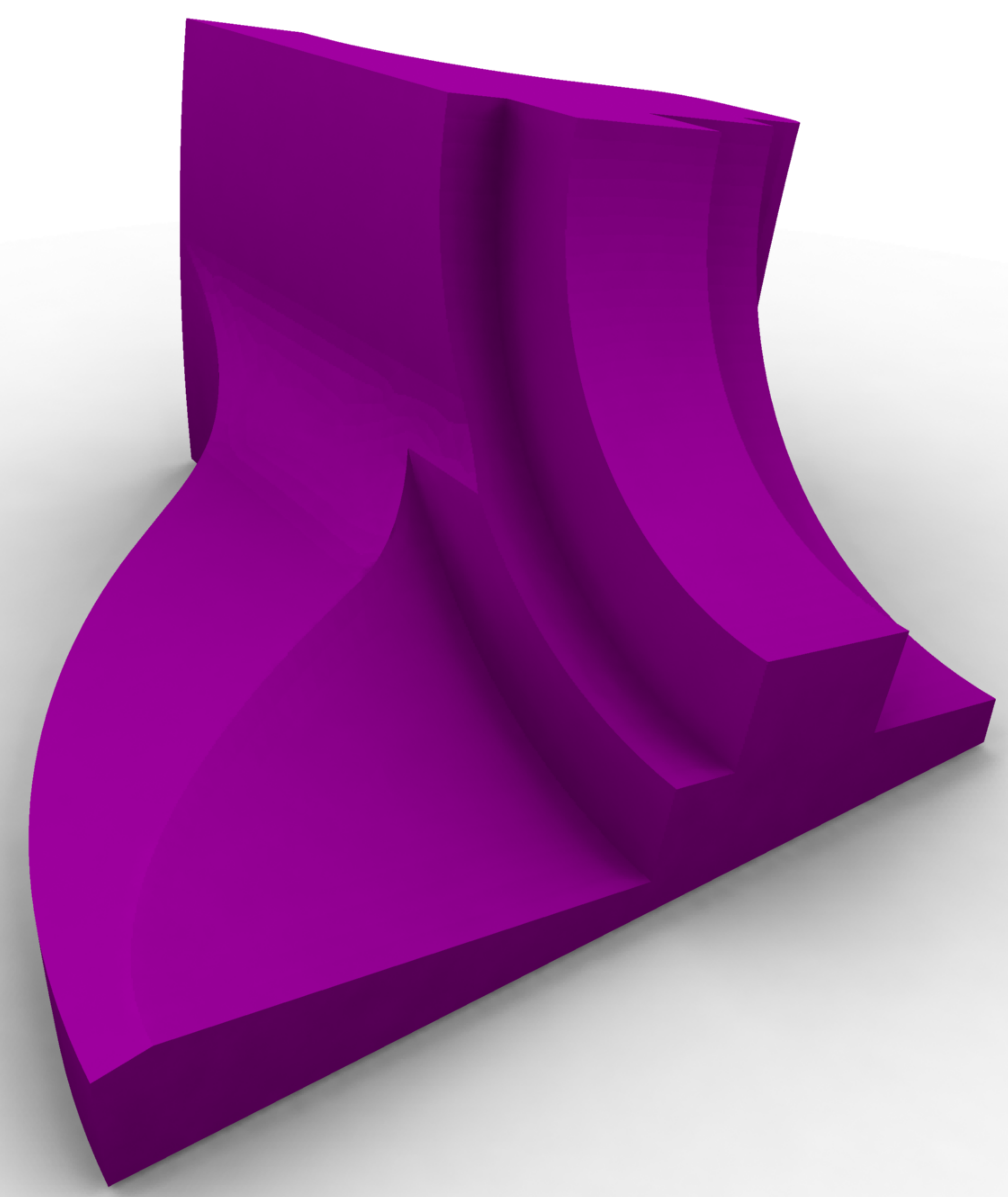} &
\includegraphics[height=.8in]{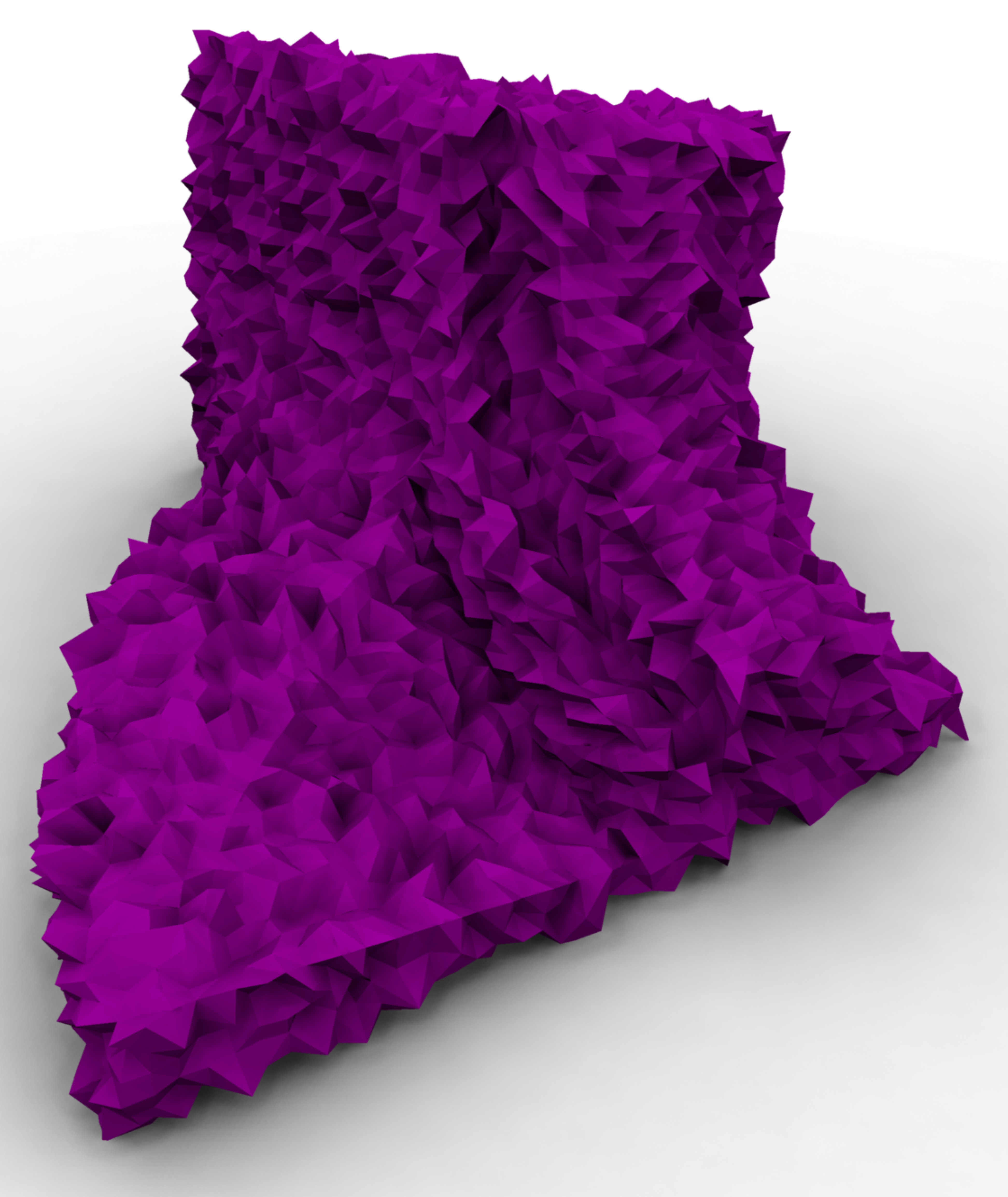} &
\includegraphics[height=.8in]{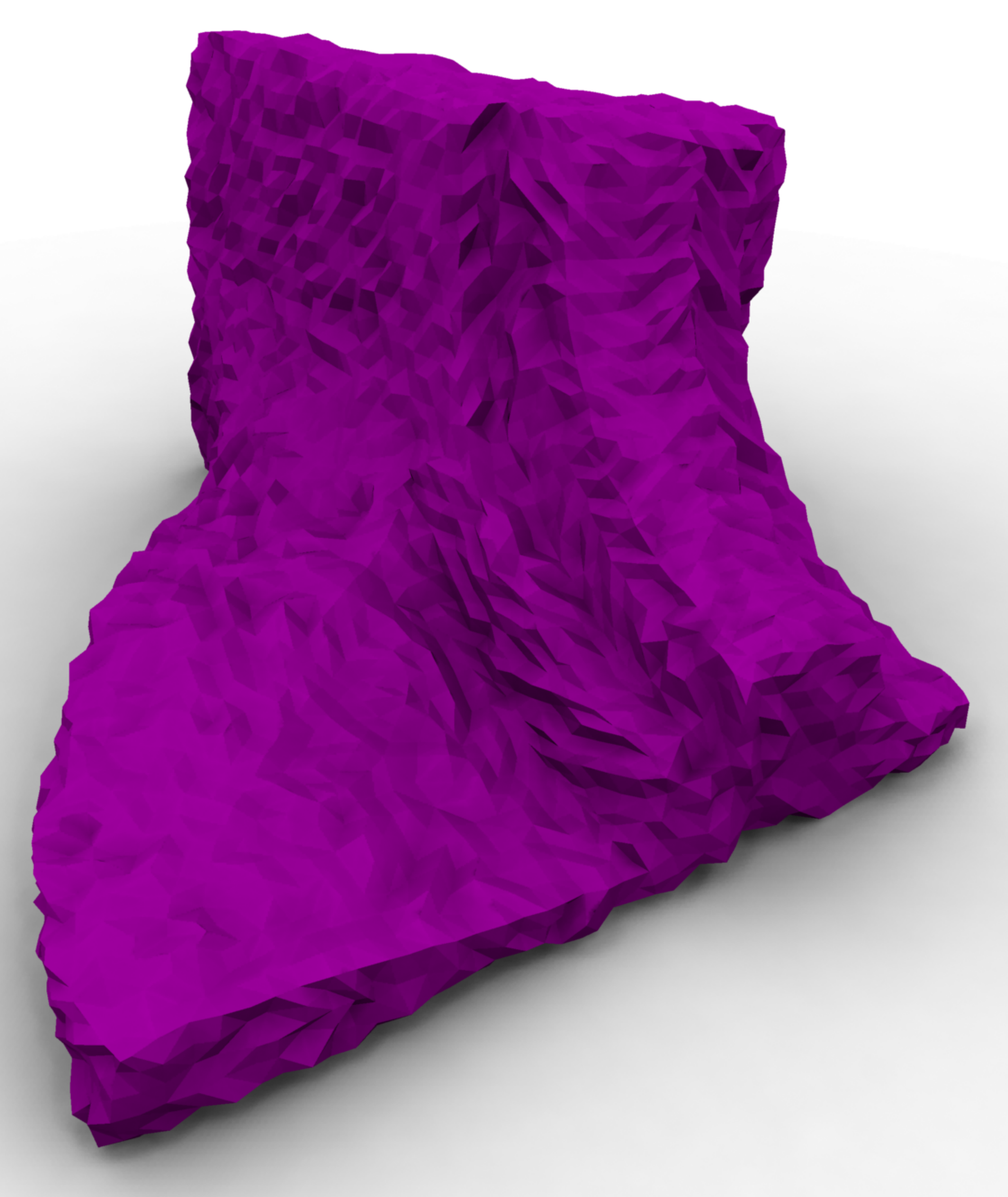} &
\includegraphics[height=.8in]{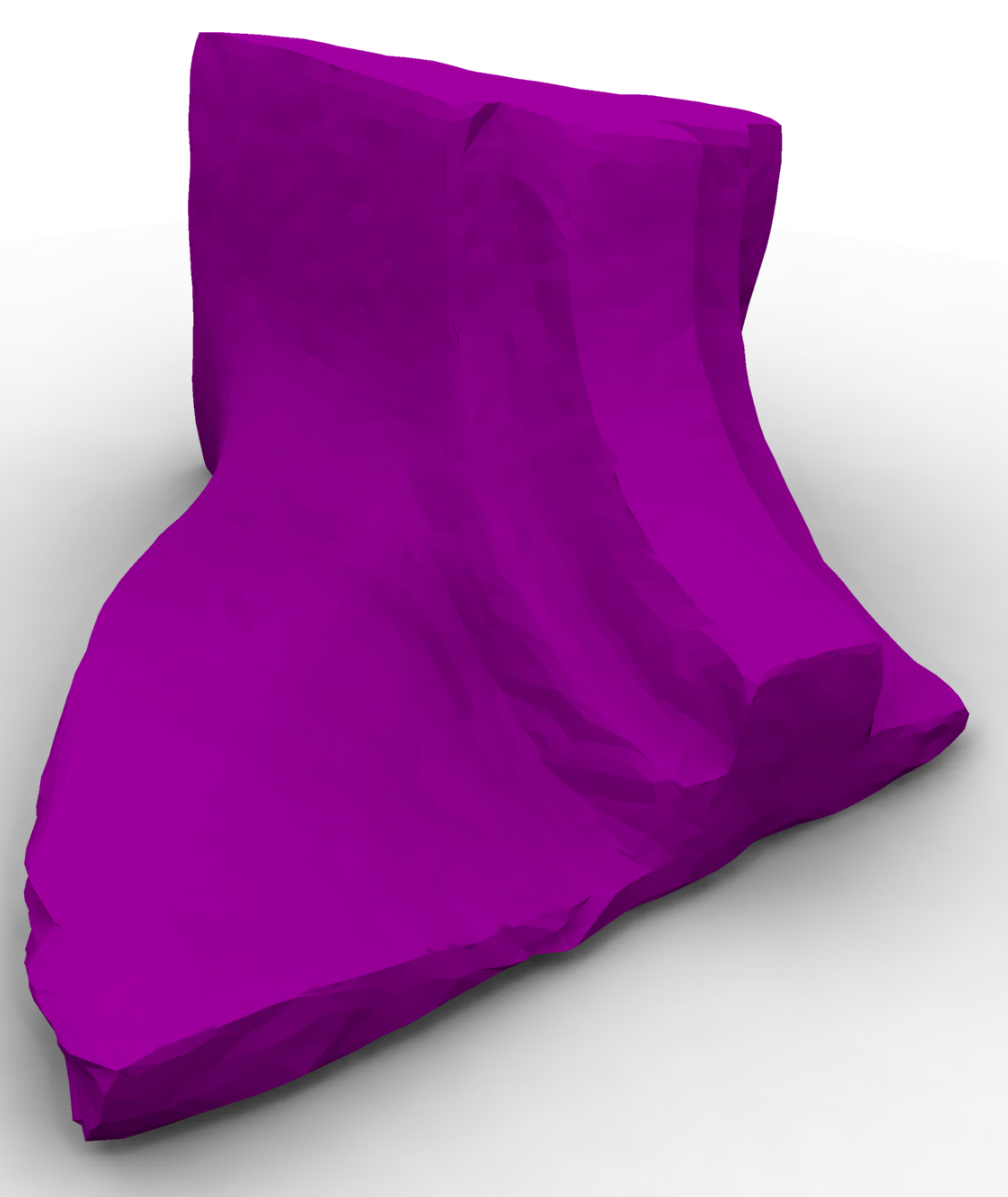} &
\includegraphics[height=.8in]{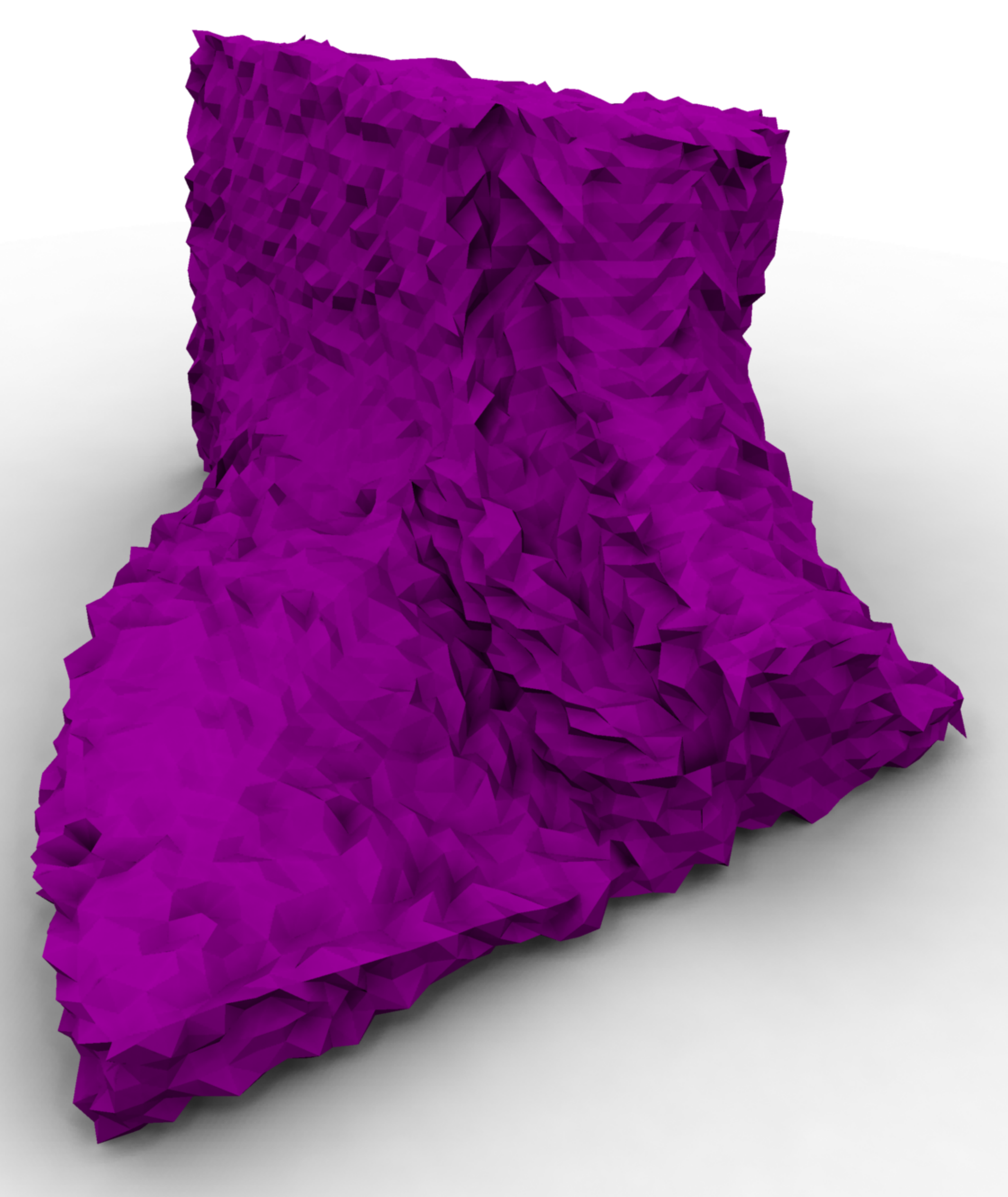} &
\includegraphics[height=.8in]{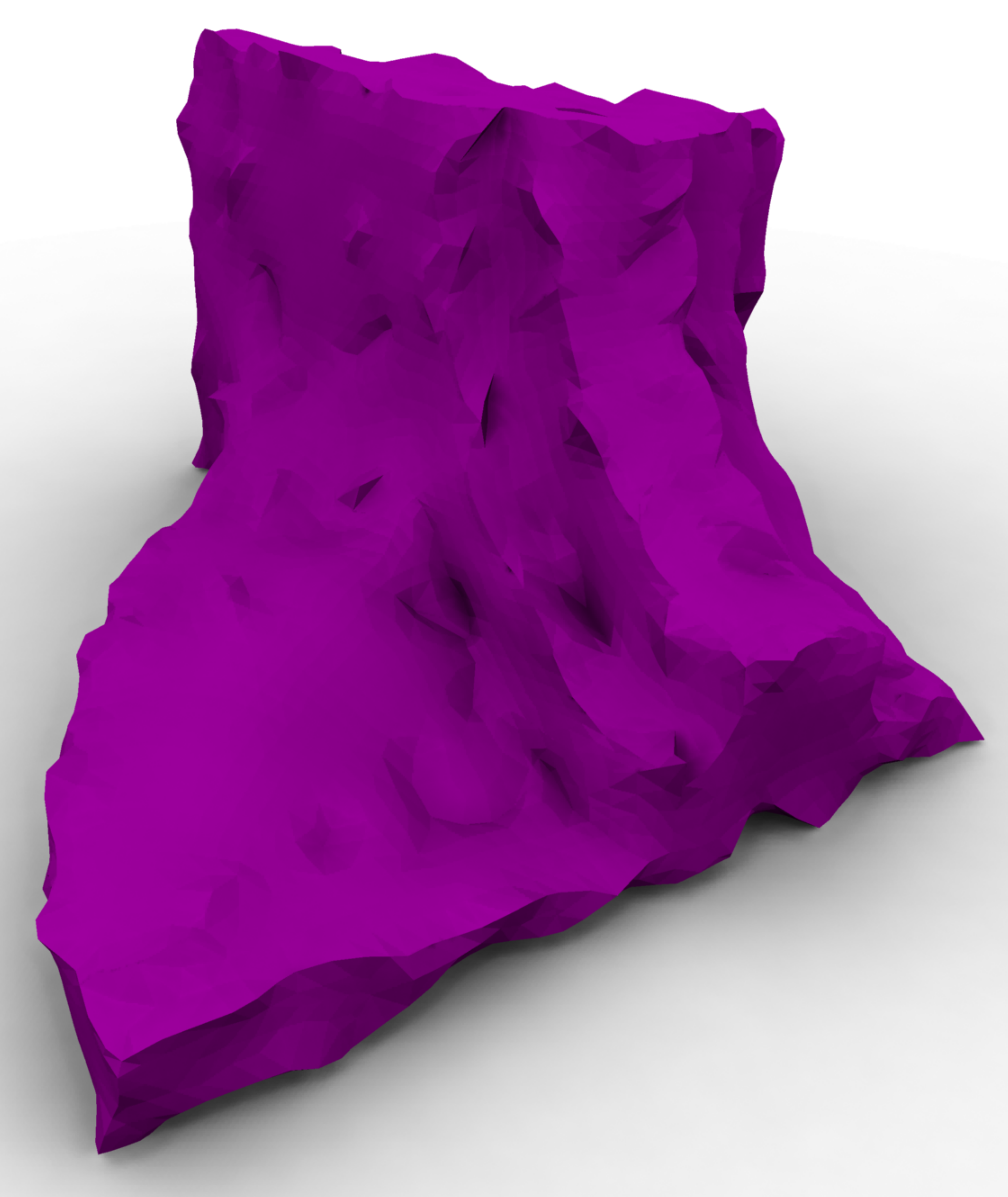} &
\includegraphics[height=.8in]{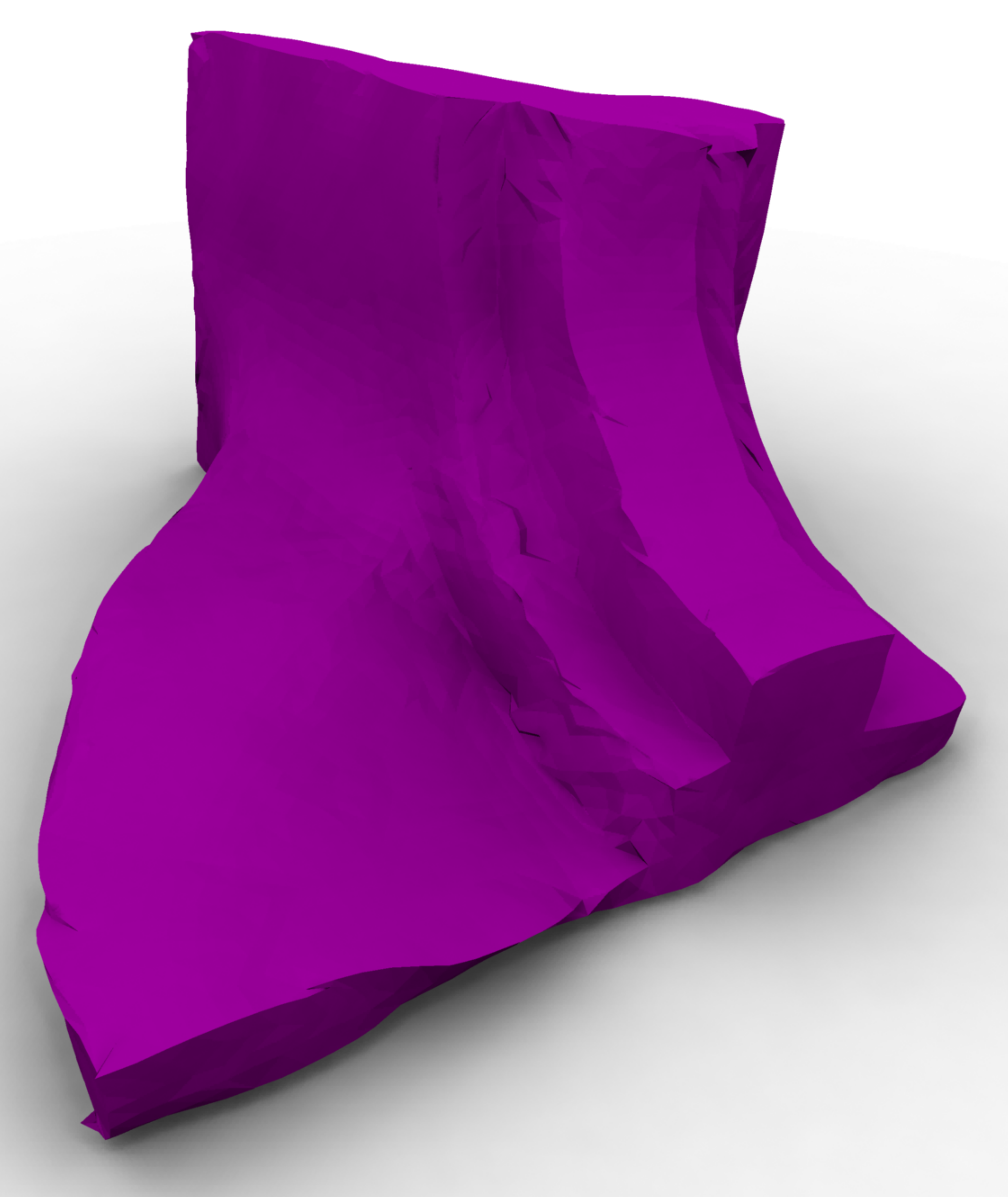}\\
(a)&(b) 0.243&(c) 0.156&(d) 0.109&(e) &(f) 0.171&(g) 0.120
\end{tabular}
}

\def\fandiskNoisy{
\begin{tabular}{c@{}c@{}c@{}c@{}c@{}c@{}c}
\includegraphics[height=.8in]{figures/mesh_smoothing/fandisk/orig_cropped.png.pdf} &
\includegraphics[height=.8in]{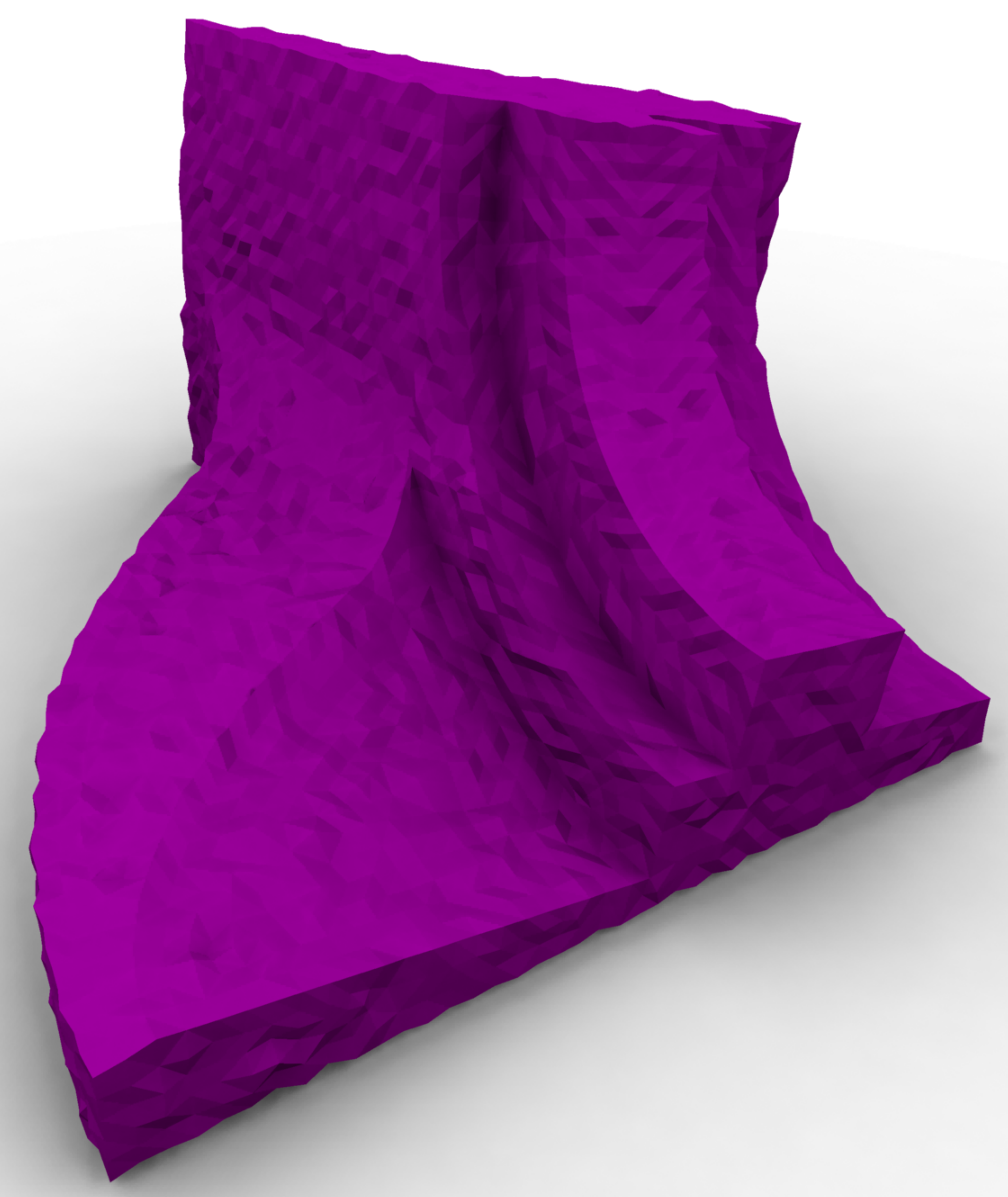} &
\includegraphics[height=.8in]{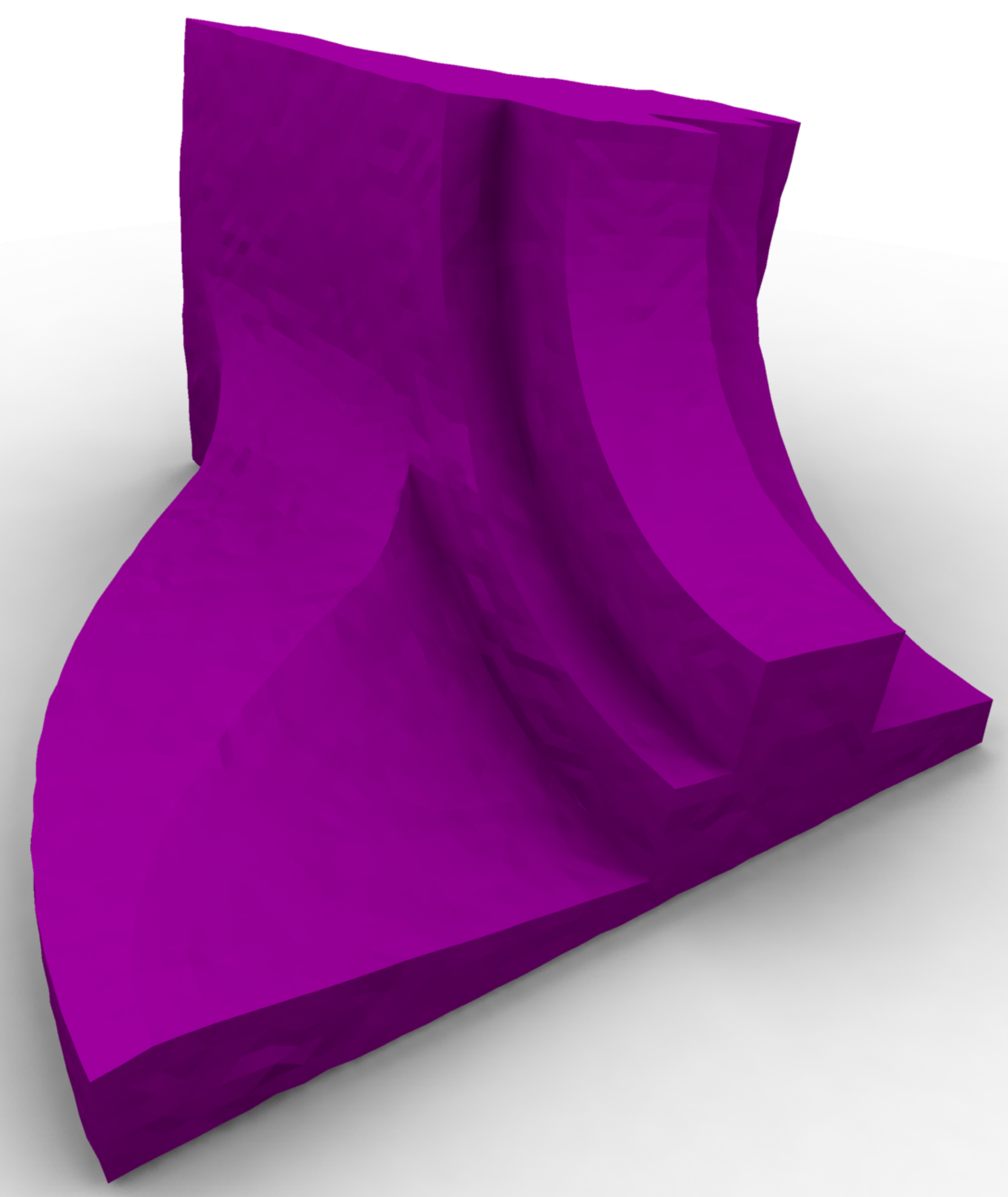} &
\includegraphics[height=.8in]{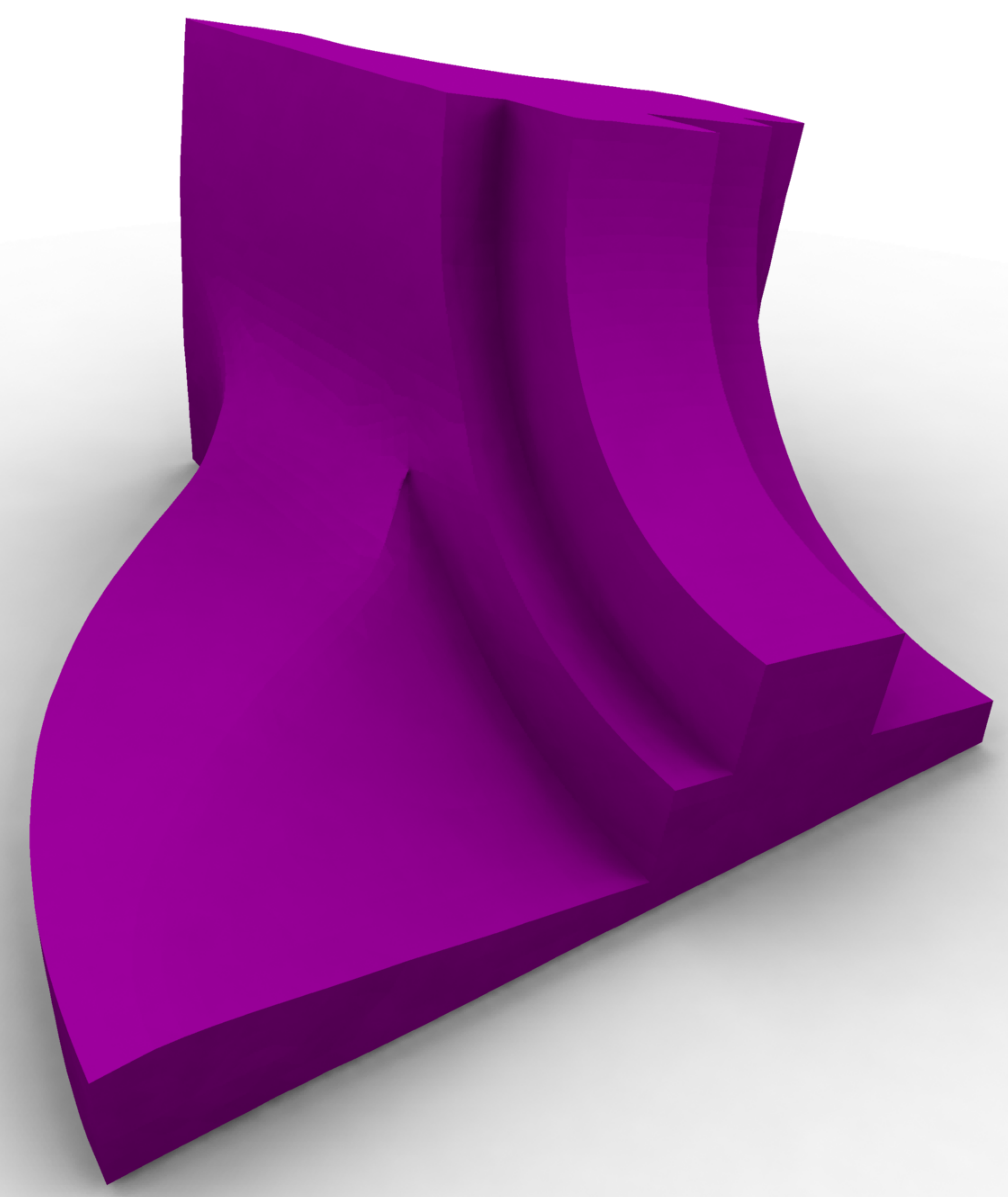} &
\includegraphics[height=.8in]{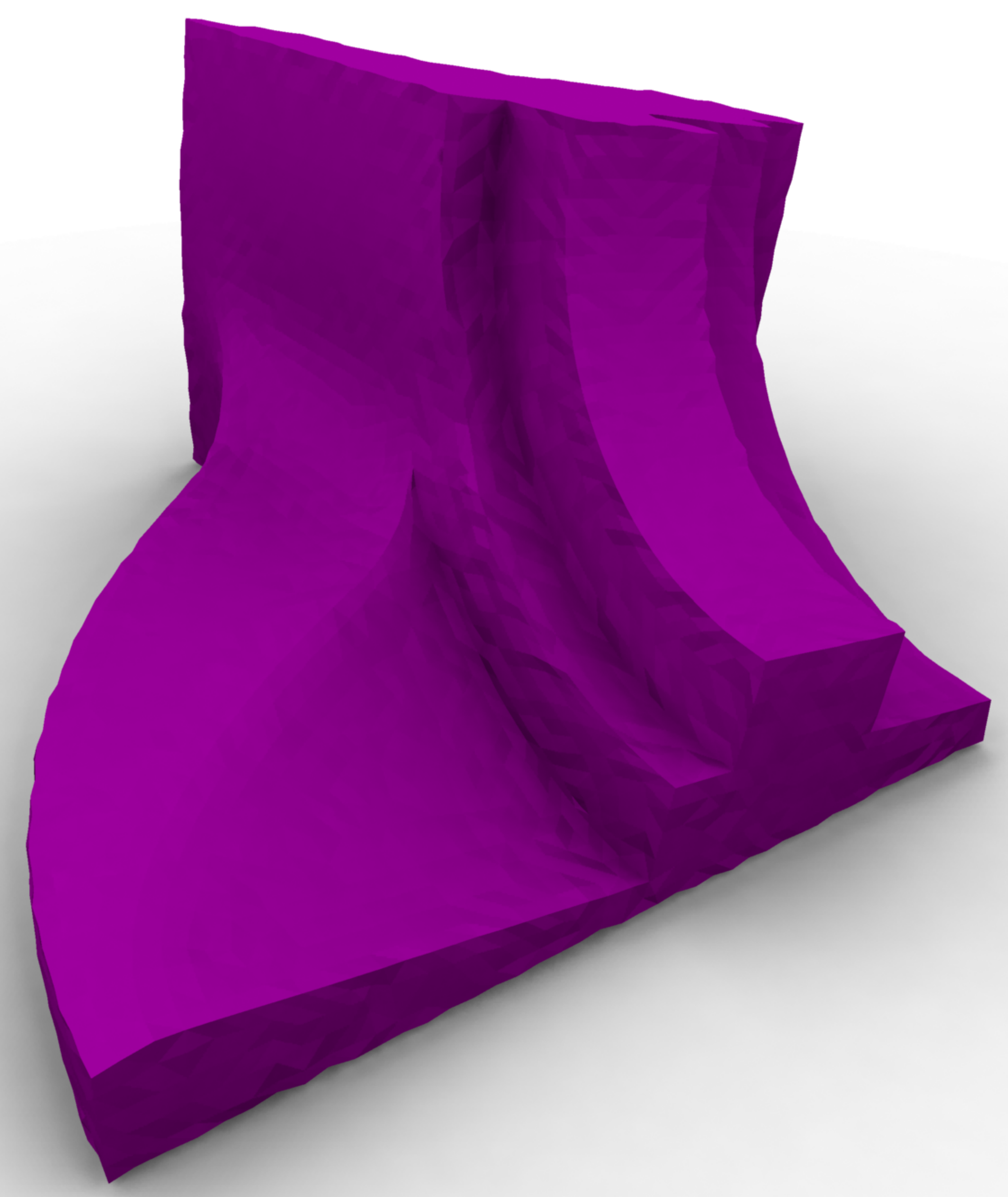} &
\includegraphics[height=.8in]{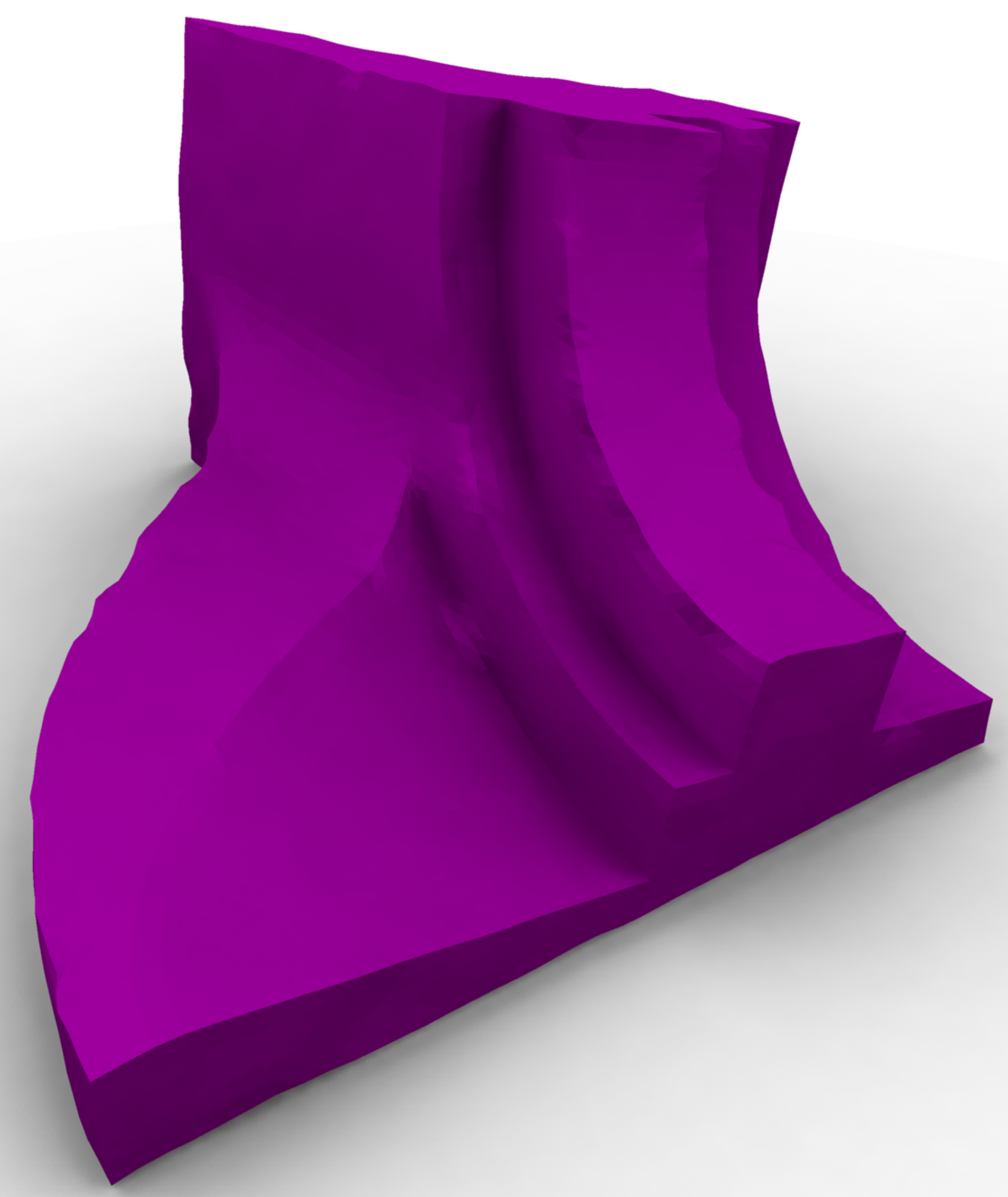} &
\includegraphics[height=.8in]{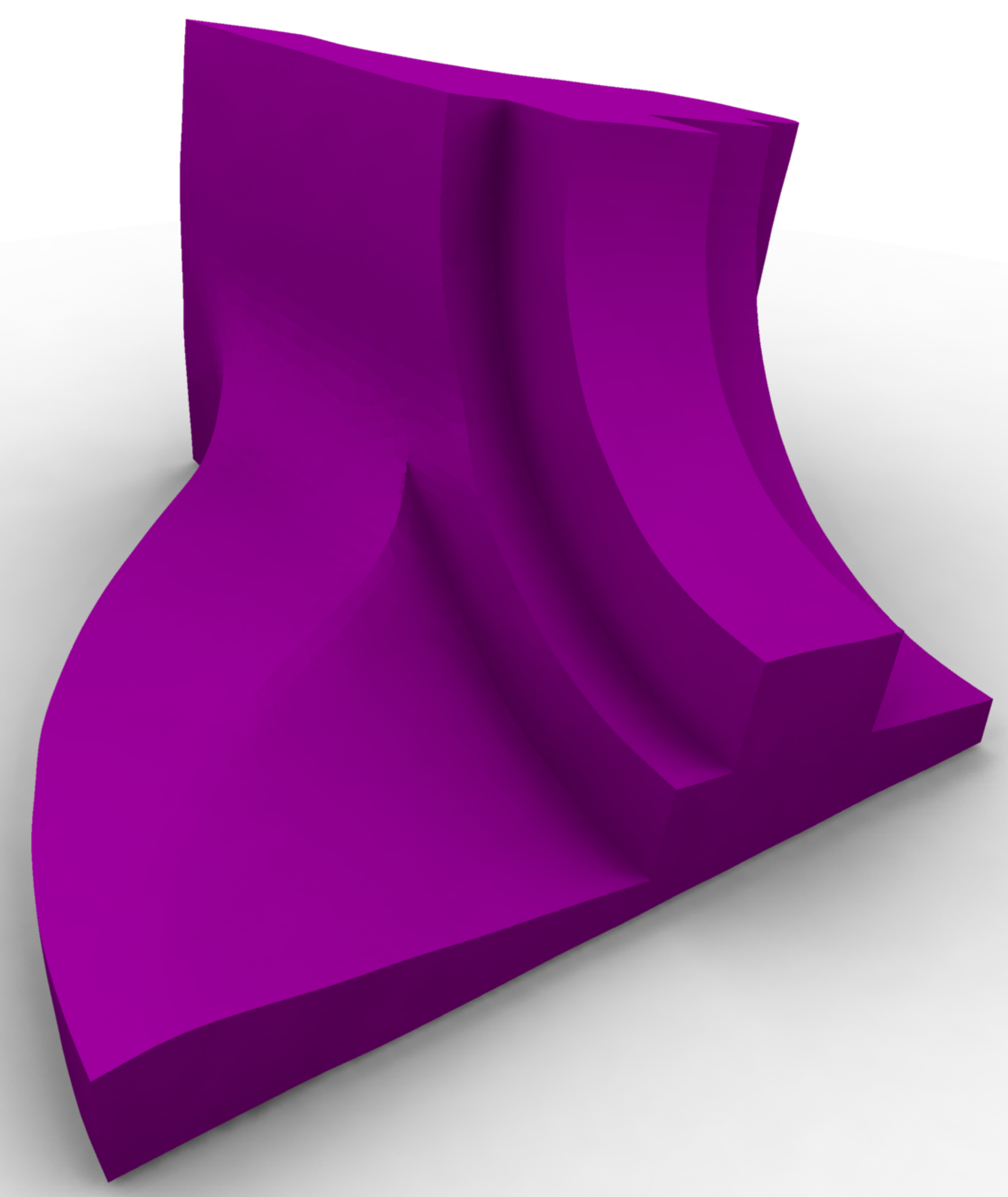}\\
(a)&(b) 0.022&(c) 0.019&(d) 0.013&(e)&(f) 0.021&(g) 0.012
\end{tabular}
}

\def\fountain{
\begin{tabular}{c@{}c@{}c@{}c@{}c@{}c@{}c}
\includegraphics[height=.8in]{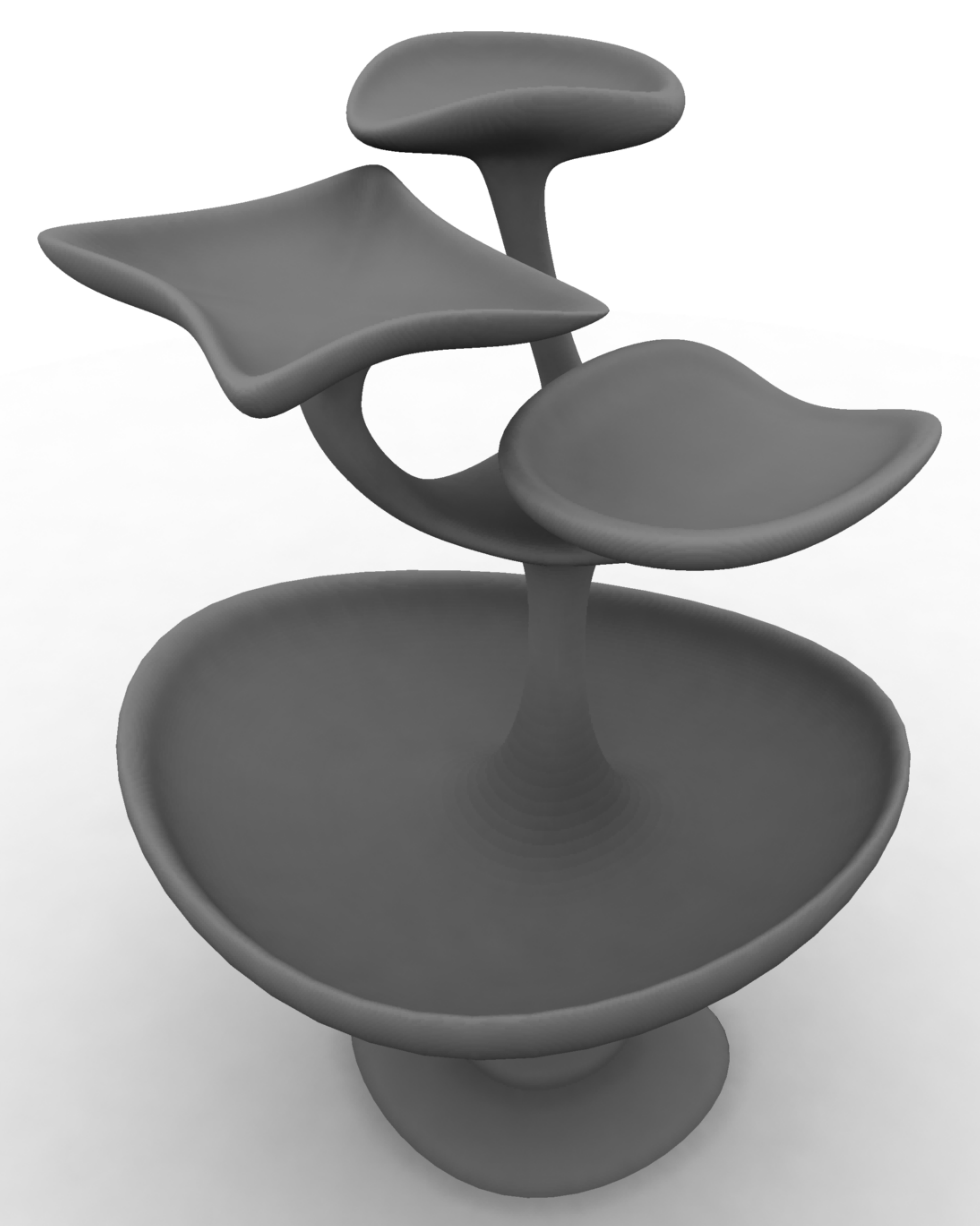} &
\includegraphics[height=.8in]{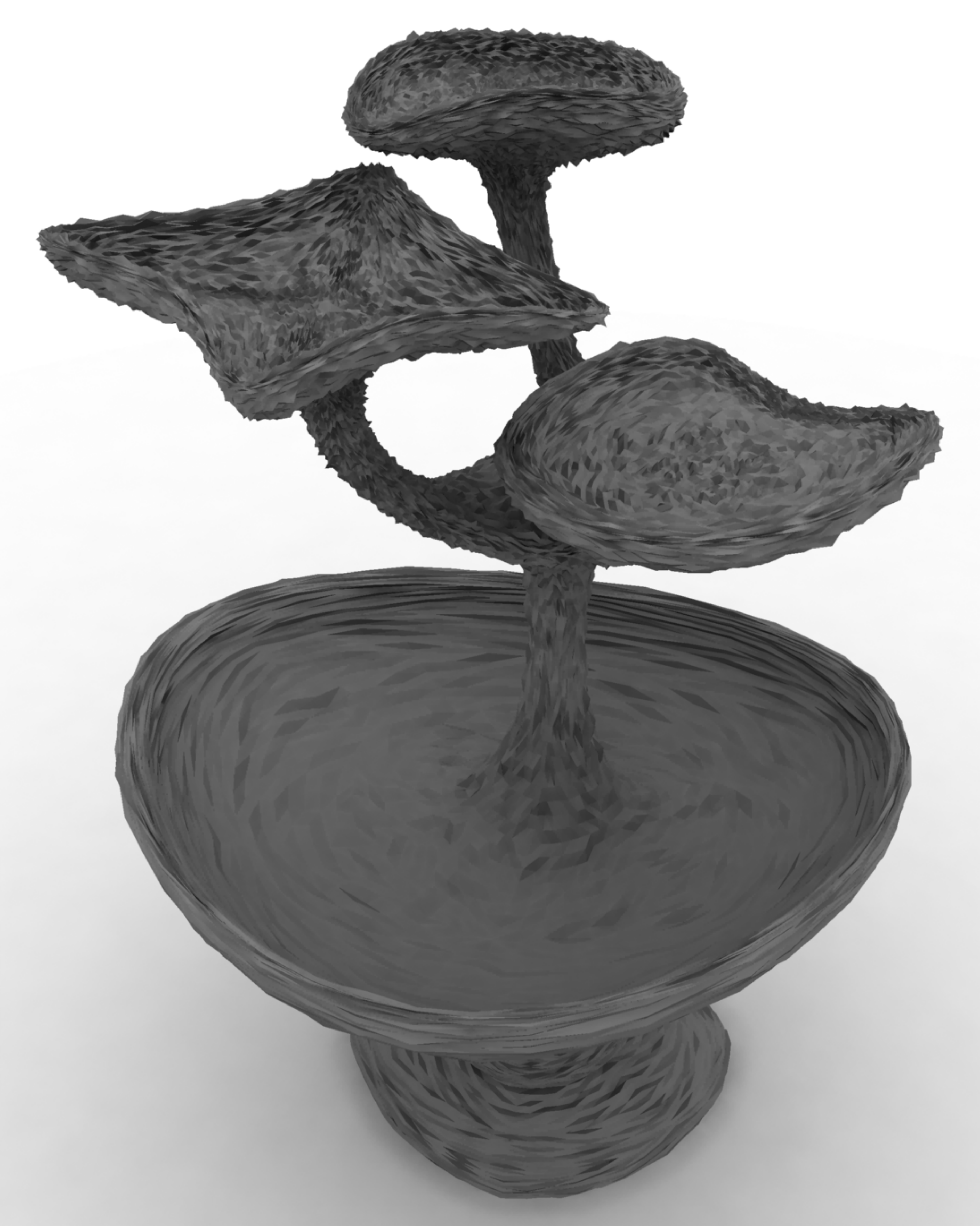} &
\includegraphics[height=.8in]{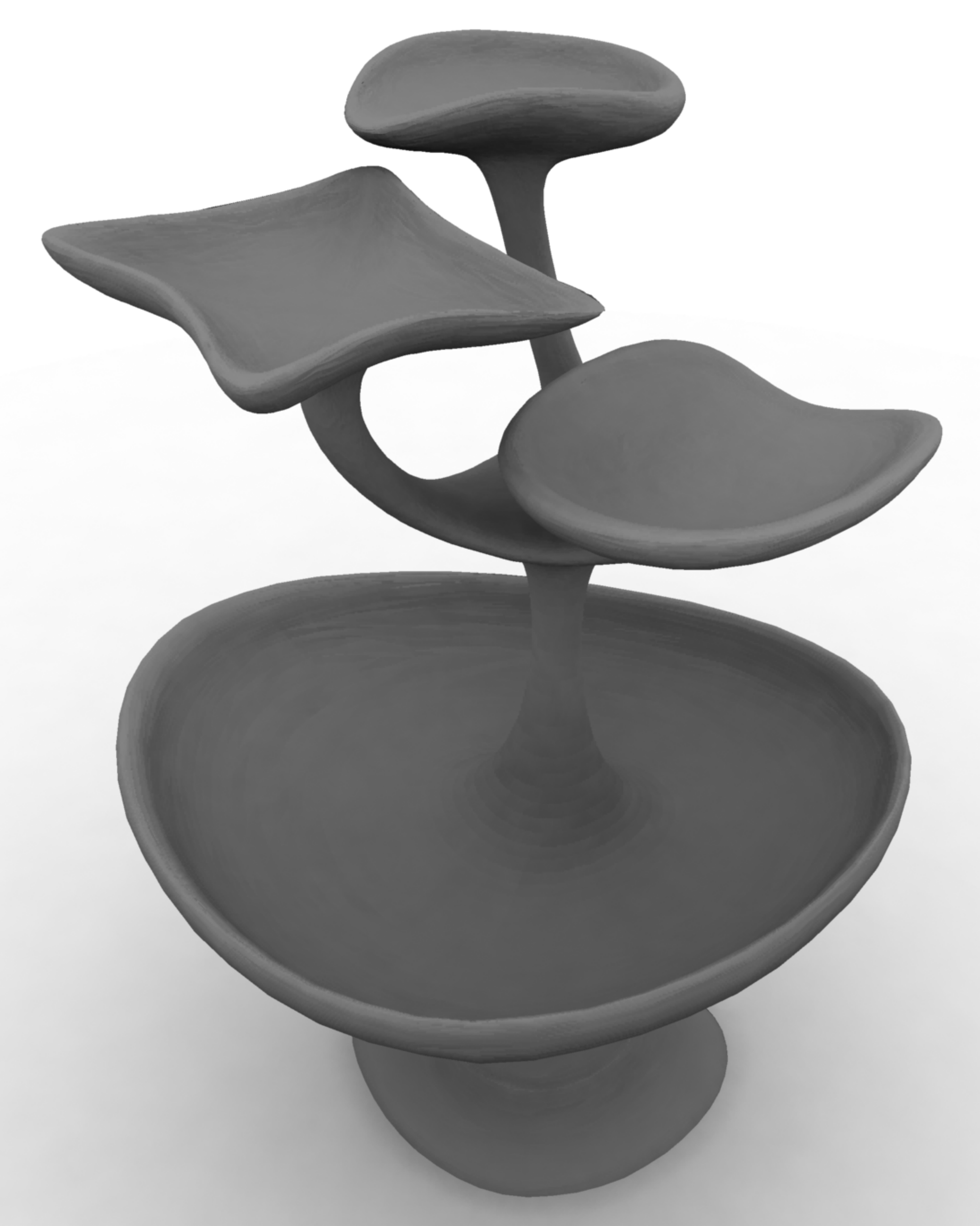} &
\includegraphics[height=.8in]{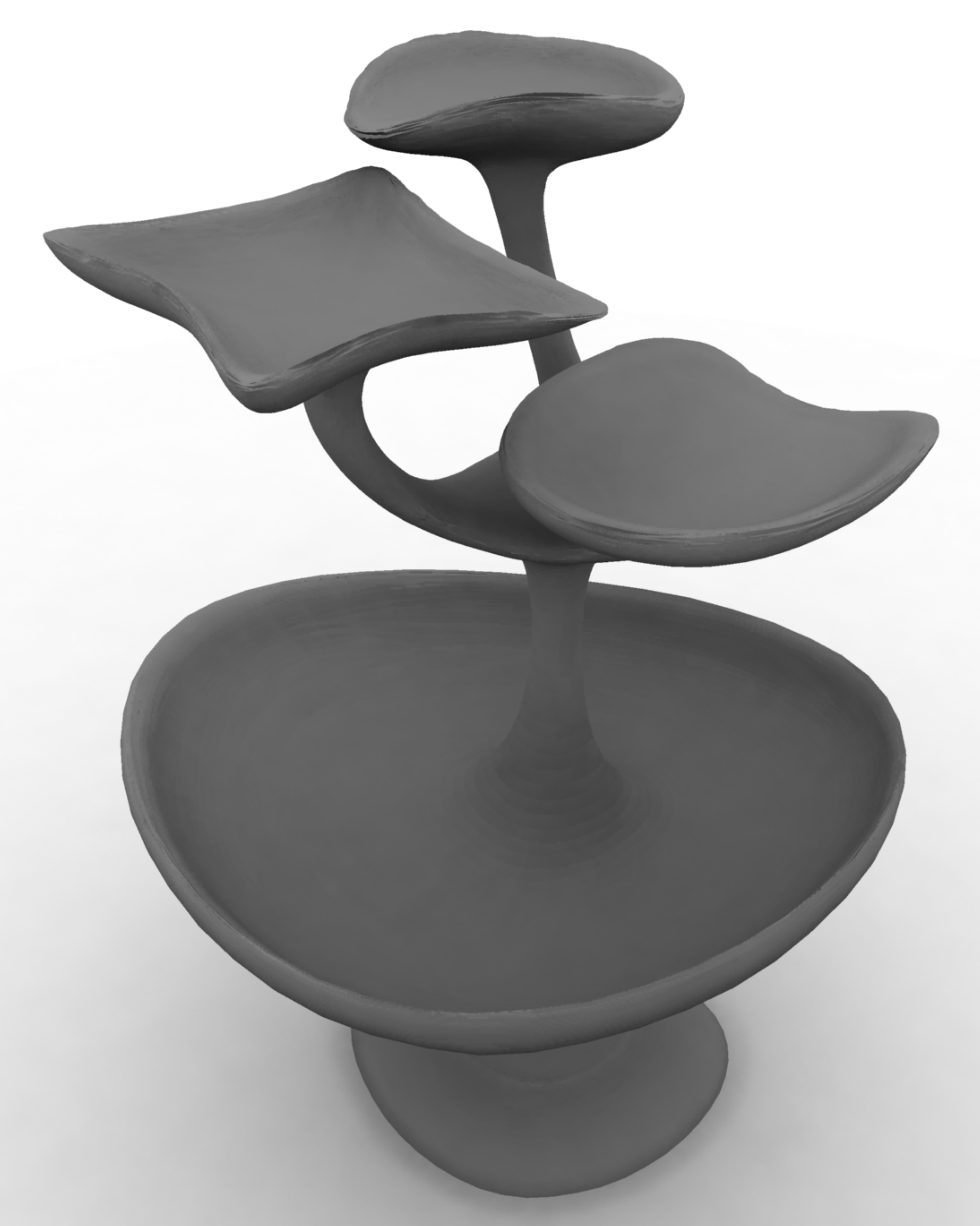} &
\includegraphics[height=.8in]{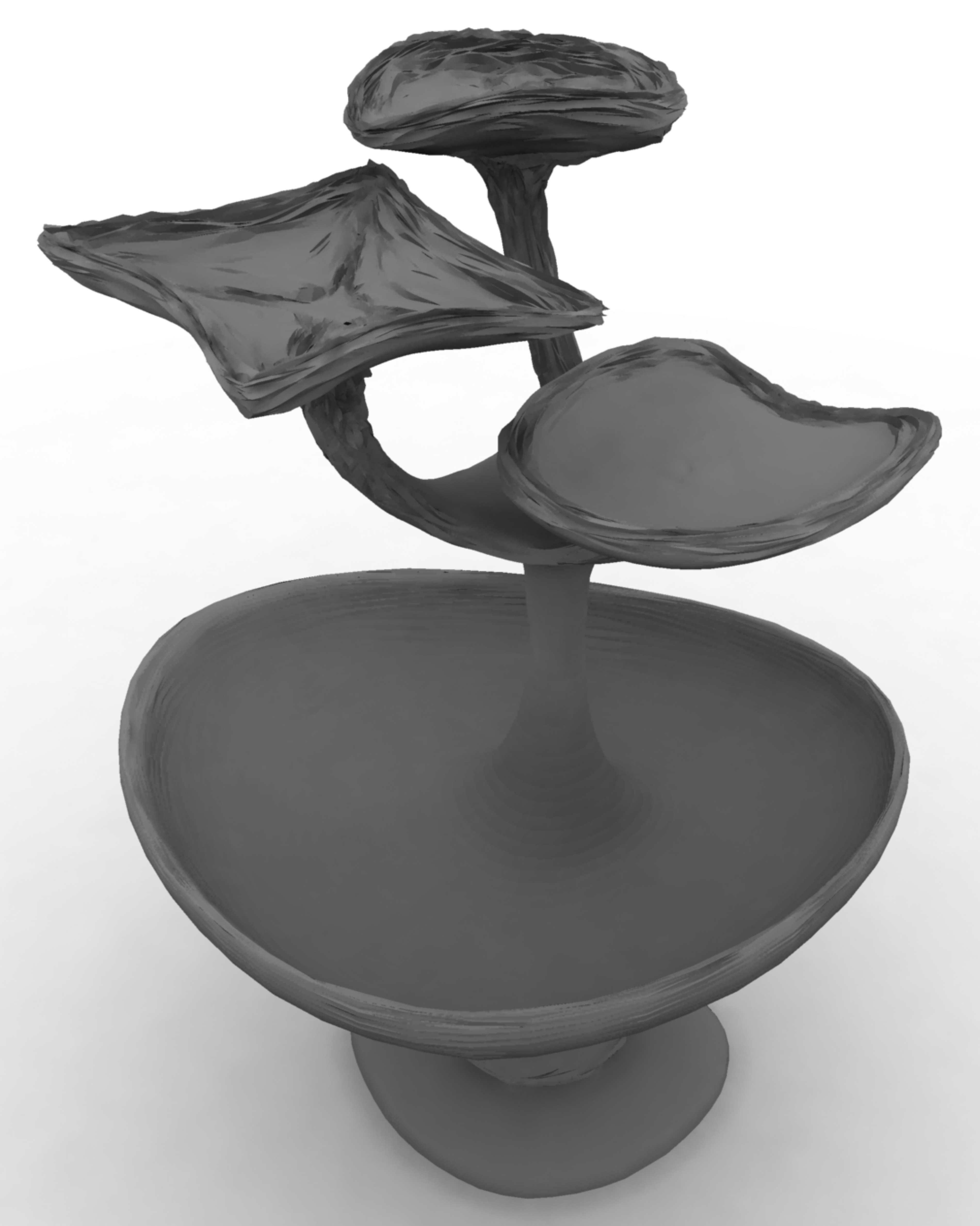}\\
(a)&(b) 0.257&(c) 0.060&(d) 0.069&(g) 0.236
\end{tabular}
}

\def\frog{
\begin{tabular}{c@{}c@{}c@{}c@{}c@{}c@{}c}
\includegraphics[height=0.8in]{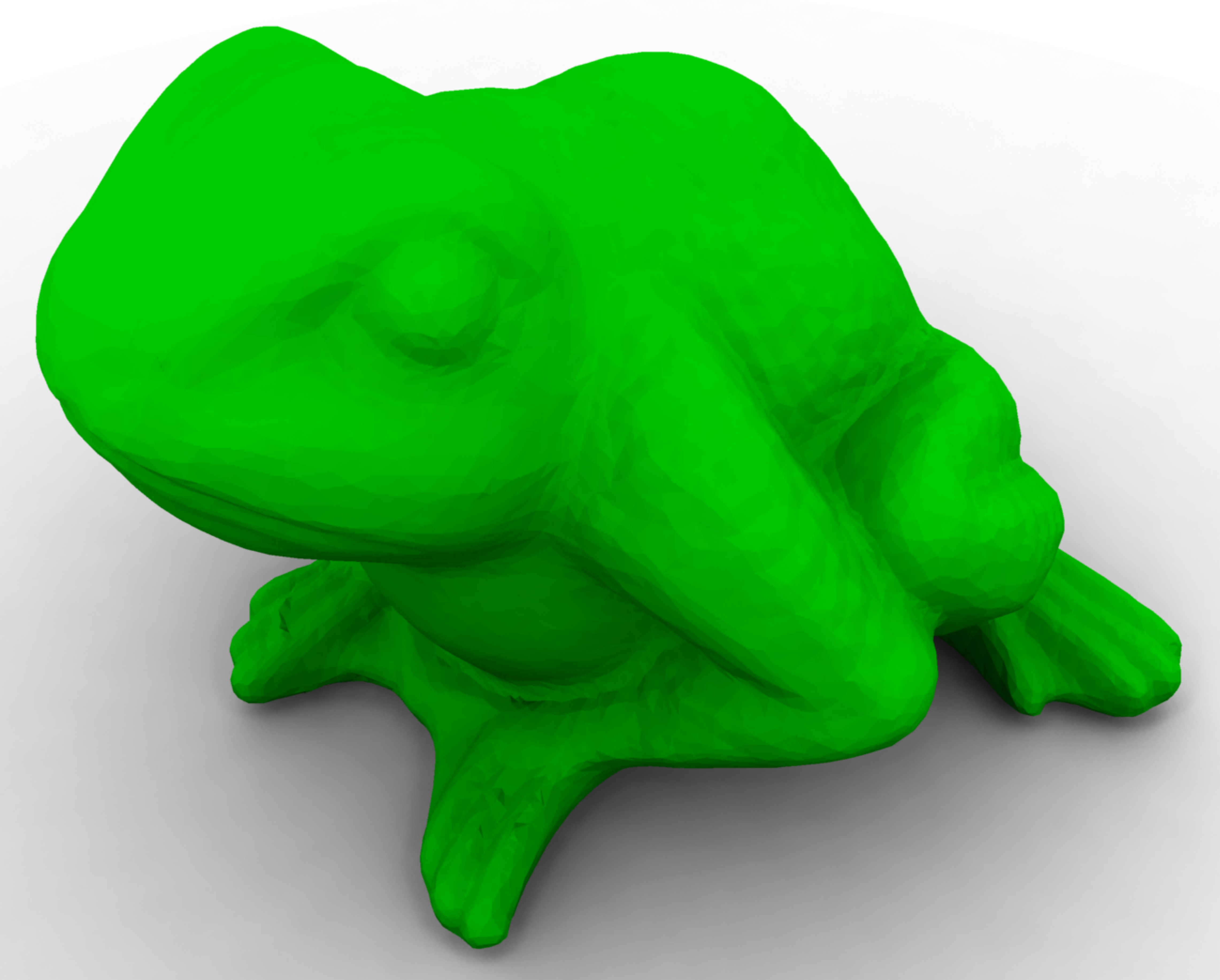} &
\includegraphics[height=0.8in]{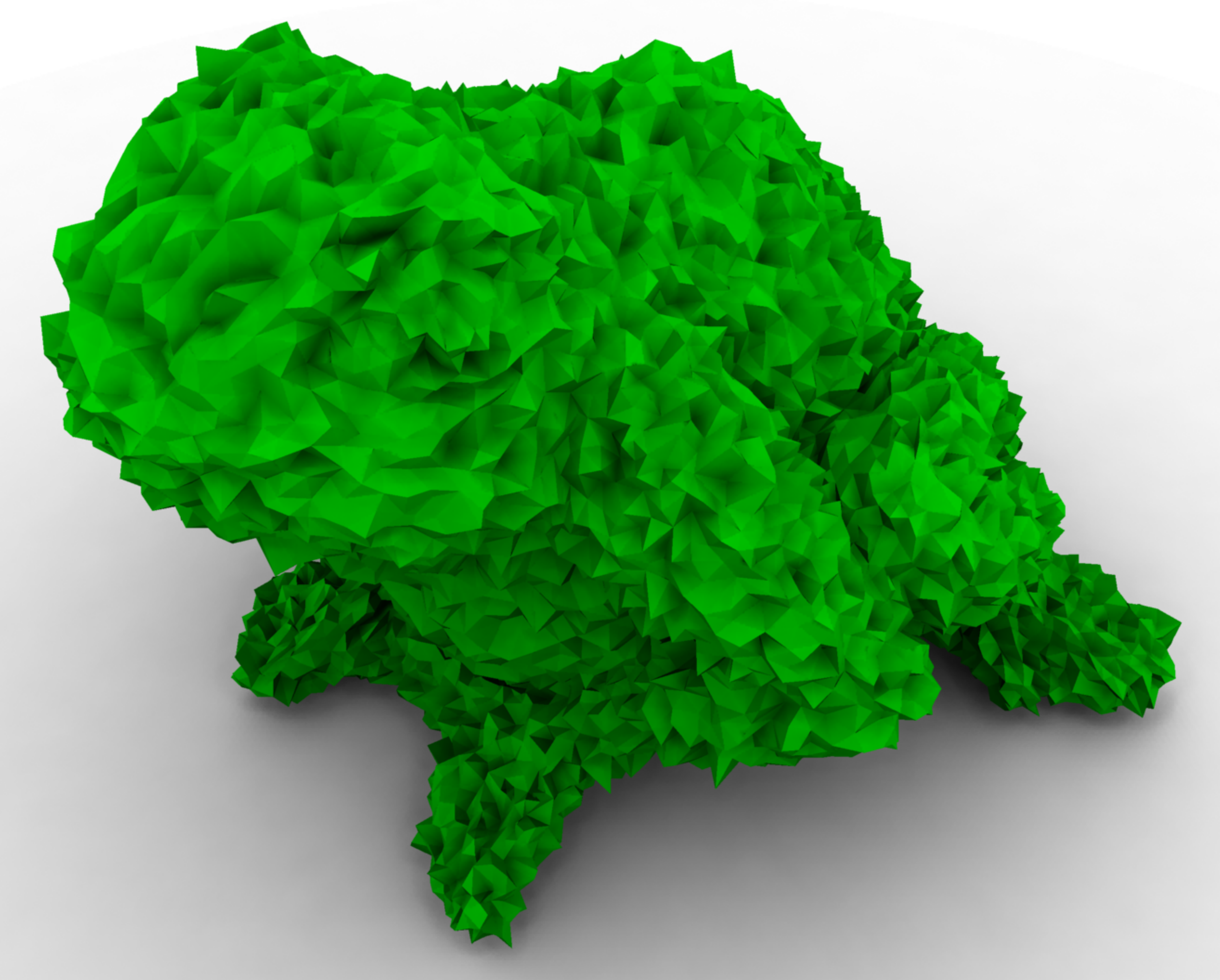} &
\includegraphics[height=0.8in]{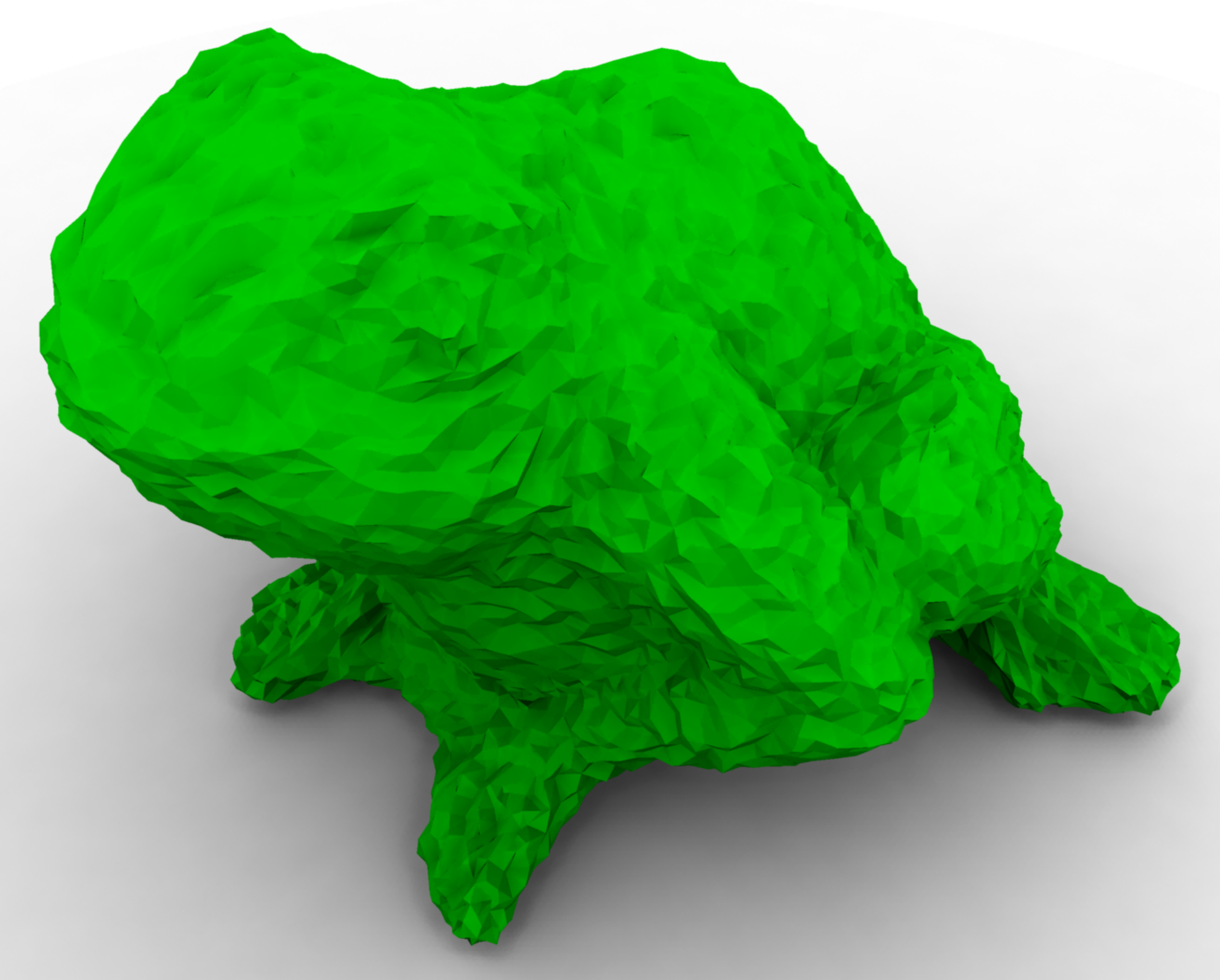} &
\includegraphics[height=0.8in]{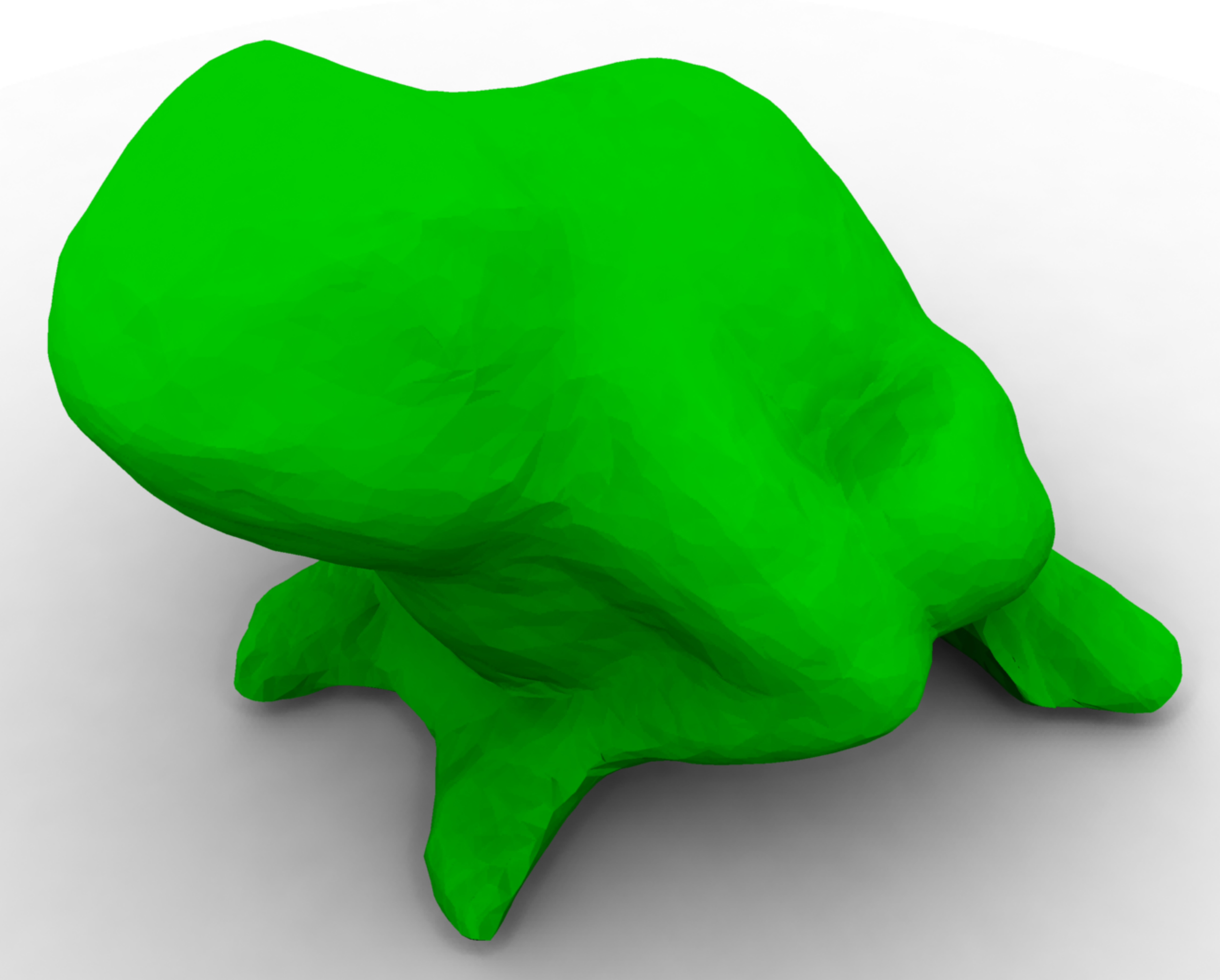} &
\includegraphics[height=0.8in]{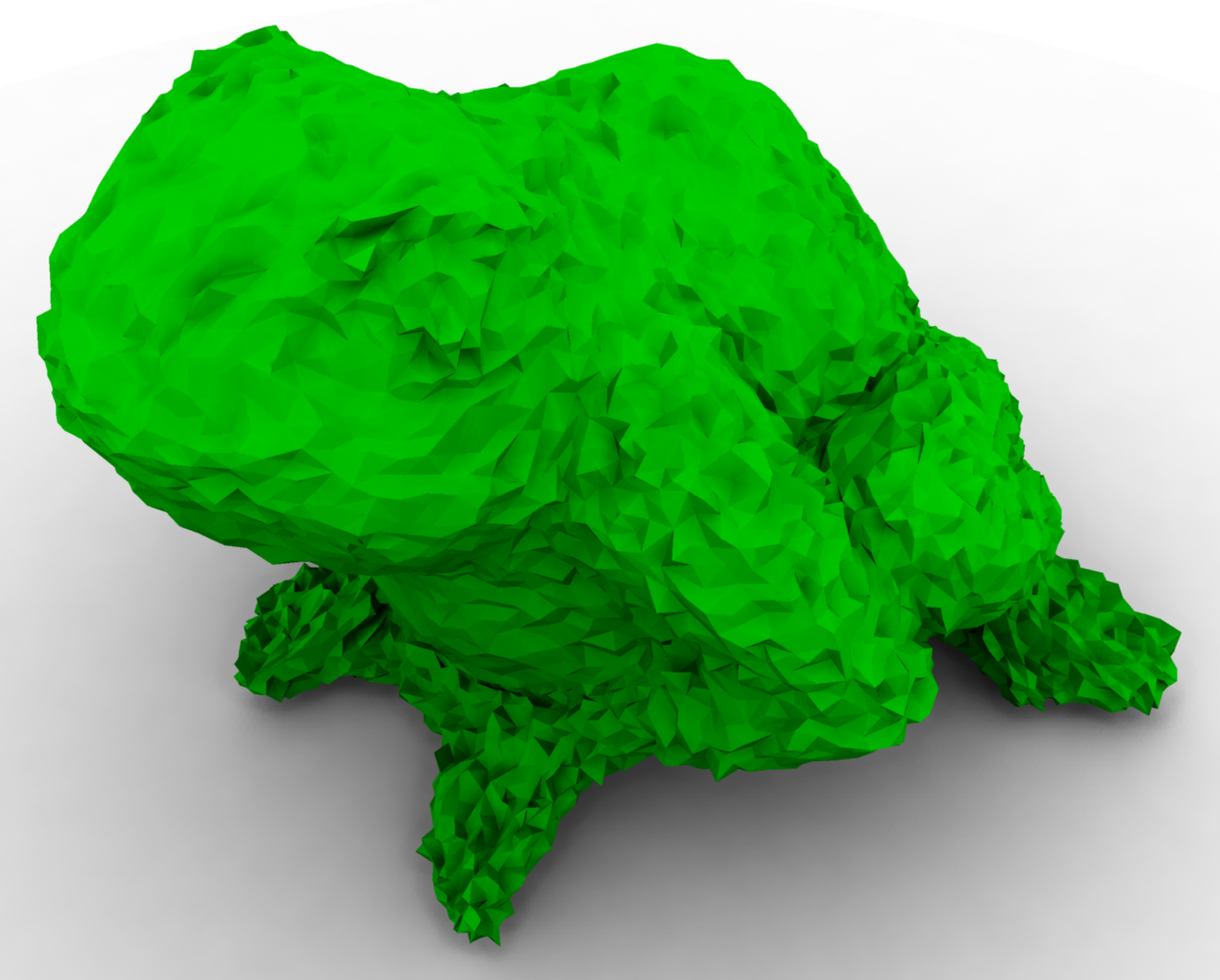} &
\includegraphics[height=0.8in]{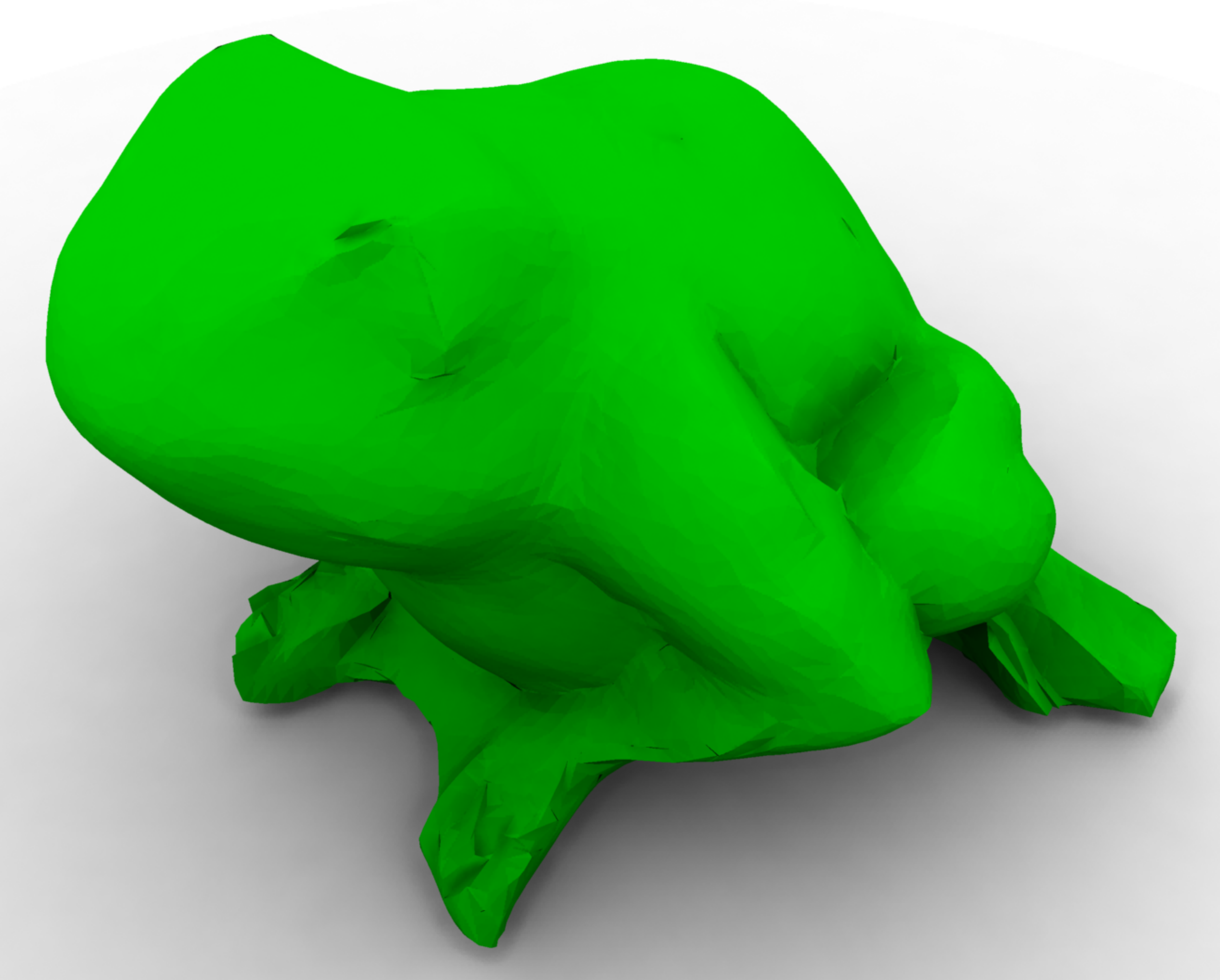}\\
(a)&(b) 0.281&(c) 0.174&(d) 0.121&(e)&(g) 0.166
\end{tabular}
}

\def\ramessesWeakNoise{
\begin{tabular}{c@{}c@{}c@{}c@{}c@{}c@{}c}
\includegraphics[height=.8in]{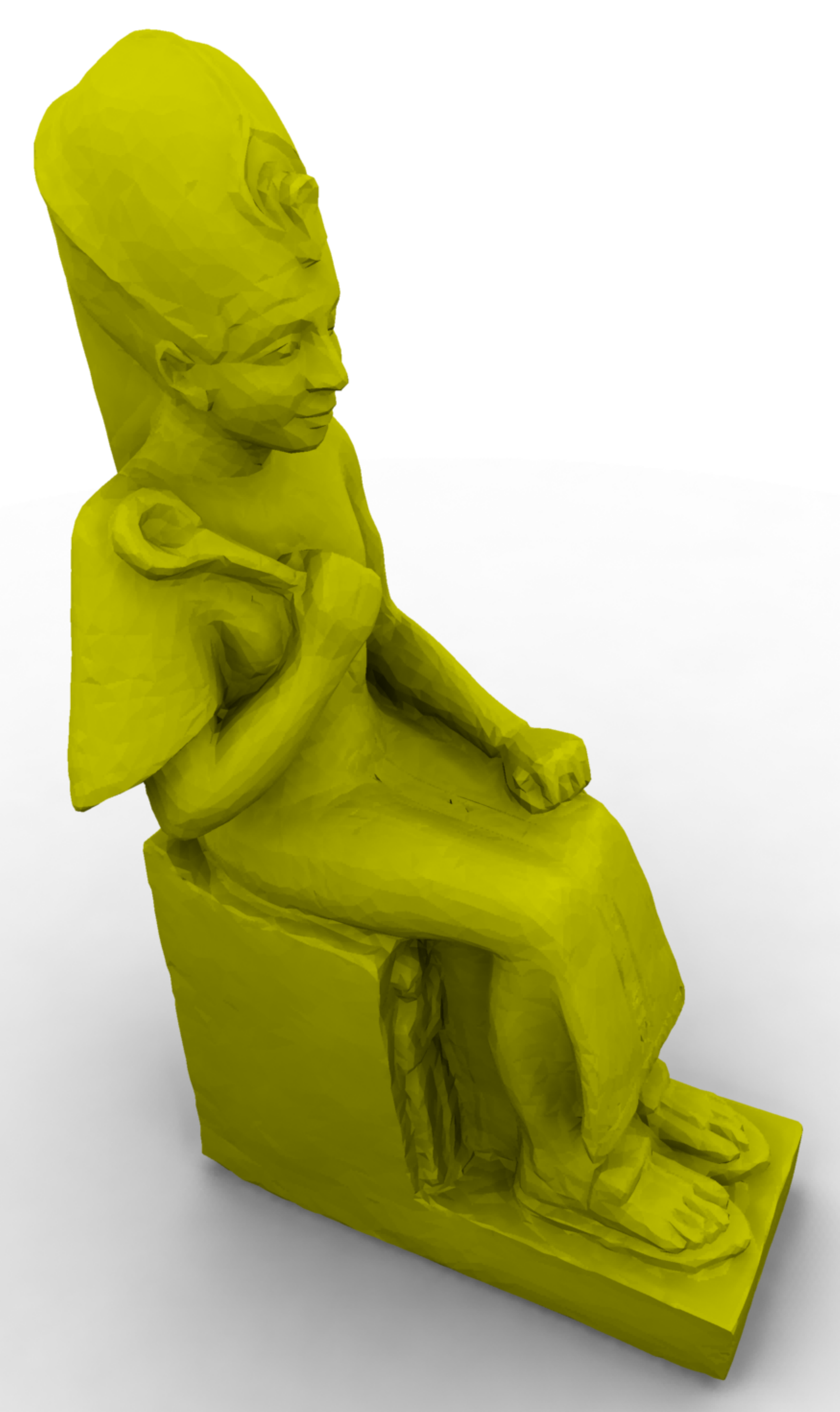} &
\includegraphics[height=.8in]{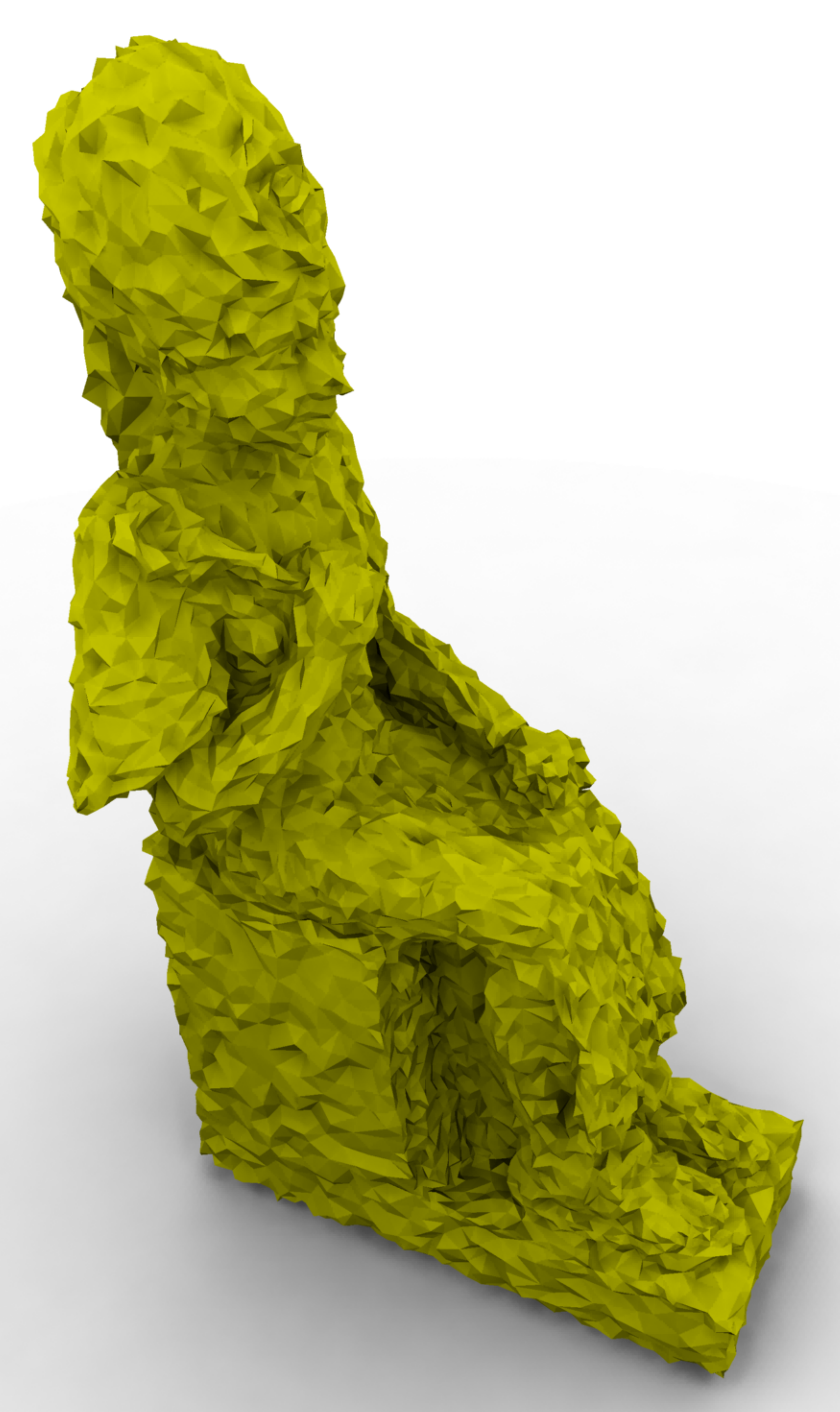} &
\includegraphics[height=.8in]{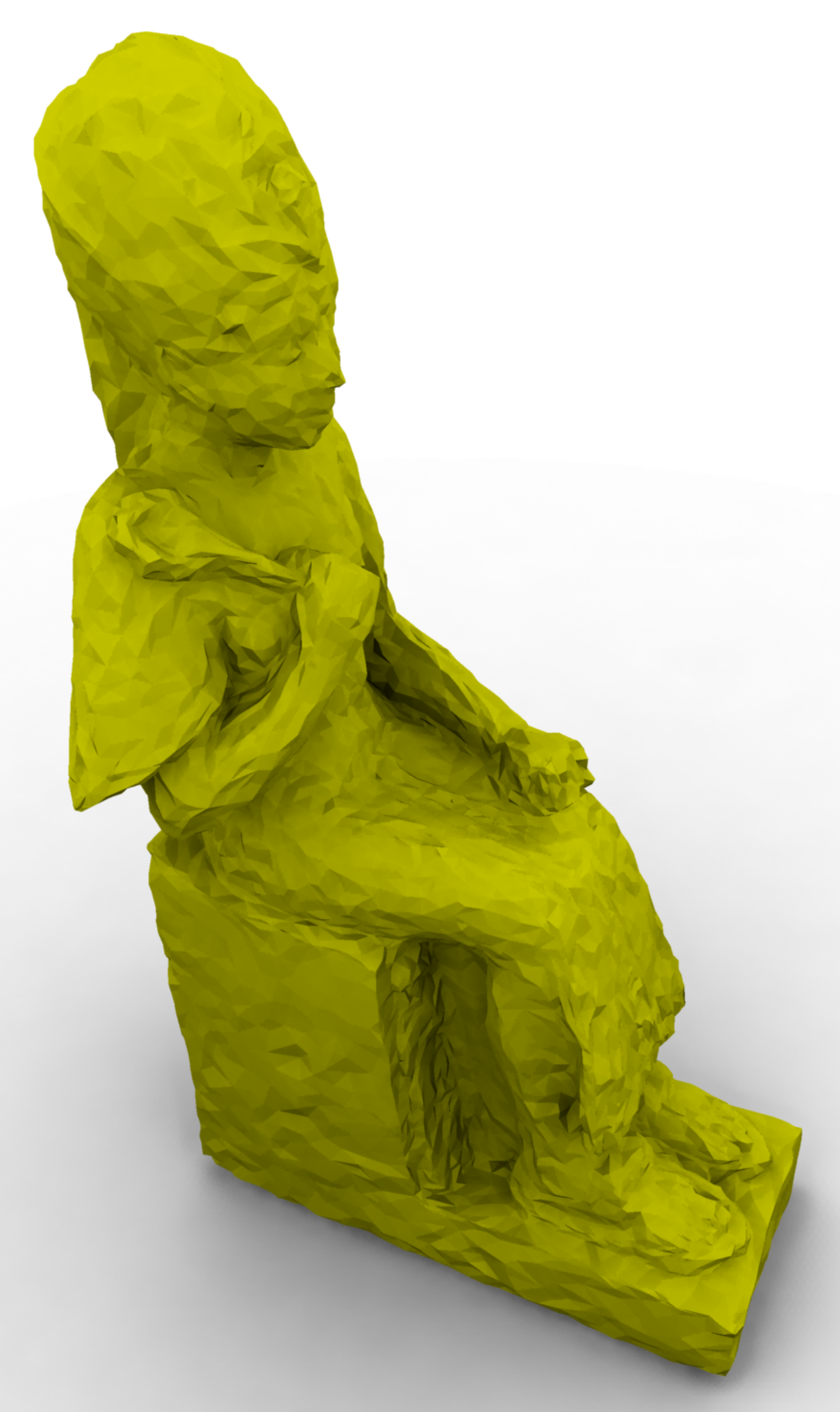} &
\includegraphics[height=.8in]{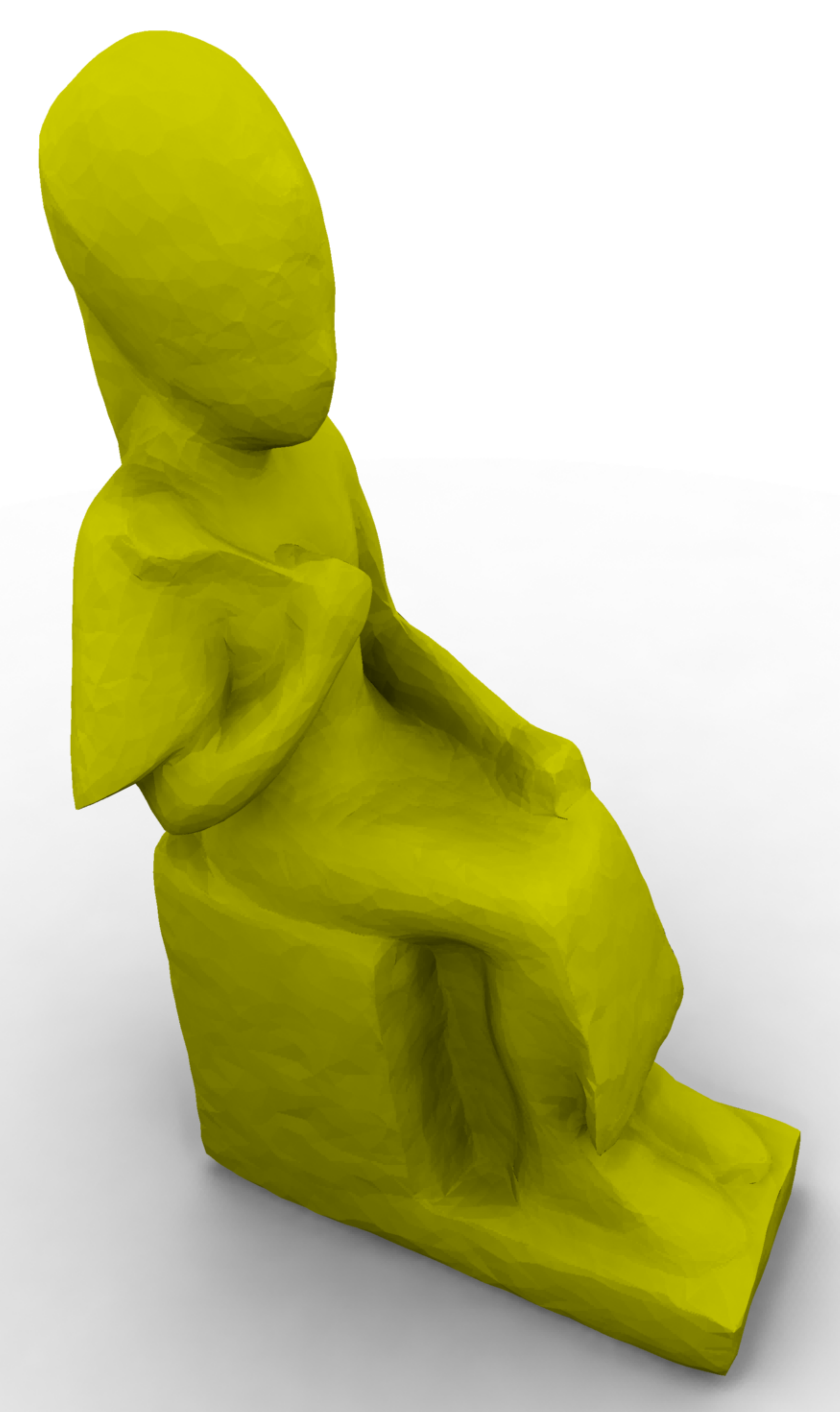} &
\includegraphics[height=.8in]{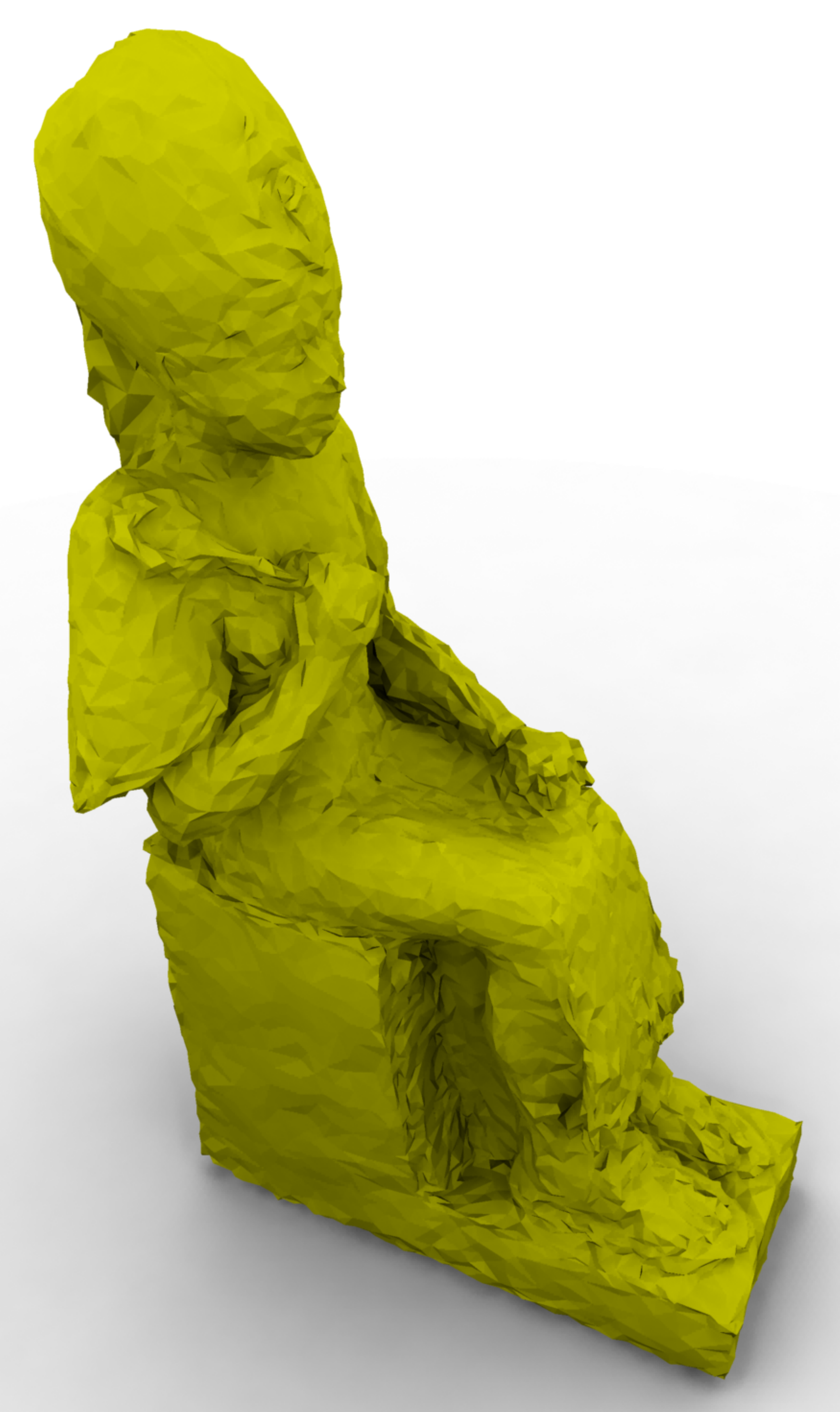} &
\includegraphics[height=.8in]{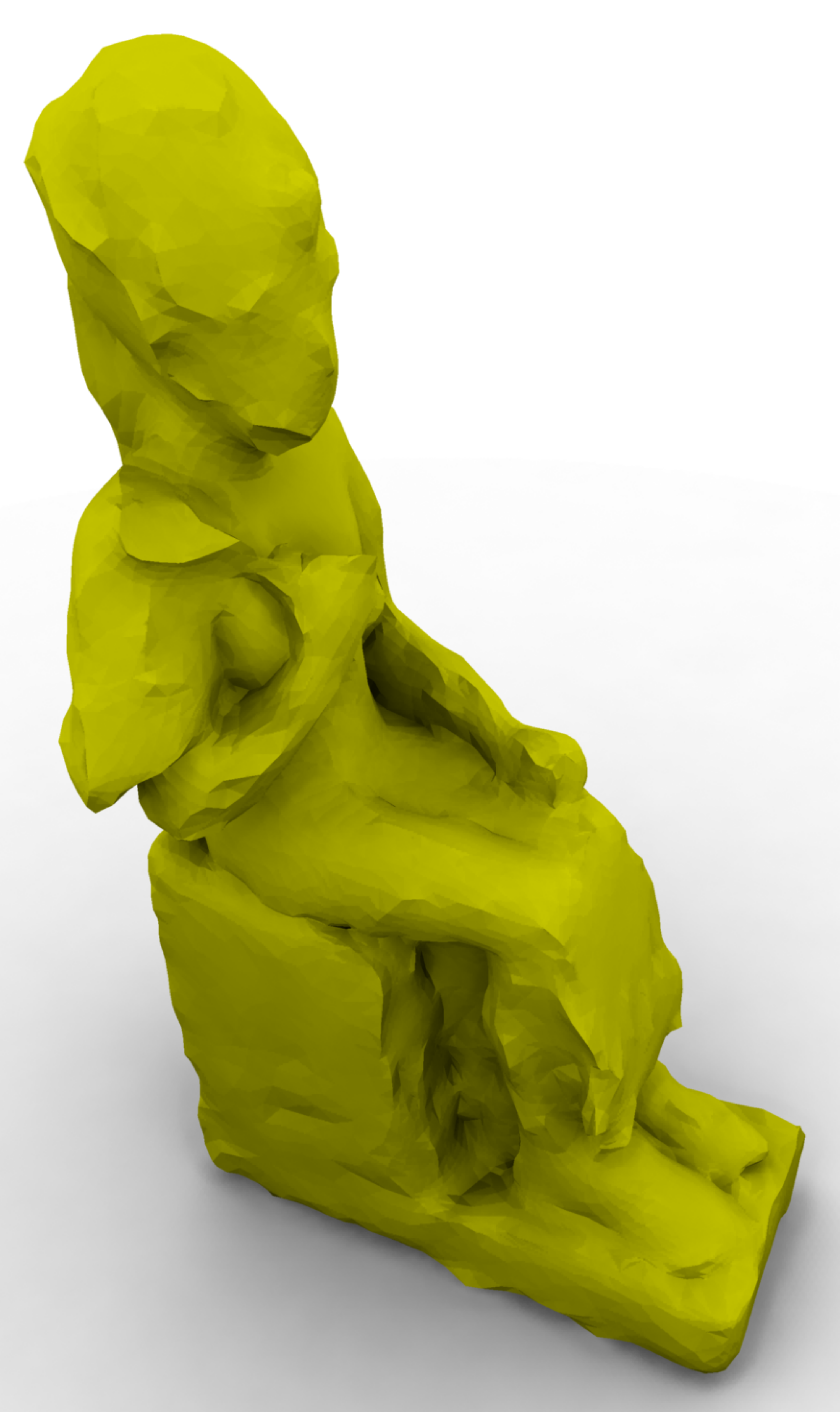} &
\includegraphics[height=.8in]{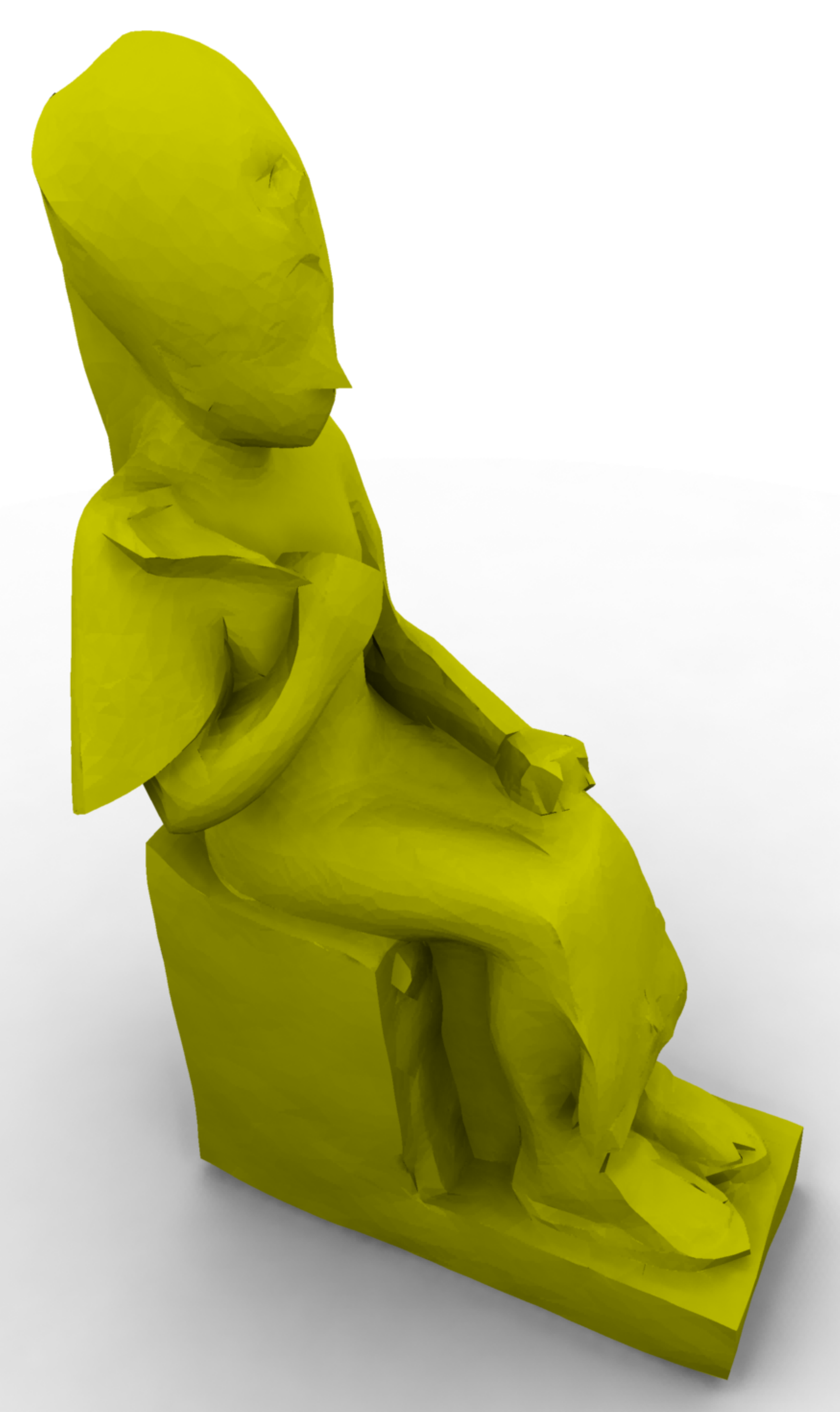}\\
(a)&(b) 0.120&(c) 0.119&(d) 0.111&(e)&(f) 0.163&(g) 0.139
\end{tabular}
}

\def\ramessesStrongNoise{
\begin{tabular}{c@{}c@{}c@{}c@{}c@{}c@{}c}
\includegraphics[height=.8in]{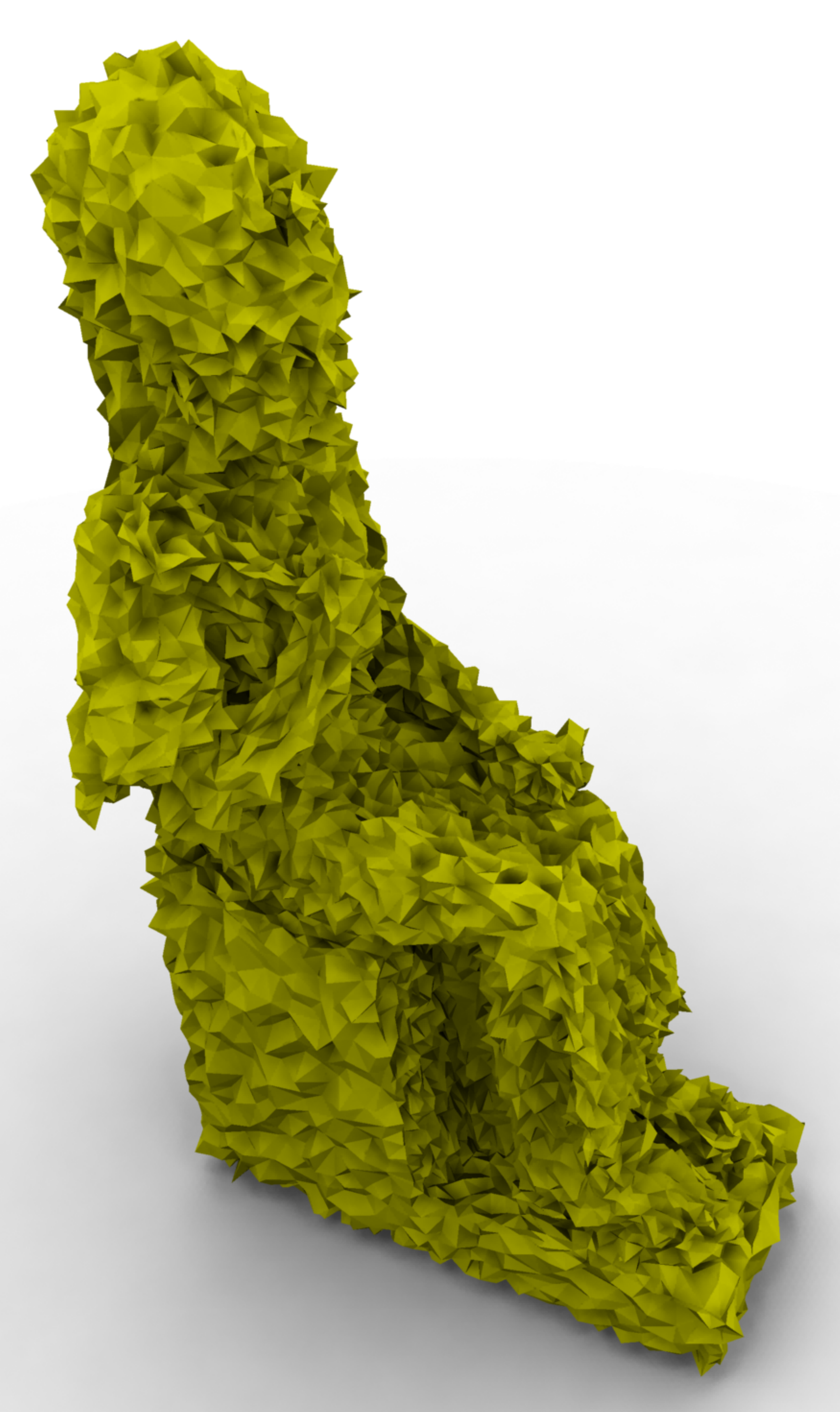} &
\includegraphics[height=.8in]{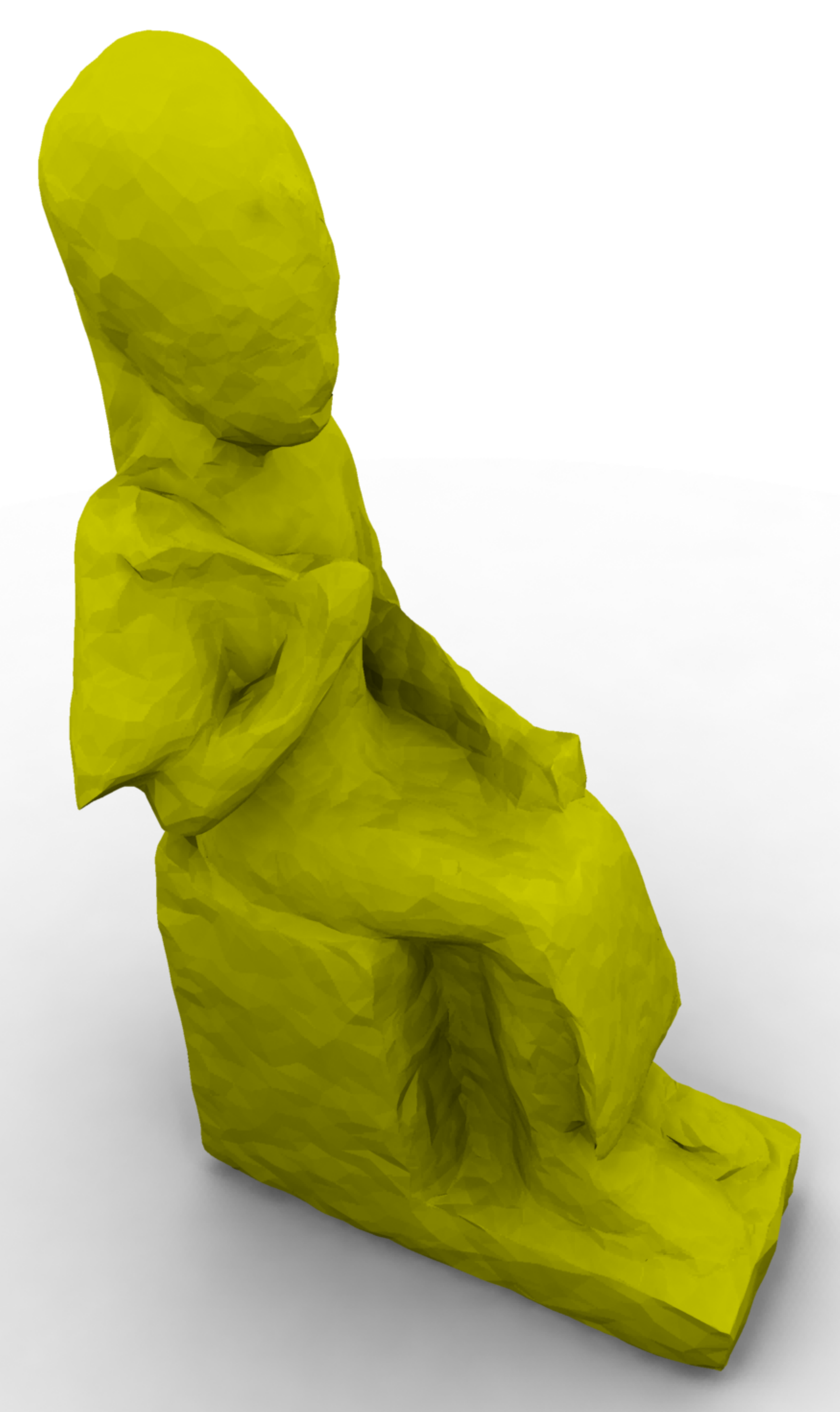} &
\includegraphics[height=.8in]{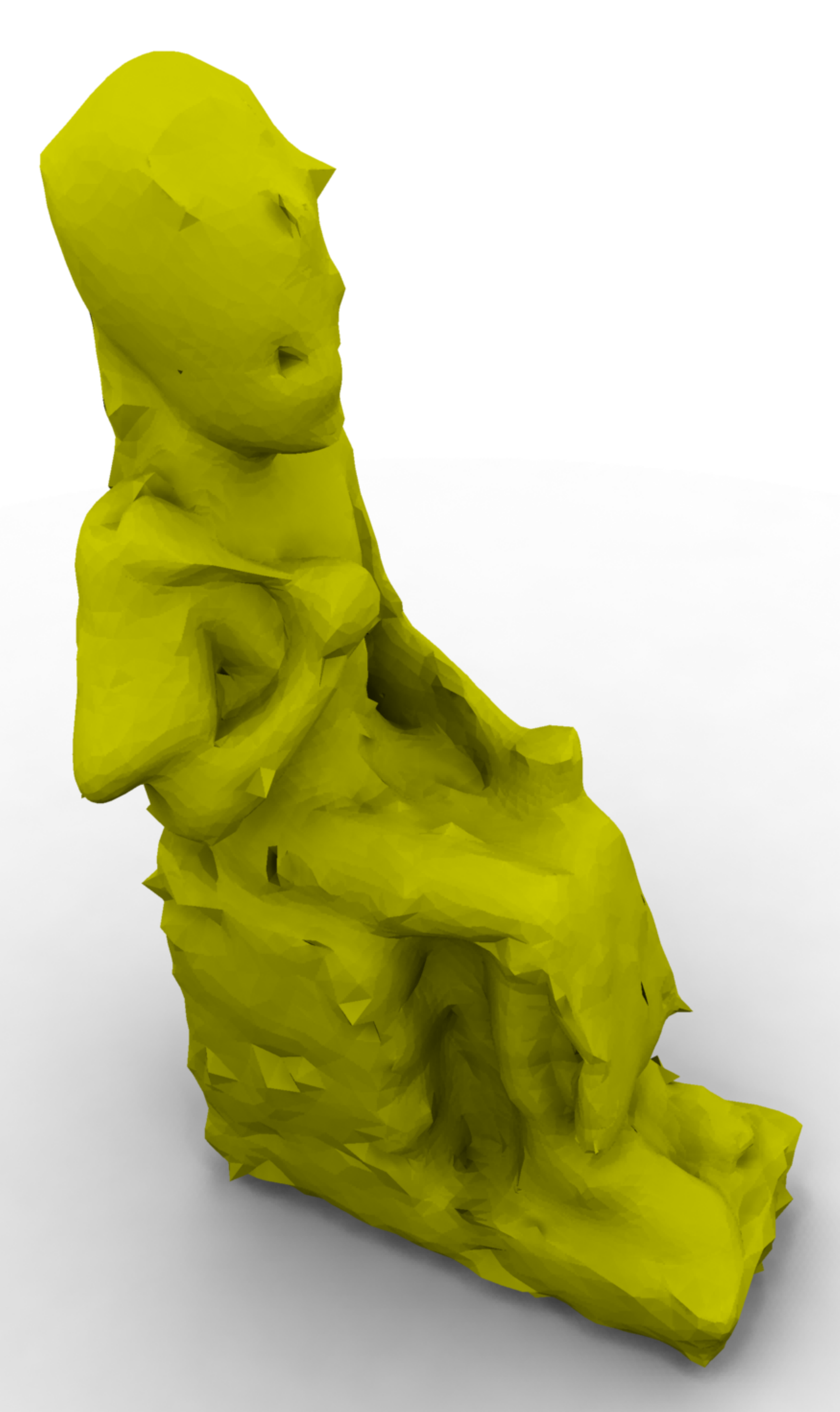} &
\includegraphics[height=.8in]{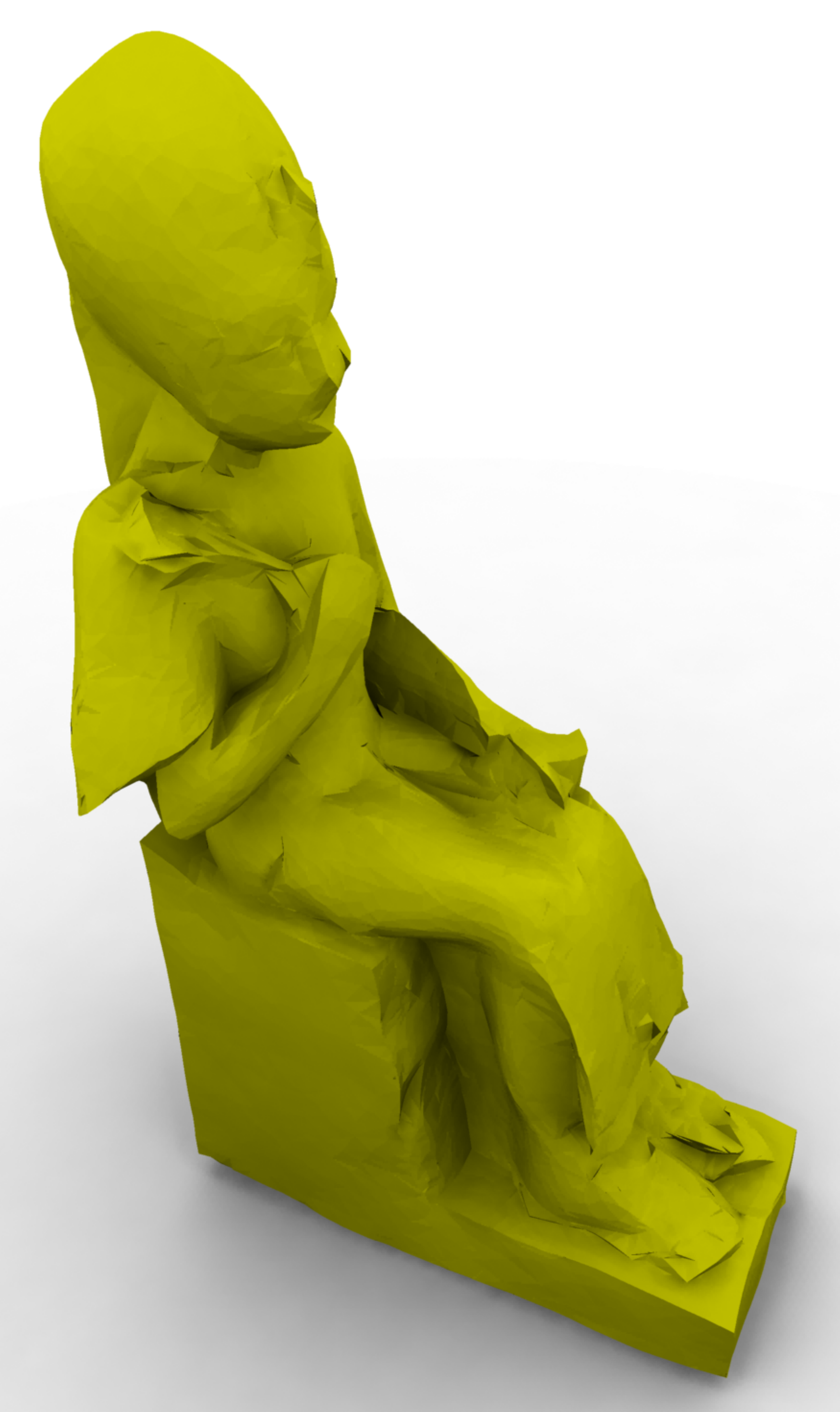}\\
(b) 0.318&(d) 0.174&(f) 0.213&(g) 0.232
\end{tabular}
}

\def\block{
\begin{tabular}{c@{}c@{}c@{}c@{}c}
\includegraphics[height=0.6in]{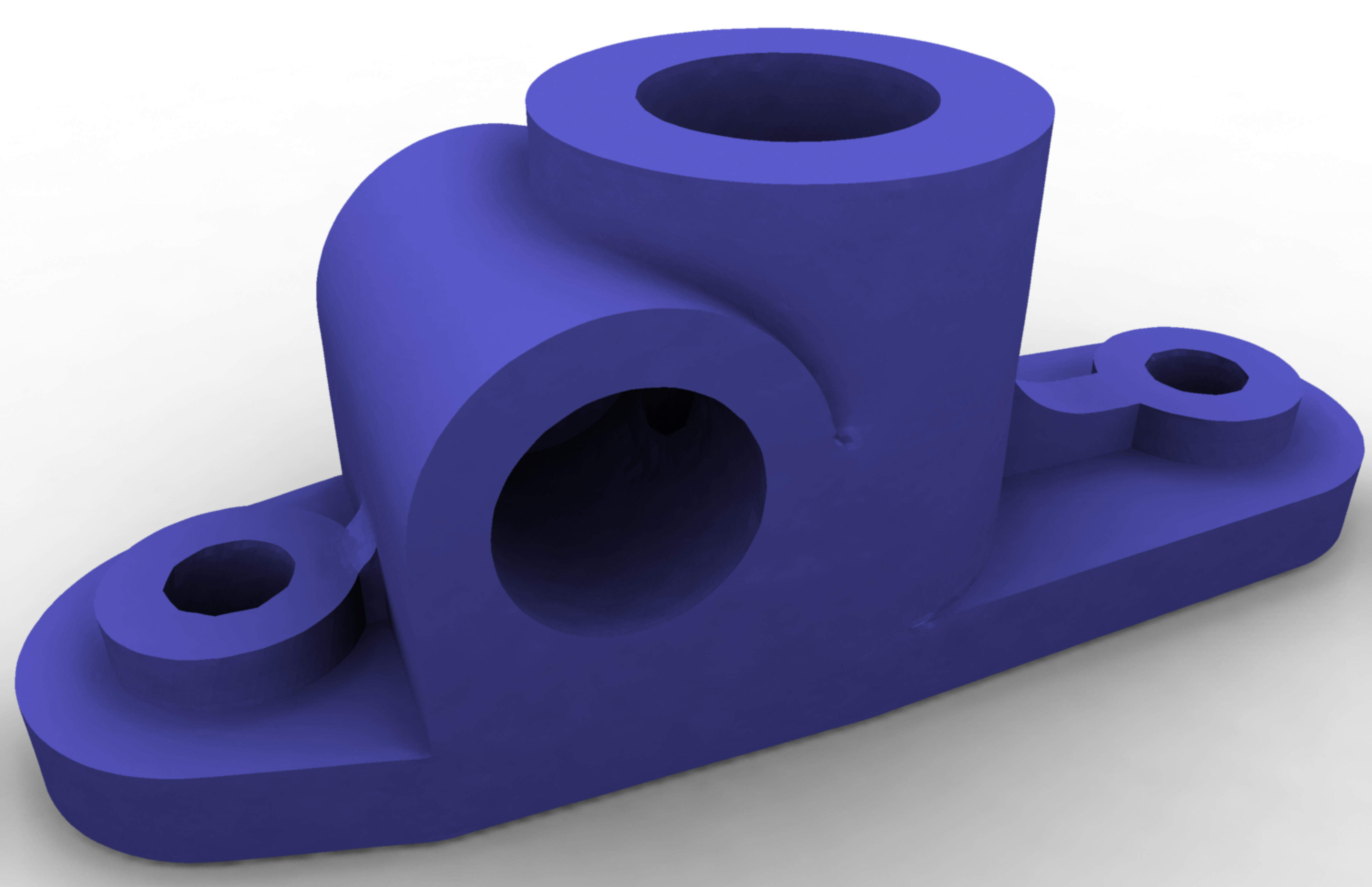} &
\includegraphics[height=0.6in]{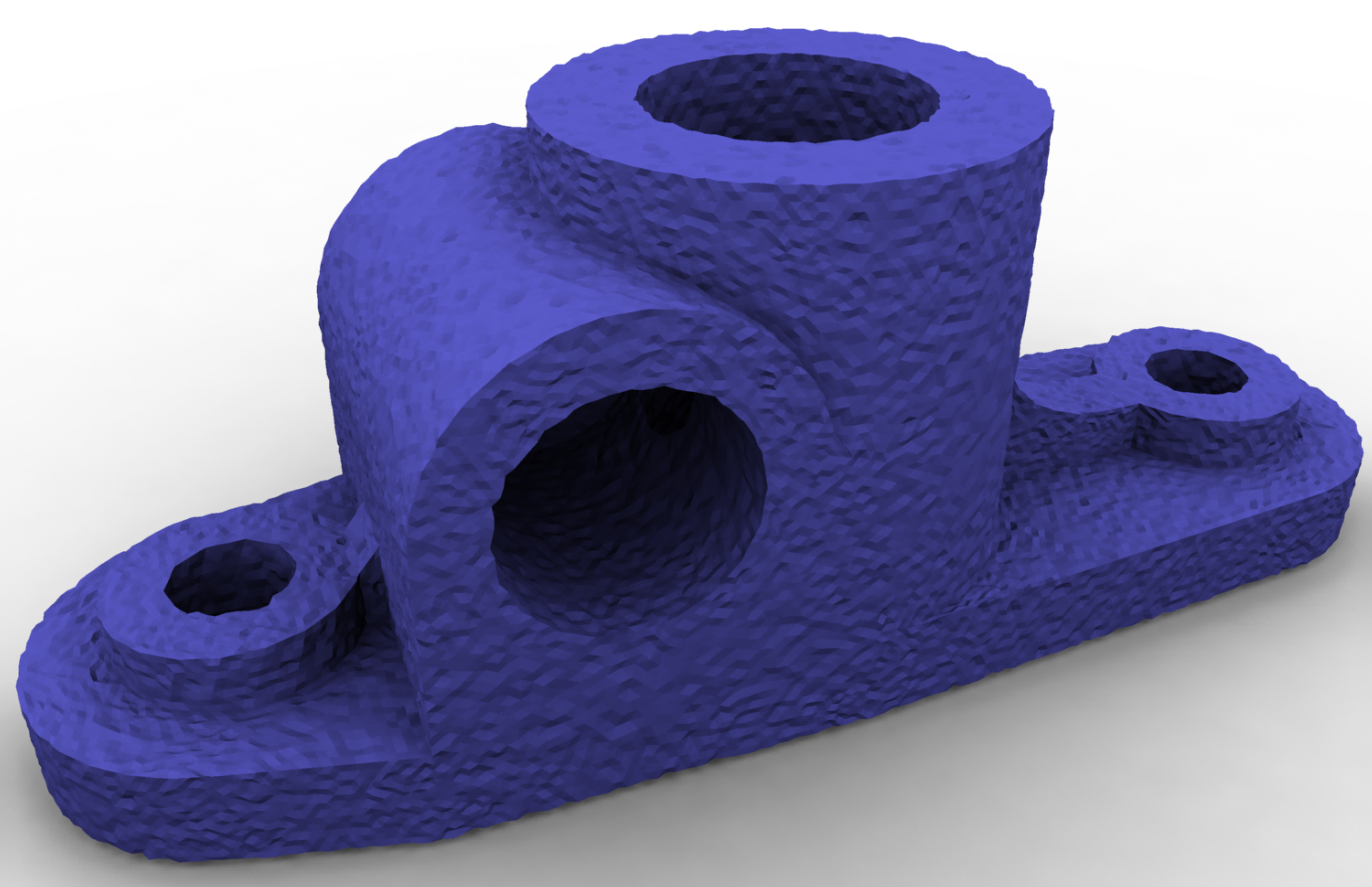} &
\includegraphics[height=0.6in]{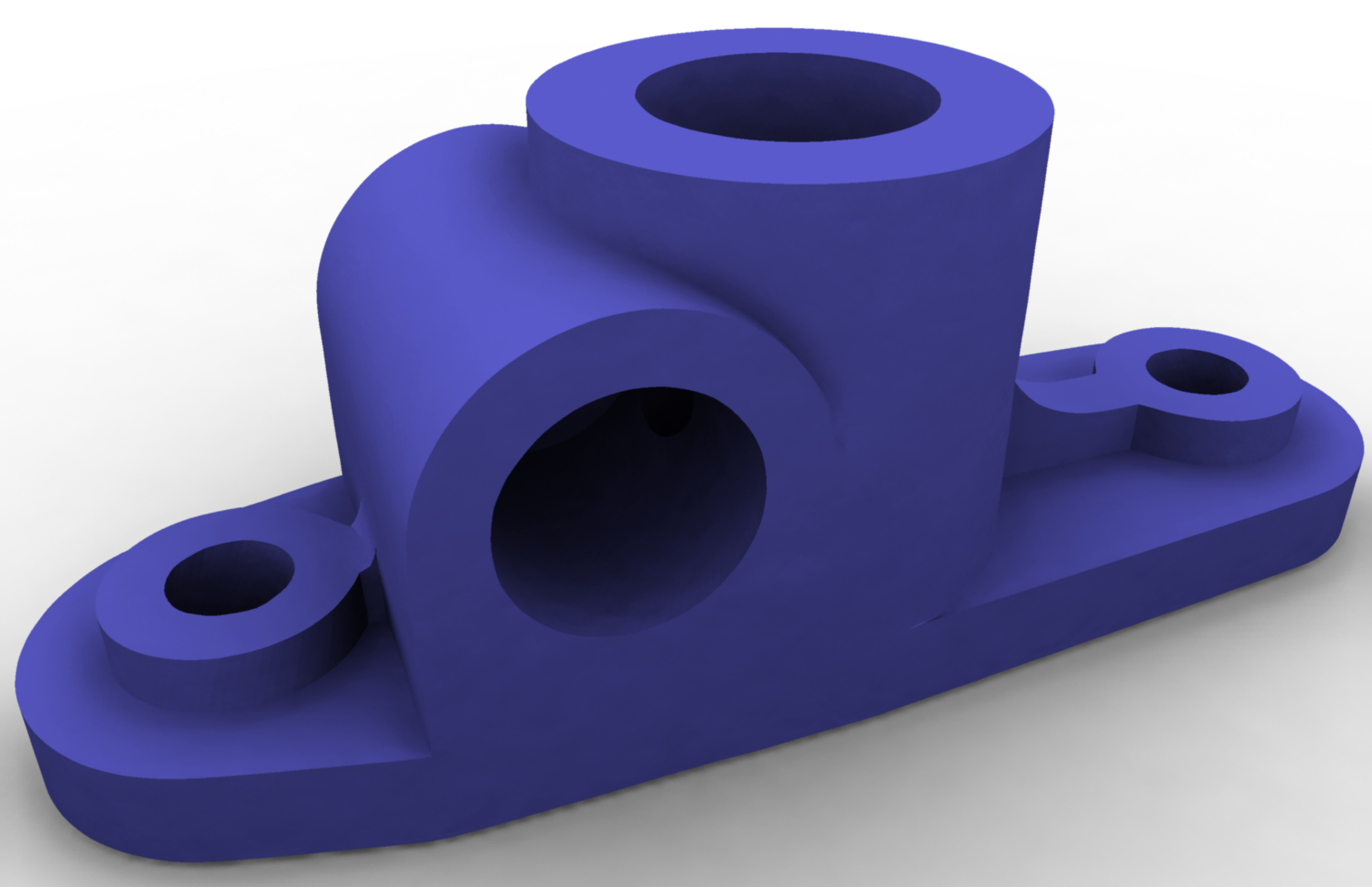} &
\includegraphics[height=0.6in]{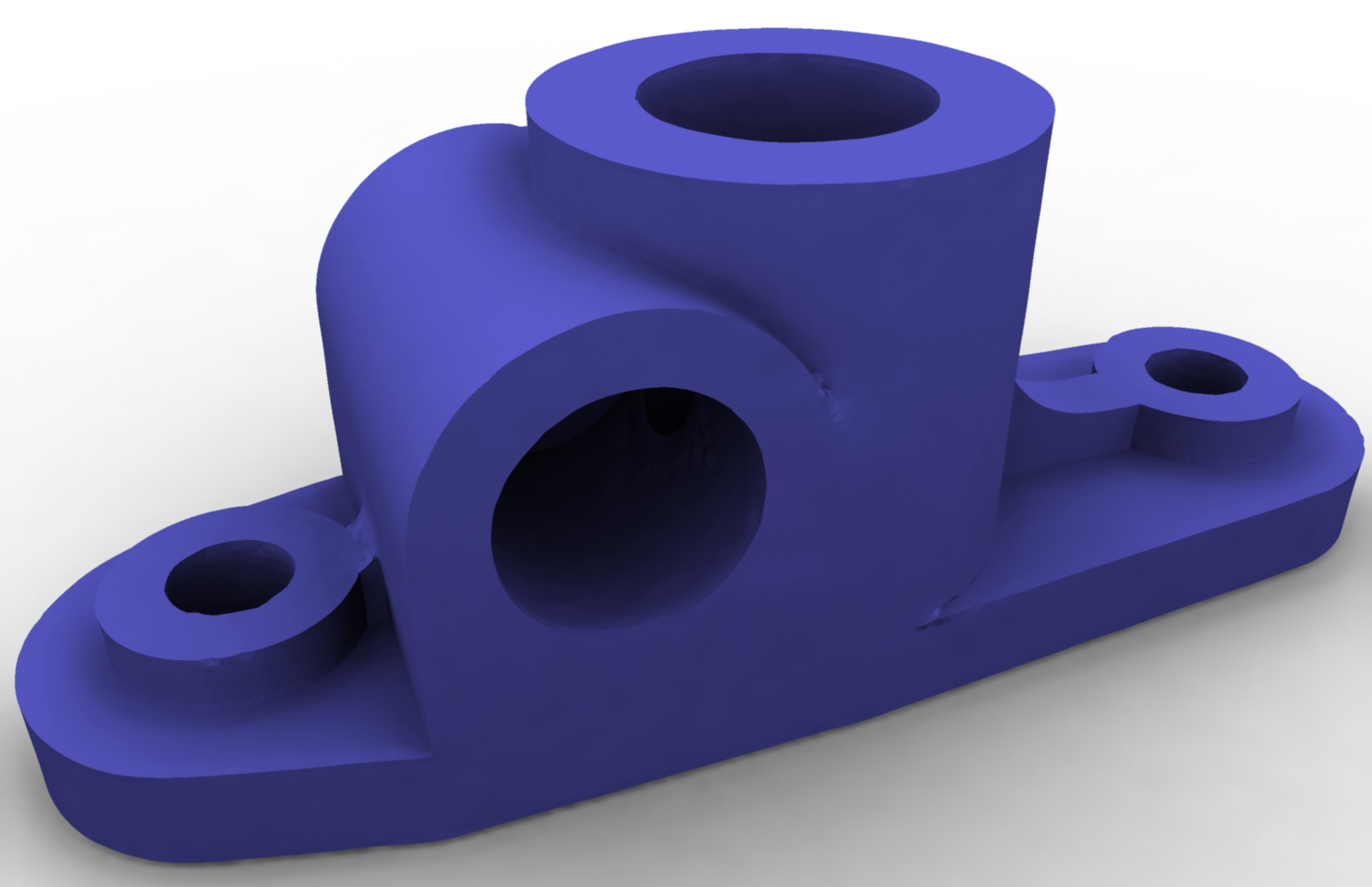}\\
(a)&(b) 0.024 &(d) 0.011&(h) 0.010
\end{tabular}
}

\def\bunny{
\begin{tabular}{c@{}c@{}c@{}c@{}c}
\includegraphics[height=0.6in]{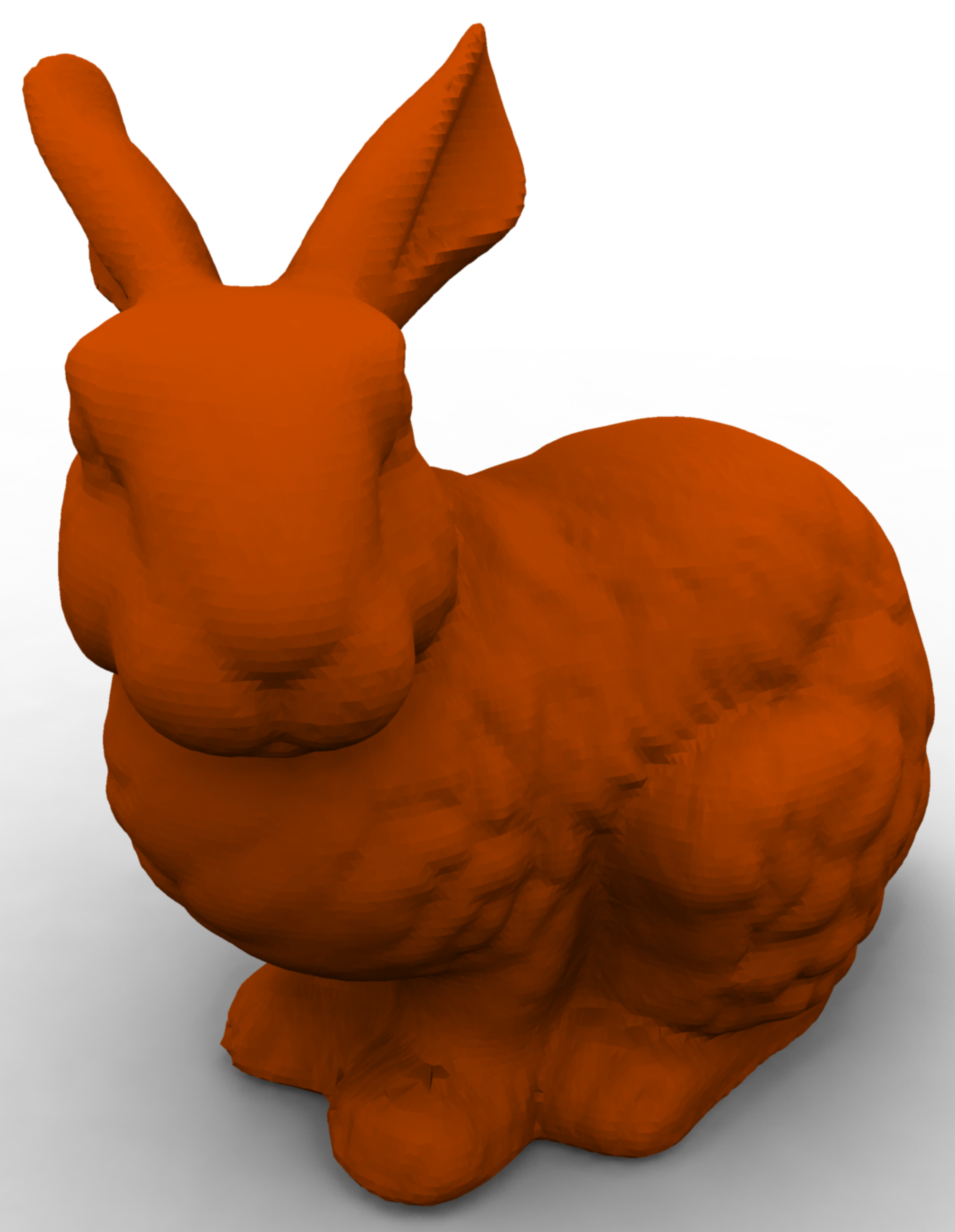} &
\includegraphics[height=0.6in]{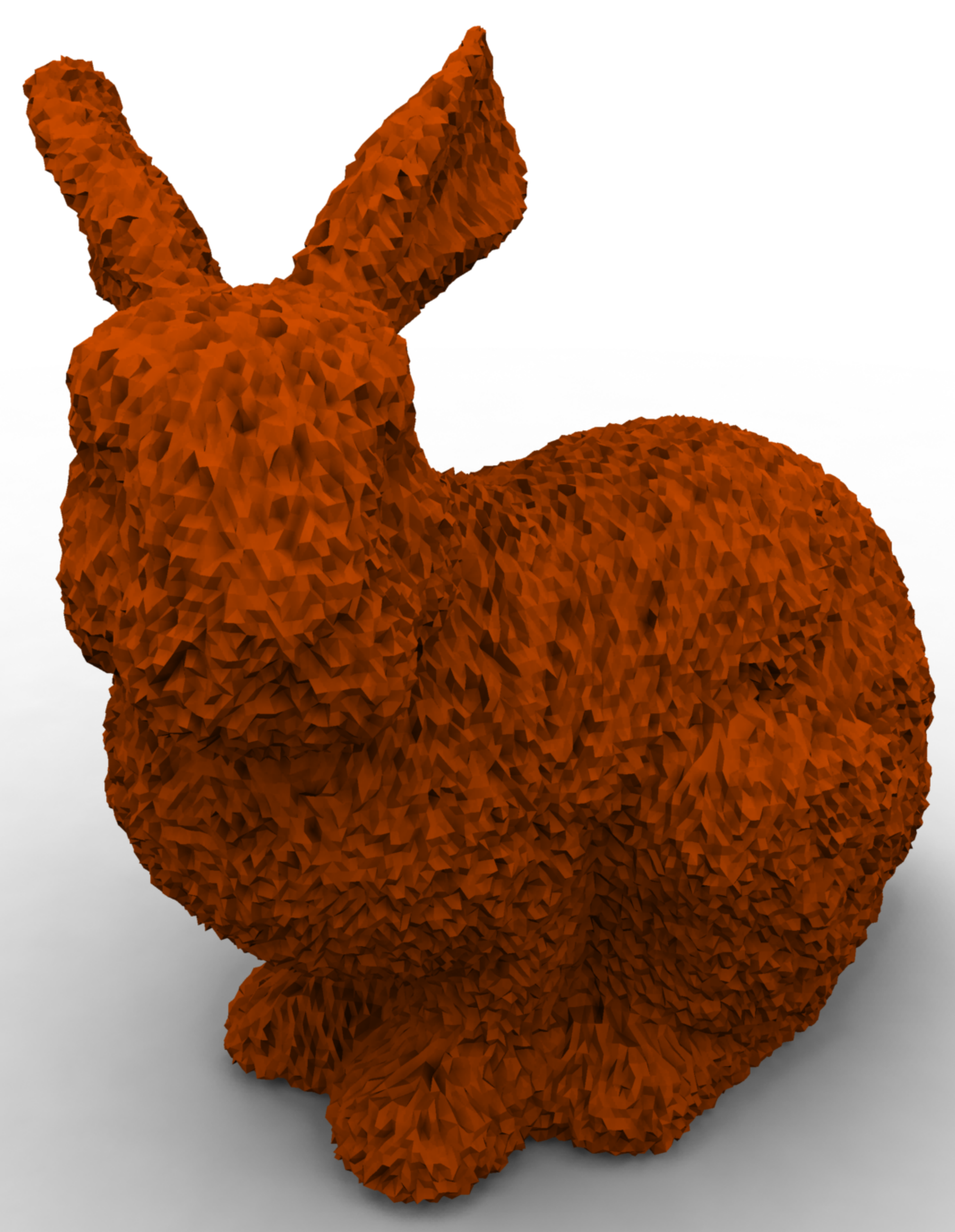} &
\includegraphics[height=0.6in]{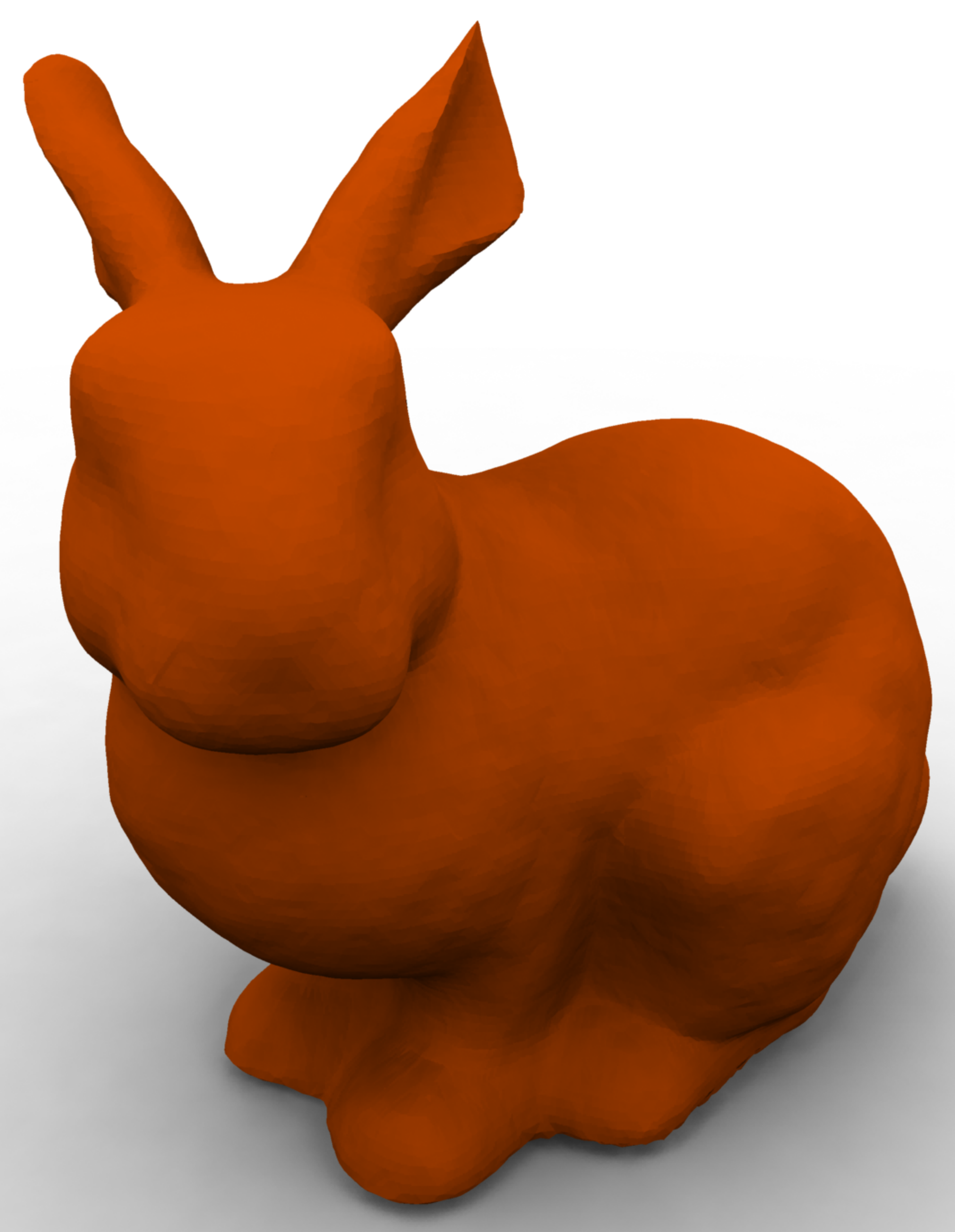} &
\includegraphics[height=0.6in]{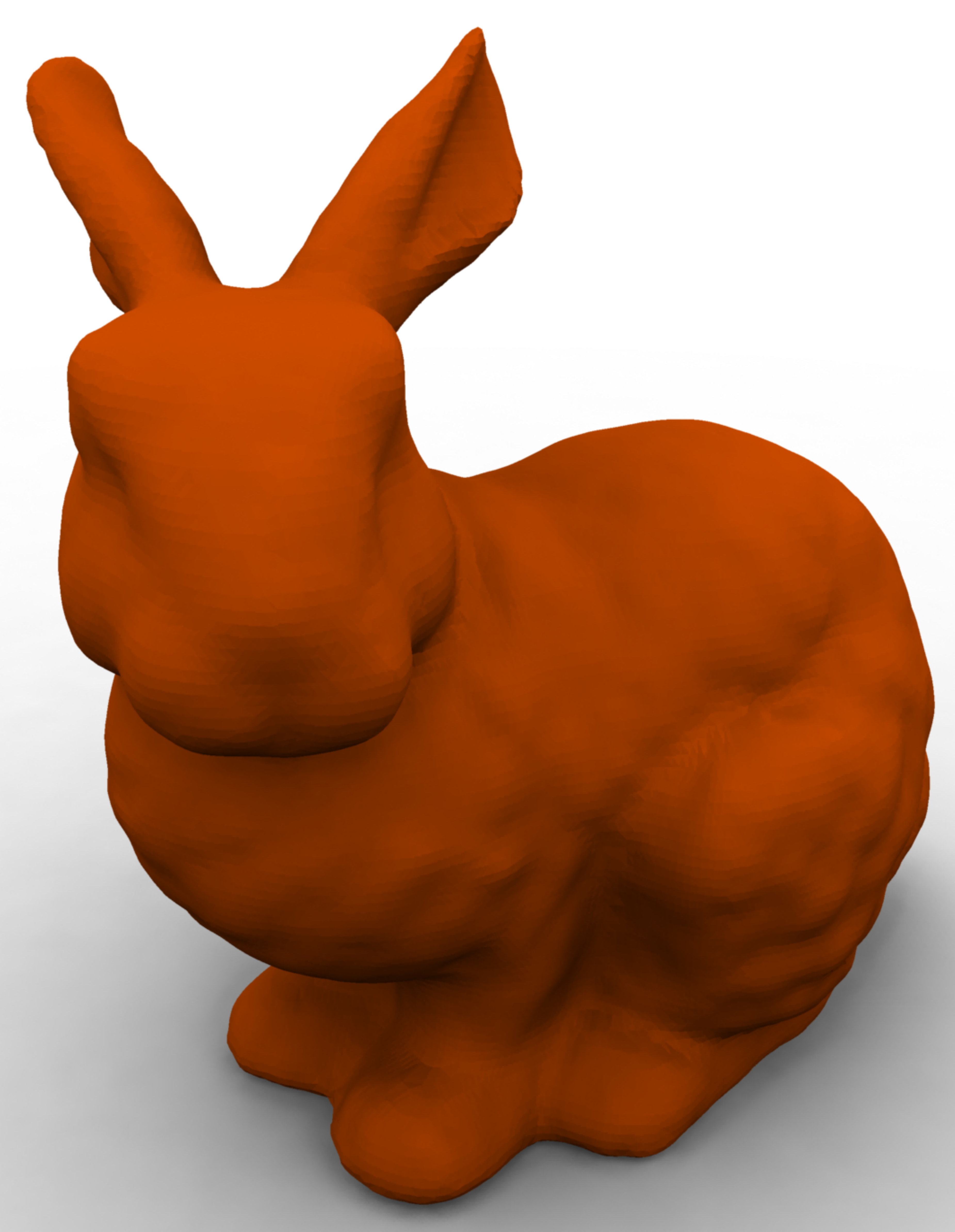}\\
(a)&(b) 0.163&(d) 0.087&(h)
\end{tabular}
}

\begin{figure*}
\centering
\bustStrongNoise\bustWeakNoise\doubleTorus\fandiskNoisy\fandiskNoisier\fountain\circularBox\frog\ramessesWeakNoise\ramessesStrongNoise\block\bunny
\caption{Noise is added to (a) to generate test case (b).  We smooth using our bilateral (c) and mean shift (d) filters and provide comparisons with~\protect\cite{jones03} (e),~\protect\cite{hildebrandt04} (f),~\protect\cite{zheng11} (g), and~\protect\cite{fan10} (h).  Perceptual STED distance~\protect\cite{vasa11_2} from original non-noisy surface is shown underneath when computable and relevant.  Noise is generated by randomly displacing mesh vertices under a uniform distribution, except for the bottom row, which uses tests from~\protect\cite{fan10}.  Data for Figure~\ref{fig:bilateral_reconstruction} is from real-world scans; here we opt to generate synthetic noise to enable use of the STED metric.}\label{fig:denoising}
\end{figure*}

\begin{figure}
\centering
\includegraphics[width=\linewidth]{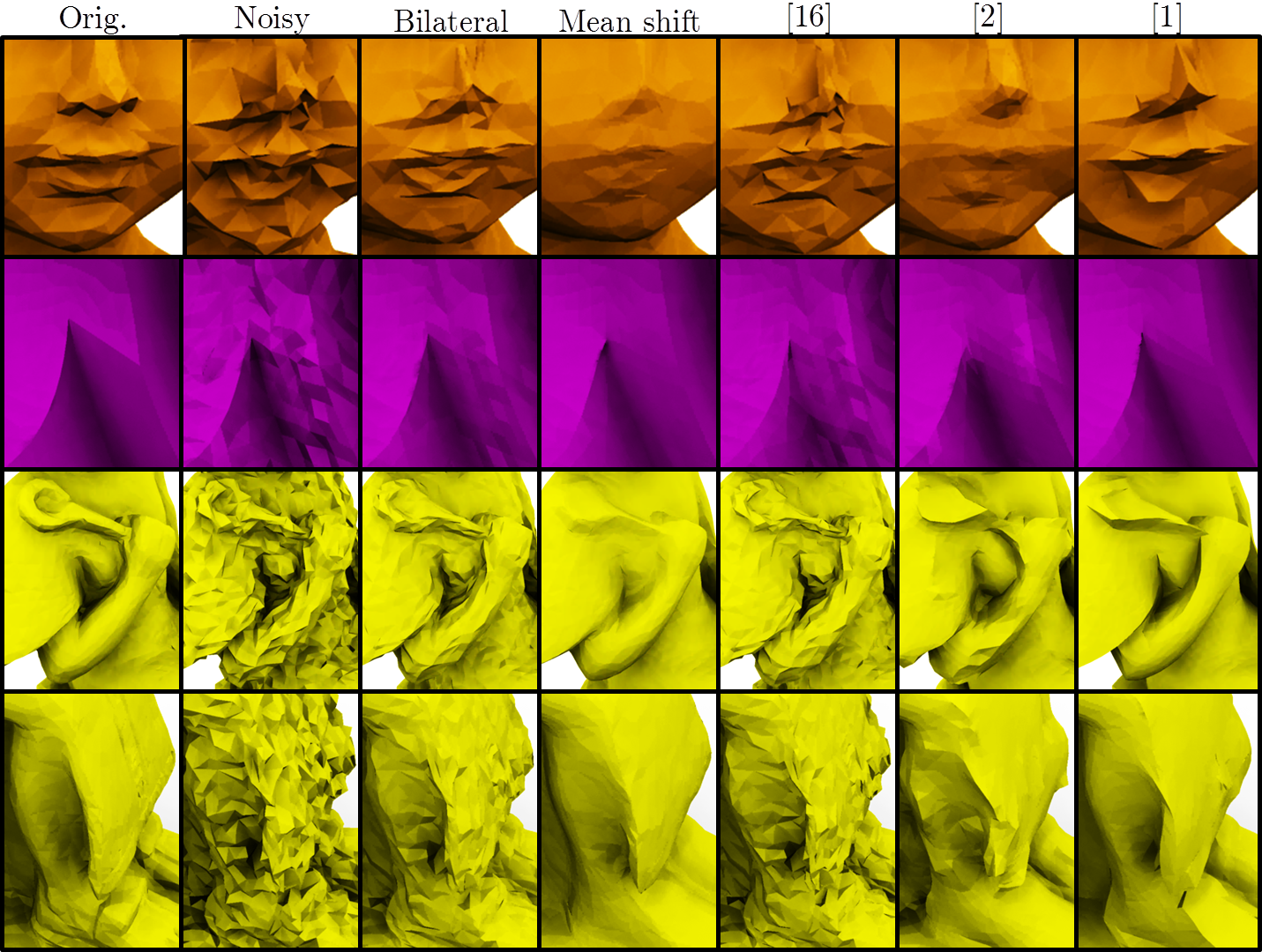}\\
\caption{Closer views of some examples from Figure~\ref{fig:denoising} with increased contrast (Figure~\ref{fig:denoising} is rendered with per-face Lambertian shading for simplicity).}\label{fig:zoom}
\end{figure}

If we filter $\N:\Sigma\rightarrow S^2$ itself, we obtain a denoised normal field over $\Sigma$; this step evaluates the normal vector bilateral proposed in~\cite{zheng11}, although their method resorts to a somewhat severe approximation effective for small blending radii.  As in~\cite{zheng11} and others, we subsequently adjust $\Sigma$ to match the denoised normals using the method in~\cite{sun07}.  While~\cite{sun07} is presented in discrete terms, it simply is solving a Poisson-type equation to recover a nearby surface with the adjusted normals; it is designed not to induce shrinkage and other artifacts.  Figure~\ref{fig:denoising} compares denoising results of the normal bilateral and mean shift filters with those of some previous methods; of course, the choice of reconstruction methods is independent of our filter and can be replaced if desired.

\section{Additional Applications}\label{sec:additional_applications}

Here we provide some applications of our method outside of mesh processing.  These show its broad variety of applications for smoothing and other signal processing tasks.

\subsection{Oriented Point Clouds}\label{sec:oriented_cloud}

Algorithms like~\cite{kazhdan06} for surface reconstruction rely on \emph{oriented} point clouds, which contain both sample points and their normals, to generate meshes; the normals help decipher tangent directions, orientation, and connectivity.  Methods for obtaining or computing orientations often yield noisy normals at best, which, combined with already noisy point clouds, can lead to topological and geometric reconstruction errors that can be difficult to correct \emph{a posteriori}.

\begin{figure*}
\centering
\begin{tabular}{ccccc}
\begin{tabular}{c@{}c@{}c}
\includegraphics[height=.76in]{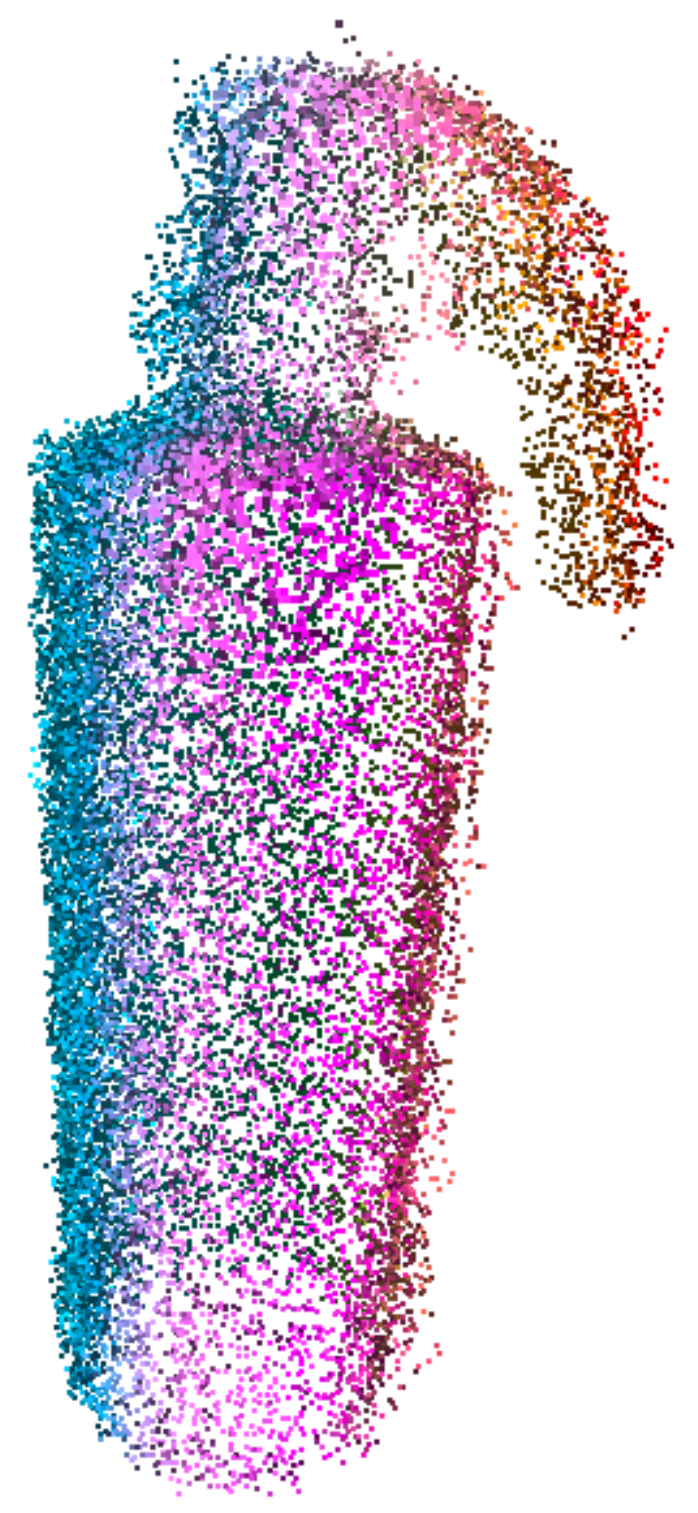} &
\includegraphics[height=.76in]{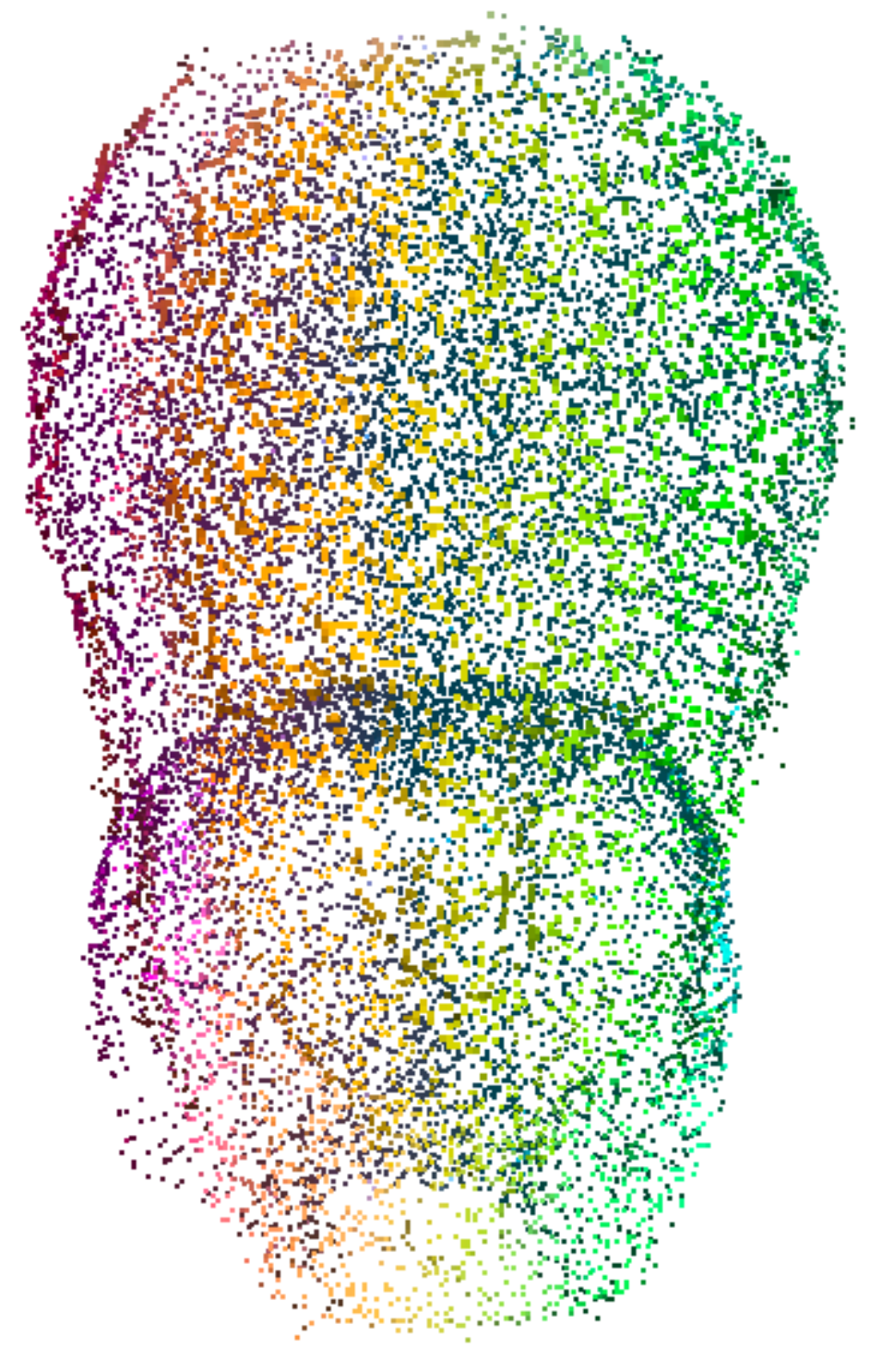} &
\includegraphics[height=.76in]{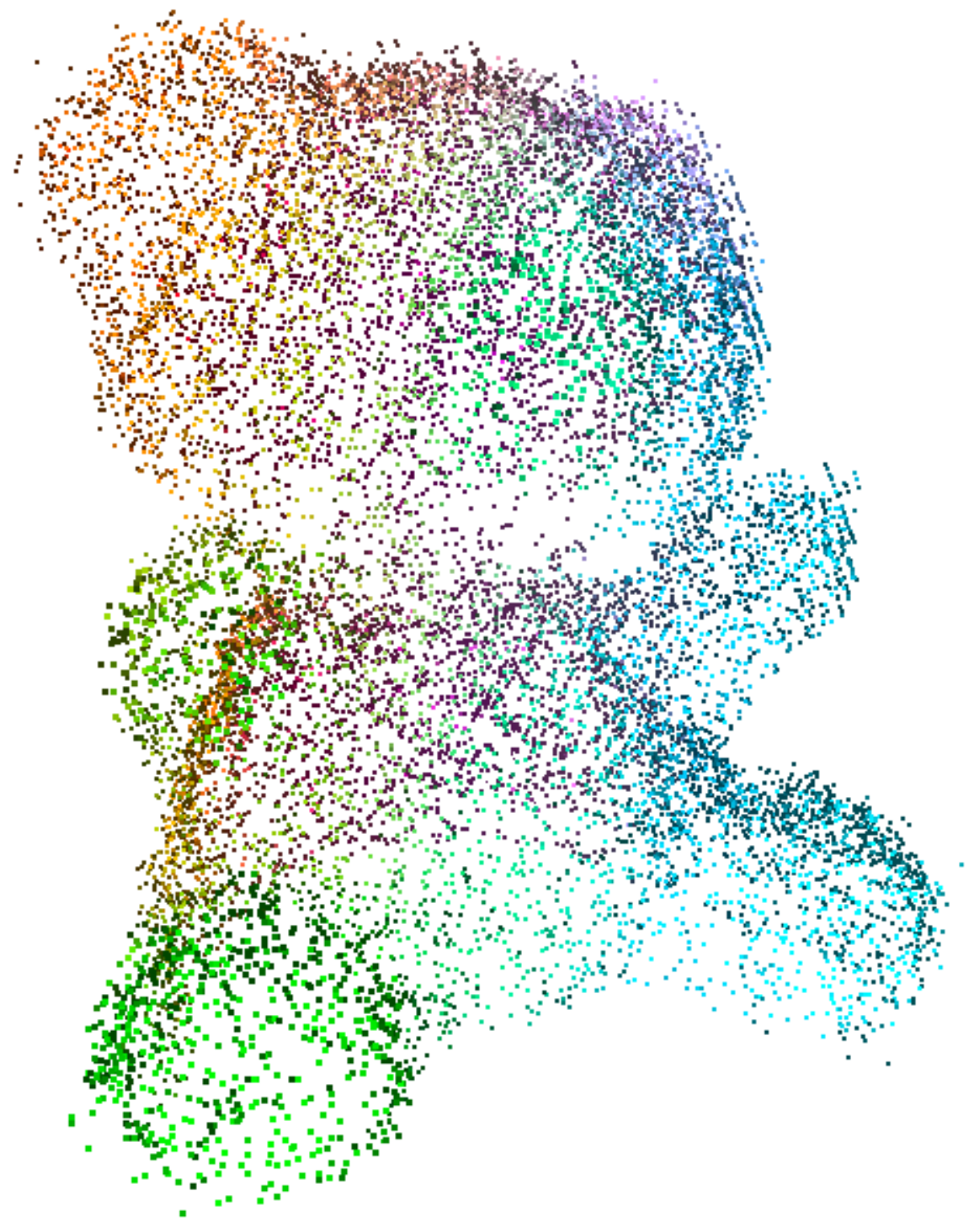}
\end{tabular} &\hspace{-.35in}
\begin{tabular}{c@{}c}
\includegraphics[height=.76in]{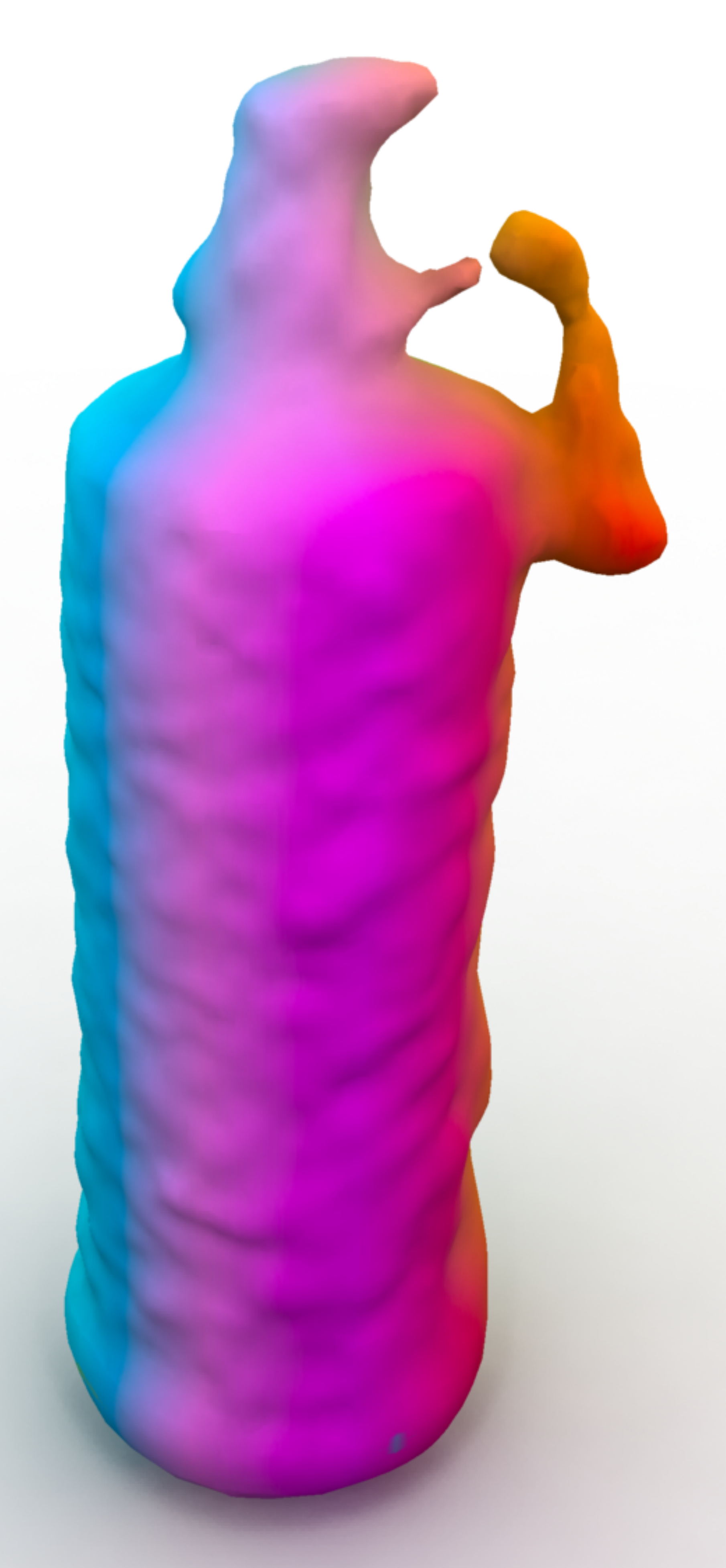} &
\includegraphics[height=.76in]{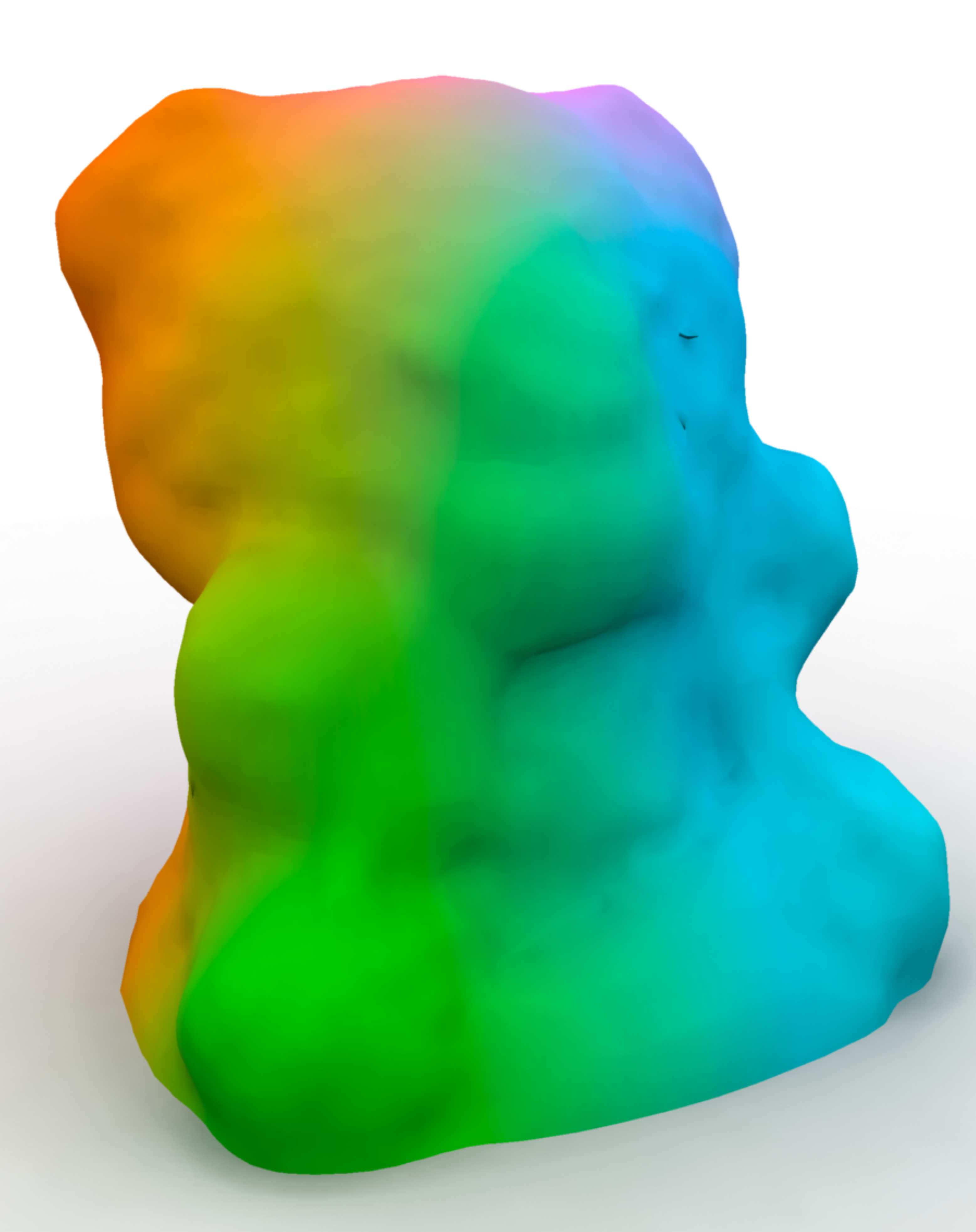}
\end{tabular} &\hspace{-.35in}
\begin{tabular}{c@{}c@{}c}
\includegraphics[height=.76in]{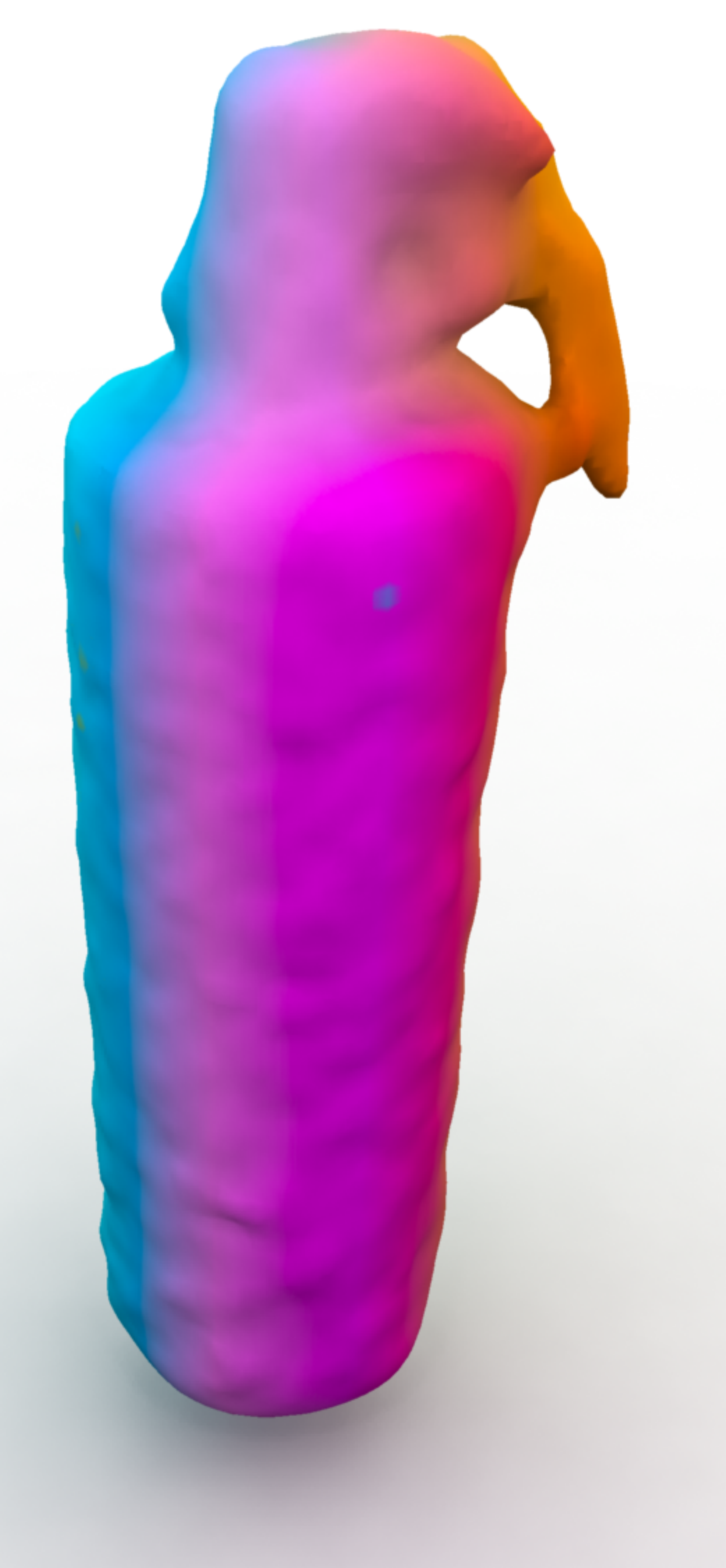} &
\includegraphics[height=.76in]{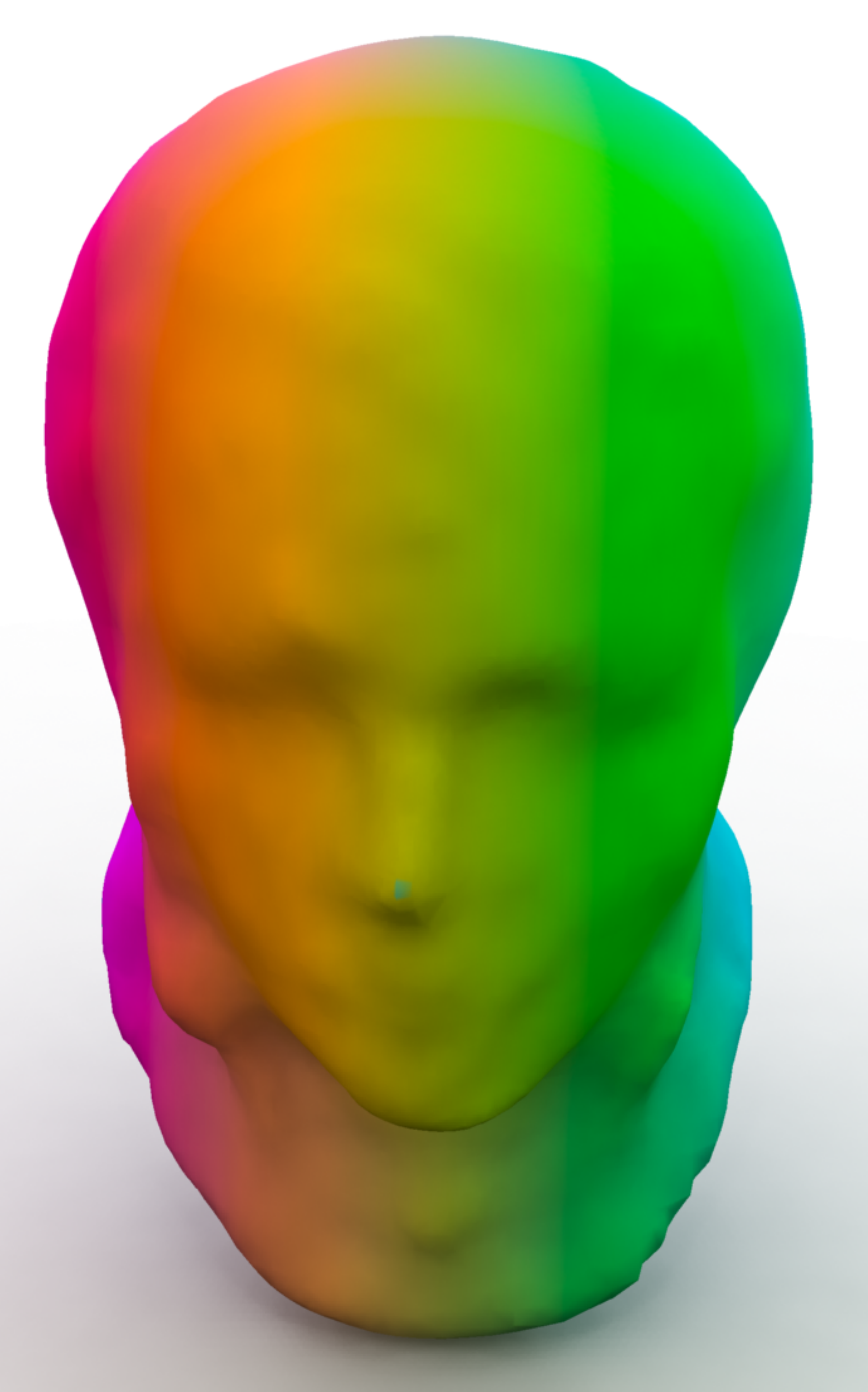} &
\includegraphics[height=.76in]{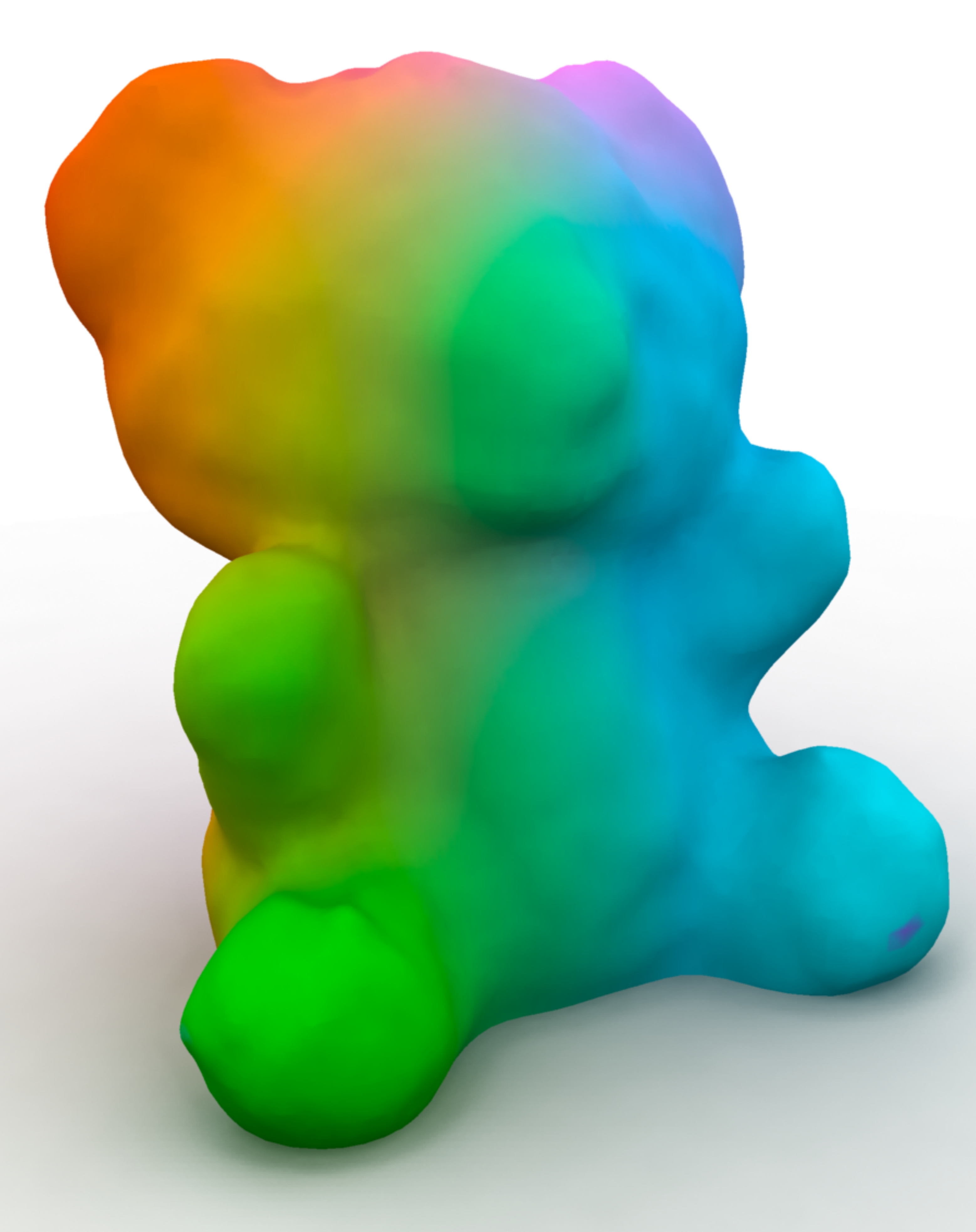}
\end{tabular}&\hspace{-.35in}
\begin{tabular}{c@{}c@{}c}
\includegraphics[height=.76in]{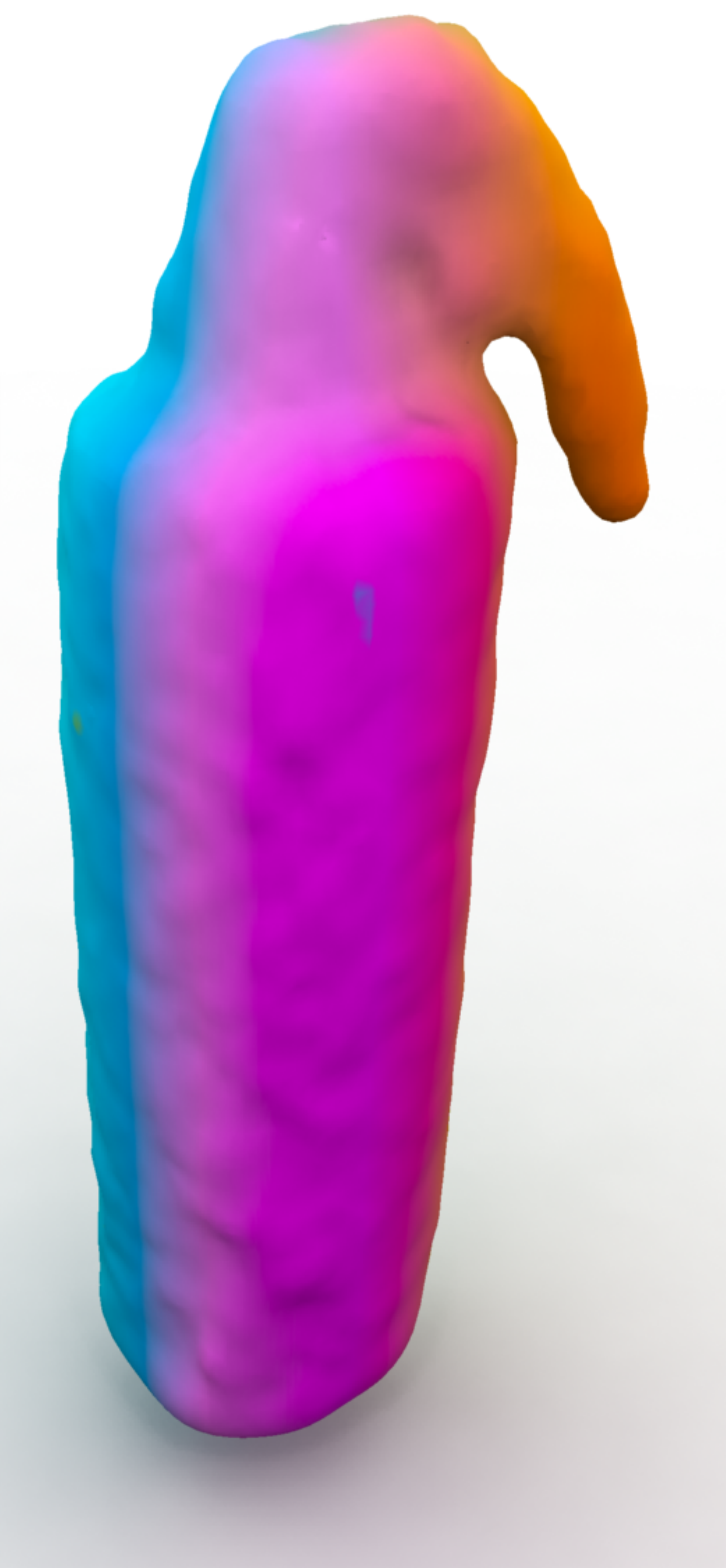} &
\includegraphics[height=.76in]{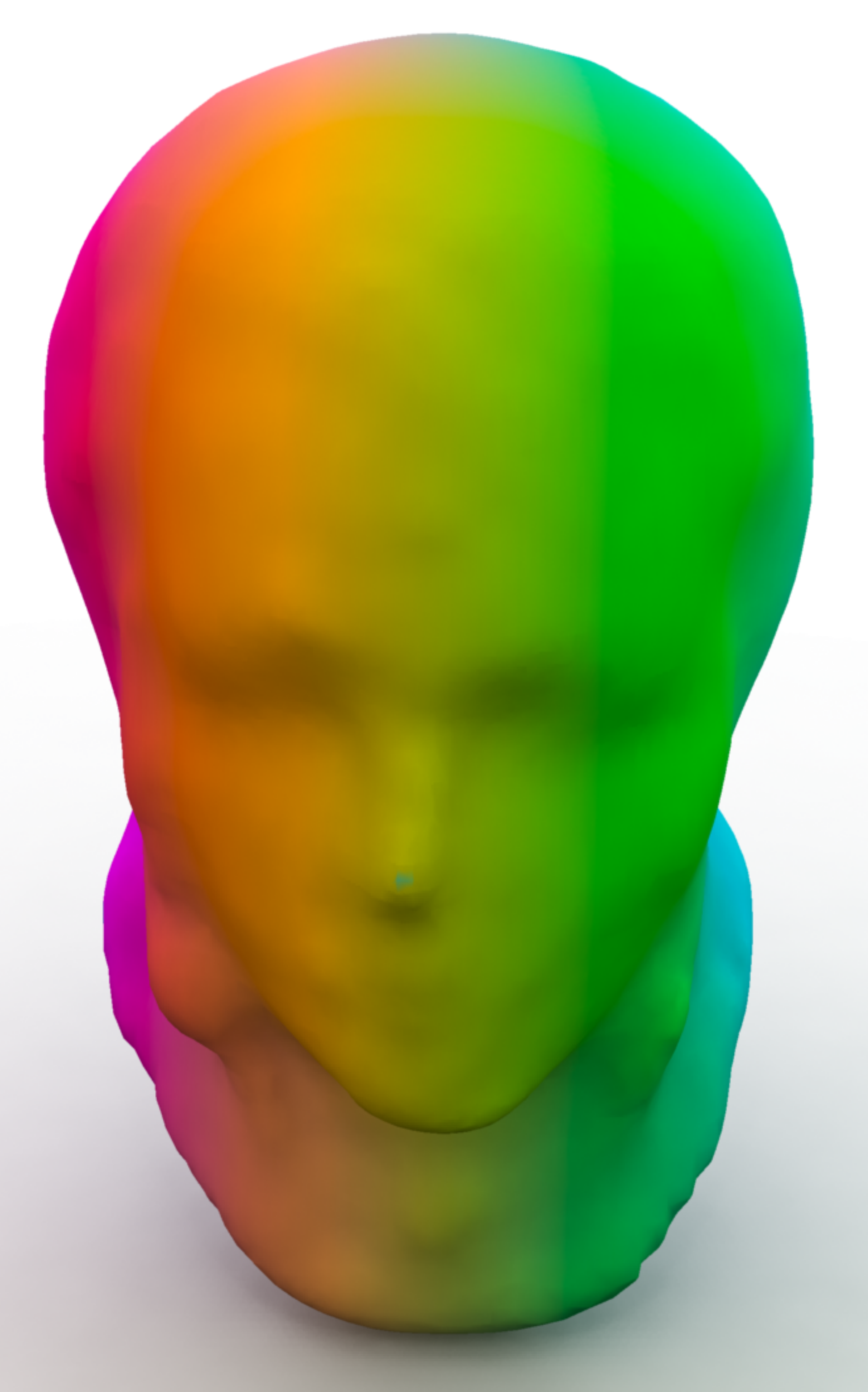} &
\includegraphics[height=.76in]{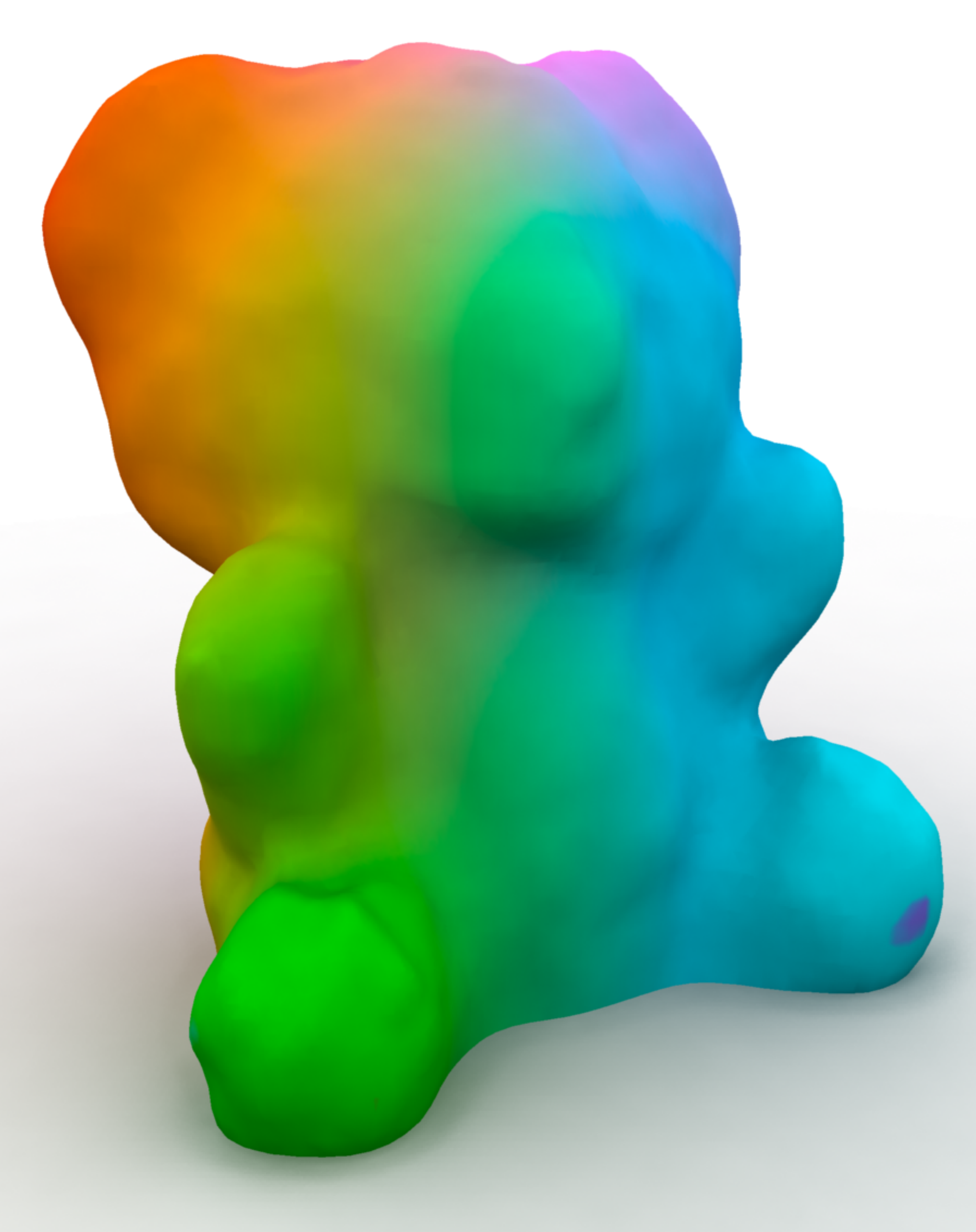}
\end{tabular}&\hspace{-.35in} 
\begin{tabular}{c@{}}
\includegraphics[height=.76in]{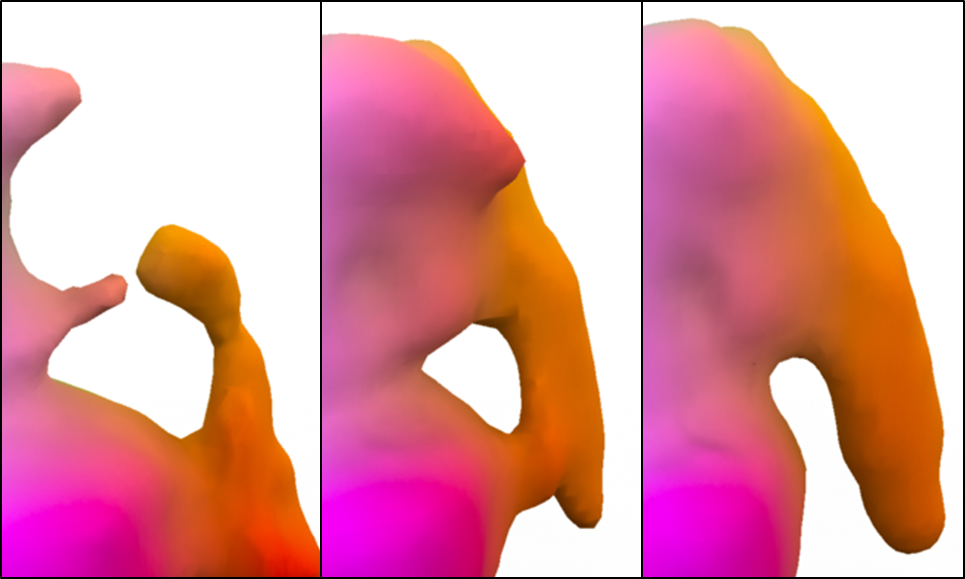}
\end{tabular}\\
(a)&(b)&(c)&(d)&(e)
\end{tabular}
\caption{Surface reconstruction from the oriented point cloud (a; rendered using normal vectors for lighting with hue chosen by position), with original normals (b; bust case fails), bilaterally-smoothed normals (c), and mean-shifted normals (d).  Bilateral and mean shift filtering create considerably better reconstruction results; even in the difficult case of the fire extinguisher cloud, mean shift filter is able to generate normals that separate the handle from the body of the extinguisher (e).}\label{fig:bilateral_reconstruction}
\end{figure*}

Fortunately,~\cite{belkin09} introduces a Laplacian for signals on point clouds with provable convergence.  Laplacian heat diffusion along with the bilateral term ensures that edges are preserved and that surface topology is respected while combining ``nearby'' normals.  Figure~\ref{fig:bilateral_reconstruction} shows examples of reconstruction using~\cite{kazhdan06} with and without bilateral normal filtering on point clouds from~\cite{reinhardt12}.

\subsection{Bilateral and Mean Shift on Other Signals}

The filters we discuss above are by no means the only ones that fit in our framework.  Additional domains and signals to which we could apply Algorithm~\ref{alg:generalized} include:
\begin{itemize}
\item Textures equipped with a blurring operator from MIP maps or a Laplacian pulled back from the mesh
\item Signals on polygonal meshes using the Laplacian from~\cite{alexa11} for diffusion
\item Point clouds with skeletons as in~\cite{cao10}, so points are combined when they are close on the skeleton and with respect to point cloud Laplacian heat flow
\item Quadric surface approximations as in some works in Table~\ref{table:other_papers}, with cross bilateral signals suggested here or in the original papers
\item Graphs with discrete Laplacian diffusion
\item Range images with RGB or normals for the cross bilateral
\item Volumetric signals with heat flow using $f_1$ as a density
\item Simplicial complexes with combinatorial Laplacian flow
\end{itemize}
Many of these applications are outside computer graphics; others may not benefit as much from a bilateral filter as from related techniques suggested by our method, like that for computing local histograms below.

\subsection{Local Histograms}\label{sec:histograms}

\cite{kass10} suggests that the histogram $h_{\x}(\p)$ in~\eqref{eq:local_histogram} has value for understanding signals on images; in particular, they use this function to understand the distribution of intensities in some smoothly-weighted neighborhood of each pixel.  An \emph{identical} formulation applies to our more general setting.  In particular, evaluation of $h(\p_i;\x)\ \forall\x\in\Sigma$ occurs while computing the samples in the denominator in Algorithm~\ref{alg:generalized}.  Thus, we can efficiently extract local histograms of signals $f:\Sigma\rightarrow\Gamma$ using the same partition of unity approach.  This allows for the direct evaluation of the filters in~\cite{kass10} applied to scalar functions on surfaces and other domains.

\begin{figure}[t]
\centering
\begin{tabular}{c@{}c}
\includegraphics[height=.25\linewidth]{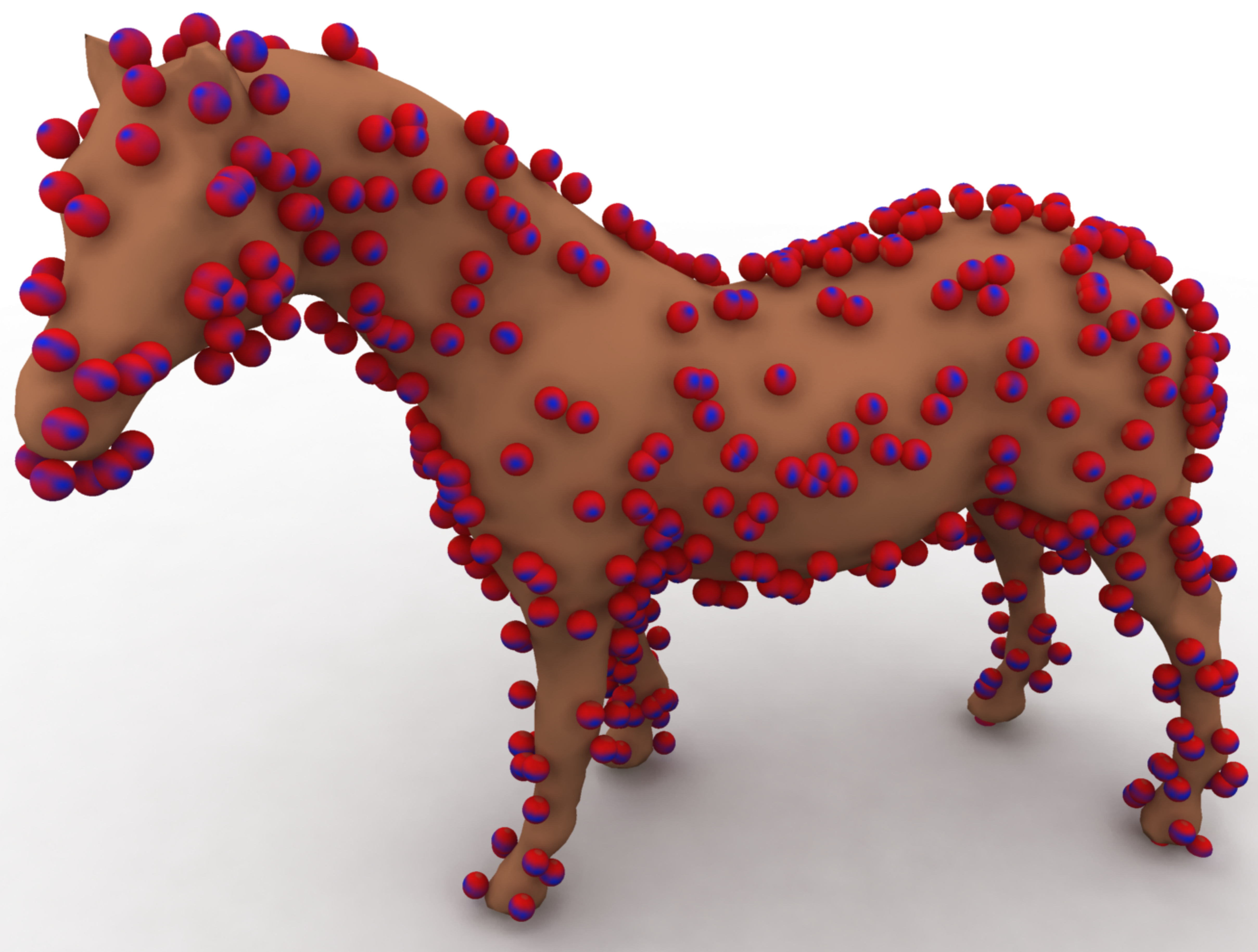} &
\includegraphics[height=.25\linewidth]{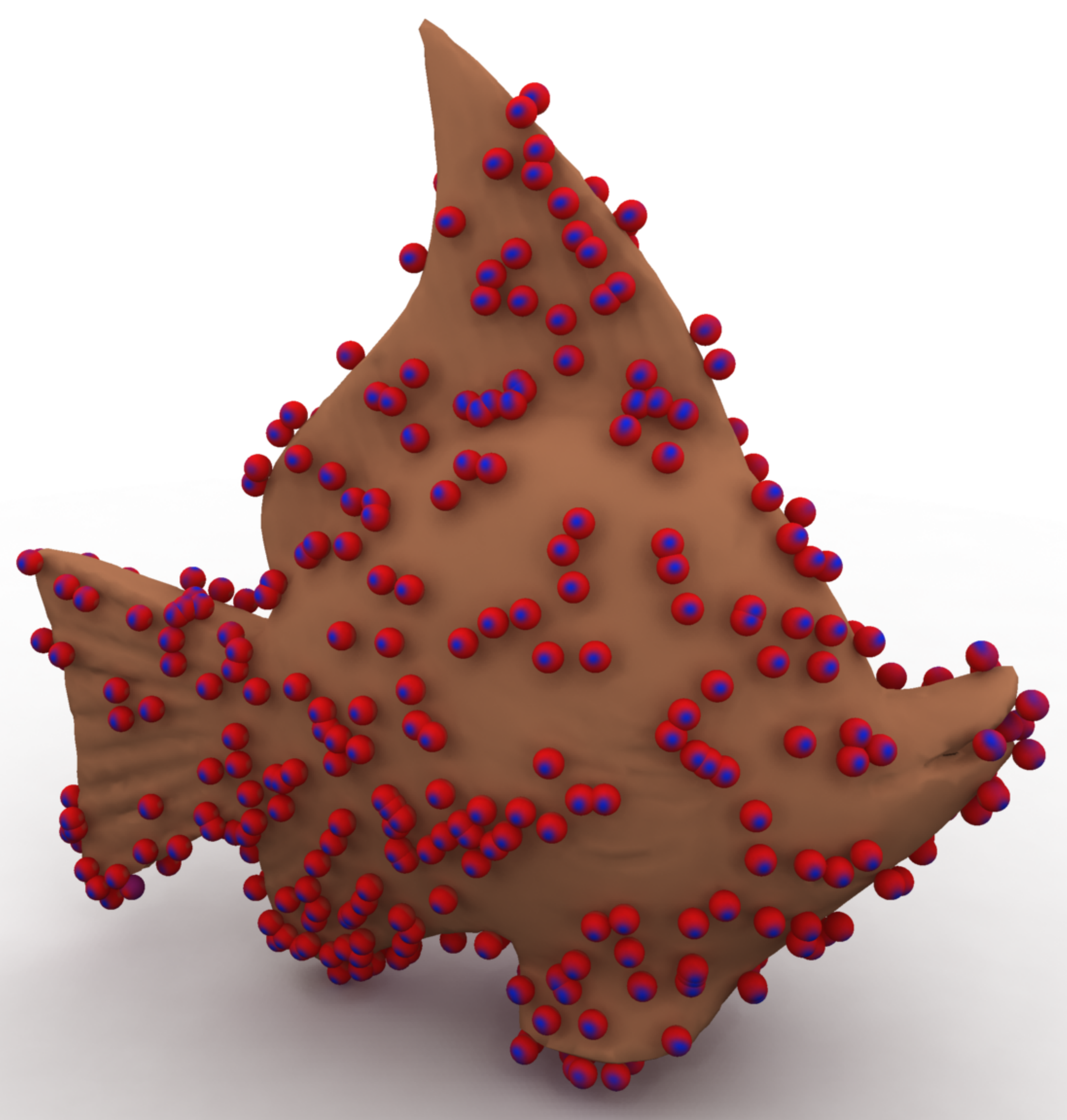}\\
\begin{tabular}{cc}
\includegraphics[height=.25\linewidth]{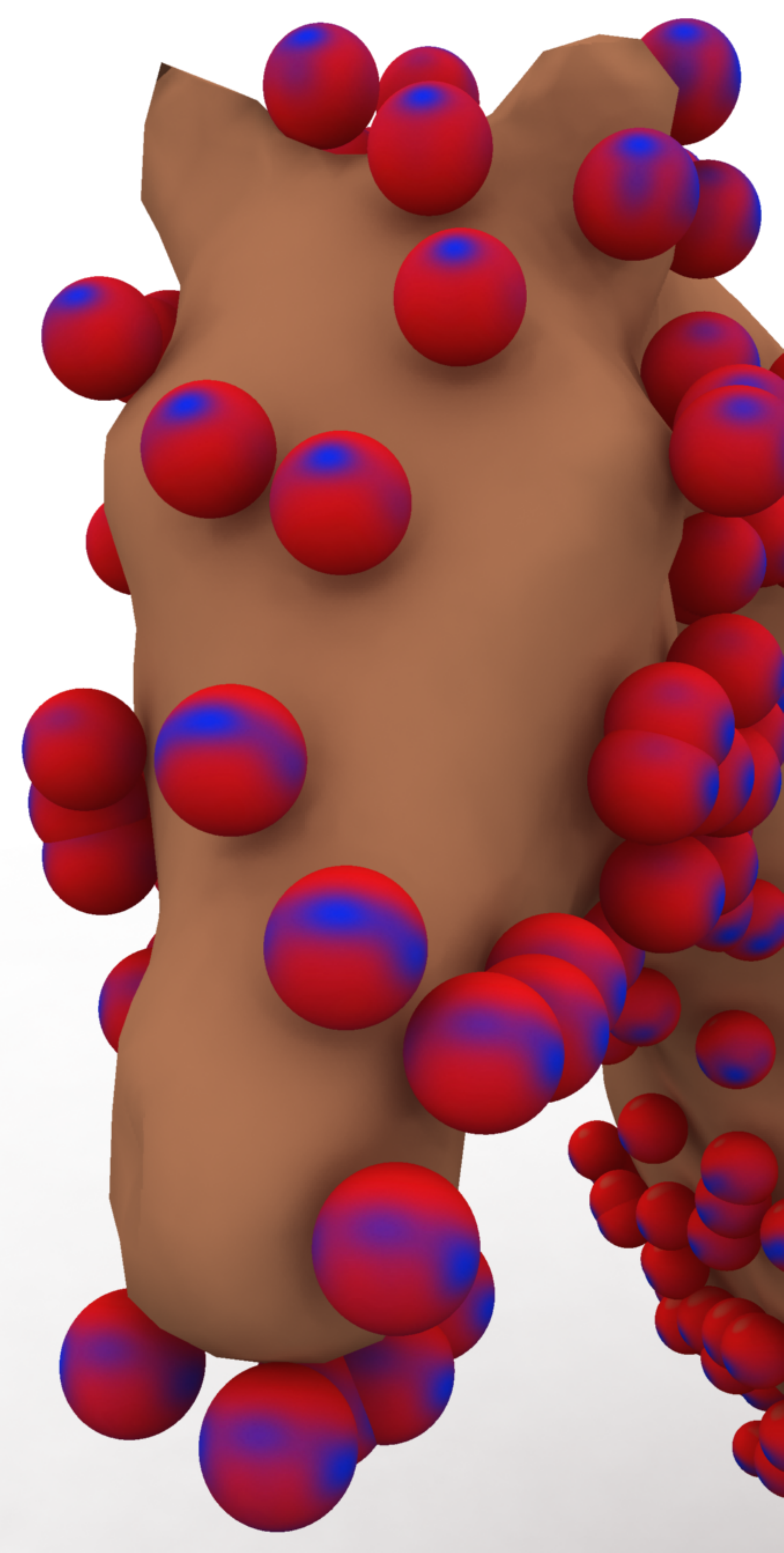} &
\includegraphics[height=.25\linewidth]{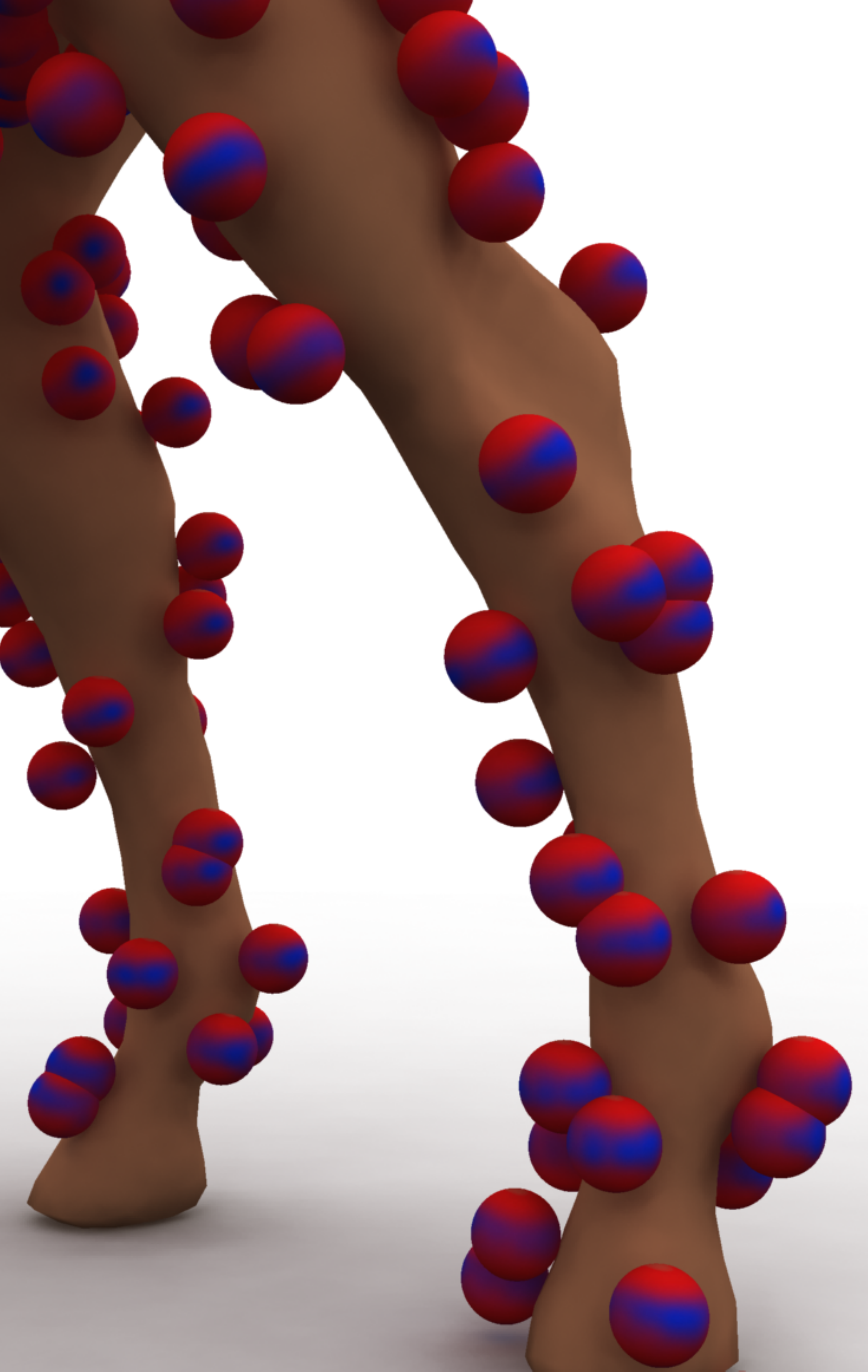}
\end{tabular}
&
\begin{tabular}{cc}
\includegraphics[height=.25\linewidth]{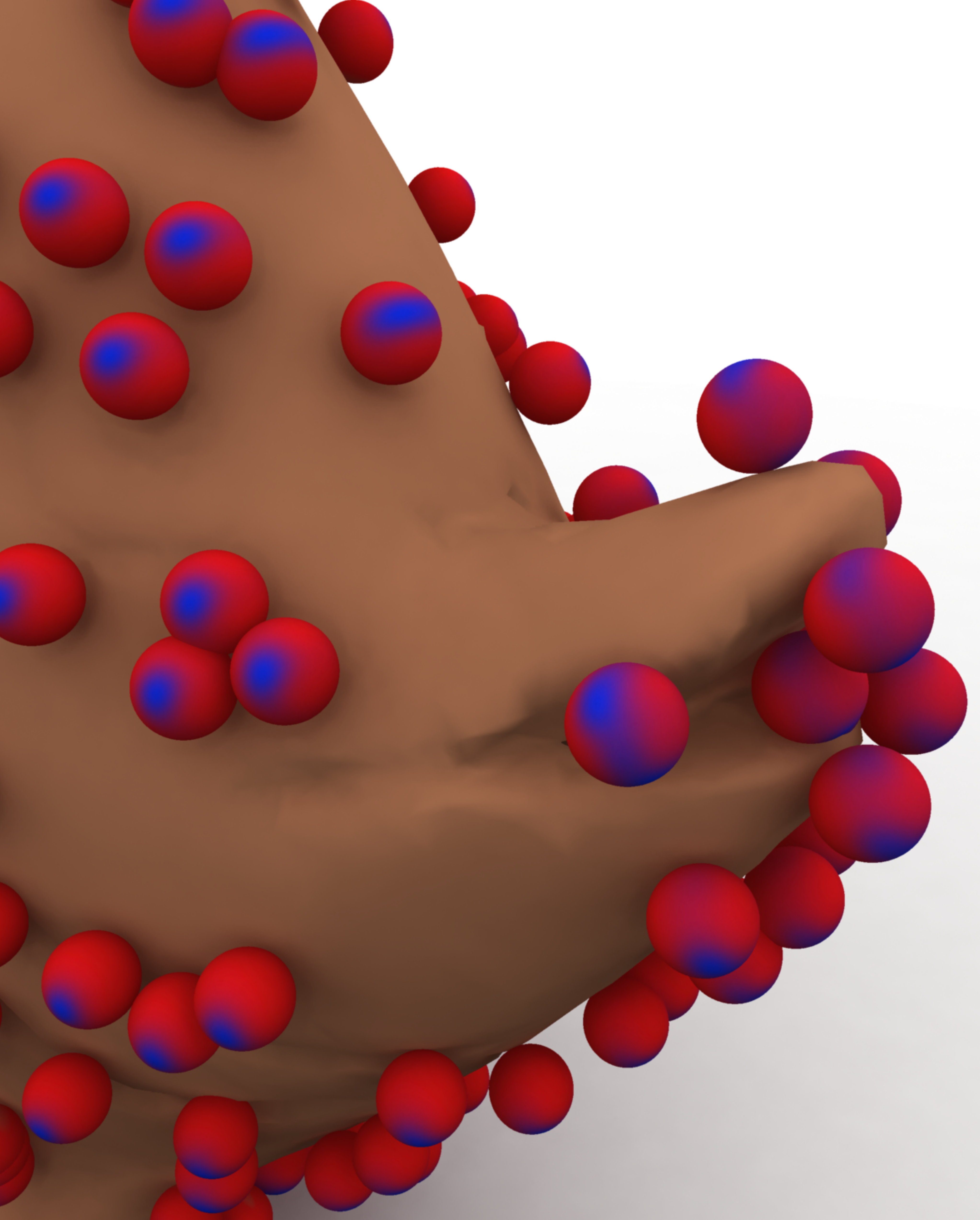} &
\includegraphics[height=.25\linewidth]{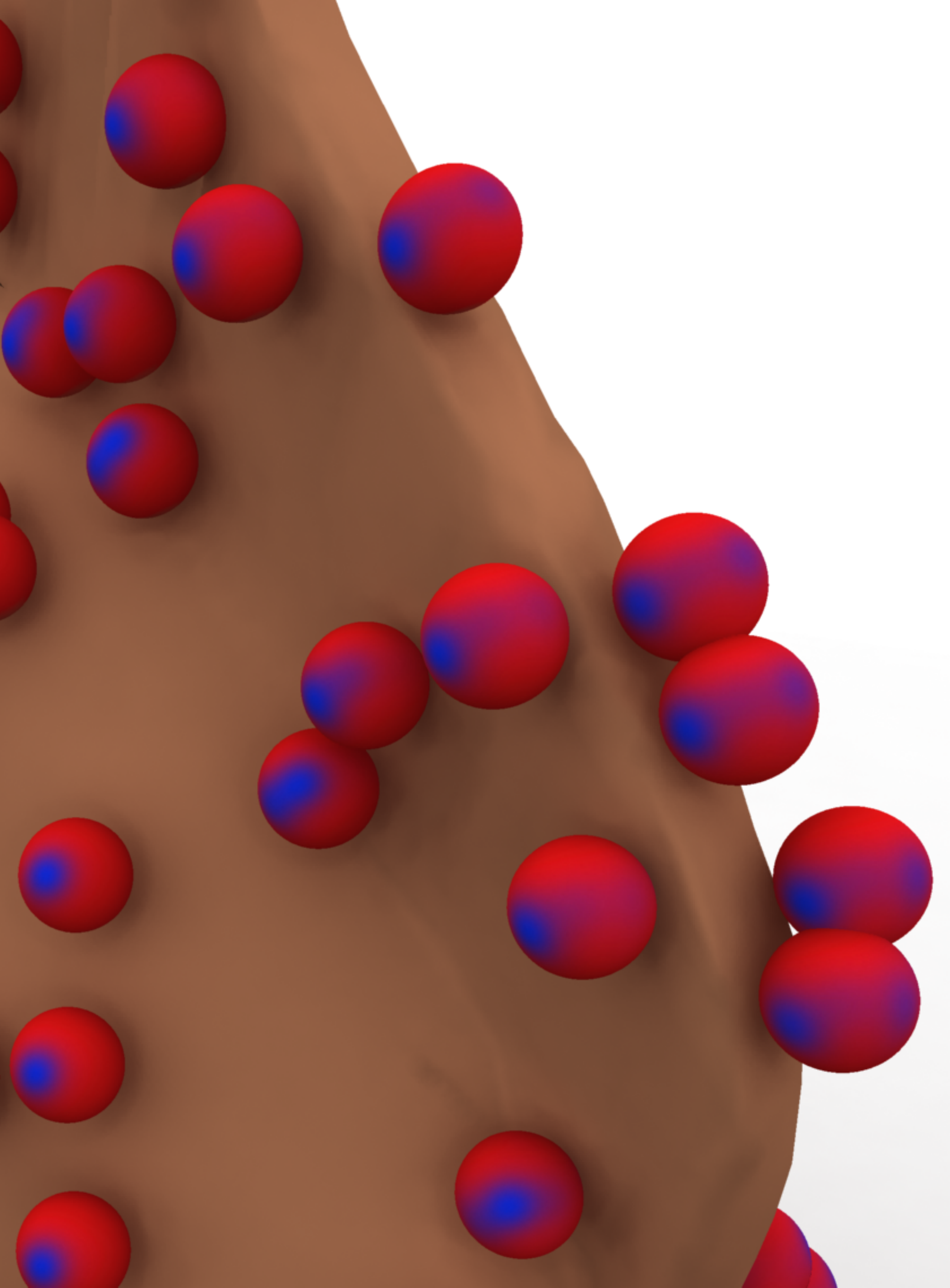}
\end{tabular}
\end{tabular}
\caption{Local normal histograms describing the distribution of normals near a given face, color-coded on the unit sphere; histograms are shown at a random set of faces.}\label{fig:local_histograms}
\end{figure}

The method at our level of generality, however, can be applied to a much wider array of signals.  For example, once again taking $f:\Sigma\rightarrow\Gamma=S^2$ to be the normal vector signal on $\Sigma$, the histogram $h_{\x}(\p)$ at a fixed $\x\in\Sigma$ now represents the distribution over $S^2$ of normal vectors to $\Sigma$ near $\x$.  This distribution can be viewed (after suitable rotation) as a version of the SHOT descriptor introduced in~\cite{tombari10} with smoothly varying, intrinsic heat kernel weights on $\Sigma$ rather than extrinsic distance weights, with straightforward regularization control by changing blurring radii on $\Sigma$ and $\Gamma$.  Figure~\ref{fig:local_histograms} shows some examples of normal vector histograms computed using this technique.  These images show that our histograms of normals are equally informative to the SHOT descriptor; viewed as probability distributions on the unit sphere, these histograms also suggest the possibility of applying filtering techniques such as~\cite{hadwiger12} to meshed domains.

\subsection{Feature-Preserving Filters}

We have gone a long way toward pushing the bilateral filter to a maximal of generality.  One additional avenue for flexibility, however, is in the choice of kernels $K_\Sigma$ and $K_\Gamma$.

The most obvious potential change in $K_\Sigma$ or $K_\Gamma$ might be in the choice of smoothing kernels.  We implicitly have made use of this flexibility by suggesting that a single implicit time step of the heat equation suffices for bilateral filtering on meshes.  In practice, we find that any reasonable choice of smoothing kernel behaves in a qualitatively similar fashion for most bilateral and mean shift applications.

Even more generally, heat flow is a member of a huge class of linear operators used in mesh processing.  Band-pass, high-pass, unsharp mask, and other filters can be applied to signals on a surface using analogs of Fourier theory and a discretization of the Laplacian.  Even if these filters are described using some sort of local operation, their linearity implies the existence of an operator matrix containing kernel values $K_\Sigma : \R^{|V|}\times\R^{|V|}\rightarrow\R$; the theory of Schwartz kernels can be used to prove a similar statement in the continuous limit~\cite{hormander90}.  The bilateral simply reweights these linear kernels to respect signal edges.

\begin{figure}[t]
\centering
\begin{tabular}{cc}
\includegraphics[width=.45\linewidth]{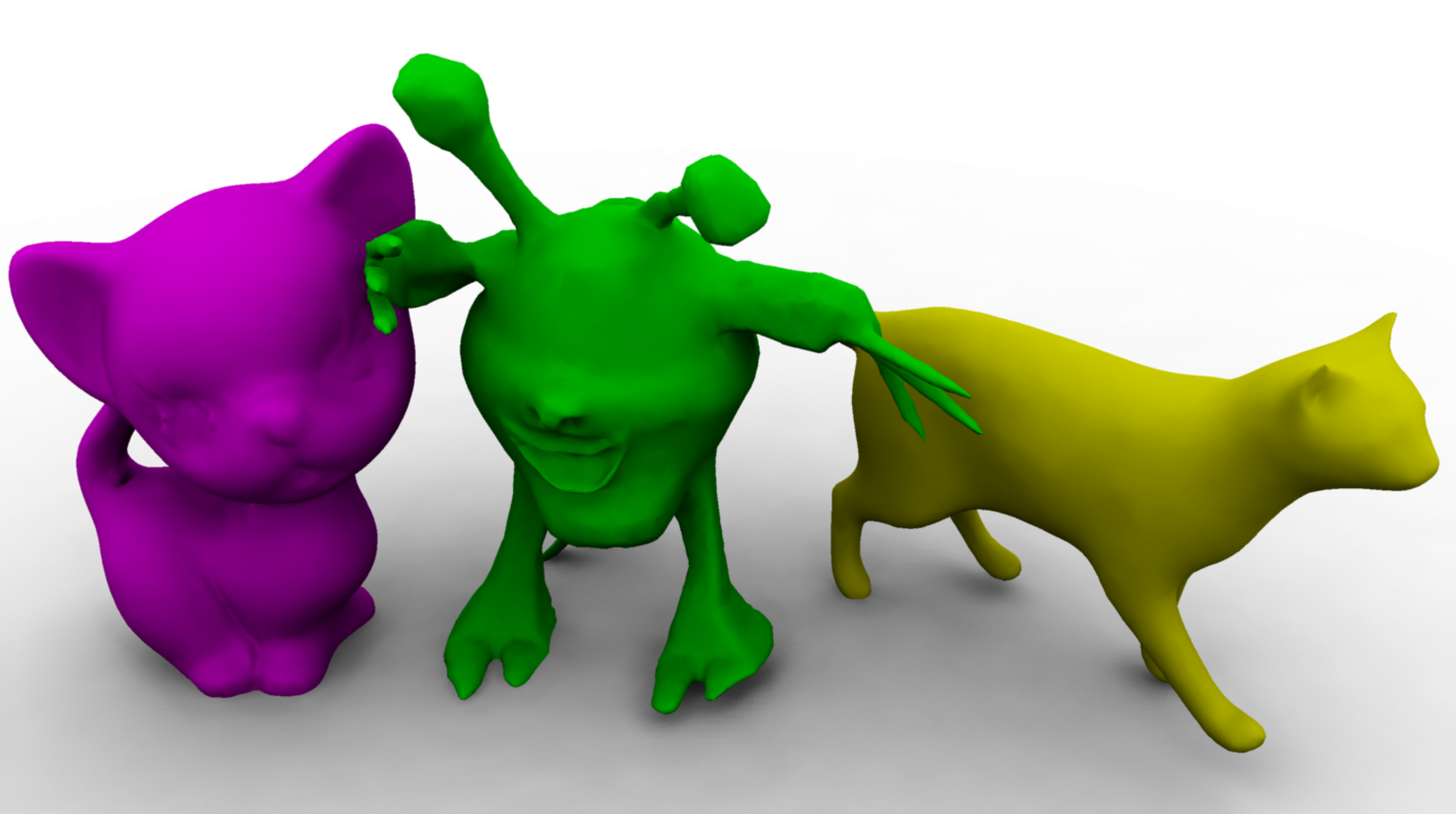}&
\includegraphics[width=.45\linewidth]{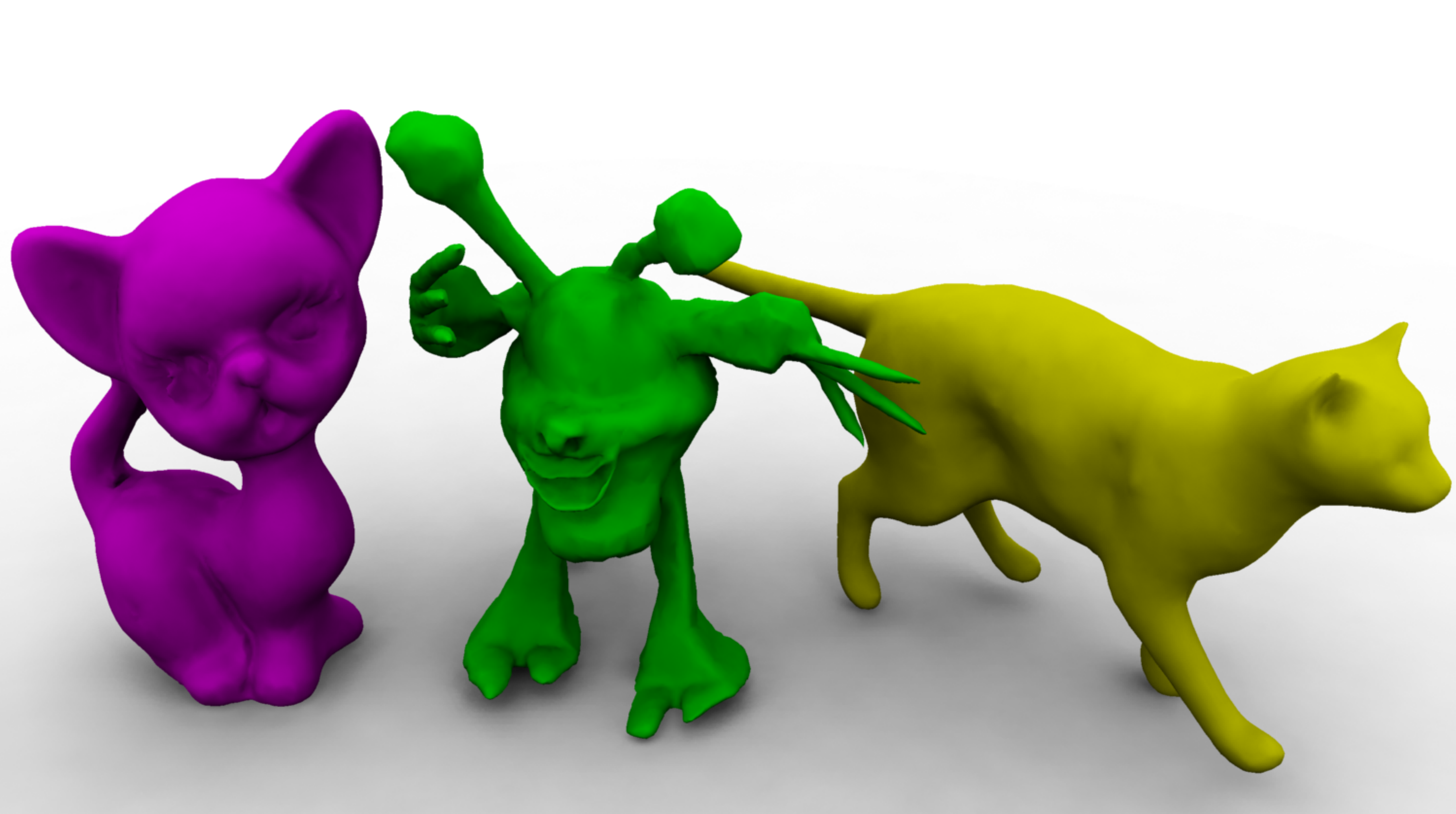}\\
(a)&(b)
\end{tabular}
\caption{Examples of non-blurring bilateral filters to achieve interesting edge-preserving shape deformations.}\label{fig:non_smoothing}
\end{figure}

Although fully exploring the domain of feature-preserving mesh operations is worthy of a larger study, Figure~\ref{fig:non_smoothing} shows examples of the application of our bilateral where the kernel $K_\Sigma$ has been replaced with the kernels of other linear operators.  In particular, we use the unsharp mask for $K_\Sigma$ while keeping $K_\Gamma$ Gaussian in mesh normals.  The resulting filter is applied to mean curvature normals, yielding meshes with exaggerated curvature while avoiding artifacts like ringing near sharp corners.

\section{Discussion}

We have written an implementation of our algorithm in C++, taking advantage of templates to encode Algorithm~\ref{alg:generalized} in full generality; we use OpenMP directives to achieve parallel evaluation of the blurs needed for each sample $\p_i$.  On a four-core 2.40 GHz Intel Xeon machine, this na\"ive implementation can apply bilateral filters to mesh normals on 12946 faces in 2.72 seconds using 42 sample points on $S^2$.  Subsequent iterations for the mean shift are even faster, since they can reuse the same prefactored heat flow matrix; this method converges in as few as five to ten iterations.

Faster run times could be achieved with an optimized implementation and faster linear solvers.  Our runtime is limited by the time it takes to blur $2m$ signals using $K_\Sigma$, so fewer samples $\p_i\in S^2$ make for better timings; we can cut our number of samples to half of the ones listed here with reasonable effect but slight visible artifact in exchange for a faster filter.

Figure~\ref{fig:denoising} compares against recent work on mesh smoothing; larger image of representative examples are shown in Figure~\ref{fig:zoom}.  We apply uniform noise of varying sizes to mesh vertices and then apply our and other smoothing methods to recover the original shape.  We show the perceptual ``STED'' distance between the filtered signals and the original~\cite{vasa11_2,vasa11}.  In general, we find that our algorithm behaves comparably with state-of-the-art, yielding small STED distances to the original meshes even when compared to the results of more specialized papers.

\subsection{Limitations}

While the theoretical and practical properties of our generalized bilateral filter make it an obvious choice in a variety of circumstances, it is important to note tasks for which our construction is not as well-suited.  In particular, we require $\Gamma$ to be compact (possibly with boundary) and to admit a partition of unity; this assumption is fairly weak for signals such as mesh normals, which live on $S^2$, but makes it difficult to consider signals like the tangent plane projections in~\cite{jones03} that can take values within a large part of $\mathbb{R}^n$.

One property exhibited by mesh smoothing algorithms making use of geometric flows rather than integral operators like the bilateral is that they somehow ``directly'' filter the geometry rather than treating it as a signal.  In fact, our method as-is actually can deal with geometry in at least two ways.  First, as proposed in Section~\ref{sec:normal_signal}, we can use normals to process geometry indirectly.  This approach has the advantage that edges in the geometry become discontinuities in the signal, whereas $xyz$ positions on a mesh are continuous everywhere.  Given the reconstruction method in~\cite{sun07}, one can view the normal signal as an alternative non-Euclidean expression of geometry that can be processed like any other embedding.  Second, our bilateral could be applied directly to $xyz$ positions as the signal on $\Sigma$ using normals on $\Gamma=S^2$.  This alternative better mimics flows, but we found it less effective than normal processing and omitted the results.  Normal processing has been shown repeatedly to be a highly-effective denoising technique, so we are hardly the first to come to this conclusion~\cite{lee05,sun07,zheng11}.  We leave the interpretation of our filter as an anisotropic flow as in~\cite{barash02} for images for future research.

A related issue that will require additional study is the effect of the reconstruction in~\cite{sun07} on the convergence properties of our normal-based mesh processing technique.  Nonetheless, consistency for signals on fixed irregular domains is a valuable feature of our method, and one that is not guaranteed by any existing method.

\section{Conclusion}\label{sec:conclusion}

The sheer number of attempts to discretize bilateral filtering on non-image domains illustrated in Table~\ref{table:other_papers} demonstrates the elusiveness and importance of a generalized bilateral filter.  Expressions for the bilateral, whether for images as in~\eqref{eq:cross_bilateral} or in the more general sense as in~\eqref{eq:generalized_bilateral}, are easy to state and understand and have only a few intuitive parameters.  The bilateral's behavior is well-understood and forms the basis for more complex methods such as the mean shift.  It has withstood the test of time and remains a foundational tool used to construct state-of-the-art algorithms in diverse parts of image processing, vision, and graphics.

Our new discretization makes the process of defining a bilateral filter on a given domain and signal straightforward.  Feature-preserving filters can be achieved on arbitrary domains simply by choosing domains $\Sigma,\Gamma$ and kernels $K_\Sigma,K_\Gamma$, with the assumption that $\Gamma$ can be sampled reasonably.  This process has an easily-understood continuous limit~\eqref{eq:generalized_bilateral} and can even be extended to tasks like histogram computation and shape editing.  The speed of the filter simply depends on the number of samples in $\Gamma$ and the time it takes to apply $K_\Sigma$, the latter of which often boils down to a simple pre-factored linear solve.

While we have illustrated only a few applications of our method within the domain of geometry processing, we hope that its simplicity and effectiveness will lead to its application in other settings.  For instance, in image processing, some results show that distances between signatures for commonly-used cross bilateral signals may not be measured using the Euclidean metric but rather along some underlying manifold~\cite{carlsson08,peyre09}; this type of relationship can be encoded in our framework by defining $\Sigma$ to be a part of the image plane and $\Gamma$ to be the cross bilateral manifold in question.  As another example, local histograms may be useful for understanding structure and local information in graphs, using Laplacian heat flow to evaluate proximity.  These broad applications and many others are no harder to implement or understand than the ones we have suggested in this paper, and they begin to reveal the exciting potential implications of a reliable generalized bilateral filtering technique.

\ifCLASSOPTIONcompsoc
  \section*{Acknowledgments}
\else
  \section*{Acknowledgment}
\fi

The authors would like to thank Andrew Adams, Leonidas Guibas, Abe Davis, Michael Kass, Andy Nguyen, and others for discussing ideas from the paper at various stages of its creation.

\ifCLASSOPTIONcaptionsoff
  \newpage
\fi

\bibliographystyle{IEEEtran}
\bibliography{IEEEabrv,MeshBilateral}

\begin{IEEEbiography}[{\includegraphics[width=1in,height=1.25in,clip,keepaspectratio]{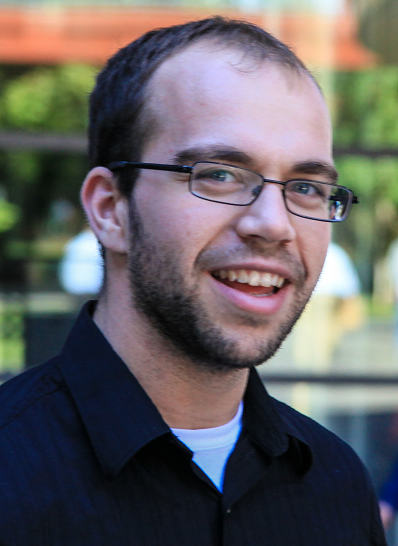}}]{\href{http://www.stanford.edu/~justso1}{Justin Solomon}} is a PhD candidate in the Geometric Computing Group of Stanford University's Department of Computer Science.  He also received a BS in Mathematics and Computer Science (2010) and an MS in Computer Science (2012) at Stanford.  His areas of study include geometry processing, computer graphics, and numerical methods with a focus on understanding geometric data.  He is supported by the Hertz, NDSEG, and NSF graduate fellowships.
\end{IEEEbiography}

\begin{IEEEbiography}[{\includegraphics[width=1in,height=1.25in,clip,keepaspectratio]{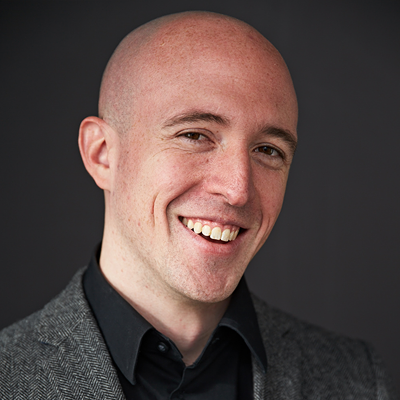}}]{\href{http://users.cms.caltech.edu/~keenan/}{Keenan Crane}} is a PhD student in the Department of Computing and Mathematical Sciences at Caltech.  He received a BS in Computer Science from the University of Illinois at Urbana Champaign in 2006.  He is the recipient of a Google PhD Fellowship and a National Science Foundation Mathematical Sciences Postdoctoral Research Fellowship.  His current research focuses on discrete differential geometry with applications in digital geometry processing.
\end{IEEEbiography}

\begin{IEEEbiography}[{\includegraphics[width=1in,height=1.25in,clip,keepaspectratio]{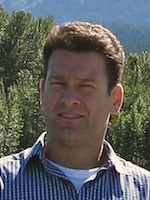}}]{Adrian Butscher} is a senior research scientist with the Max Planck Institute for Computer Science.  He received his PhD in mathematics at Stanford University in 2000.  His current research interests include discrete and continuous differential geometry with applications in digital geometry processing.
\end{IEEEbiography}

\begin{IEEEbiography}[{\includegraphics[width=1in,height=1.25in,clip,keepaspectratio]{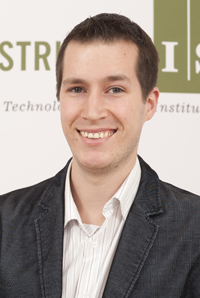}}]{\href{http://pub.ist.ac.at/group_wojtan/}{Chris Wojtan}} received his B.S. in Computer Science in 2004 from the University of Illinois in Urbana Champaign and his Ph.D. in Computer Graphics from the Georgia Institute of Technology in 2010. He was awarded a National Science Foundation Graduate Research Fellowship, the Georgia Tech Sigma Xi Best Ph.D. Thesis Award, and the Microsoft Visual Computing Award. Chris is currently an Assistant Professor at the Institute of Science and Technology Austria (IST Austria), and his research interests are physically-based animation and geometry processing.
\end{IEEEbiography}

%
\vfill
%
%
%

\end{document}